\documentclass[apj]{emulateapj}
\shorttitle{Mass Outflows in the NLR of Mrk 34}
\shortauthors{Revalski et al.}
\slugcomment{Accepted for Publication in ApJ}
\usepackage{graphics}
\usepackage{microtype}
\usepackage{color}
\usepackage{subfigure}
\usepackage{float}
\usepackage{epstopdf}
\usepackage{comment}
\usepackage[colorlinks=true, citecolor=blue, linkcolor=blue, urlcolor=blue]{hyperref}
\usepackage{etoolbox}
\usepackage[normalem]{ulem}

\makeatletter
\patchcmd{\NAT@citex}
  {\@citea\NAT@hyper@{%
     \NAT@nmfmt{\NAT@nm}%
     \hyper@natlinkbreak{\NAT@aysep\NAT@spacechar}{\@citeb\@extra@b@citeb}%
     \NAT@date}}
  {\@citea\NAT@nmfmt{\NAT@nm}%
   \NAT@aysep\NAT@spacechar\NAT@hyper@{\NAT@date}}{}{}
\patchcmd{\NAT@citex}
  {\@citea\NAT@hyper@{%
     \NAT@nmfmt{\NAT@nm}%
     \hyper@natlinkbreak{\NAT@spacechar\NAT@@open\if*#1*\else#1\NAT@spacechar\fi}%
       {\@citeb\@extra@b@citeb}%
     \NAT@date}}
  {\@citea\NAT@nmfmt{\NAT@nm}%
   \NAT@spacechar\NAT@@open\if*#1*\else#1\NAT@spacechar\fi\NAT@hyper@{\NAT@date}}
  {}{}
\makeatother

\begin{document}
\title{\vspace{-15pt} Quantifying Feedback from Narrow Line Region Outflows in Nearby Active Galaxies. II. Spatially Resolved Mass Outflow Rates for the QSO2 Markarian 34$^{\dagger \star}$}

\author{M. Revalski\altaffilmark{1,9}, D. Dashtamirova\altaffilmark{1}, D. M. Crenshaw\altaffilmark{1}, S. B. Kraemer\altaffilmark{2}, T. C. Fischer\altaffilmark{3,10}, H. R. Schmitt\altaffilmark{4}, C. L. Gnilka\altaffilmark{1}, J. Schmidt\altaffilmark{5}, M. Elvis\altaffilmark{6}, G. Fabbiano\altaffilmark{6}, T. Storchi-Bergmann\altaffilmark{6, 7}, W. P. Maksym\altaffilmark{6}, P. Gandhi\altaffilmark{8}}
\altaffiltext{1}{Department of Physics and Astronomy, Georgia State University, 25 Park Place, Suite 605, Atlanta, GA 30303, USA; revalski@astro.gsu.edu}
\altaffiltext{2}{Institute for Astrophysics and Computational Sciences, Department of Physics, The Catholic University of America, Washington, DC 20064, USA}
\altaffiltext{3}{Astrophysics Science Division, Goddard Space Flight Center, Code 665, Greenbelt, MD 20771, USA}
\altaffiltext{4}{Naval Research Laboratory, Washington, DC 20375, USA}
\altaffiltext{5}{Designer/Developer for the Astrophysics Source Code Library}
\altaffiltext{6}{Harvard-Smithsonian Center for Astrophysics, 60 Garden St., Cambridge, MA 02138, USA}
\altaffiltext{7}{Departamento de Astronomia, Universidade Federal do Rio Grande do Sul, IF, CP 15051, 91501-970 Porto Alegre, RS, Brazil}
\altaffiltext{8}{Department of Physics and Astronomy, University of Southampton, Southampton SO17 3RT, UK}
\altaffiltext{9}{National Science Foundation Graduate Research Fellow (DGE-1550139)}
\altaffiltext{10}{{\it James Webb Space Telescope} NASA Postdoctoral Program Fellow}
\altaffiltext{$\dagger$}{Based on observations made with the NASA/ESA Hubble Space Telescope, obtained from the Data Archive at the Space Telescope Science Institute, which is operated by the Association of Universities for Research in Astronomy, Inc., under NASA contract NAS 5-26555. These observations are associated with program \# 14360.}
\altaffiltext{$\star$}{Based in part on observations obtained with the Apache Point Observatory 3.5-meter telescope, which is owned and operated by the Astrophysical Research Consortium.}

\begin{abstract}
We present spatially resolved mass outflow rate measurements ($\dot M_{out}$) for the narrow line region of Markarian 34, the nearest Compton-thick type 2 quasar (QSO2). Spectra obtained with the {\it Hubble Space Telescope} and at {\it Apache Point Observatory} reveal complex kinematics, with distinct signatures of outflow and rotation within 2 kpc of the nucleus. Using multi-component photoionization models, we find that the outflow contains a total ionized gas mass of $M \approx 1.6 \times 10^6 M_{\odot}$. Combining this with the kinematics yields a peak outflow rate of $\dot M_{out} \approx$ 2.0 $\pm$ 0.4 $M_{\odot}$ yr$^{-1}$ at a distance of 470 pc from the nucleus, with a spatially integrated kinetic energy of $E \approx 1.4 \times 10^{55}$ erg. These outflows are more energetic than those observed in Mrk 573 and NGC 4151, supporting a correlation between luminosity and outflow strength even though they have similar peak outflow rates. The mix of rotational and outflowing components suggests that spatially resolved observations are required to determine accurate outflow parameters in systems with complex kinematics. (\textcolor{red}{{\it See appended erratum for updated values.}})
\end{abstract}

\keywords{galaxies: active -- galaxies: individual (Mrk 34) -- galaxies: kinematics and dynamics -- galaxies: Seyfert -- ISM: jets and outflows \vspace{-5pt}}

\section{Introduction}

\subsection{Feedback from Mass Outflows in Active Galaxies}

Accreting supermassive black holes (SMBHs) are thought to be the central engines that power luminous Active Galactic Nuclei (AGN). As material from an accretion disk falls onto the SMBH, it releases ionizing radiation into the host galaxy that interacts with the interstellar medium. This feeding and feedback process drives outflows of ionized and molecular gas that may regulate the SMBH accretion rate and clear the galaxy bulge of star forming gas, thereby linking the evolution of SMBHs and their galaxies \citep{ciotti2001, hopkins2005, kormendy2013, heckman2014, fiore2017, harrison2018}. This feedback model is often used to explain the observed scaling relationships between SMBHs and galaxy bulge properties, including the mass and stellar velocity dispersion ($M_{\bullet} - \sigma_{\star}$, \citealp{ferrarese2000, gebhardt2000, batiste2017}).

Outflows that are capable of delivering feedback on the kiloparsec (kpc) scale of galaxy bulges often reside in the narrow emission line region (NLR), which is composed of ionized gas at distances of $\sim$ 1 $-$ 1000 pc from the AGN with densities of $n_\mathrm{H}$ $\approx 10^2 - 10^6$ cm$^{-3}$ \citep{peterson1997}. AGN ionized gas at larger distances that exhibits primarily rotational kinematics is referred to as the extended narrow line region (ENLR, \citealp{unger1987}).

The impact of outflows must be quantified to determine whether or not they deliver effective feedback capable of altering star formation rates and evacuating gas reservoirs \citep{harrison2017}. This can be accomplished by characterizing the outflow kinetic energy ($E = 1/2 Mv^2$), mass outflow rate ($\dot M = Mv/\delta r$), and related energetic measurements. Determining these quantities is greatly aided by spatially resolved observations that constrain the physical size and location of the outflowing material, making nearby AGN prime targets, as observations can resolve structures on scales of tens of parsecs (pc).

\begin{figure*}
\vspace{-14pt}
\centering
\subfigure{
\includegraphics[scale=0.523]{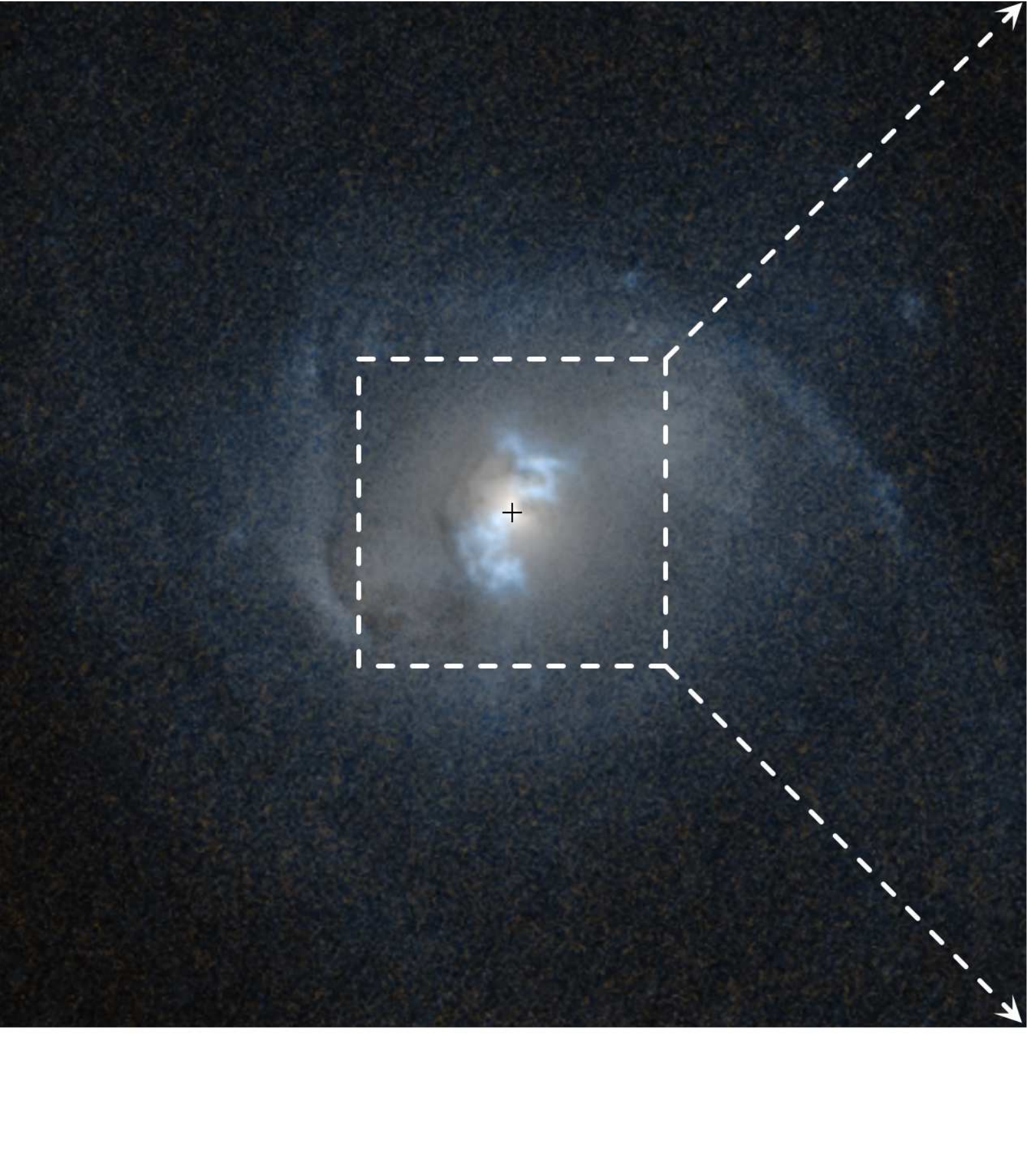}}
\subfigure{
\includegraphics[scale=0.618]{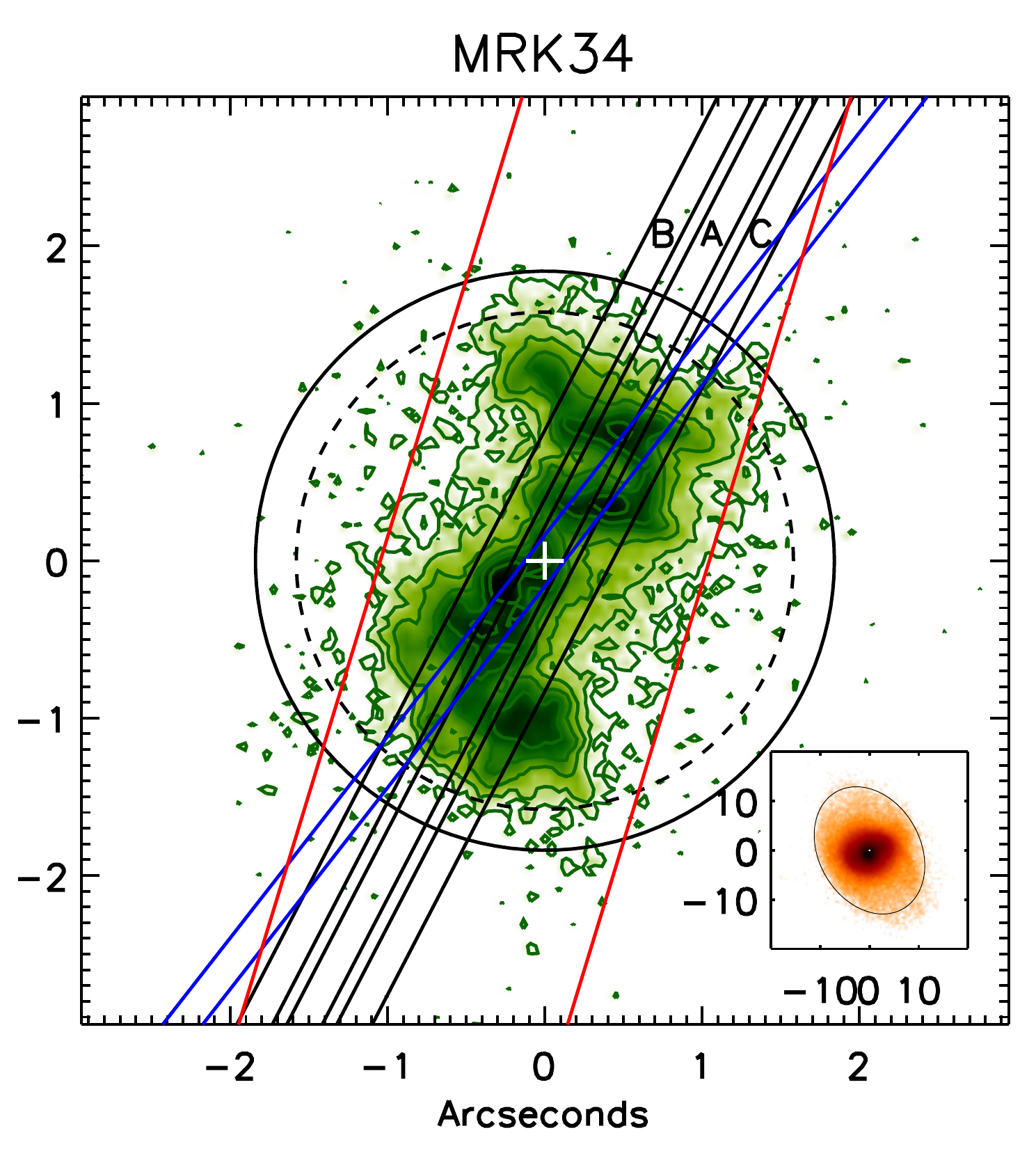}}
\vspace{-10pt}
\caption{The left panel is a 20$\arcsec$ x 20$\arcsec$ composite image of Mrk 34 with a 6$\arcsec$ x 6$\arcsec$ inset square. The individual color channels are composed of {\it HST} WFC3/UVIS F814W (red) and WFPC2/PC F547M (blue) images, which are dominated by galaxy continuum and [O~III] emission, respectively. The right panel shows an [O~III] flux contour map of the NLR of Mrk 34 modified from \cite{fischer2018}, where the dashed and solid circles are the outflow radius and 3$\sigma$ flux detection limit, respectively. The solid lines delineate the locations and slit widths of the observations, with {\it HST} STIS using the G430M grating and 0$\farcs$2 slit in black, {\it HST} STIS using the G430L/G750M gratings and 0$\farcs$2 slit in blue, and our APO DIS observations along PA = 163$\degr$ with a 2$\farcs$0 slit in red. North is up and east is to the left in both panels. This image was created with the help of the ESA/ESO/NASA FITS Liberator.}
\label{structure}
\end{figure*}

Studies of NLR outflows have traditionally measured ``global'' outflow rates that characterize the outflows with a single mass, velocity, and outflow extent (e.g. \citealp{forster2014, genzel2014, harrison2014, mcelroy2015, kakkad2016, karouzos2016, villar2016, bae2017, bischetti2017, leung2017}). This has the advantage of quickly determining the outflow parameters; however, global measurements may suffer from large uncertainties if the sizes and kinematic profiles of the NLRs are not well constrained \citep{kang2018} and/or the proper techniques for determining masses are not employed \citep{revalski2018}. Spatially resolved observations can be used to probe the assumptions of global techniques and understand how outflows manifest at different distances from the nucleus (e.g., the recent detailed study by \citealp{venturi2018}).

A pilot study to quantify NLR outflows as a function of spatial position was conducted by \cite{crenshaw2015} for the nearby Seyfert 1 galaxy NGC 4151 ($z = 0.00332, D = 13.3$ Mpc). They found that the peak outflow rate exceeded the SMBH mass accretion rate, as well as the outflow rates for the UV/X-ray absorbers, which motivated us to quantify feedback from NLR outflows.

In \cite{revalski2018}, hereafter Paper I, we presented a refined methodology, and results for the Seyfert 2 galaxy Mrk 573. In this second paper of the series we continue our detailed investigations of individual AGN using high spatial resolution spectroscopy and imaging to quantify the impact of NLR outflows. We follow a similar procedure to \cite{revalski2018}, presenting our observations (\S2), spectral and image analyses (\S3), photoionization modeling (\S4), calculations (\S5), results (\S6), discussion (\S7), and conclusions (\S8).

\setlength{\tabcolsep}{0.012in} 
\tabletypesize{\scriptsize}
\begin{deluxetable*}{cccccccccccccc}
\tablenum{1}
\vspace{-26pt}
\tablecaption{Summary of Observations}
\tablehead{
\colhead{Observing} & \colhead{Instrument} & \colhead{Observation} & \colhead{Program} & \colhead{Date} & \colhead{Exposure} & \colhead{Grating} & \colhead{Spectral} & \colhead{Wavelength} & \colhead{Wavelength} & \colhead{Spatial} & \colhead{Position} & \colhead{Mean} &\colhead {Seeing}\\
\colhead{Facility} & \colhead{Name} & \colhead{ID} & \colhead{ID} & \colhead{(UT)} & \colhead{Time} & \colhead{/ Filter} & \colhead{Dispersion} & \colhead{Centroid} & \colhead{Range} & \colhead{Scale} & \colhead{Angle} & \colhead{Air Mass} &\colhead {}\\
\colhead{} & \colhead{} & \colhead{} & \colhead{} & \colhead{} & \colhead{(s)} & \colhead{} & \colhead{(\AA~pix$^{-1}$)} & \colhead{(\AA)} & \colhead{(\AA)} & \colhead{($\arcsec$~pix$^{-1}$)} & \colhead{(deg)} & \colhead{} &\colhead {($\arcsec$)}
}
\startdata
HST & STIS CCD & O5G404010 & 8253 & 2000-02-17 & 1500 & G430M & 0.28 & 5216 & 5076-5357 & 0.051 & 152.5 & ... & ...\\
HST & STIS CCD & O5G404030 & 8253 & 2000-02-17 & 1460 & G430M & 0.28 & 5216 & 5076-5357 & 0.051 & 152.5$^\star$ & ... & ...\\
HST & STIS CCD & O5G404040 & 8253 & 2000-02-18 & 1460 & G430M & 0.28 & 5216 & 5076-5357 & 0.051 & 152.5$^\star$ & ... & ...\\
 HST & STIS CCD & OD2501010 & 14360 & 2016-02-24 & 437 & G430L & 2.73 & 4300 & 2900-5700 & 0.051 & 143.1 & ... & ...\\
HST & STIS CCD & OD2501020 & 14360 & 2016-02-24 & 437 & G430L & 2.73 & 4300 & 2900-5700 & 0.051 & 143.1& ... & ...\\
HST & STIS CCD & OD2501030 & 14360 & 2016-02-24 & 437 & G430L & 2.73 & 4300 & 2900-5700 & 0.051 & 143.1 & ... & ...\\
HST & STIS CCD & OD2501040 & 14360 & 2016-02-24 & 143 & G750L & 4.92 & 7751 & 5240-10270 & 0.051 & 143.1 & ... & ...\\
HST & STIS CCD & OD2501050 & 14360 & 2016-02-24 & 143 & G750L & 4.92 & 7751 & 5240-10270 & 0.051 & 143.1 & ... & ...\\
HST & STIS CCD & OD2501060 & 14360 & 2016-02-24 & 143 & G750L & 4.92 & 7751 & 5240-10270 & 0.051 & 143.1 & ... & ...\\
HST & WFC3/UVIS & IBY80U020 & 12903 & 2013-10-09 & 75 & F814W & ... & 8030 & 1536 & 0.04 & ... & ... & ...\\
HST & WFC3/UVIS & IBY80U030 & 12903 & 2013-10-09 & 75 & F814W & ... & 8030 & 1536 & 0.04 & ... & ... & ...\\
HST & WFPC2/PC & U9PM0201M & 10873 & 2007-06-07 & 1300 & F547M & ... & 5468 & 483 & 0.046 & ... & ... & ...\\
HST & WFPC2/PC & U9PM0202M & 10873 & 2007-06-07 & 1300 & F547M & ... & 5468 & 483 & 0.046 & ... & ... & ...\\
HST & WFPC2/PC & U9PM0203M & 10873 & 2007-06-07 & 1300 & F547M & ... & 5468 & 483 & 0.046 & ... & ... & ...\\
HST & WFPC2/PC & U9PM0204M & 10873 & 2007-06-07 & 1300 & F547M & ... & 5468 & 483 & 0.046 & ... & ... & ...\\
HST & WFPC2/PC & U9PM0207M & 10873 & 2007-06-07 & 1300 & F547M & ... & 5468 & 483 & 0.046 & ... & ... & ...\\
HST & WFPC2/PC & U9PM0208M & 10873 & 2007-06-07 & 1300 & F547M & ... & 5468 & 483 & 0.046 & ... & ... & ...\\
HST & WFPC2/PC & U9PM0205M & 10873 & 2007-06-07 & 1300 & F467M & ... & 5468 & 483 & 0.046 & ... & ... & ...\\
HST & WFPC2/PC & U9PM0206M & 10873 & 2007-06-07 & 1300 & F467M & ... & 5468 & 483 & 0.046 & ... & ... & ...\\
HST & WFPC2/PC & U9PM0209M & 10873 & 2007-06-07 & 1300 & F467M & ... & 5468 & 483 & 0.046 & ... & ... & ...\\
HST & WFPC2/PC & U9PM020AM & 10873 & 2007-06-07 & 1300 & F467M & ... & 5468 & 483 & 0.046 & ... & ... & ...\\
APO & DIS & ... & ... & 2017-01-31 & 900 & B1200 & 0.615 & 4898 & 4257-5517 & 0.42 & 73 & 1.13 & 1.22\\
APO & DIS & ... & ... & 2017-01-31 & 900 & R1200 & 0.580 & 6600 & 6006-7192 & 0.40 & 73 & 1.13 & 1.22\\
APO & DIS & ... & ... & 2017-04-23 & 900 & B1200 & 0.615 & 4900 & 4273-5545 & 0.42 & 118 & 1.14 & 1.80\\
APO & DIS & ... & ... & 2017-04-23 & 900 & R1200 & 0.580 & 6600 & 5996-7183 & 0.40 & 118 & 1.14 & 1.80\\
APO & DIS & ... & ... & 2017-01-31 & 900 & B1200 & 0.615 & 4898 & 4257-5517 & 0.42 & 163 & 1.17 & 1.22\\
APO & DIS & ... & ... & 2017-01-31 & 900 & R1200 & 0.580 & 6600 & 6005-7191 & 0.40 & 163 & 1.17 & 1.22\\
APO & DIS & ... & ... & 2014-11-19 & 900 & B1200 & 0.615 & 5380 & 4734-5584 & 0.42 & 208 & 1.20 & 1.13\\
APO & DIS & ... & ... & 2014-11-19 & 900 & R1200 & 0.580 & 6582 & 5990-7179 & 0.40 & 208 & 1.20 & 1.13
\enddata
\tablecomments{A summary of the observations and data used in this study. The columns list the observing facility, instrument, MAST archive dataset name, {\it HST} Program ID, observation date, exposure time, grating (for spectra) or filter (for imaging), spectral dispersion, wavelength centroid, wavelength range (for spectra) or bandpass (for imaging), spatial resolution, position angle (for spectra), air mass, and seeing. All {\it HST} values are defined in their respective instrument handbooks \citep{stisihb, wfpc2ihb, wfc3ihb}, with the exact STIS spatial scale quoted as 0.05078$\arcsec$~pix$^{-1}$. $^\star$These observations are spatially offset from the nucleus by $\pm0\farcs28$ (see Figure \ref{structure}).}
\end{deluxetable*}

\subsection{Physical Characteristics of Markarian 34}

We selected the type 2 quasar (QSO2) Markarian 34 (Mrk 34, SDSS J103408.58+600152.1, MCG+10-15-104) to extend our sample to higher redshifts, bolometric luminosities, and outflow radii, complementing our studies in \cite{crenshaw2015} and \cite{revalski2018}. The central AGN resides in an Sa-type galaxy with spiral arms and a weak bar structure \citep{nair2010} that are visible in Figure \ref{structure}. Using our observations, we derive (\S 3.3) a new heliocentric redshift of $z = 0.05080$, a recessional velocity of 14,843 km s$^{-1}$, a Hubble distance of 209 Mpc, and a spatial scale of 1014 pc arcsec$^{-1}$, assuming $H_0 = 71$ km s$^{-1}$ Mpc$^{-1}$.

In \cite{fischer2018} we found that the host galaxy major axis is along a position angle (PA) of $\approx$ 30$\degr$, with an ellipticity of $e = 1-b/a = 0.25$, and an inclination of $i = cos^{-1}(b/a) = 41\degr$, in agreement with previous studies \citep{haniff1988}. The NLR major axis is along PA $\approx$ 150$\degr$, which is similar to the observed radio jet that displays a double-lobed structure, and enhanced emission in regions of low gas excitation \citep{ulvestad1984, unger1987, baum1993, falcke1998, nagar1999}. The ionized gas displays multiple kinematic components, and emission has been traced to radial distances of $\sim$ 12 kpc \citep{whittle1988}. These details are explored extensively in \cite{rosario2007}.

As discussed by \cite{gandhi2014}, Mrk 34 is the nearest Compton-thick QSO2, and a firm black hole mass estimate is not currently available. A water maser has been detected \citep{henkel2005, liu2017}, but a resolved velocity map has not yet been obtained. Estimates from the $M_{\bullet} - \sigma_{\star}$ method using proxies for $\sigma_{\star}$ \citep{wang2007, oh2011} yield a mass range of $M_{\bullet} \approx 10^{6.8-7.5} M_{\odot}$, resulting in a wide range of Eddington ratios. We adopt a bolometric luminosity of Log(L$_{\mathrm{bol}}$/erg s$^{-1}$) = 46.2 $\pm$ 0.4 (\S3.3). The host galaxy morphology and ionized gas distribution within the NLR are shown in Figure \ref{structure}.

\section{Observations}

\subsection{Hubble Space Telescope (HST)}

The {\it Hubble Space Telescope} ({\it HST}) spectroscopy and imaging used in this study were obtained with the Space Telescope Imaging Spectrograph (STIS), Wide Field Camera 3 (WFC3), and Wide Field and Planetary Camera 2 (WFPC2/PC). We employ low and medium-dispersion spectra to characterize the physical conditions and kinematics of the emission line gas, as well as [O~III] imaging to determine the ionized gas mass. The calibrated data were retrieved from the Mikulski Archive at the Space Telescope Science Institute, and the multiple spatially dithered exposures were combined using the Interactive Data Language (IDL). These data are summarized in Table 1, and all spectral observations employed the 52$\arcsec$~x~0$\farcs$2 slit.

\begin{figure*}
\vspace{-10pt}
\centering
\includegraphics[scale=0.55]{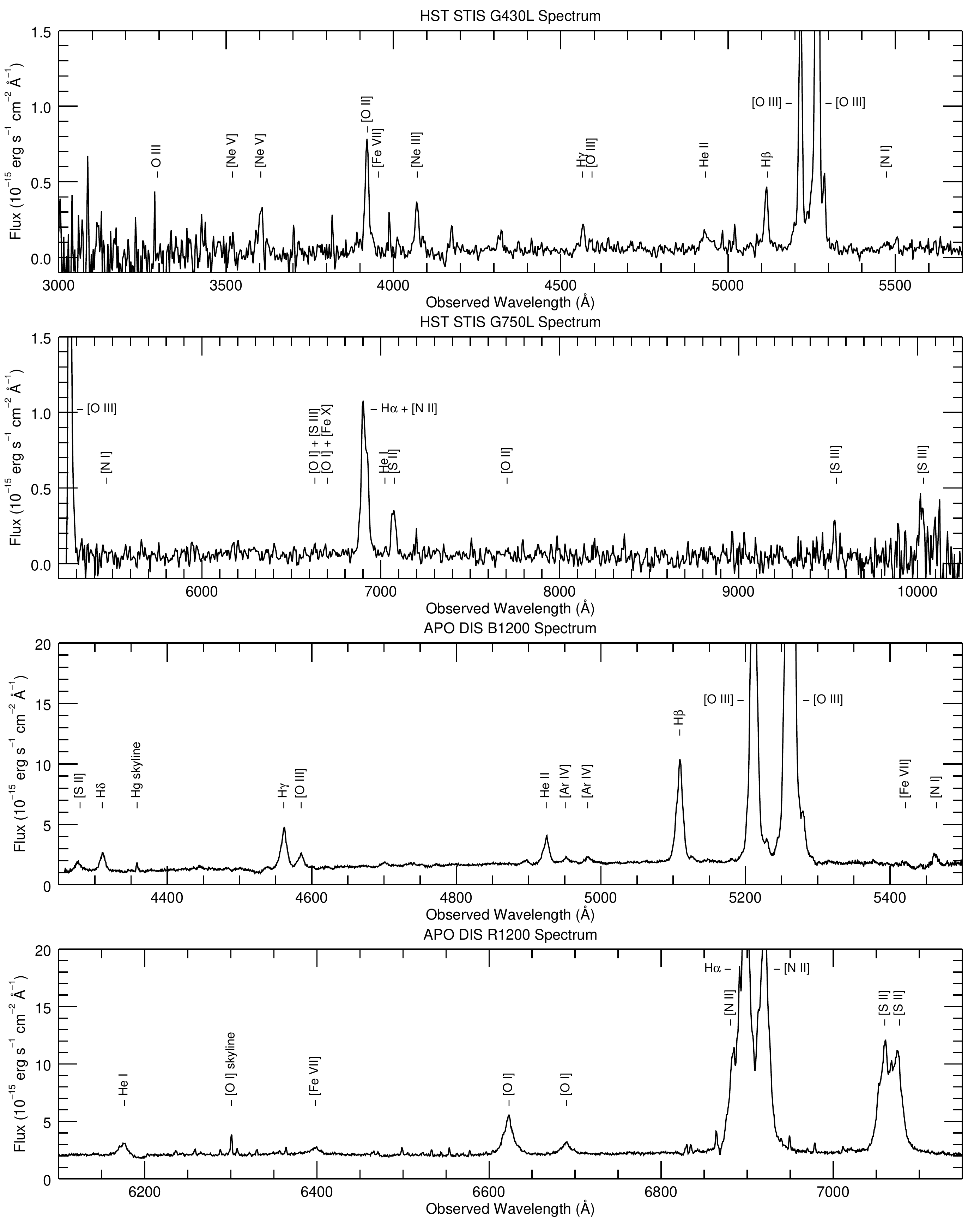}
\caption{Spectral traces centered on the nucleus and spatially summed over $\sim 1\farcs3$ and $\sim 2\farcs9$ for the {\it HST} and APO data, respectively, with the positions of common emission lines labeled. The spectra are shown at observed wavelengths, and from top to bottom are: {\it HST} STIS G430L, {\it HST} STIS G750L, APO DIS B1200, and APO DIS R1200.}
\label{spectra}
\end{figure*}

In \cite{fischer2018} we used the G430M observations with a resolving power of $R \approx 9400$ to characterize the NLR kinematics of Mrk 34. These slits are labeled A, B, and C in Figure \ref{structure}, and additional details are given in \cite{fischer2013, fischer2018}. To characterize the physical conditions of the gas, we obtained new STIS spectroscopy with a larger wavelength range and resolving powers of $R \approx 900-1000$ using the G430L/G750L gratings under Program ID 14360 (PI: M. Elvis). The position of these observations is represented by a blue slit in Figure \ref{structure}, and extracted spectra are shown in Figure \ref{spectra}. The corollary X-ray observations obtained with {\it Chandra} to investigate shocks associated with the outflows will be presented in a future paper (Fischer et al. in prep).

To examine the NLR structure, we created a color-composite image using {\it HST} WFC3/UVIS images with the F814W and F547M filters. An ArcSinh(x) stretch function was applied to reveal faint details in each image, and then they were combined into a single color average. The ``Z-shaped'' NLR structure and its continuation into the spiral arms are visible in the left panel of Figure \ref{structure}. The [O~III] emission line image used to calculate the ionized gas mass was taken from our study in \cite{fischer2018}, which consists of the F547M images discussed previously, in combination with an F467M image for continuum subtraction.

\subsection{Apache Point Observatory (APO)}

We obtained additional spectra using the Dual Imaging Spectrograph (DIS) on the Astrophysical Research Consortium's Apache Point Observatory (ARC's APO) 3.5 meter telescope in Sunspot, New Mexico. The DIS uses a dichroic element to split light into blue and red channels, allowing for simultaneous data collection in the H$\beta$ and H$\alpha$ regions of the spectrum. The spectral resolution is 1.23 $\AA$ in the blue and 1.16 $\AA$ in the red, corresponding to resolving powers of $R \approx$ 3400 -- 6200. These spectra have lower spatial resolution than the {\it HST} data, but allow us to search for outflow signatures outside of the narrow {\it HST} slits, probe the ENLR kinematics, and detect important diagnostic emission lines with greater S/N out to larger distances from the nucleus.  We obtained observations at four evenly spaced position angles of $\sim 73\degr$, $118\degr$, $163\degr$, and $208\degr$, using a 2$\farcs$0 slit at low air masses (Table 1). While the slits were not at the ideal parallactic angle, we compared with available SDSS spectra and found no evidence for loss of blue light due to atmospheric refraction.

\begin{figure*}
\vspace{-8pt}
\centering
\subfigure{
\includegraphics[scale=0.682]{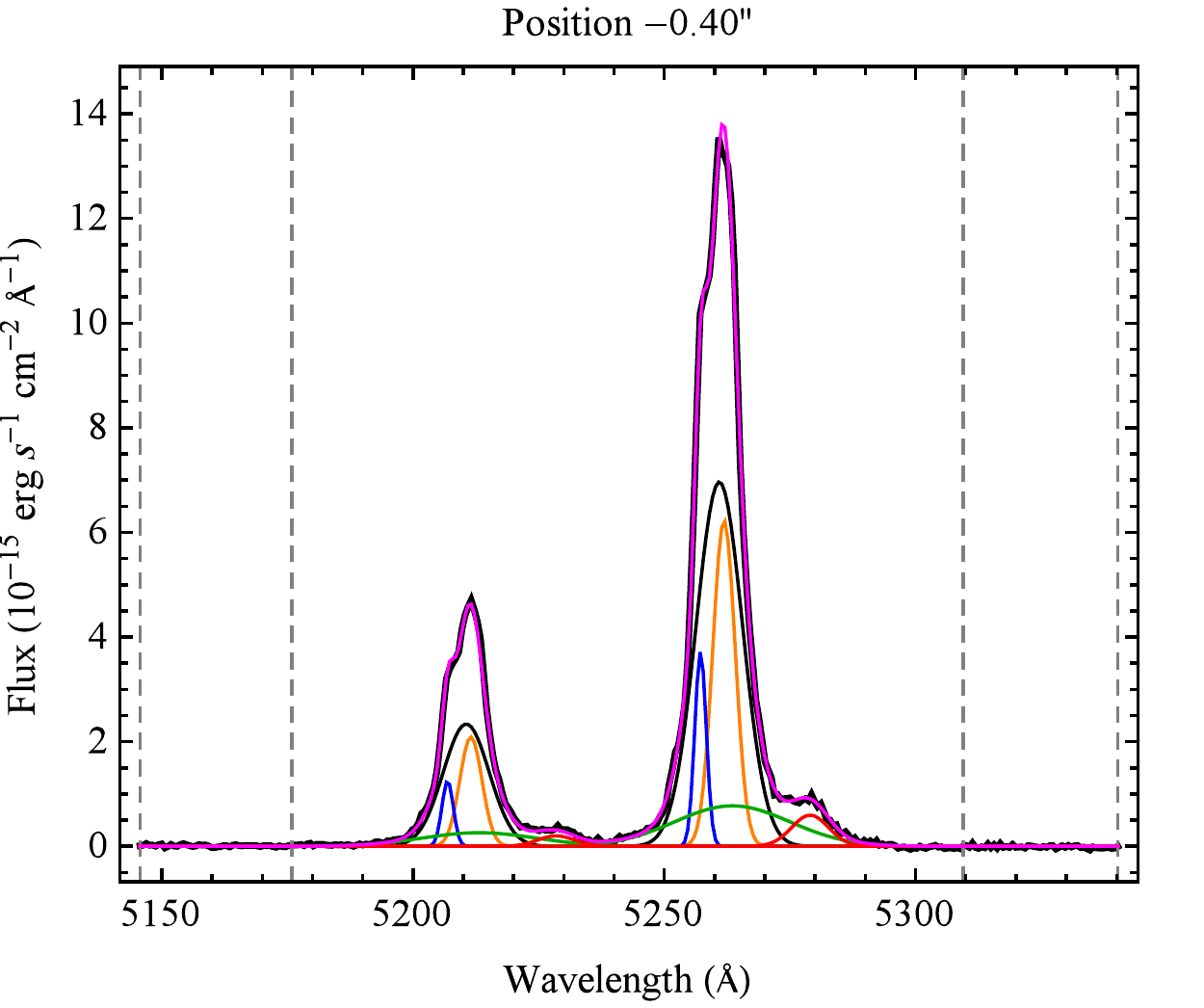}}
\subfigure{
\includegraphics[scale=0.435]{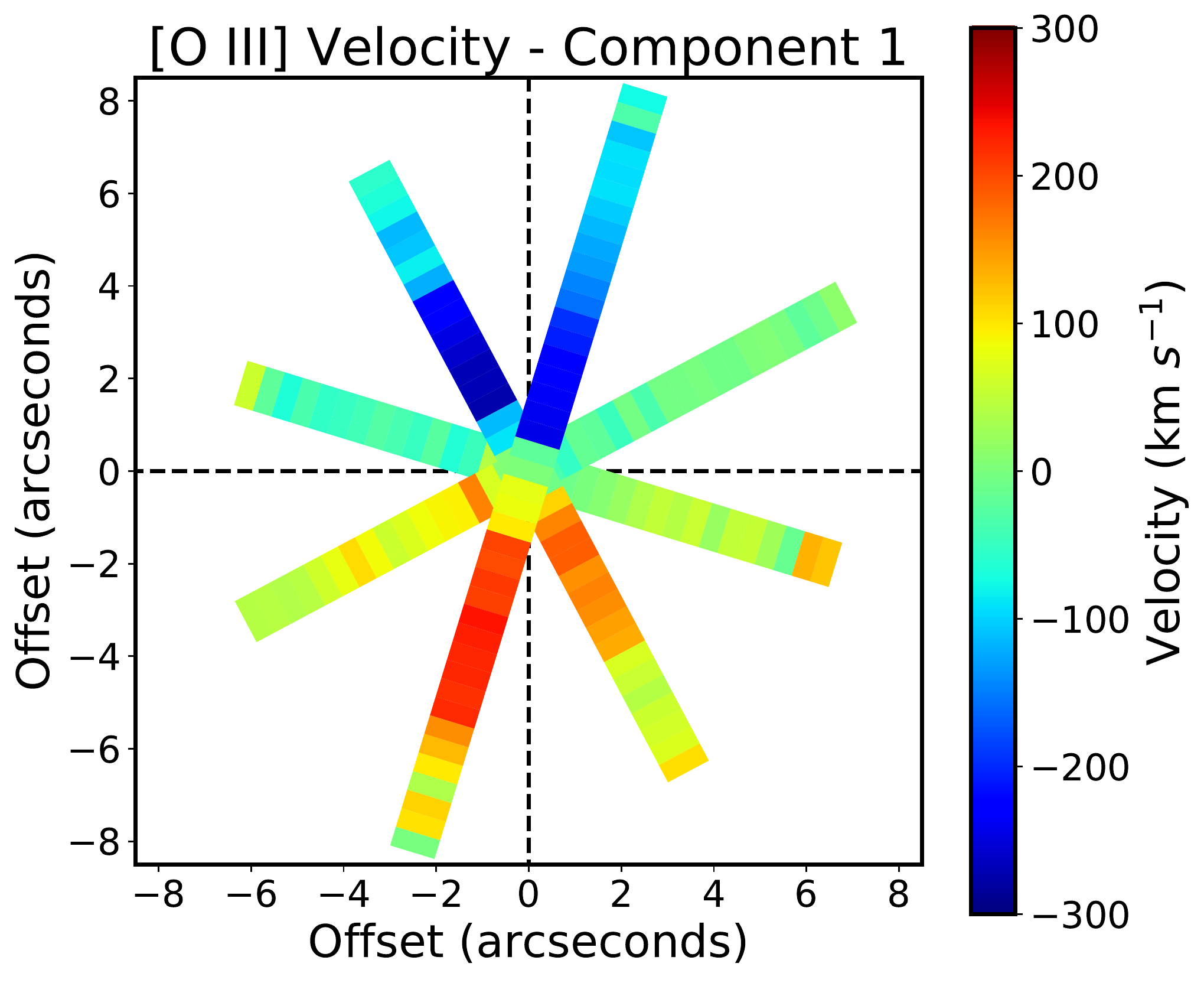}}
\subfigure{
\includegraphics[scale=0.20]{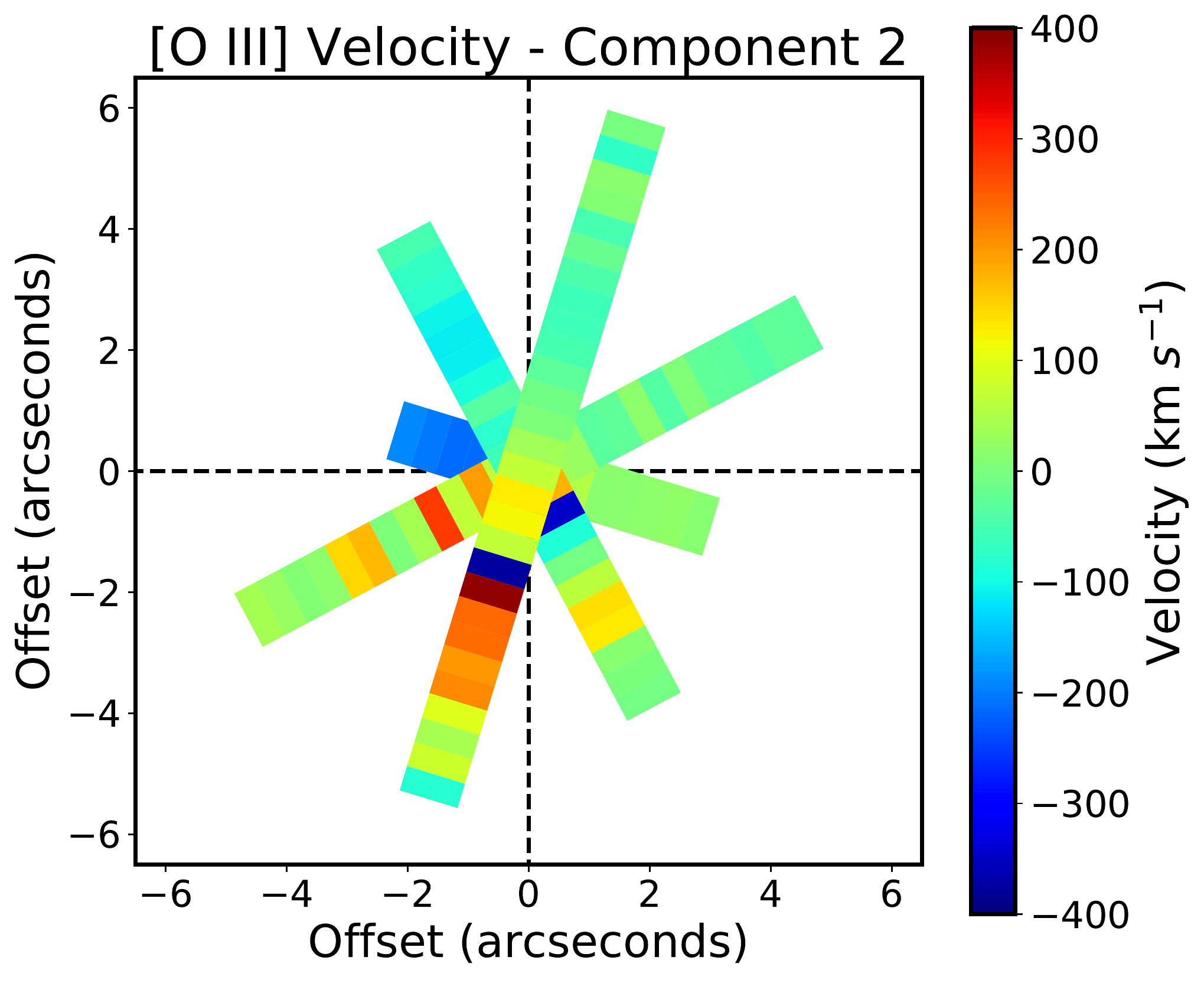}}
\subfigure{
\includegraphics[scale=0.20]{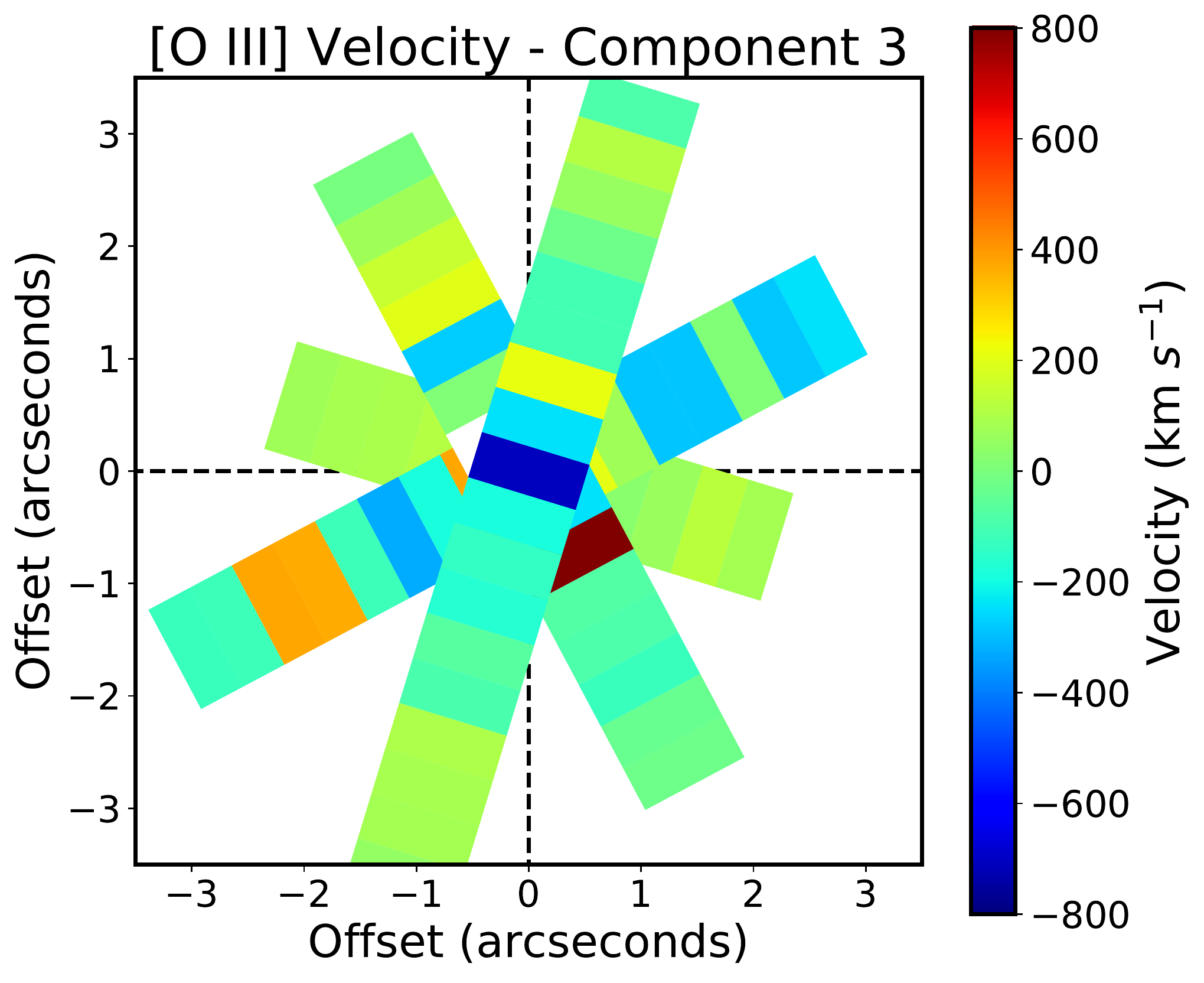}}
\subfigure{
\includegraphics[scale=0.20]{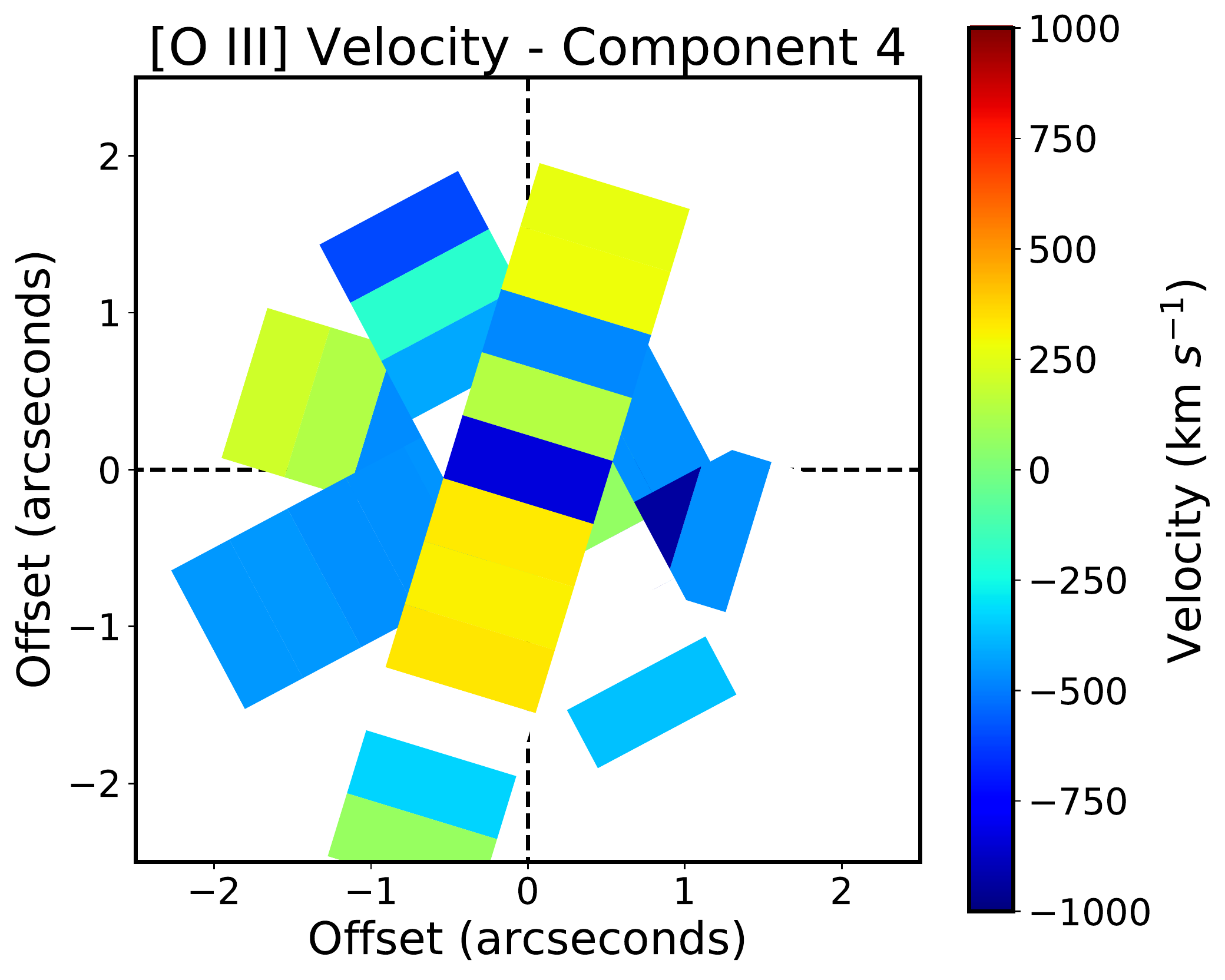}}
\subfigure{
\includegraphics[scale=0.20]{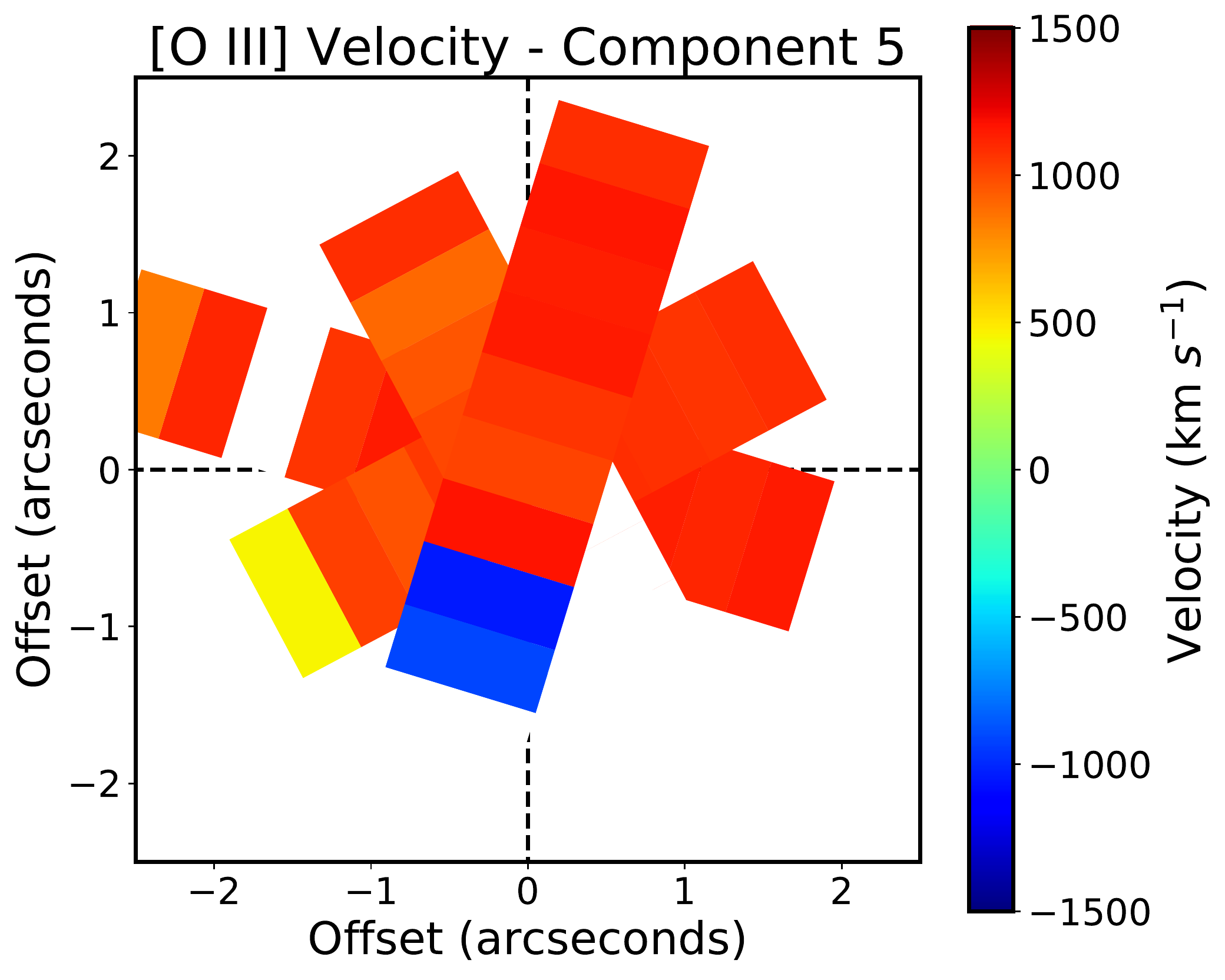}}
\caption{The left panel is a spectral trace of the [O~III] doublet from our APO DIS observations along PA = 163$\degr$ at a position $0\farcs4$ SE of the nucleus, overlaid with a multi-component fit. The data are represented by a thick black line, and the composite model by magenta. The individual kinematic components are color-coded from strongest to weakest peak flux in the following order: black, orange, blue, green, and red. See the Appendix for additional fits. The right panel shows the observed velocities at each position for the strongest [O~III] $\lambda$5007 emission line component in each of the four APO observations. The redshifts and blueshifts are a signature of galactic rotation. The lower row shows the observed velocities for the second, third, fourth, and fifth components. Note the different spatial range displayed in each panel. The observations employed a 2$\arcsec$ wide slit and are represented by 1$\arcsec$ wide rectangles for visibility. North is up, and East is to the left.}
\label{fitting}
\end{figure*}

We reduced the data using IRAF \citep{tody1986, tody1993} \footnote{IRAF is distributed by the National Optical Astronomy Observatories, which are operated by the Association of Universities for Research in Astronomy, Inc., under cooperative agreement with the National Science Foundation.} following the standard techniques of bias subtraction, image trimming, bad pixel replacement, flat-fielding, Laplacian edge cosmic ray removal \citep{vandokkum2001}, image combining, and sky line subtraction. Wavelength calibration was completed using comparison lamp images taken before the science exposures, and velocities were converted to heliocentric values. Flux calibration was completed using Oke standard stars \citep{oke1990} and the air mass at midexposure. The DIS dispersion and spatial axes are not perpendicular, so we fit a line to the galaxy continuum and resampled the data to ensure that measurements of emission lines from the same pixel row sample the same spatial location. We focus on our observations along PA = $163\degr$, which is closest to the NLR major axis. Some of the extracted spectra are shown in Figure \ref{spectra}.

\section{Analysis}

\subsection{Spectral Fitting}

We fit Gaussian profiles to all emission lines in our spectra to obtain the gas kinematics and emission line flux ratios for comparison with photoionization models. We employ a Bayesian fitting routine developed for our recent kinematic studies \citep{fischer2017, fischer2018} based on the Importance Nested Sampling algorithm in MultiNest\footnote{\url{https://ccpforge.cse.rl.ac.uk/gf/project/multinest/}} \citep{feroz2008, feroz2009, feroz2013, buchner2014}, and a detailed description is given in the Appendix of \cite{fischer2017}. We require an emission line to have a height signal-to-noise ratio (S/N) of $>$~2 in our {\it HST} spectra, and S/N $>$~3 in our APO spectra, for a positive detection.

As in Paper I, we use a spectral template method by first fitting the strong [O~III] emission line, and then use the Gaussian fit parameters to calculate the centroids and widths for all other lines at each location. This ensures that we are sampling the same kinematic components in each line, although minor differences in the intrinsic line widths may be neglected (see \S3.1 of Paper I). The widths are scaled to maintain the same intrinsic velocity and account for the instrumental line-spread functions. The height of each component is then allowed to vary to encompass the total line flux.

\begin{figure*}
\vspace{-8pt}
\centering
\subfigure{
\includegraphics[scale=0.458]{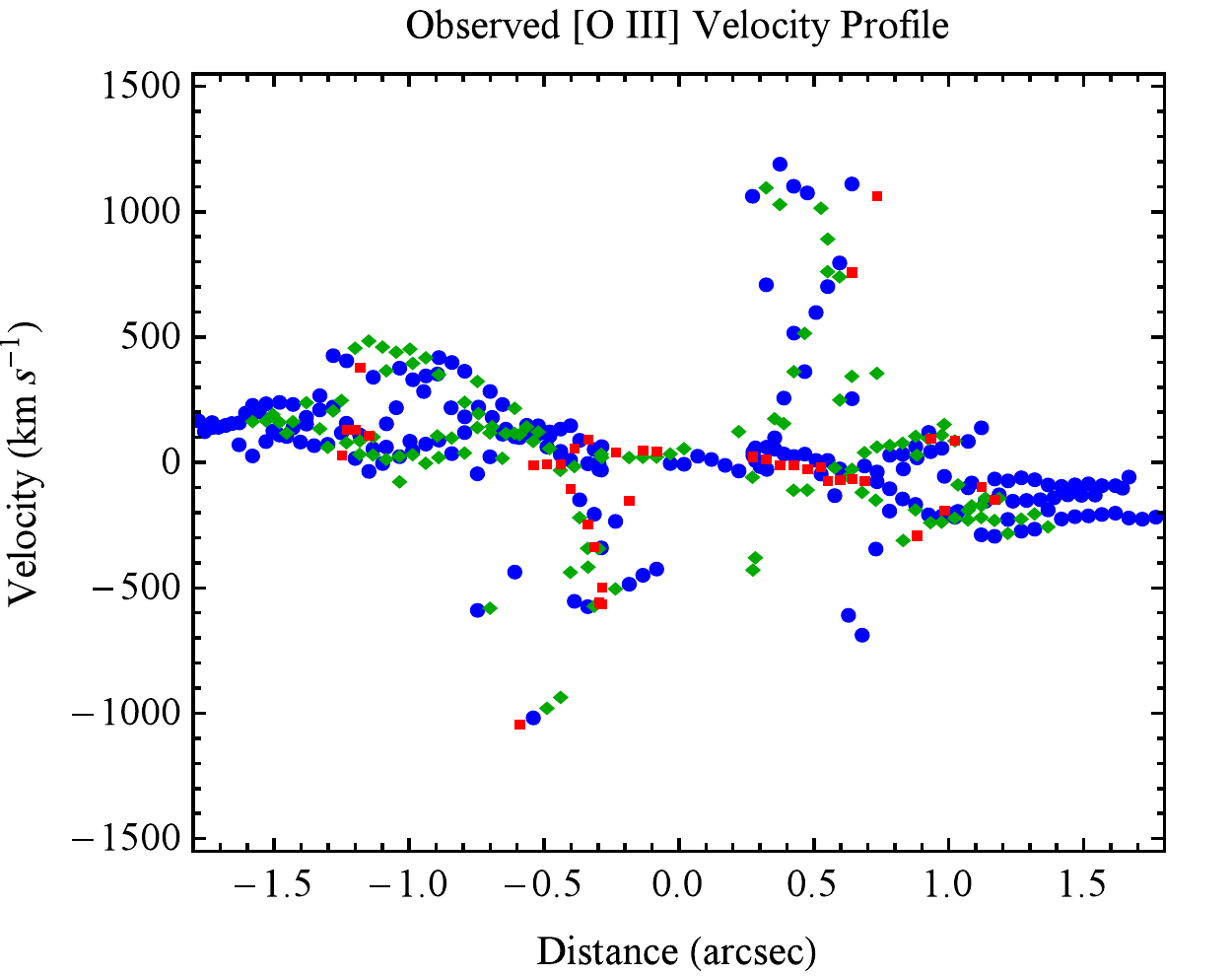}}
\subfigure{
\includegraphics[scale=0.635]{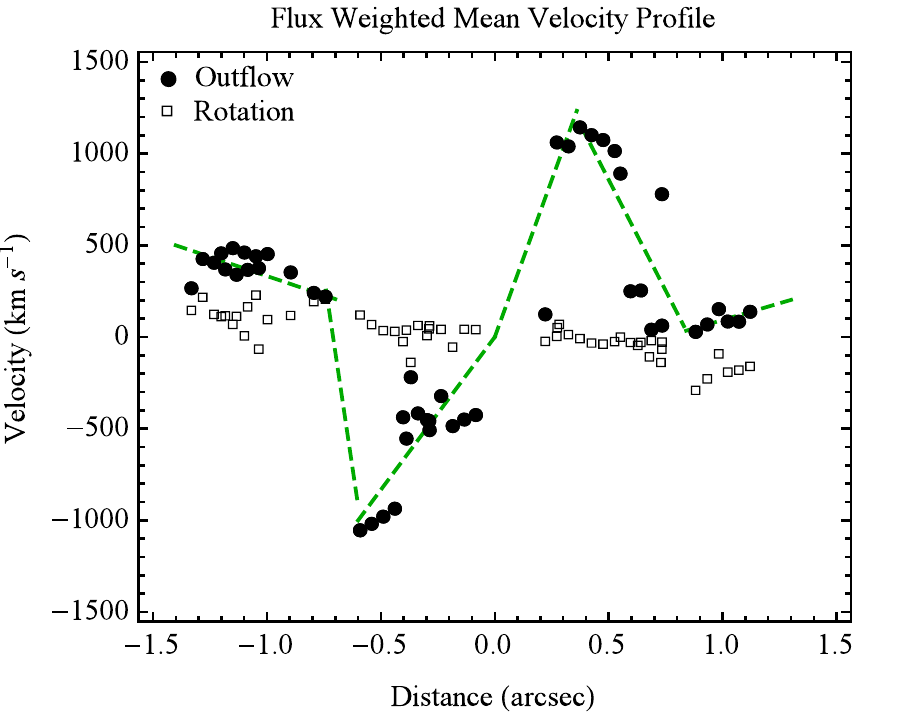}}
\subfigure{
\includegraphics[scale=0.601]{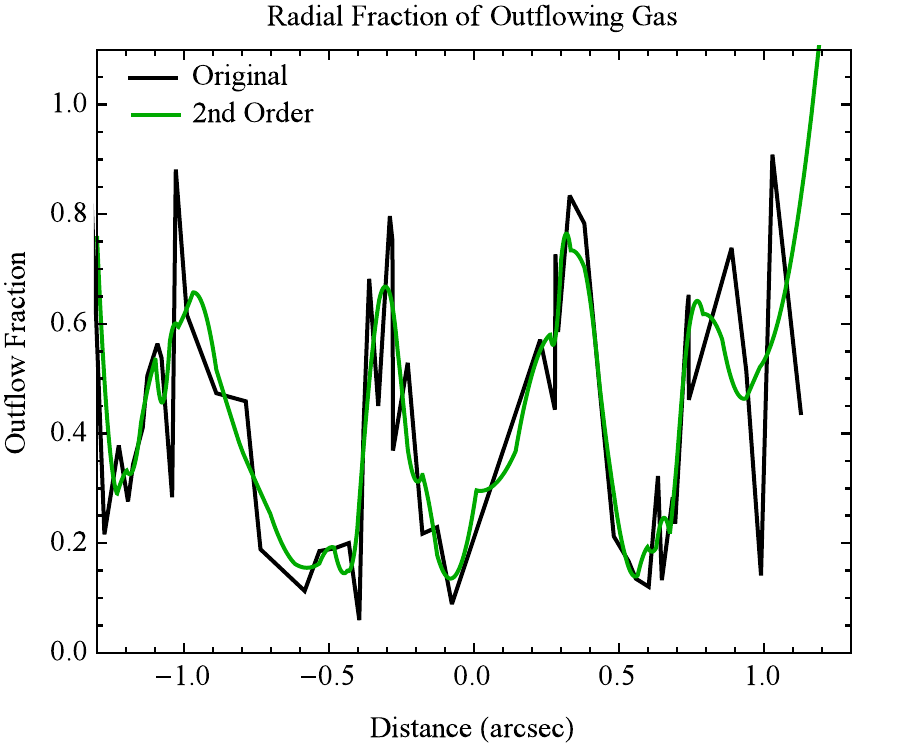}}
\vspace{-8pt}
\caption{Left: The measured kinematics from the three parallel G430M slits \citep{fischer2018} after spatial alignment. The components are sorted from largest to smallest flux as blue circles, green diamonds, and red squares, respectively. Middle: The flux-weighted average velocity profile with components classified as rotation (open squares) or outflow (filled black circles), with the best linear fit in dashed green. Right: The ratio of the outflowing flux to the total flux as a function of distance, with the data in black and a second-order fit in green. Southeast is to the left and Northwest is to the right in all figures.}
\label{velocitylaw}
\end{figure*}

We further improve the fitting procedure by fixing the relative height ratios of doublet lines to their theoretical values \citep{osterbrock2006}. Specifically, [O~III] $\lambda \lambda$5007/4959 = 3.01, [O~I] $\lambda \lambda$6300/6363 = 3.0, and [N~II] $\lambda \lambda$6584/6548 = 2.95. The resulting fits to key emission lines are shown in Figure \ref{fitting} and the Appendix.

The uncertainty in flux for each line is calculated from the residuals between the data and fit. As we detect multiple kinematic components, we scale the uncertainty for each component based on its fractional contribution. If $F = \sum f_i$ is the total flux of all $N$ Gaussians, each of flux $f_i$, and $\delta F$ is the difference in flux between the data and fit, then the fractional flux uncertainty for each $i$-th component is
\begin{equation}
\sigma f_i = \left(\frac{\delta F}{f_i \sqrt{N}}\right).
\end{equation}

\subsection{Ionized Gas Kinematics}

The observed gas kinematics are significantly more complex than those seen in Mrk 573 (Paper I). In that target, we observed a single outflowing component that transitioned to rotation at a radial distance of $\sim$~600 pc. Here, we observe a combination of rotational, disturbed, and outflow kinematics at all radial distances $\lesssim 1\farcs6$ from the nucleus. To derive a velocity law that describes the intrinsic outflow velocity at each position, we created a flux-weighted mean velocity profile from the three parallel {\it HST} STIS G430M observations \citep{rosario2008, fischer2013, fischer2018}. This yields a better average velocity profile over the spatial extent of the NLR than was available with our single slit position for Mrk 573. As the two offset observations do not pass through the nucleus, there is no zero point, and the nuclear distance of each pixel is a function of position along the slit. The projected nuclear distance in arcseconds for each pixel is then represented by Pythagorean's theorem,
\begin{equation}
D = \sqrt{(\delta N \cdot S)^2 + (\delta R)^2}
\end{equation}
\noindent
where $\delta N$ is the number of pixels from the pixel passing closest to the nucleus as defined by the continuum peak, $S$ is the STIS pixel scale of $0\farcs05078$ pixel$^{-1}$, and $\delta R = 0\farcs28$ is the offset distance of the parallel slits. The results of this procedure are shown in the left panel of Figure \ref{velocitylaw}. Next, we grouped the resulting kinematic components at each distance into either rotation or outflow based on their velocities, and created a single flux-weighted velocity profile with one rotational and outflow component at each distance, shown in the middle panel of Figure \ref{velocitylaw}. This also allows us to derive the fraction of flux in the outflowing component relative to the total flux. Finally, the velocities and positions were corrected for projection effects using Equations (1) and (2) in Paper I, with $i = 41\degr$ and $\varphi = 77\degr$. This technique assumes that the outflows are moving radially outward along the galactic disk as suggested by \cite{fischer2017}. The resulting maximum deprojected velocities are $\sim 2000$ km s$^{-1}$. This provides us with the intrinsic outflow velocity and the fraction of flux in outflow that are needed to calculate the mass outflow rates and energetics. Further discussion of the kinematics is presented in \cite{fischer2018}.

\begin{figure*}
\centering
\subfigure{
\includegraphics[scale=0.34]{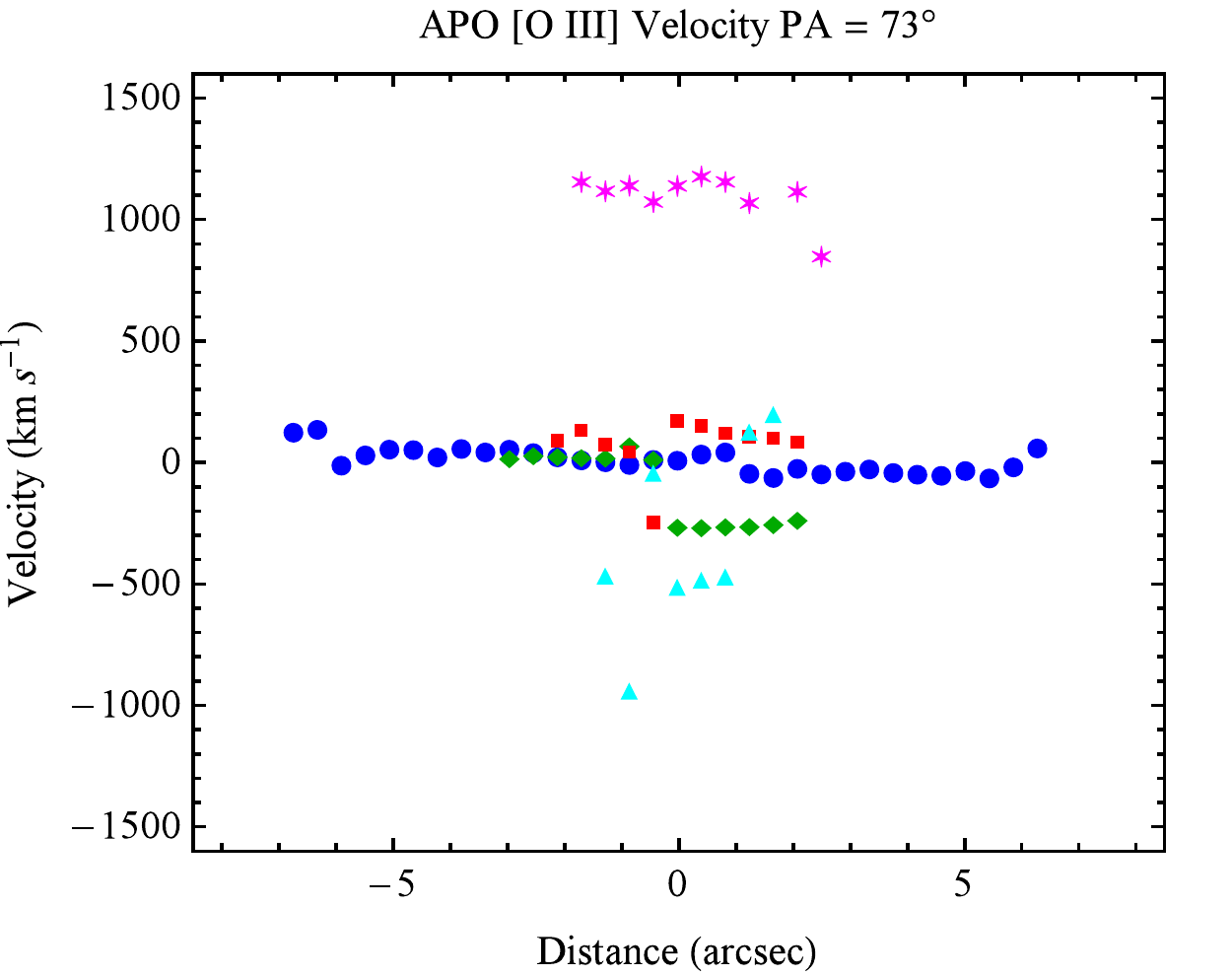}}
\vspace{-6pt}
\subfigure{
\includegraphics[scale=0.34]{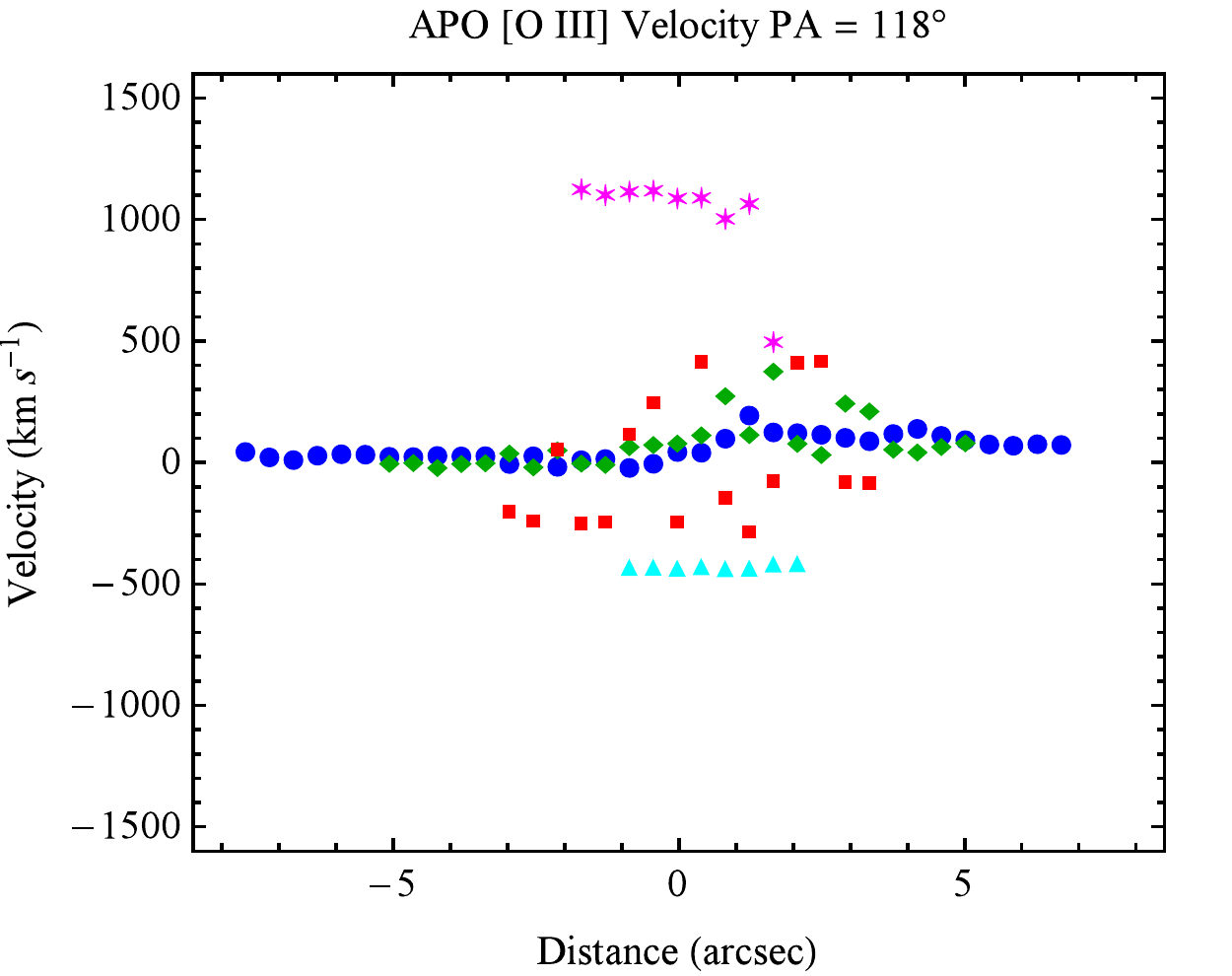}}
\subfigure{
\includegraphics[scale=0.34]{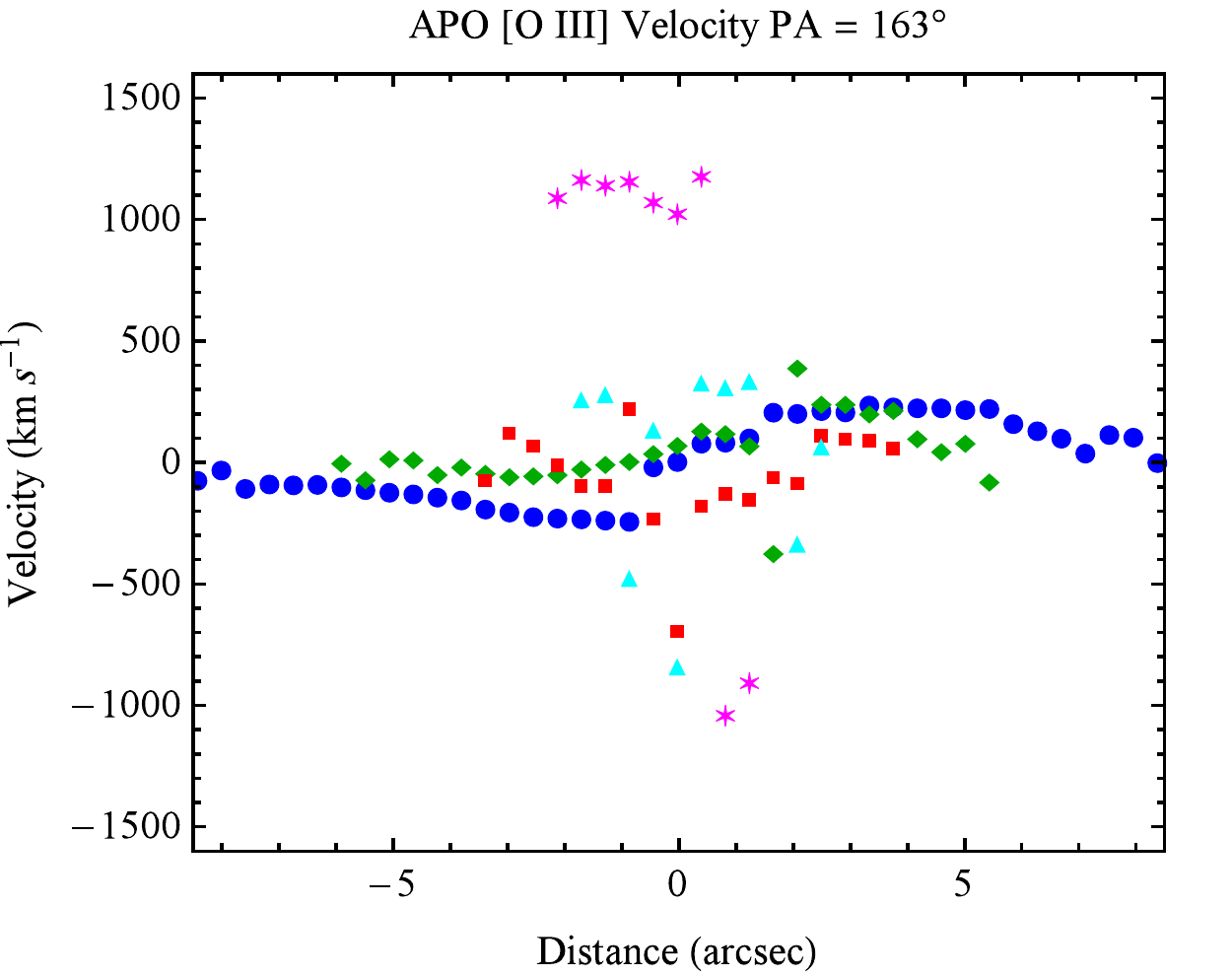}}
\subfigure{
\includegraphics[scale=0.34]{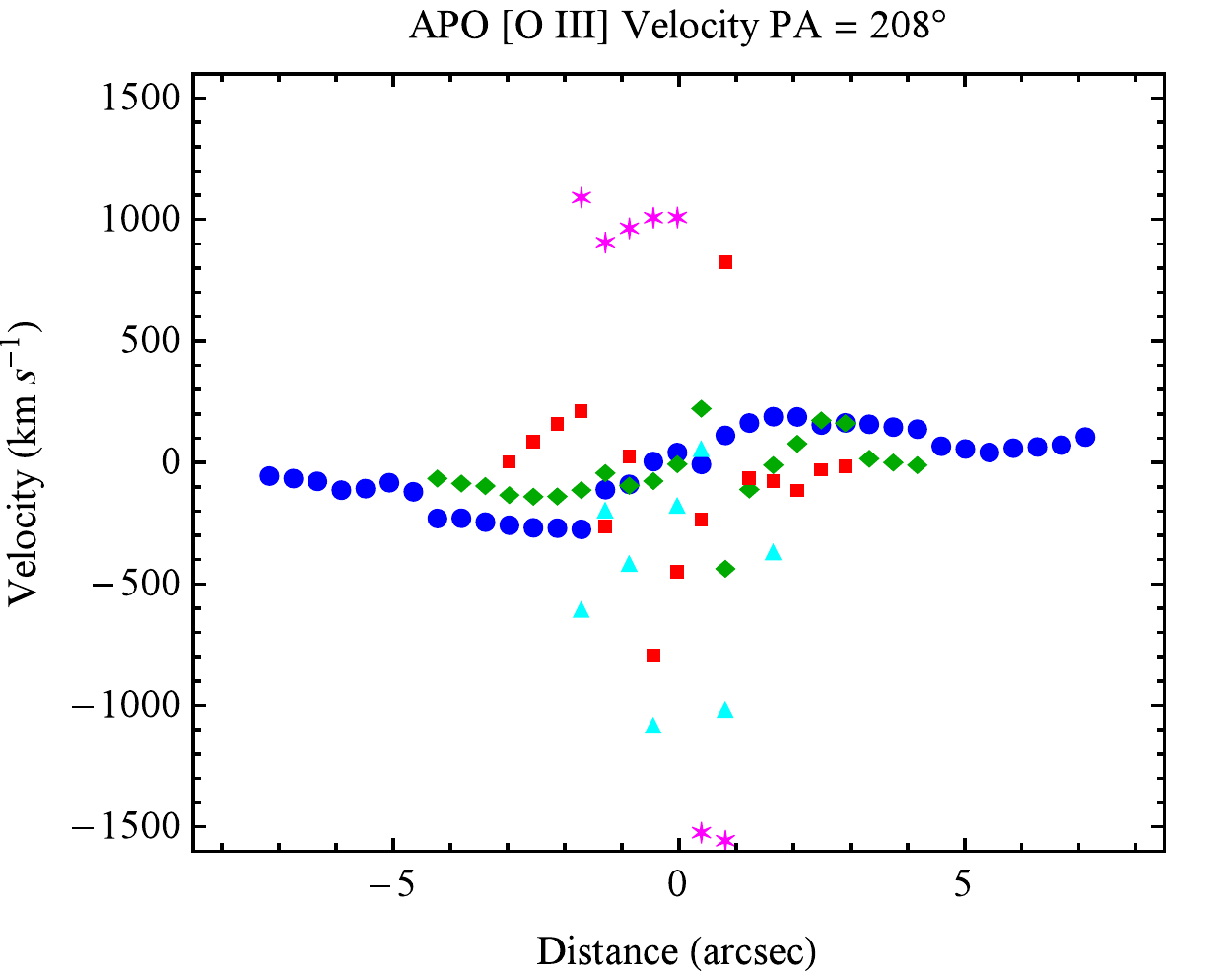}}
\vspace{-6pt}
\subfigure{
\includegraphics[scale=0.34]{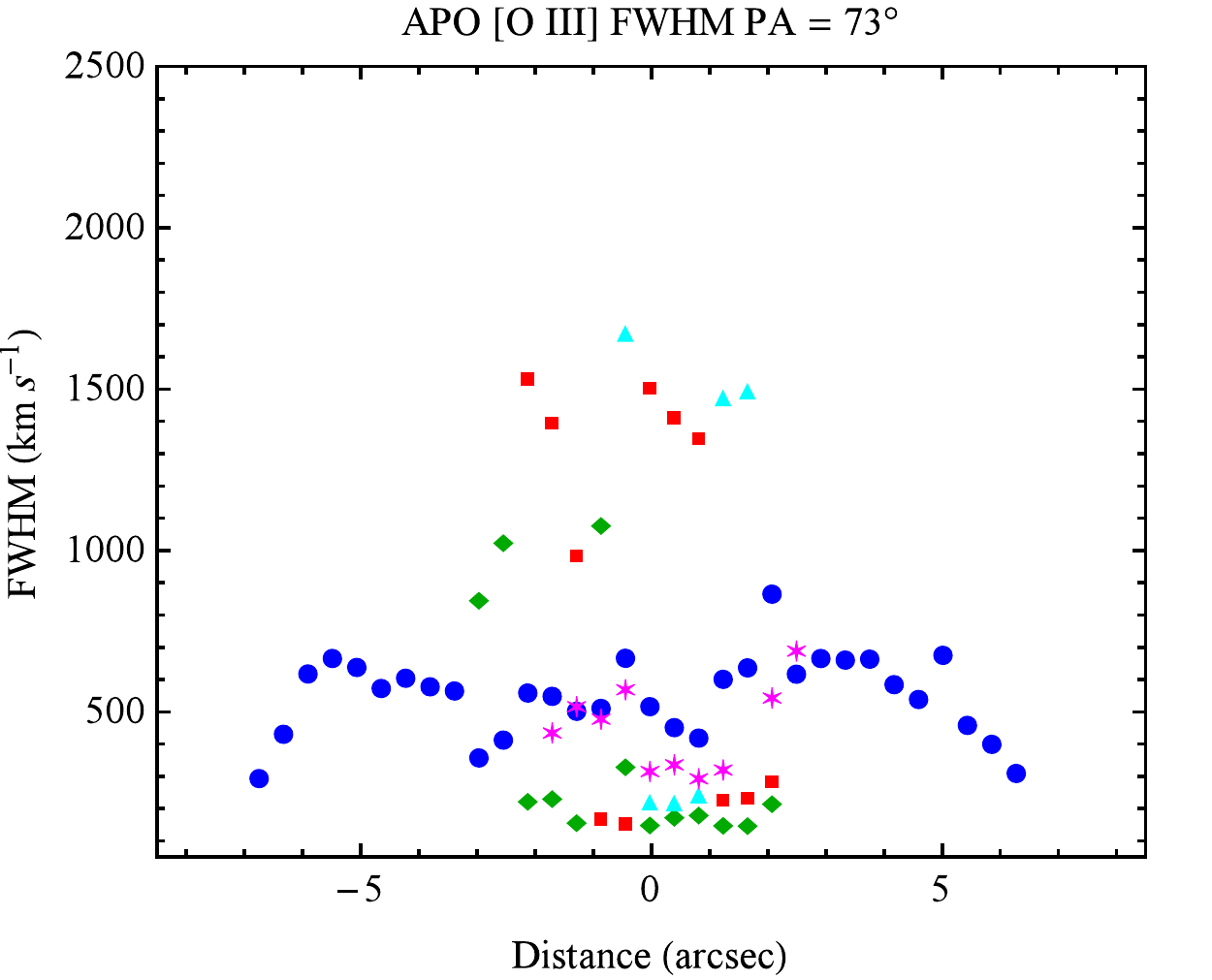}}
\subfigure{
\includegraphics[scale=0.34]{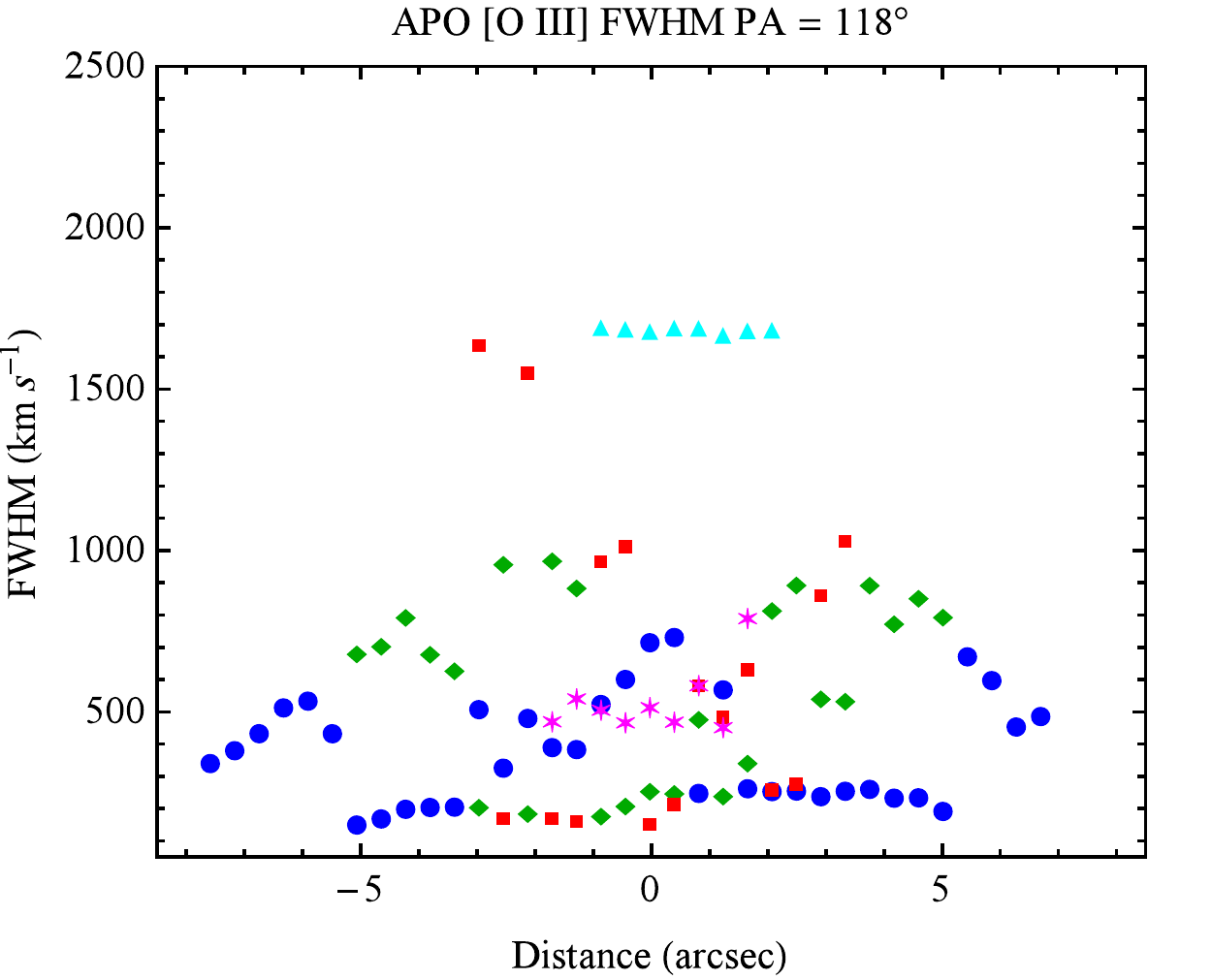}}
\subfigure{
\includegraphics[scale=0.34]{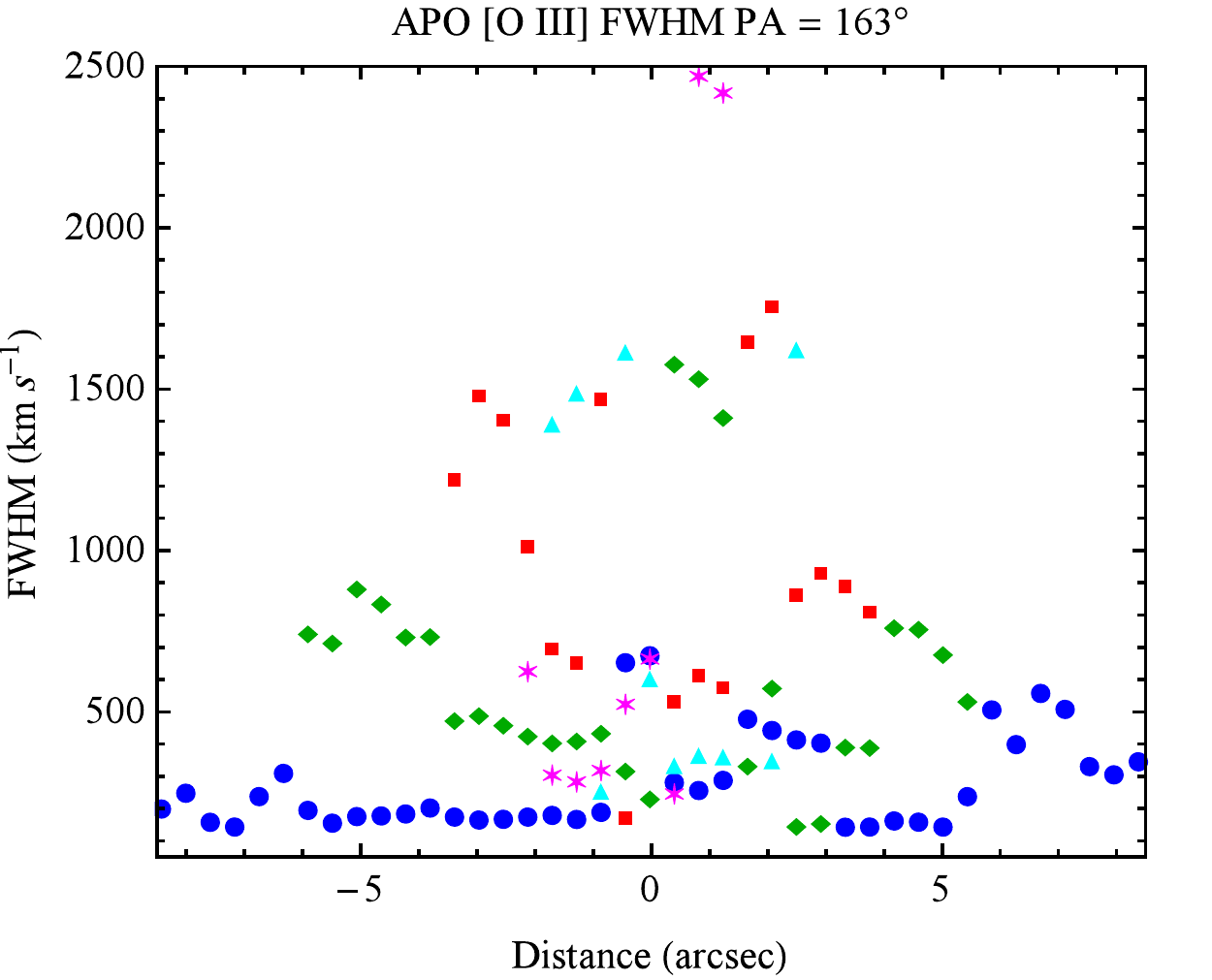}}
\subfigure{
\includegraphics[scale=0.34]{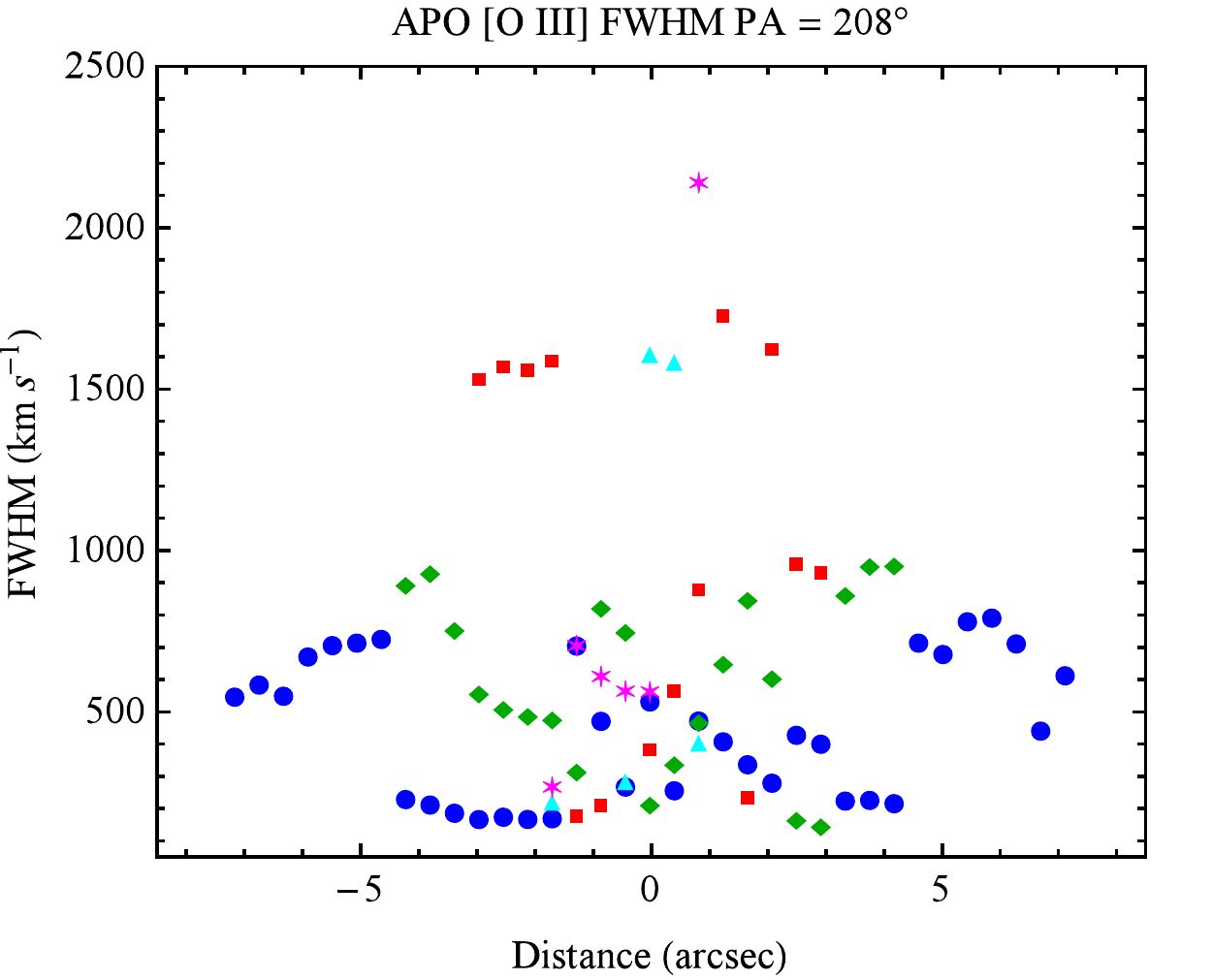}}
\subfigure{
\includegraphics[scale=0.34]{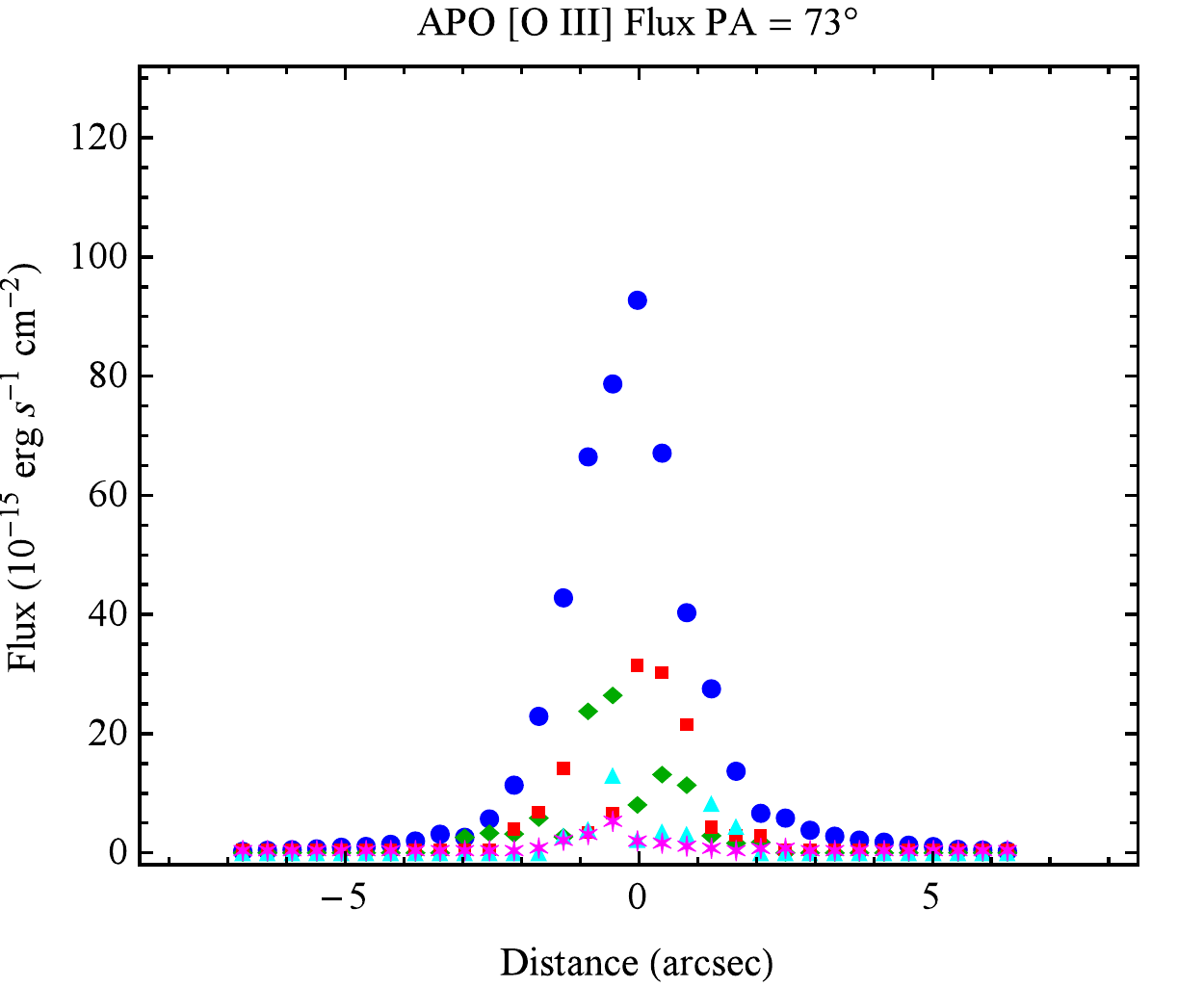}}
\subfigure{
\includegraphics[scale=0.34]{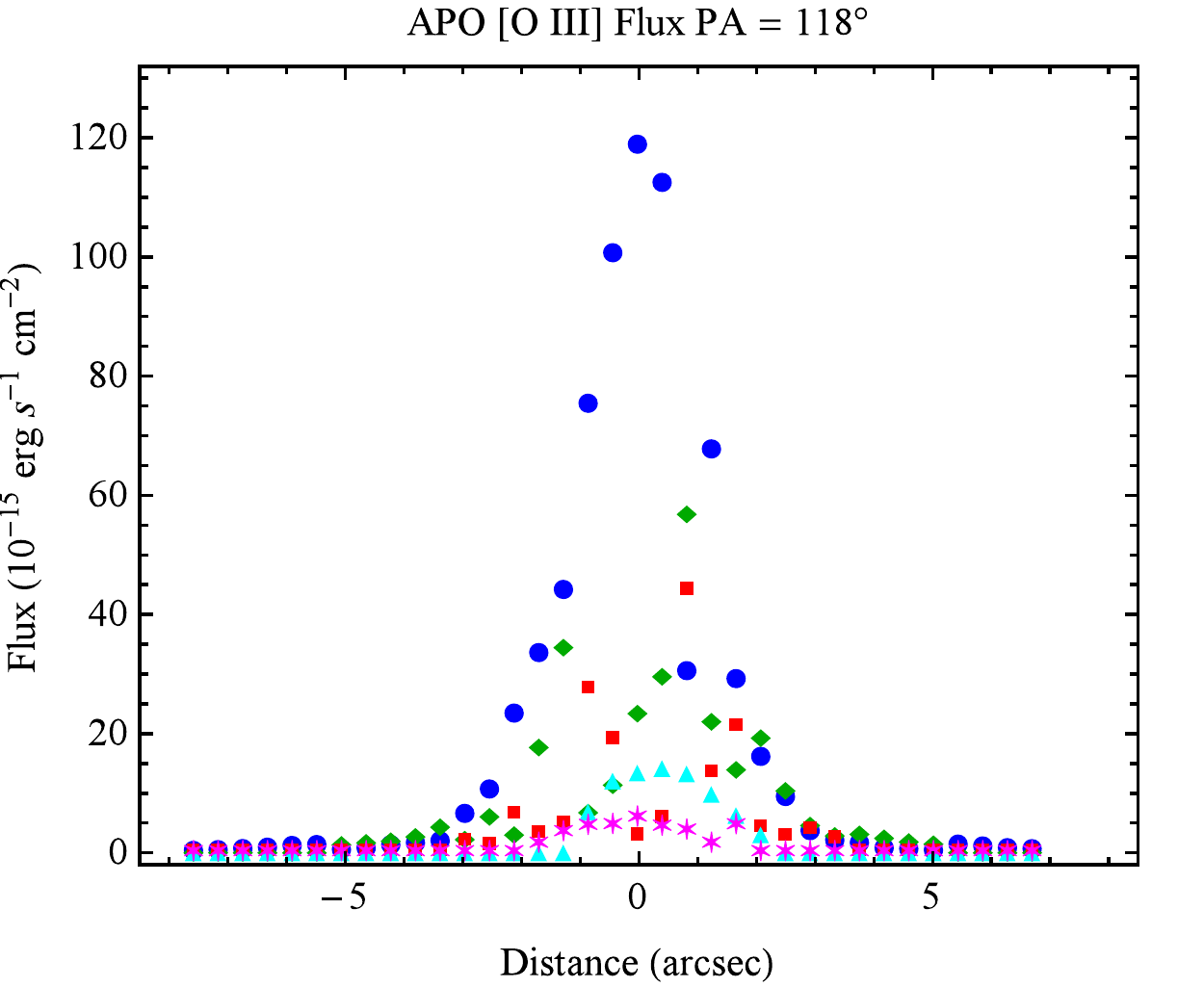}}
\subfigure{
\includegraphics[scale=0.34]{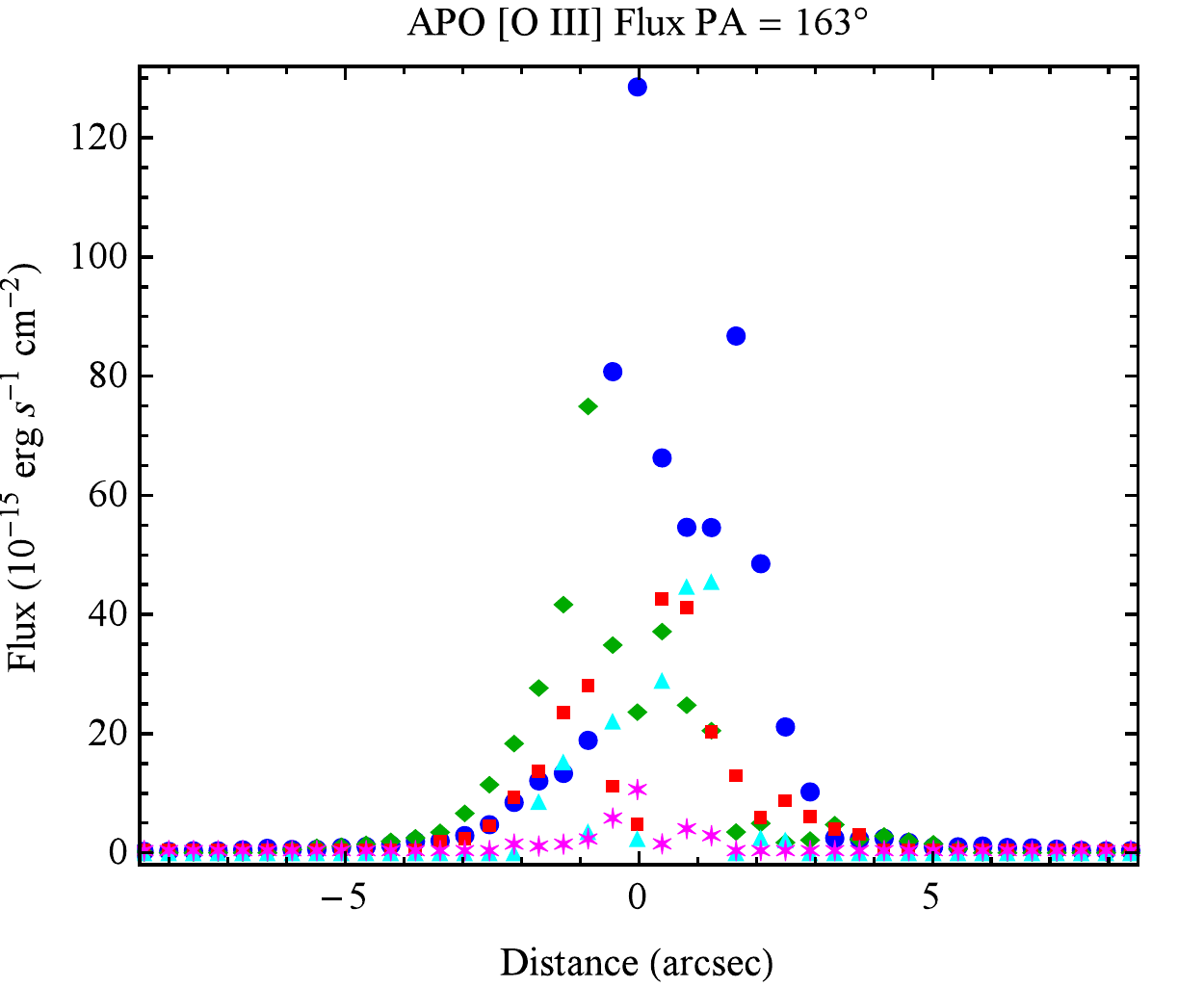}}
\subfigure{
\includegraphics[scale=0.34]{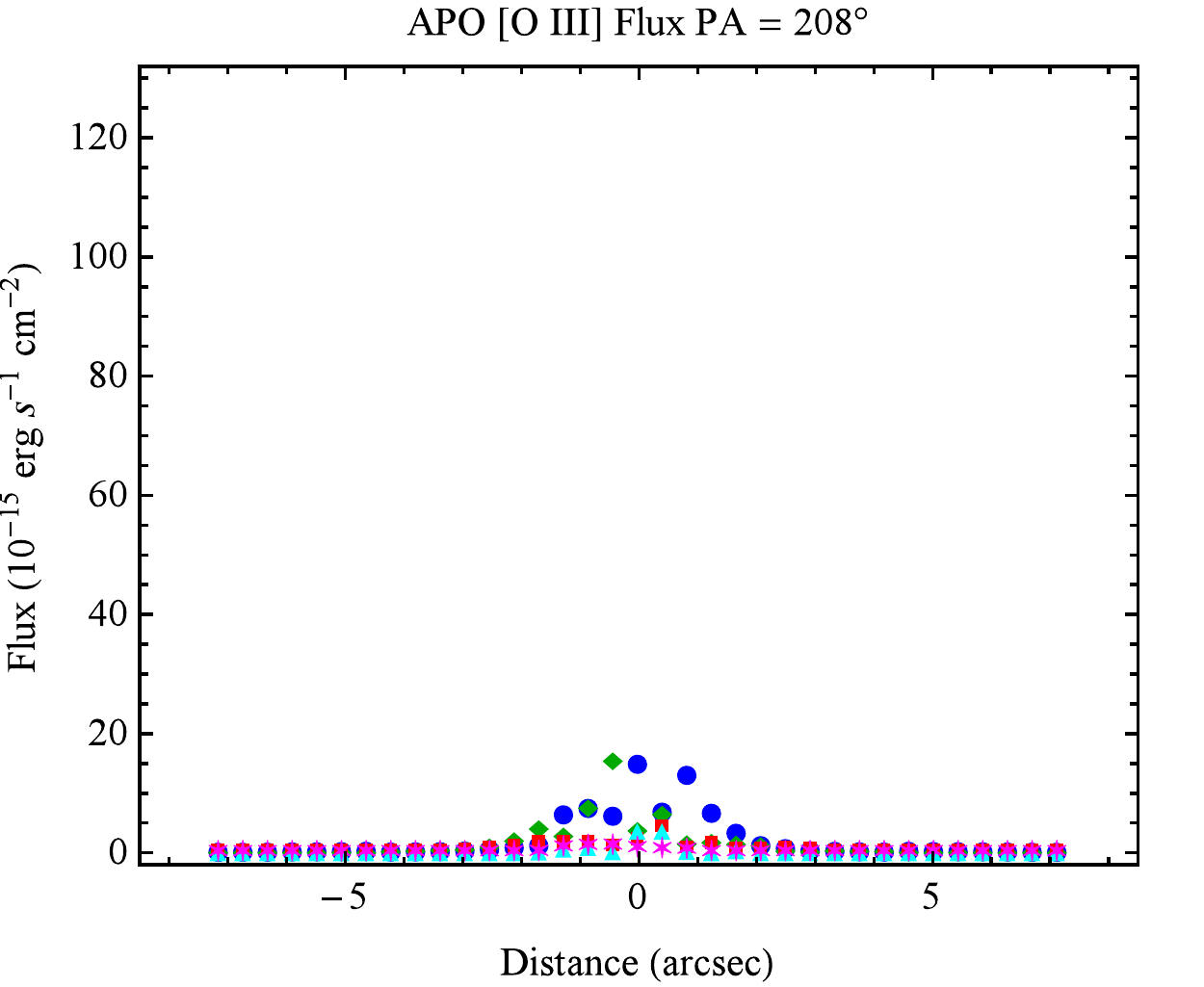}}
\caption{The observed velocities (top), FWHM (middle), and integrated line fluxes (bottom) for the [O~III] $\lambda$5007 emission line in each of the four long-slit APO observations. The points are color-coded from strongest to weakest flux in the order: blue circles, green diamonds, red squares, cyan triangles, and magenta stars. Left to right are position angles 73$\degr$, 118$\degr$, 163$\degr$, and 208$\degr$, East of North.}
\label{kinematics}
\end{figure*}

We also applied our fitting routine to the multiple APO DIS long-slit observations to trace the kinematics and physical conditions of the gas at larger radii. We require up to five Gaussian components to match the observed line profiles near the nucleus, in agreement with the study by \cite{whittle1988}. Assigning a physical meaning to each component must be done cautiously, as a wide, low-flux component may represent a superposition of the non-Gaussian emission line wings, rather than a physical component of gas with a unique velocity \citep{peterson1997}. An example of our multi-component emission line fitting is shown in Figure \ref{fitting}, with additional fits provided in the Appendix. In Figure \ref{fitting} we also show the observed velocities for the strongest emission line component along each of our APO DIS observations in the form of a pseudo-integral field unit (IFU) velocity field, which shows characteristic redshifts and blueshifts indicating that the underlying galactic rotation is traced by at least one emission line component at all radii.

We present the observed velocities, full width at half maxima (FWHM), and integrated line fluxes for all components in our four APO DIS observations in Figure \ref{kinematics}. Due to the $2\arcsec$ wide slit, the inner regions of all four slits sample the same gas and kinematic components. The maximum rotational velocity is seen for PA = 208$\degr$, along the host galaxy major axis. The large velocity amplitude of the strongest component along PA = 163$\degr$, combined with the weak but highly redshifted $\sim$ 1000 km s$^{-1}$ component seen in all slits, suggests that the outflows extend to larger distances than seen in the {\it HST} data.

Using our APO observations, we derive a new heliocentric redshift, recessional velocity, Hubble distance, and spatial scale for Mrk 34. These are based on the centroids of the rotational component of the [O~III] emission line from the central pixel of each PA. The quoted errors are purely instrumental, and the true uncertainty in distance and spatial scale will be dominated by peculiar motion of the galaxy relative to the Hubble flow, which can be up to $\sim$ 600 km s$^{-1}$. For $H_0 = 71$ km s$^{-1}$ Mpc$^{-1}$,
\begin{equation}
z = \left(\frac{\lambda - \lambda_0}{\lambda_0} \right) =  0.05080 \pm 0.00005,
\end{equation}
\begin{equation}
\frac{v}{c} = \frac{(1+z)^2-1}{(1+z)^2+1} = 14843 \pm 21~\mathrm{km/s},
\end{equation}
\begin{equation}
D = \left(\frac{v}{H_0}\right) = 209.1 \pm 0.3~\mathrm{Mpc},~~~~~~~~
\end{equation}
\begin{equation}
R = \left(\frac{D (pc)}{206265}\right) = 1013.5 \pm 1.4~\mathrm{pc/}\arcsec,~
\end{equation}
\noindent
where $\lambda$ is the observed wavelength, and $\lambda_0$ is the rest wavelength of [O~III] $\lambda$5007 (5006.843~$\AA$ in air, 5008.240~$\AA$ in vacuum). This uses the relativistic velocity expression, and the mean Hubble velocity ($v \approx cz$) and distance are 15,229 km s$^{-1}$ and 214.5 Mpc, respectively, consistent with the literature \citep{whittle1988}. Finally, we derive the bolometric luminosity using $L_{\mathrm{bol}} = 3500 \times L_{\mathrm{[O~III]}}$, with a scatter of 0.38 dex \citep{heckman2004}. Using an observed [O~III] luminosity\footnote{Published values span Log($L_{\mathrm{[O~III]}}$) $ \approx 42.39-42.83$ erg s$^{-1}$ (\citealp{reyes2008, heckman2005}, this work). The result is higher but consistent with the relationship of \cite{netzer2009}, which with a reddening correction yields Log(L$_{\mathrm{bol}}$) = 45.8 $\pm$ 0.4 erg s$^{-1}$.} of Log($L_{\mathrm{[O~III]}}$) = 42.64 $\pm$ 0.24 erg s$^{-1}$, we find Log(L$_{\mathrm{bol}}$) = 46.2 $\pm$ 0.4 erg s$^{-1}$.


\setlength{\tabcolsep}{0.015in} 
\tabletypesize{\tiny}
\begin{deluxetable*}{l|c|c|c|c|c|c|c|c|c|c|c|c|c|}
\tablenum{2}
\tablecaption{Observed Emission Line Ratios - Markarian 34 HST STIS Spectrum - Sum of All Components}
\tablewidth{0pt}
\tablehead{
\colhead{Emission Line} & \colhead{--0$\farcs$66} & \colhead{--0$\farcs$56} & \colhead{--0$\farcs$46} & \colhead{--0$\farcs$36} & \colhead{--0$\farcs$25} & \colhead{--0$\farcs$15} & \colhead{--0$\farcs$05} & \colhead{+0$\farcs$05} & \colhead{+0$\farcs$15} & \colhead{+0$\farcs$46} & \colhead{+0$\farcs$56} & \colhead{+0$\farcs$66}}
\startdata
[Ne V] $\lambda$3426	& ... $\pm$ ...	& 0.67 $\pm$ 0.25	& 0.98 $\pm$ 0.43	& 0.80 $\pm$ 0.35	& 1.21 $\pm$ 0.74	& 1.13 $\pm$ 0.50	& 0.70 $\pm$ 0.22	& ... $\pm$ ...	& ... $\pm$ ...	& ... $\pm$ ...	& ... $\pm$ ...	& ... $\pm$ ...	\\ \relax
[O II] $\lambda$3728	& 2.92 $\pm$ 1.20	& 1.96 $\pm$ 0.73	& 2.02 $\pm$ 0.89	& 1.61 $\pm$ 0.71	& 1.19 $\pm$ 0.73	& 1.28 $\pm$ 0.56	& 1.61 $\pm$ 0.51	& 1.30 $\pm$ 0.44	& 1.95 $\pm$ 1.28	& 1.14 $\pm$ 0.89	& 1.34 $\pm$ 0.74	& 1.96 $\pm$ 0.62	\\ \relax
[Ne III] $\lambda$3870	& 1.50 $\pm$ 0.62	& 0.91 $\pm$ 0.34	& 0.96 $\pm$ 0.42	& 0.95 $\pm$ 0.42	& 0.75 $\pm$ 0.46	& 0.80 $\pm$ 0.35	& 0.63 $\pm$ 0.20	& 0.36 $\pm$ 0.12	& 0.68 $\pm$ 0.45	& 0.79 $\pm$ 0.61	& 0.62 $\pm$ 0.34	& 1.29 $\pm$ 0.41	\\ \relax
[Ne III] $\lambda$3968	& 0.63 $\pm$ 0.26	& 0.27 $\pm$ 0.10	& 0.31 $\pm$ 0.13	& 0.39 $\pm$ 0.17	& 0.17 $\pm$ 0.11	& 0.39 $\pm$ 0.17	& 0.34 $\pm$ 0.11	& 0.30 $\pm$ 0.10	& ... $\pm$ ...	& 0.19 $\pm$ 0.15	& 0.18 $\pm$ 0.10	& 0.31 $\pm$ 0.10	\\ \relax
H$\delta$ $\lambda$4102	& ... $\pm$ ...	& 0.16 $\pm$ 0.06	& 0.25 $\pm$ 0.11	& 0.25 $\pm$ 0.11	& ... $\pm$ ...	& ... $\pm$ ...	& 0.18 $\pm$ 0.06	& ... $\pm$ ...	& 0.52 $\pm$ 0.34	& ... $\pm$ ...	& ... $\pm$ ...	& 0.32 $\pm$ 0.10	\\ \relax
H$\gamma$ $\lambda$4341	& 0.31 $\pm$ 0.13	& 0.46 $\pm$ 0.17	& 0.70 $\pm$ 0.31	& 0.29 $\pm$ 0.13	& 0.24 $\pm$ 0.15	& 0.47 $\pm$ 0.20	& 0.20 $\pm$ 0.07	& 0.26 $\pm$ 0.09	& ... $\pm$ ...	& 0.27 $\pm$ 0.21	& 0.22 $\pm$ 0.12	& 0.38 $\pm$ 0.12	\\ \relax
[O III] $\lambda$4364	& 0.27 $\pm$ 0.11	& 0.07 $\pm$ 0.02	& ... $\pm$ ...	& ... $\pm$ ...	& ... $\pm$ ...	& ... $\pm$ ...	& ... $\pm$ ...	& ... $\pm$ ...	& ... $\pm$ ...	& 0.39 $\pm$ 0.31	& ... $\pm$ ...	& ... $\pm$ ...	\\ \relax
He II $\lambda$4687	& 0.35 $\pm$ 0.14	& 0.45 $\pm$ 0.17	& 0.44 $\pm$ 0.19	& 0.26 $\pm$ 0.12	& 0.42 $\pm$ 0.26	& ... $\pm$ ...	& 0.20 $\pm$ 0.06	& ... $\pm$ ...	& 0.52 $\pm$ 0.35	& 0.28 $\pm$ 0.22	& 0.41 $\pm$ 0.23	& 0.36 $\pm$ 0.11	\\ \relax
H$\beta$ $\lambda$4862	& 1.00 $\pm$ 0.29	& 1.00 $\pm$ 0.26	& 1.00 $\pm$ 0.31	& 1.00 $\pm$ 0.31	& 1.00 $\pm$ 0.43	& 1.00 $\pm$ 0.31	& 1.00 $\pm$ 0.23	& 1.00 $\pm$ 0.24	& 1.00 $\pm$ 0.47	& 1.00 $\pm$ 0.55	& 1.00 $\pm$ 0.39	& 1.00 $\pm$ 0.22	\\ \relax
[O III] $\lambda$4960	& 6.96 $\pm$ 2.87	& 5.28 $\pm$ 1.96	& 5.07 $\pm$ 2.23	& 4.42 $\pm$ 1.96	& 4.31 $\pm$ 2.65	& 6.06 $\pm$ 2.65	& 4.44 $\pm$ 1.42	& 3.15 $\pm$ 1.08	& 4.90 $\pm$ 3.23	& 3.53 $\pm$ 2.75	& 4.34 $\pm$ 2.40	& 5.51 $\pm$ 1.74	\\ \relax
[O III] $\lambda$5008	& 20.94 $\pm$ 8.63	& 15.89 $\pm$ 5.91	& 15.26 $\pm$ 6.71	& 13.32 $\pm$ 5.90	& 12.96 $\pm$ 7.97	& 18.24 $\pm$ 7.99	& 13.37 $\pm$ 4.28	& 9.47 $\pm$ 3.24	& 14.74 $\pm$ 9.71	& 10.63 $\pm$ 8.26	& 13.08 $\pm$ 7.23	& 16.58 $\pm$ 5.22	\\ \relax
[O I] $\lambda$6302	& ... $\pm$ ...	& ... $\pm$ ...	& ... $\pm$ ...	& ... $\pm$ ...	& ... $\pm$ ...	& ... $\pm$ ...	& ... $\pm$ ...	& ... $\pm$ ...	& ... $\pm$ ...	& ... $\pm$ ...	& 0.25 $\pm$ 0.14	& ... $\pm$ ...	\\ \relax
[N II] $\lambda$6549	& 0.99 $\pm$ 0.41	& 0.94 $\pm$ 0.35	& 0.75 $\pm$ 0.33	& 1.08 $\pm$ 0.48	& 0.68 $\pm$ 0.42	& 0.69 $\pm$ 0.30	& 0.71 $\pm$ 0.23	& 0.62 $\pm$ 0.21	& 1.57 $\pm$ 1.03	& ... $\pm$ ...	& 0.54 $\pm$ 0.30	& 0.81 $\pm$ 0.26	\\ \relax
H$\alpha$ $\lambda$6564	& 5.28 $\pm$ 2.18	& 3.50 $\pm$ 1.30	& 2.91 $\pm$ 1.28	& 2.36 $\pm$ 1.04	& 3.45 $\pm$ 2.12	& 4.39 $\pm$ 1.92	& 3.98 $\pm$ 1.27	& 3.91 $\pm$ 1.34	& 8.49 $\pm$ 5.59	& 3.10 $\pm$ 2.41	& 2.41 $\pm$ 1.33	& 3.75 $\pm$ 1.18	\\ \relax
[N II] $\lambda$6585	& 2.92 $\pm$ 1.20	& 2.76 $\pm$ 1.03	& 3.07 $\pm$ 1.35	& 3.18 $\pm$ 1.41	& 2.02 $\pm$ 1.24	& 3.39 $\pm$ 1.49	& 2.11 $\pm$ 0.67	& 2.29 $\pm$ 0.78	& 4.63 $\pm$ 3.05	& 2.04 $\pm$ 1.59	& 2.03 $\pm$ 1.12	& 2.40 $\pm$ 0.76	\\ \relax
[S II] $\lambda$6718	& 0.52 $\pm$ 0.22	& 0.65 $\pm$ 0.24	& 0.66 $\pm$ 0.29	& 0.55 $\pm$ 0.25	& 0.66 $\pm$ 0.40	& 0.70 $\pm$ 0.31	& 0.82 $\pm$ 0.26	& 0.77 $\pm$ 0.26	& ... $\pm$ ...	& 0.58 $\pm$ 0.45	& 0.84 $\pm$ 0.47	& 0.41 $\pm$ 0.13	\\ \relax
[S II] $\lambda$6732	& 1.01 $\pm$ 0.42	& 0.94 $\pm$ 0.35	& 0.79 $\pm$ 0.35	& 0.78 $\pm$ 0.35	& 0.90 $\pm$ 0.56	& 1.19 $\pm$ 0.52	& 0.51 $\pm$ 0.16	& 0.66 $\pm$ 0.23	& ... $\pm$ ...	& 0.49 $\pm$ 0.38	& 0.59 $\pm$ 0.33	& 0.53 $\pm$ 0.17	\\ \hline
F(H$\beta$) $\times$10$^{-15}$ & 	0.28 $\pm$ 0.08 & 	0.49 $\pm$ 0.13 & 	0.46 $\pm$ 0.14 & 	0.61 $\pm$ 0.19 & 	0.56 $\pm$ 0.24 & 	0.30 $\pm$ 0.09 & 	0.59 $\pm$ 0.13 & 	0.78 $\pm$ 0.19 & 	0.21 $\pm$ 0.10 & 	0.38 $\pm$ 0.21 & 	0.55 $\pm$ 0.22 & 	0.42 $\pm$ 0.09
\enddata
\tablecomments{{\it HST} STIS observed emission line ratios relative to H$\beta$ with the flux of all kinematic components summed together at each spatial distance from the nucleus in arcseconds. Positive is toward the NW and negative is to the SE. Emission lines were fit using widths and centroids calculated from fits to [O~III] $\lambda$5007 and error bars are the quadrature sum of the fractional flux uncertainty in H$\beta$ and each respective line. Rows marked with ``...	$\pm$	...'' represent nondetections. Wavelengths are approximate vacuum values, and the last row lists the observed H$\beta$ flux in units of $10^{-15}$ erg s$^{-1}$ cm$^{-2}$.}
\end{deluxetable*}

\setlength{\tabcolsep}{0.015in} 
\tabletypesize{\tiny}
\begin{deluxetable*}{l|c|c|c|c|c|c|c|c|c|c|c|c|c|}
\tablenum{3}
\tablecaption{Reddening-Corrected Emission Line Ratios - Markarian 34 HST STIS Spectrum - Sum of All Components}
\tablewidth{0pt}
\tablehead{
\colhead{Emission Line} & \colhead{--0$\farcs$66} & \colhead{--0$\farcs$56} & \colhead{--0$\farcs$46} & \colhead{--0$\farcs$36} & \colhead{--0$\farcs$25} & \colhead{--0$\farcs$15} & \colhead{--0$\farcs$05} & \colhead{+0$\farcs$05} & \colhead{+0$\farcs$15} & \colhead{+0$\farcs$46} & \colhead{+0$\farcs$56} & \colhead{+0$\farcs$66}}
\startdata
[Ne V] $\lambda$3426	& ... $\pm$ ...	& 0.82 $\pm$ 0.43	& 0.98 $\pm$ 0.67	& 0.80 $\pm$ 0.55	& 1.46 $\pm$ 1.42	& 1.75 $\pm$ 1.03	& 0.98 $\pm$ 0.41	& ... $\pm$ ...	& ... $\pm$ ...	& ... $\pm$ ...	& ... $\pm$ ...	& ... $\pm$ ...	\\ \relax
[O II] $\lambda$3728	& 4.67 $\pm$ 2.03	& 2.27 $\pm$ 1.01	& 2.03 $\pm$ 1.17	& 1.61 $\pm$ 0.94	& 1.37 $\pm$ 1.09	& 1.77 $\pm$ 0.87	& 2.06 $\pm$ 0.74	& 1.64 $\pm$ 0.64	& 4.51 $\pm$ 3.23	& 1.20 $\pm$ 1.40	& 1.34 $\pm$ 1.01	& 2.40 $\pm$ 0.88	\\ \relax
[Ne III] $\lambda$3870	& 2.26 $\pm$ 0.92	& 1.04 $\pm$ 0.44	& 0.96 $\pm$ 0.53	& 0.95 $\pm$ 0.52	& 0.85 $\pm$ 0.63	& 1.06 $\pm$ 0.49	& 0.78 $\pm$ 0.27	& 0.45 $\pm$ 0.17	& 1.42 $\pm$ 0.92	& 0.83 $\pm$ 0.89	& 0.62 $\pm$ 0.43	& 1.54 $\pm$ 0.53	\\ \relax
[Ne III] $\lambda$3968	& 0.91 $\pm$ 0.36	& 0.31 $\pm$ 0.13	& 0.31 $\pm$ 0.16	& 0.39 $\pm$ 0.21	& 0.19 $\pm$ 0.14	& 0.51 $\pm$ 0.23	& 0.41 $\pm$ 0.14	& 0.37 $\pm$ 0.13	& ... $\pm$ ...	& 0.20 $\pm$ 0.20	& 0.18 $\pm$ 0.12	& 0.37 $\pm$ 0.12	\\ \relax
H$\delta$ $\lambda$4102	& ... $\pm$ ...	& 0.17 $\pm$ 0.07	& 0.25 $\pm$ 0.13	& 0.25 $\pm$ 0.13	& ... $\pm$ ...	& ... $\pm$ ...	& 0.22 $\pm$ 0.07	& ... $\pm$ ...	& 0.92 $\pm$ 0.53	& ... $\pm$ ...	& ... $\pm$ ...	& 0.37 $\pm$ 0.12	\\ \relax
H$\gamma$ $\lambda$4341	& 0.39 $\pm$ 0.14	& 0.50 $\pm$ 0.19	& 0.70 $\pm$ 0.34	& 0.29 $\pm$ 0.14	& 0.25 $\pm$ 0.16	& 0.55 $\pm$ 0.23	& 0.23 $\pm$ 0.07	& 0.29 $\pm$ 0.10	& ... $\pm$ ...	& 0.28 $\pm$ 0.24	& 0.22 $\pm$ 0.13	& 0.42 $\pm$ 0.13	\\ \relax
[O III] $\lambda$4364	& 0.34 $\pm$ 0.12	& 0.07 $\pm$ 0.03	& ... $\pm$ ...	& ... $\pm$ ...	& ... $\pm$ ...	& ... $\pm$ ...	& ... $\pm$ ...	& ... $\pm$ ...	& ... $\pm$ ...	& 0.40 $\pm$ 0.35	& ... $\pm$ ...	& ... $\pm$ ...	\\ \relax
He II $\lambda$4687	& 0.38 $\pm$ 0.15	& 0.46 $\pm$ 0.17	& 0.44 $\pm$ 0.19	& 0.26 $\pm$ 0.12	& 0.43 $\pm$ 0.26	& ... $\pm$ ...	& 0.21 $\pm$ 0.07	& ... $\pm$ ...	& 0.60 $\pm$ 0.35	& 0.29 $\pm$ 0.22	& 0.41 $\pm$ 0.23	& 0.37 $\pm$ 0.11	\\ \relax
H$\beta$ $\lambda$4862	& 1.00 $\pm$ 0.29	& 1.00 $\pm$ 0.26	& 1.00 $\pm$ 0.31	& 1.00 $\pm$ 0.31	& 1.00 $\pm$ 0.43	& 1.00 $\pm$ 0.31	& 1.00 $\pm$ 0.23	& 1.00 $\pm$ 0.24	& 1.00 $\pm$ 0.47	& 1.00 $\pm$ 0.55	& 1.00 $\pm$ 0.39	& 1.00 $\pm$ 0.22	\\ \relax
[O III] $\lambda$4960	& 6.68 $\pm$ 2.87	& 5.21 $\pm$ 1.97	& 5.07 $\pm$ 2.23	& 4.42 $\pm$ 1.97	& 4.26 $\pm$ 2.65	& 5.90 $\pm$ 2.66	& 4.35 $\pm$ 1.42	& 3.08 $\pm$ 1.08	& 4.56 $\pm$ 3.23	& 3.52 $\pm$ 2.76	& 4.34 $\pm$ 2.41	& 5.41 $\pm$ 1.74	\\ \relax
[O III] $\lambda$5008	& 19.71 $\pm$ 8.67	& 15.59 $\pm$ 5.94	& 15.26 $\pm$ 6.75	& 13.32 $\pm$ 5.93	& 12.74 $\pm$ 8.02	& 17.50 $\pm$ 8.03	& 12.95 $\pm$ 4.30	& 9.19 $\pm$ 3.26	& 13.23 $\pm$ 9.77	& 10.56 $\pm$ 8.34	& 13.08 $\pm$ 7.27	& 16.15 $\pm$ 5.25	\\ \relax
[O I] $\lambda$6302	& ... $\pm$ ...	& ... $\pm$ ...	& ... $\pm$ ...	& ... $\pm$ ...	& ... $\pm$ ...	& ... $\pm$ ...	& ... $\pm$ ...	& ... $\pm$ ...	& ... $\pm$ ...	& ... $\pm$ ...	& 0.25 $\pm$ 0.19	& ... $\pm$ ...	\\ \relax
[N II] $\lambda$6549	& 0.54 $\pm$ 0.47	& 0.78 $\pm$ 0.47	& 0.75 $\pm$ 0.49	& 1.08 $\pm$ 0.71	& 0.58 $\pm$ 0.61	& 0.46 $\pm$ 0.37	& 0.52 $\pm$ 0.29	& 0.46 $\pm$ 0.27	& 0.54 $\pm$ 1.13	& ... $\pm$ ...	& 0.54 $\pm$ 0.46	& 0.63 $\pm$ 0.33	\\ \relax
H$\alpha$ $\lambda$6564	& 2.90 $\pm$ 2.54	& 2.90 $\pm$ 1.75	& 2.90 $\pm$ 1.91	& 2.36 $\pm$ 1.56	& 2.90 $\pm$ 3.10	& 2.90 $\pm$ 2.38	& 2.90 $\pm$ 1.61	& 2.90 $\pm$ 1.71	& 2.91 $\pm$ 6.14	& 2.90 $\pm$ 4.31	& 2.41 $\pm$ 2.08	& 2.90 $\pm$ 1.52	\\ \relax
[N II] $\lambda$6585	& 1.59 $\pm$ 1.40	& 2.28 $\pm$ 1.38	& 3.06 $\pm$ 2.03	& 3.18 $\pm$ 2.12	& 1.69 $\pm$ 1.82	& 2.24 $\pm$ 1.85	& 1.53 $\pm$ 0.85	& 1.69 $\pm$ 1.00	& 1.57 $\pm$ 3.35	& 1.91 $\pm$ 2.86	& 2.03 $\pm$ 1.76	& 1.85 $\pm$ 0.98	\\ \relax
[S II] $\lambda$6718	& 0.28 $\pm$ 0.25	& 0.53 $\pm$ 0.33	& 0.65 $\pm$ 0.45	& 0.55 $\pm$ 0.38	& 0.55 $\pm$ 0.61	& 0.45 $\pm$ 0.39	& 0.58 $\pm$ 0.34	& 0.56 $\pm$ 0.34	& ... $\pm$ ...	& 0.54 $\pm$ 0.86	& 0.84 $\pm$ 0.76	& 0.31 $\pm$ 0.17	\\ \relax
[S II] $\lambda$6732	& 0.53 $\pm$ 0.49	& 0.77 $\pm$ 0.48	& 0.78 $\pm$ 0.54	& 0.78 $\pm$ 0.54	& 0.75 $\pm$ 0.84	& 0.76 $\pm$ 0.66	& 0.37 $\pm$ 0.21	& 0.48 $\pm$ 0.29	& ... $\pm$ ...	& 0.46 $\pm$ 0.73	& 0.59 $\pm$ 0.54	& 0.40 $\pm$ 0.22	\\ \hline
E(B-V) & 	0.55 $\pm$ 0.40 & 	0.17 $\pm$ 0.36 & 	0.00 $\pm$ 0.43 & 	0.00 $\pm$ 0.43 & 	0.16 $\pm$ 0.65 & 	0.38 $\pm$ 0.43 & 	0.29 $\pm$ 0.30 & 	0.27 $\pm$ 0.33 & 	0.98 $\pm$ 0.72 & 	0.06 $\pm$ 0.95 & 	0.00 $\pm$ 0.57 & 	0.24 $\pm$ 0.30
\enddata
\tablecomments{Same as in Table 2, but with line ratios corrected for galactic extinction using a galactic reddening curve \citep{savage1979}. The $H\alpha / H\beta$ ratios were fixed at 2.90 and negative E(B-V) values were set to zero. Error bars are the quadrature sum of the fractional flux uncertainty in H$\beta$ and each respective line along with the reddening uncertainty. The last row lists the color excess E(B-V).}
\end{deluxetable*}

\subsection{Emission Line Ratios}

We used our Gaussian fit parameters to calculate integrated emission line fluxes and their ratios relative to H$\beta$. We determined line ratios with the fluxes of all kinematic components added together for the highest possible S/N, as well as for each individual component to probe differences in the physical conditions between rotational and outflowing gas. The observed and reddening-corrected line ratios for the sum of all components are given in Tables 2 and 3 for the {\it HST} data, and in Tables 4 and 5 for the APO data, respectively. The observed emission line ratios for the individual components with a S/N $>$~2 are given in the Appendix. The procedure for reddening correction is described in \S3.3 of Paper I.

There are a maximum of two components for the {\it HST} data, sorted into rotation and outflow based on velocity, and up to five components for the APO data, sorted from highest to lowest peak flux. The larger number of components in the APO observations can be attributed to the wider slit that encompasses additional emission line knots, and spatial blending across adjacent pixels due to atmospheric smearing. The weakest components likely encompass the non-Gaussian wings of the combined profile, and we caution against a physical interpretation for these ratios. Only emission within $\lesssim 2\arcsec$ of the nucleus displays outflow kinematics and are included in our tables. All measurements are available by request to M.R.

The {\it HST} data have detectable emission over a limited spatial extent with modest S/N and large uncertainties due to the relatively short exposure times of the G430L/G750L observations, allowing us to place some constraints on the physical conditions in the gas at very high spatial resolution. The APO observations yield emission line ratios over a larger spatial extent with higher S/N and smaller uncertainties, but at lower spatial resolution, allowing us to probe conditions on large scales.

To first order, our scale factor and the resulting ionized gas masses depend solely on the gas density and the ratio of [O~III]/H$\beta$, and the latter does not vary by more than a factor of $\sim 2$ between components at each location. The similarity between the [O~III]/H$\beta$ ratios in the {\it HST} and APO observations suggests that both sample gas with similar physical conditions and ionization states. The mildly larger [O~III]/H$\beta$ ratios in the {\it HST} observations are likely due to the noisy continuum regions surrounding H$\beta$ and/or isolating individual knots of emission. In addition, the consistency between the [O~III]/H$\beta$ ratios in all five components of the APO data suggests that modeling the combined emission line ratios yields a sufficiently accurate representation of the conditions in the photoionized gas. 

In Paper I the {\it HST} spectra had sufficient S/N for the key diagnostic emission lines, and we did not model the APO data, as they offered similar insights at lower spatial resolution. For Mrk 34, the larger uncertainties of the {\it HST} observations would yield models with limited constraints on the conditions in the gas; however, the similarity of the ratios between the {\it HST} and APO observations means we can be confident that photoionization models of the APO data will yield an equivalent scale factor for converting [O~III] flux to mass, while the higher spatial resolution {\it HST} observations help to constrain the density profile using the [S~II] doublet.


\setlength{\tabcolsep}{0.034in} 
\tabletypesize{\tiny}
\begin{deluxetable*}{l|c|c|c|c|c|c|c|c|c|c|c|c|c|c|c|c|c|c|}
\tablenum{4}
\tablecaption{Observed Emission Line Ratios - Markarian 34 APO DIS Spectrum - Sum of All Components}
\tablehead{
\colhead{Emission Line} & \colhead{--2$\farcs$0} & \colhead{--1$\farcs$6} & \colhead{--1$\farcs$2} & \colhead{--0$\farcs$8} & \colhead{--0$\farcs$4} & \colhead{0$\farcs$0} & \colhead{+0$\farcs$4} & \colhead{+0$\farcs$8} & \colhead{+1$\farcs$2} & \colhead{+1$\farcs$6}  & \colhead{+2$\farcs$0}}
\startdata
[S II] $\lambda$4072	& 0.17 $\pm$ 0.03	& 0.14 $\pm$ 0.02	& 0.12 $\pm$ 0.01	& 0.10 $\pm$ 0.01	& 0.08 $\pm$ 0.01	& 0.07 $\pm$ 0.01	& 0.08 $\pm$ 0.01	& 0.08 $\pm$ 0.01	& 0.10 $\pm$ 0.01	& 0.05 $\pm$ 0.01	& ... $\pm$ ...	\\ \relax
H$\delta$ $\lambda$4101	& 0.23 $\pm$ 0.04	& 0.18 $\pm$ 0.03	& 0.19 $\pm$ 0.02	& 0.16 $\pm$ 0.02	& 0.16 $\pm$ 0.02	& 0.14 $\pm$ 0.02	& 0.14 $\pm$ 0.01	& 0.14 $\pm$ 0.02	& 0.14 $\pm$ 0.02	& 0.11 $\pm$ 0.01	& ... $\pm$ ...	\\ \relax
H$\gamma$ $\lambda$4340	& 0.47 $\pm$ 0.08	& 0.44 $\pm$ 0.06	& 0.40 $\pm$ 0.04	& 0.41 $\pm$ 0.04	& 0.40 $\pm$ 0.04	& 0.30 $\pm$ 0.03	& 0.38 $\pm$ 0.04	& 0.39 $\pm$ 0.04	& 0.39 $\pm$ 0.05	& 0.39 $\pm$ 0.05	& 0.38 $\pm$ 0.07	\\ \relax
[O III] $\lambda$4363	& 0.15 $\pm$ 0.03	& 0.13 $\pm$ 0.02	& 0.14 $\pm$ 0.01	& 0.15 $\pm$ 0.02	& 0.15 $\pm$ 0.02	& 0.21 $\pm$ 0.02	& 0.14 $\pm$ 0.01	& 0.15 $\pm$ 0.02	& 0.17 $\pm$ 0.02	& 0.18 $\pm$ 0.03	& 0.20 $\pm$ 0.04	\\ \relax
He II $\lambda$4685	& 0.24 $\pm$ 0.04	& 0.23 $\pm$ 0.03	& 0.26 $\pm$ 0.03	& 0.25 $\pm$ 0.03	& 0.23 $\pm$ 0.02	& 0.22 $\pm$ 0.02	& 0.23 $\pm$ 0.02	& 0.23 $\pm$ 0.03	& 0.23 $\pm$ 0.03	& 0.24 $\pm$ 0.03	& 0.26 $\pm$ 0.05	\\ \relax
H$\beta$ $\lambda$4861	& 1.00 $\pm$ 0.13	& 1.00 $\pm$ 0.10	& 1.00 $\pm$ 0.07	& 1.00 $\pm$ 0.07	& 1.00 $\pm$ 0.08	& 1.00 $\pm$ 0.08	& 1.00 $\pm$ 0.07	& 1.00 $\pm$ 0.08	& 1.00 $\pm$ 0.08	& 1.00 $\pm$ 0.10	& 1.00 $\pm$ 0.13	\\ \relax
[O III] $\lambda$4958	& 4.01 $\pm$ 0.73	& 3.95 $\pm$ 0.57	& 3.93 $\pm$ 0.38	& 3.92 $\pm$ 0.41	& 3.80 $\pm$ 0.41	& 3.65 $\pm$ 0.41	& 3.66 $\pm$ 0.34	& 3.58 $\pm$ 0.39	& 3.51 $\pm$ 0.41	& 3.47 $\pm$ 0.49	& 3.55 $\pm$ 0.67	\\ \relax
[O III] $\lambda$5006	& 12.08 $\pm$ 2.20	& 11.88 $\pm$ 1.72	& 11.82 $\pm$ 1.13	& 11.81 $\pm$ 1.24	& 11.44 $\pm$ 1.23	& 10.99 $\pm$ 1.23	& 11.00 $\pm$ 1.03	& 10.77 $\pm$ 1.18	& 10.57 $\pm$ 1.25	& 10.44 $\pm$ 1.47	& 10.68 $\pm$ 2.02	\\ \relax
[N I] $\lambda$5199	& ... $\pm$ ...	& ... $\pm$ ...	& ... $\pm$ ...	& 0.08 $\pm$ 0.01	& 0.11 $\pm$ 0.01	& ... $\pm$ ...	& ... $\pm$ ...	& ... $\pm$ ...	& ... $\pm$ ...	& ... $\pm$ ...	& ... $\pm$ ...	\\ \relax
He I $\lambda$5875	& 0.27 $\pm$ 0.05	& 0.18 $\pm$ 0.03	& 0.17 $\pm$ 0.02	& 0.14 $\pm$ 0.01	& 0.13 $\pm$ 0.01	& 0.13 $\pm$ 0.01	& 0.11 $\pm$ 0.01	& 0.10 $\pm$ 0.01	& 0.08 $\pm$ 0.01	& 0.11 $\pm$ 0.01	& 0.11 $\pm$ 0.02	\\ \relax
[Fe VII] $\lambda$6086	& ... $\pm$ ...	& 0.09 $\pm$ 0.01	& 0.12 $\pm$ 0.01	& 0.13 $\pm$ 0.01	& 0.10 $\pm$ 0.01	& 0.10 $\pm$ 0.01	& 0.09 $\pm$ 0.01	& 0.08 $\pm$ 0.01	& 0.06 $\pm$ 0.01	& ... $\pm$ ...	& ... $\pm$ ...	\\ \relax
[O I] $\lambda$6300	& 0.60 $\pm$ 0.11	& 0.56 $\pm$ 0.08	& 0.56 $\pm$ 0.05	& 0.54 $\pm$ 0.06	& 0.54 $\pm$ 0.06	& 0.51 $\pm$ 0.06	& 0.54 $\pm$ 0.05	& 0.56 $\pm$ 0.06	& 0.62 $\pm$ 0.07	& 0.64 $\pm$ 0.09	& 0.68 $\pm$ 0.13	\\ \relax
[O I] $\lambda$6363	& 0.20 $\pm$ 0.04	& 0.19 $\pm$ 0.03	& 0.19 $\pm$ 0.02	& 0.18 $\pm$ 0.02	& 0.18 $\pm$ 0.02	& 0.17 $\pm$ 0.02	& 0.18 $\pm$ 0.02	& 0.19 $\pm$ 0.02	& 0.21 $\pm$ 0.02	& 0.21 $\pm$ 0.03	& 0.23 $\pm$ 0.04	\\ \relax
[N II] $\lambda$6548	& 1.77 $\pm$ 0.32	& 1.39 $\pm$ 0.20	& 1.23 $\pm$ 0.12	& 1.14 $\pm$ 0.12	& 1.09 $\pm$ 0.12	& 0.88 $\pm$ 0.10	& 0.89 $\pm$ 0.08	& 0.94 $\pm$ 0.10	& 0.89 $\pm$ 0.10	& 0.90 $\pm$ 0.13	& 1.00 $\pm$ 0.19	\\ \relax
H$\alpha$ $\lambda$6562	& 5.24 $\pm$ 0.95	& 4.78 $\pm$ 0.69	& 4.03 $\pm$ 0.38	& 3.60 $\pm$ 0.38	& 3.36 $\pm$ 0.36	& 3.78 $\pm$ 0.42	& 3.80 $\pm$ 0.36	& 3.54 $\pm$ 0.39	& 3.72 $\pm$ 0.44	& 3.95 $\pm$ 0.56	& 3.98 $\pm$ 0.75	\\ \relax
[N II] $\lambda$6583	& 5.21 $\pm$ 0.95	& 4.10 $\pm$ 0.59	& 3.63 $\pm$ 0.35	& 3.37 $\pm$ 0.36	& 3.21 $\pm$ 0.34	& 2.58 $\pm$ 0.29	& 2.62 $\pm$ 0.25	& 2.79 $\pm$ 0.31	& 2.61 $\pm$ 0.31	& 2.66 $\pm$ 0.37	& 2.96 $\pm$ 0.56	\\ \relax
[S II] $\lambda$6716	& 2.49 $\pm$ 0.45	& 2.20 $\pm$ 0.32	& 1.26 $\pm$ 0.12	& 1.28 $\pm$ 0.13	& 1.13 $\pm$ 0.12	& 1.18 $\pm$ 0.13	& 1.18 $\pm$ 0.11	& 1.26 $\pm$ 0.14	& 1.14 $\pm$ 0.13	& 1.09 $\pm$ 0.15	& 1.22 $\pm$ 0.23	\\ \relax
[S II] $\lambda$6730	& 3.04 $\pm$ 0.55	& 1.90 $\pm$ 0.28	& 2.14 $\pm$ 0.20	& 1.78 $\pm$ 0.19	& 1.64 $\pm$ 0.18	& 1.29 $\pm$ 0.14	& 1.12 $\pm$ 0.11	& 1.04 $\pm$ 0.11	& 1.00 $\pm$ 0.12	& 1.01 $\pm$ 0.14	& 1.05 $\pm$ 0.20	\\ \hline
F(H$\beta$) $\times$10$^{-15}$ & 	5.09 $\pm$ 0.65 & 	8.65 $\pm$ 0.89 & 	12.10 $\pm$ 0.82 & 	14.27 $\pm$ 1.06 & 	15.36 $\pm$ 1.16 & 	15.41 $\pm$ 1.22 & 	13.99 $\pm$ 0.93 & 	11.78 $\pm$ 0.91 & 	8.93 $\pm$ 0.74 & 	5.96 $\pm$ 0.59 & 	3.44 $\pm$ 0.46
\enddata
\tablecomments{The same as Table 2 for the APO DIS observed emission line ratios along PA = 163$\degr$ with the flux of all kinematic components summed together. Wavelengths are approximate air values, and the last row lists the observed H$\beta$ flux in units of $10^{-15}$ erg s$^{-1}$ cm$^{-2}$.}
\end{deluxetable*}

\setlength{\tabcolsep}{0.034in} 
\tabletypesize{\tiny}
\begin{deluxetable*}{l|c|c|c|c|c|c|c|c|c|c|c|c|c|c|c|c|c|c|}
\tablenum{5}
\tablecaption{Reddening-Corrected Emission Line Ratios - Markarian 34 APO DIS Spectrum - Sum of All Components}
\tablehead{
\colhead{Emission Line} & \colhead{--2$\farcs$0} & \colhead{--1$\farcs$6} & \colhead{--1$\farcs$2} & \colhead{--0$\farcs$8} & \colhead{--0$\farcs$4} & \colhead{0$\farcs$0} & \colhead{+0$\farcs$4} & \colhead{+0$\farcs$8} & \colhead{+1$\farcs$2} & \colhead{+1$\farcs$6}  & \colhead{+2$\farcs$0}}
\startdata
[S II] $\lambda$4072	& 0.23 $\pm$ 0.04	& 0.18 $\pm$ 0.02	& 0.14 $\pm$ 0.01	& 0.11 $\pm$ 0.01	& 0.09 $\pm$ 0.01	& 0.08 $\pm$ 0.01	& 0.09 $\pm$ 0.01	& 0.09 $\pm$ 0.01	& 0.11 $\pm$ 0.01	& 0.06 $\pm$ 0.01	& ... $\pm$ ...	\\ \relax
H$\delta$ $\lambda$4101	& 0.31 $\pm$ 0.05	& 0.24 $\pm$ 0.03	& 0.23 $\pm$ 0.02	& 0.17 $\pm$ 0.02	& 0.17 $\pm$ 0.02	& 0.16 $\pm$ 0.02	& 0.16 $\pm$ 0.01	& 0.16 $\pm$ 0.02	& 0.16 $\pm$ 0.02	& 0.13 $\pm$ 0.02	& ... $\pm$ ...	\\ \relax
H$\gamma$ $\lambda$4340	& 0.59 $\pm$ 0.09	& 0.54 $\pm$ 0.07	& 0.46 $\pm$ 0.04	& 0.45 $\pm$ 0.05	& 0.42 $\pm$ 0.05	& 0.34 $\pm$ 0.04	& 0.43 $\pm$ 0.04	& 0.42 $\pm$ 0.05	& 0.43 $\pm$ 0.05	& 0.44 $\pm$ 0.06	& 0.44 $\pm$ 0.08	\\ \relax
[O III] $\lambda$4363	& 0.18 $\pm$ 0.03	& 0.16 $\pm$ 0.02	& 0.16 $\pm$ 0.01	& 0.16 $\pm$ 0.02	& 0.16 $\pm$ 0.02	& 0.23 $\pm$ 0.03	& 0.16 $\pm$ 0.01	& 0.17 $\pm$ 0.02	& 0.19 $\pm$ 0.02	& 0.21 $\pm$ 0.03	& 0.23 $\pm$ 0.04	\\ \relax
He II $\lambda$4685	& 0.26 $\pm$ 0.04	& 0.25 $\pm$ 0.03	& 0.27 $\pm$ 0.03	& 0.25 $\pm$ 0.03	& 0.23 $\pm$ 0.02	& 0.23 $\pm$ 0.03	& 0.23 $\pm$ 0.02	& 0.24 $\pm$ 0.03	& 0.24 $\pm$ 0.03	& 0.25 $\pm$ 0.03	& 0.27 $\pm$ 0.05	\\ \relax
H$\beta$ $\lambda$4861	& 1.00 $\pm$ 0.13	& 1.00 $\pm$ 0.10	& 1.00 $\pm$ 0.07	& 1.00 $\pm$ 0.07	& 1.00 $\pm$ 0.08	& 1.00 $\pm$ 0.08	& 1.00 $\pm$ 0.07	& 1.00 $\pm$ 0.08	& 1.00 $\pm$ 0.08	& 1.00 $\pm$ 0.10	& 1.00 $\pm$ 0.13	\\ \relax
[O III] $\lambda$4958	& 3.86 $\pm$ 0.73	& 3.82 $\pm$ 0.57	& 3.84 $\pm$ 0.38	& 3.87 $\pm$ 0.41	& 3.76 $\pm$ 0.41	& 3.59 $\pm$ 0.41	& 3.59 $\pm$ 0.34	& 3.53 $\pm$ 0.39	& 3.45 $\pm$ 0.41	& 3.40 $\pm$ 0.49	& 3.47 $\pm$ 0.67	\\ \relax
[O III] $\lambda$5006	& 11.38 $\pm$ 2.21	& 11.29 $\pm$ 1.73	& 11.44 $\pm$ 1.13	& 11.56 $\pm$ 1.25	& 11.27 $\pm$ 1.23	& 10.70 $\pm$ 1.23	& 10.71 $\pm$ 1.04	& 10.56 $\pm$ 1.19	& 10.31 $\pm$ 1.25	& 10.12 $\pm$ 1.48	& 10.34 $\pm$ 2.03	\\ \relax
[N I] $\lambda$5199	& ... $\pm$ ...	& ... $\pm$ ...	& ... $\pm$ ...	& 0.08 $\pm$ 0.01	& 0.11 $\pm$ 0.01	& ... $\pm$ ...	& ... $\pm$ ...	& ... $\pm$ ...	& ... $\pm$ ...	& ... $\pm$ ...	& ... $\pm$ ...	\\ \relax
He I $\lambda$5875	& 0.18 $\pm$ 0.05	& 0.13 $\pm$ 0.03	& 0.14 $\pm$ 0.02	& 0.12 $\pm$ 0.02	& 0.12 $\pm$ 0.02	& 0.11 $\pm$ 0.02	& 0.09 $\pm$ 0.01	& 0.09 $\pm$ 0.01	& 0.07 $\pm$ 0.01	& 0.09 $\pm$ 0.02	& 0.09 $\pm$ 0.02	\\ \relax
[Fe VII] $\lambda$6086	& ... $\pm$ ...	& 0.06 $\pm$ 0.01	& 0.09 $\pm$ 0.01	& 0.11 $\pm$ 0.02	& 0.09 $\pm$ 0.01	& 0.08 $\pm$ 0.01	& 0.07 $\pm$ 0.01	& 0.07 $\pm$ 0.01	& 0.05 $\pm$ 0.01	& ... $\pm$ ...	& ... $\pm$ ...	\\ \relax
[O I] $\lambda$6300	& 0.36 $\pm$ 0.12	& 0.36 $\pm$ 0.09	& 0.42 $\pm$ 0.06	& 0.44 $\pm$ 0.07	& 0.47 $\pm$ 0.07	& 0.40 $\pm$ 0.07	& 0.43 $\pm$ 0.06	& 0.47 $\pm$ 0.08	& 0.49 $\pm$ 0.09	& 0.49 $\pm$ 0.11	& 0.51 $\pm$ 0.15	\\ \relax
[O I] $\lambda$6363	& 0.12 $\pm$ 0.04	& 0.12 $\pm$ 0.03	& 0.14 $\pm$ 0.02	& 0.15 $\pm$ 0.02	& 0.16 $\pm$ 0.02	& 0.13 $\pm$ 0.02	& 0.14 $\pm$ 0.02	& 0.16 $\pm$ 0.03	& 0.16 $\pm$ 0.03	& 0.16 $\pm$ 0.04	& 0.17 $\pm$ 0.05	\\ \relax
[N II] $\lambda$6548	& 0.98 $\pm$ 0.37	& 0.85 $\pm$ 0.24	& 0.89 $\pm$ 0.14	& 0.92 $\pm$ 0.15	& 0.94 $\pm$ 0.15	& 0.67 $\pm$ 0.12	& 0.68 $\pm$ 0.10	& 0.78 $\pm$ 0.13	& 0.69 $\pm$ 0.13	& 0.66 $\pm$ 0.16	& 0.73 $\pm$ 0.24	\\ \relax
H$\alpha$ $\lambda$6562	& 2.90 $\pm$ 1.09	& 2.90 $\pm$ 0.81	& 2.90 $\pm$ 0.47	& 2.90 $\pm$ 0.49	& 2.90 $\pm$ 0.48	& 2.90 $\pm$ 0.53	& 2.90 $\pm$ 0.45	& 2.90 $\pm$ 0.50	& 2.90 $\pm$ 0.56	& 2.90 $\pm$ 0.69	& 2.90 $\pm$ 0.94	\\ \relax
[N II] $\lambda$6583	& 2.87 $\pm$ 1.09	& 2.48 $\pm$ 0.70	& 2.61 $\pm$ 0.43	& 2.71 $\pm$ 0.46	& 2.76 $\pm$ 0.46	& 1.98 $\pm$ 0.37	& 1.99 $\pm$ 0.31	& 2.28 $\pm$ 0.40	& 2.03 $\pm$ 0.39	& 1.95 $\pm$ 0.47	& 2.15 $\pm$ 0.70	\\ \relax
[S II] $\lambda$6716	& 1.33 $\pm$ 0.52	& 1.29 $\pm$ 0.38	& 0.89 $\pm$ 0.15	& 1.01 $\pm$ 0.18	& 0.97 $\pm$ 0.16	& 0.89 $\pm$ 0.17	& 0.88 $\pm$ 0.14	& 1.02 $\pm$ 0.18	& 0.87 $\pm$ 0.17	& 0.79 $\pm$ 0.19	& 0.87 $\pm$ 0.29	\\ \relax
[S II] $\lambda$6730	& 1.61 $\pm$ 0.64	& 1.12 $\pm$ 0.33	& 1.51 $\pm$ 0.26	& 1.41 $\pm$ 0.25	& 1.40 $\pm$ 0.24	& 0.97 $\pm$ 0.19	& 0.84 $\pm$ 0.13	& 0.84 $\pm$ 0.15	& 0.76 $\pm$ 0.15	& 0.73 $\pm$ 0.18	& 0.74 $\pm$ 0.25	\\ \hline
E(B-V) & 	0.54 $\pm$ 0.17 & 	0.46 $\pm$ 0.13 & 	0.30 $\pm$ 0.09 & 	0.20 $\pm$ 0.10 & 	0.14 $\pm$ 0.10 & 	0.24 $\pm$ 0.10 & 	0.25 $\pm$ 0.09 & 	0.18 $\pm$ 0.10 & 	0.23 $\pm$ 0.11 & 	0.28 $\pm$ 0.13 & 	0.29 $\pm$ 0.17
\enddata
\tablecomments{The same as in Table 4, but with line ratios corrected for galactic extinction using a galactic reddening curve \citep{savage1979}. The $H\alpha / H\beta$ ratios were fixed at 2.90 and negative E(B-V) values were set to zero. Error bars are the quadrature sum of the fractional flux uncertainty in H$\beta$ and each respective line along with the reddening uncertainty. Wavelengths are approximate air values, and the last row lists the color excess E(B-V).}
\end{deluxetable*}

\subsection{Emission Line Diagnostics}

\begin{figure*}
\centering
\subfigure{
\includegraphics[scale=0.46]{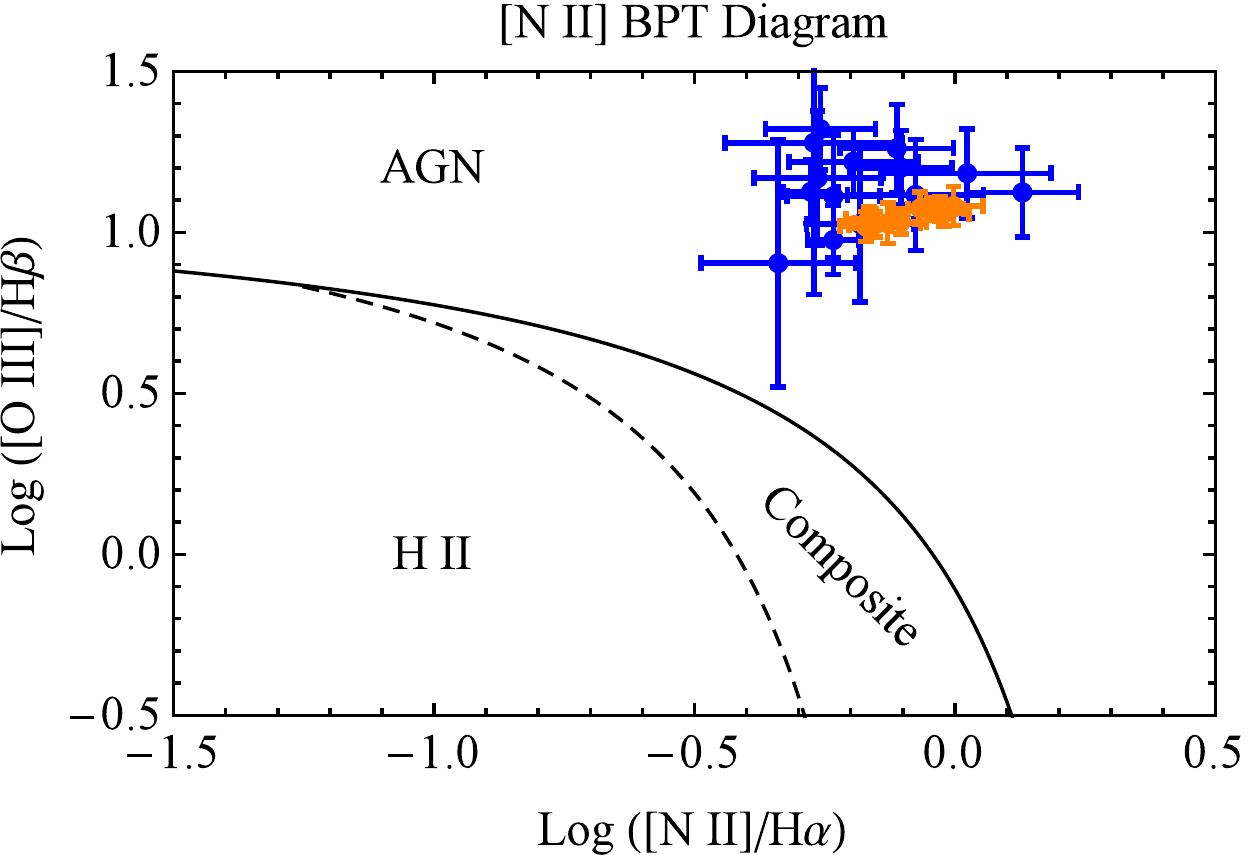}}
\subfigure{
\includegraphics[scale=0.46]{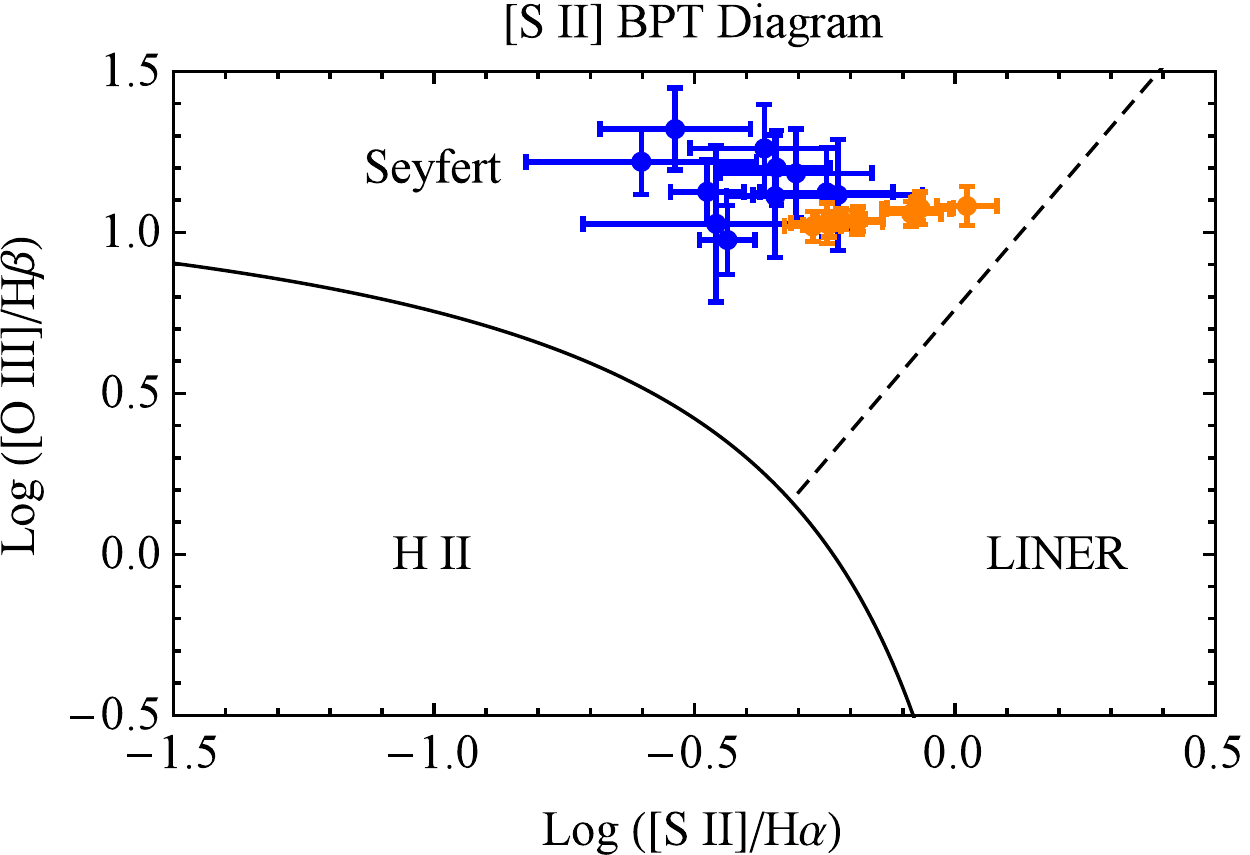}}
\vspace{-4pt}
\subfigure{
\includegraphics[scale=0.46]{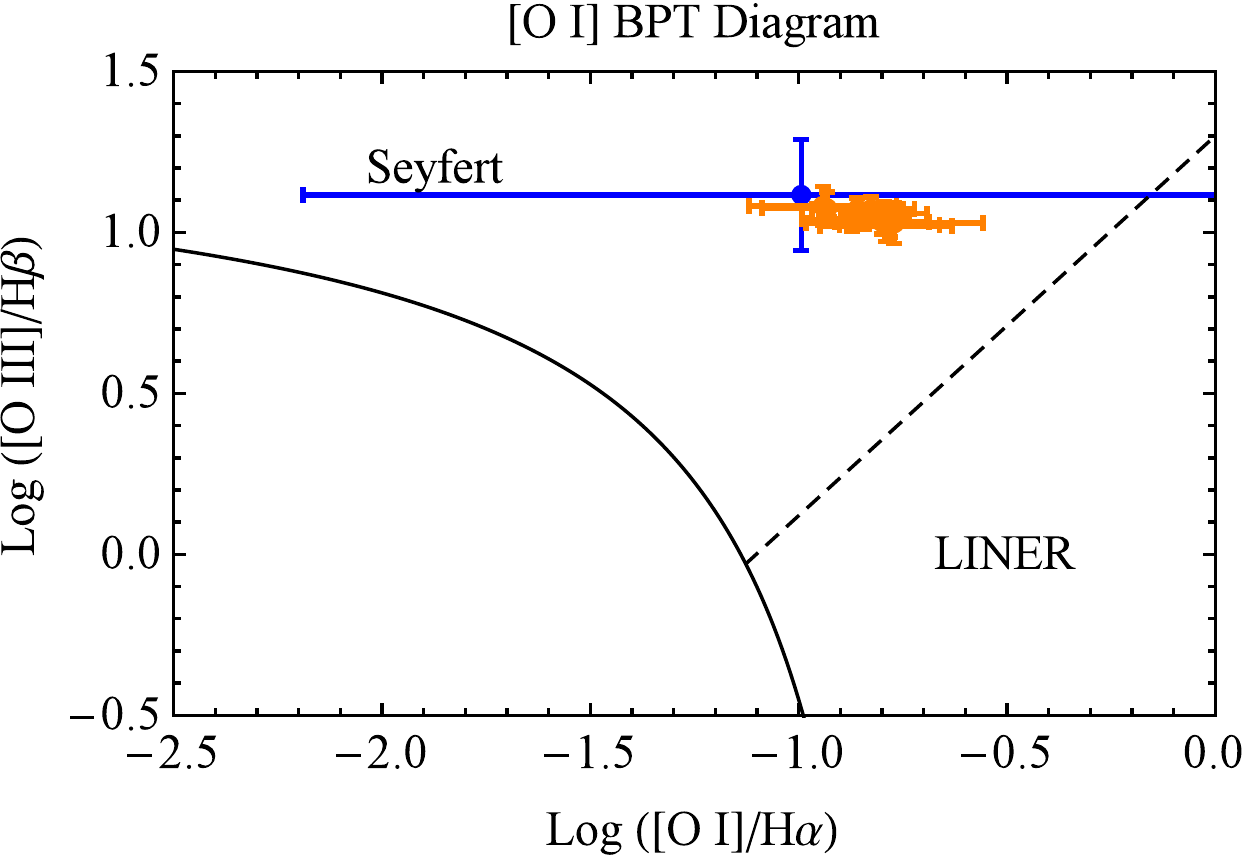}}
\subfigure{
\includegraphics[scale=0.64]{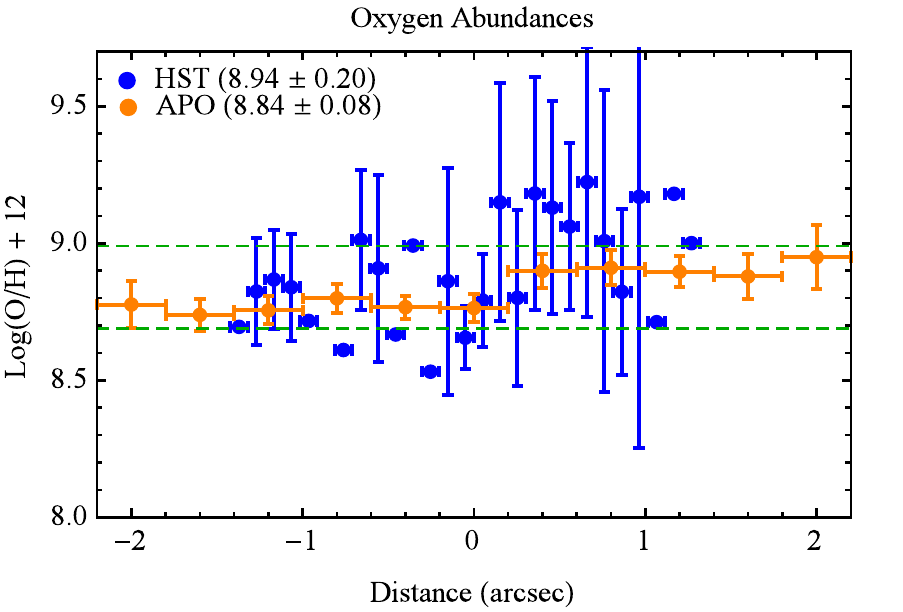}}
\subfigure{
\includegraphics[scale=0.64]{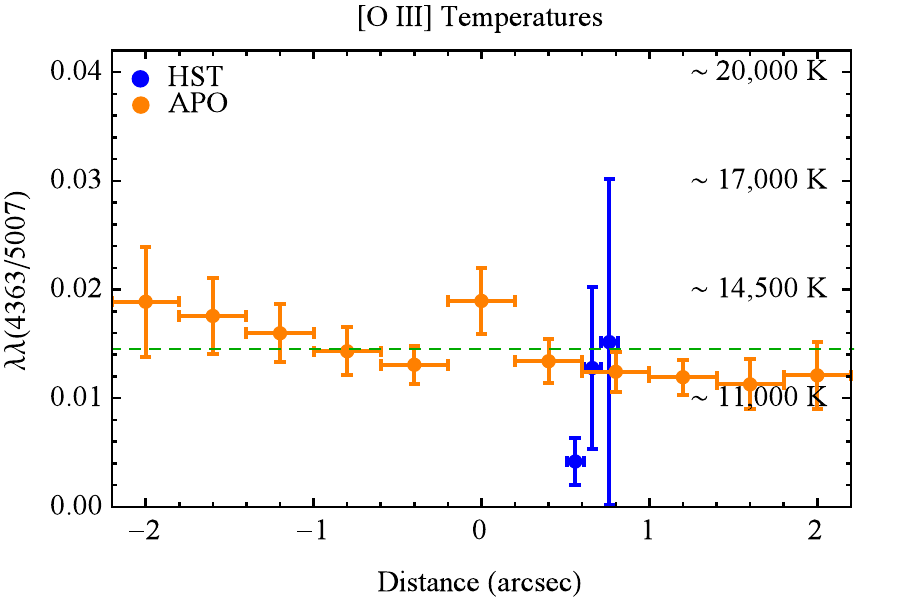}}
\subfigure{
\includegraphics[scale=0.64]{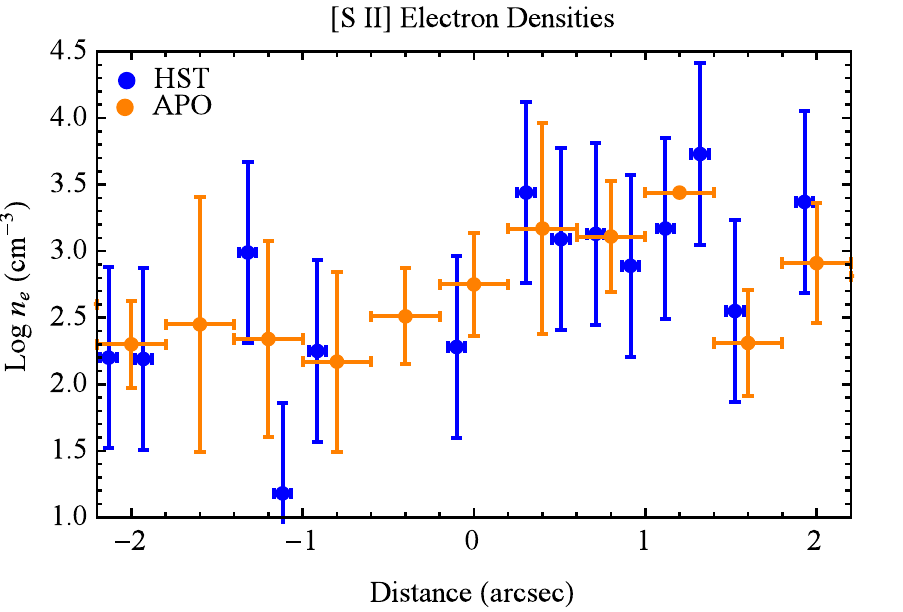}}
\vspace{-4pt}
\caption{Top: BPT ionization diagrams for [N~II], [S~II], and [O~I] using ratios calculated from the fluxes of all kinematic components summed together (Tables 3 and 5). {\it HST} STIS and APO DIS points are shown with blue and orange circles, respectively. The demarcation lines for distinguishing ionization mechanisms are from \cite{kewley2001, kewley2006, kauffmann2003}. Diagrams for the individual kinematic components and the other APO slit positions are given in the appendix. Bottom: The left panel shows the derived oxygen abundances using the method of \cite{storchibergmann1998}. The lower and upper dotted green lines represent one and two times solar abundances, respectively, with the average values provided in parentheses. The middle panel shows the electron temperatures, with the dotted green line representing the mean. The right panel shows the electron densities as derived from [S~II], with uncertainties propagated from the individual errors in the line ratios for all figures.}
\label{bpt}
\end{figure*}

As in Paper I, we use the dereddened emission line ratios to constrain the physical conditions in the emission line gas, including the ionization mechanism, elemental abundances, temperature, and electron density. In Figure \ref{bpt} and the Appendix, we present Baldwin-Phillips-Terlevich (BPT) diagrams \citep{baldwin1981, veilleux1987} that confirm the gas sampled by our observations is ionized by the central AGN\footnote{Note that there is a minor error in the y-intercept of the diagonal Seyfert/LINER line on the [S~II] diagram in Figure 4 of Paper I. The displayed intercept is 1.30, and the correct value is 0.76. This shifts the line minorly to the right, but does not affect the results or interpretation of that study.}, in agreement with \cite{stoklasov2009}. The AGN ionized gas extends to at least 8 kpc, with emission detected out to 12 kpc from the nucleus \citep{whittle1988}.

We determined the elemental abundances in gas phase using Equation (2) of \cite{storchibergmann1998}, and adopt a solar abundance of Log(O/H)+12 = 8.69 from \cite{asplund2009}. We find a mean oxygen abundance of Log(O/H)+12 = 8.84~$\pm$~0.08, or $Z = 1.40~\pm~0.26~Z_\odot$, with the spatial distribution shown in Figure \ref{bpt}. This value is $\sim 0.4$ dex higher than that found by \cite{castro2017}; however, our spectra do not contain the required emission lines for a direct comparison with that method.

We also derived the electron temperature and density, and the results are shown in Figure \ref{bpt}. We find typical NLR temperatures of $\sim 10,000 - 15,000$ K, in agreement with \cite{koski1978}. Unlike Mrk 573, we do not observe a strong, centrally peaked density profile with a characteristic power-law index. The abundance, temperature, and density profiles display a curious dichotomy across the nucleus, following a systematic decrease in [O~III]/H$\beta$ across the NLR from the SE to the NW.

\subsection{{\normalfont [O~III]} Image Analysis}

We use an [O~III] emission line image of the entire NLR to account for mass outside of our spectral slit observations. To ensure proper flux calibration, we compared our integrated [O~III] emission line fluxes to extracted regions of the image covering the same area. Owing to the excellent calibration of {\it HST} observations, the [O~III] fluxes agree to better than 5\% after scaling the image by the filter bandpass.

We determined the total [O~III] flux as a function of distance from the nucleus using the Elliptical Panda routine within the SAOImage DS9 software \citep{joye2003}. We divide the image into two concentric semi-ellipses to better account for NLR asymmetries in the flux, density, and velocity profiles. The semi-ellipses are centered on the nucleus, with spacings equal to the spatial sampling of our extracted {\it HST} observations for determining the mass profile. The ring ellipticity is calculated from the adopted inclination of $i = 41\degr$ using $b/a = cos(i)$ where $a$ and $b$ are the major and minor-axis lengths, respectively.

\begin{figure*}
\vspace{-4pt}
\centering
\subfigure{
\includegraphics[scale=0.485]{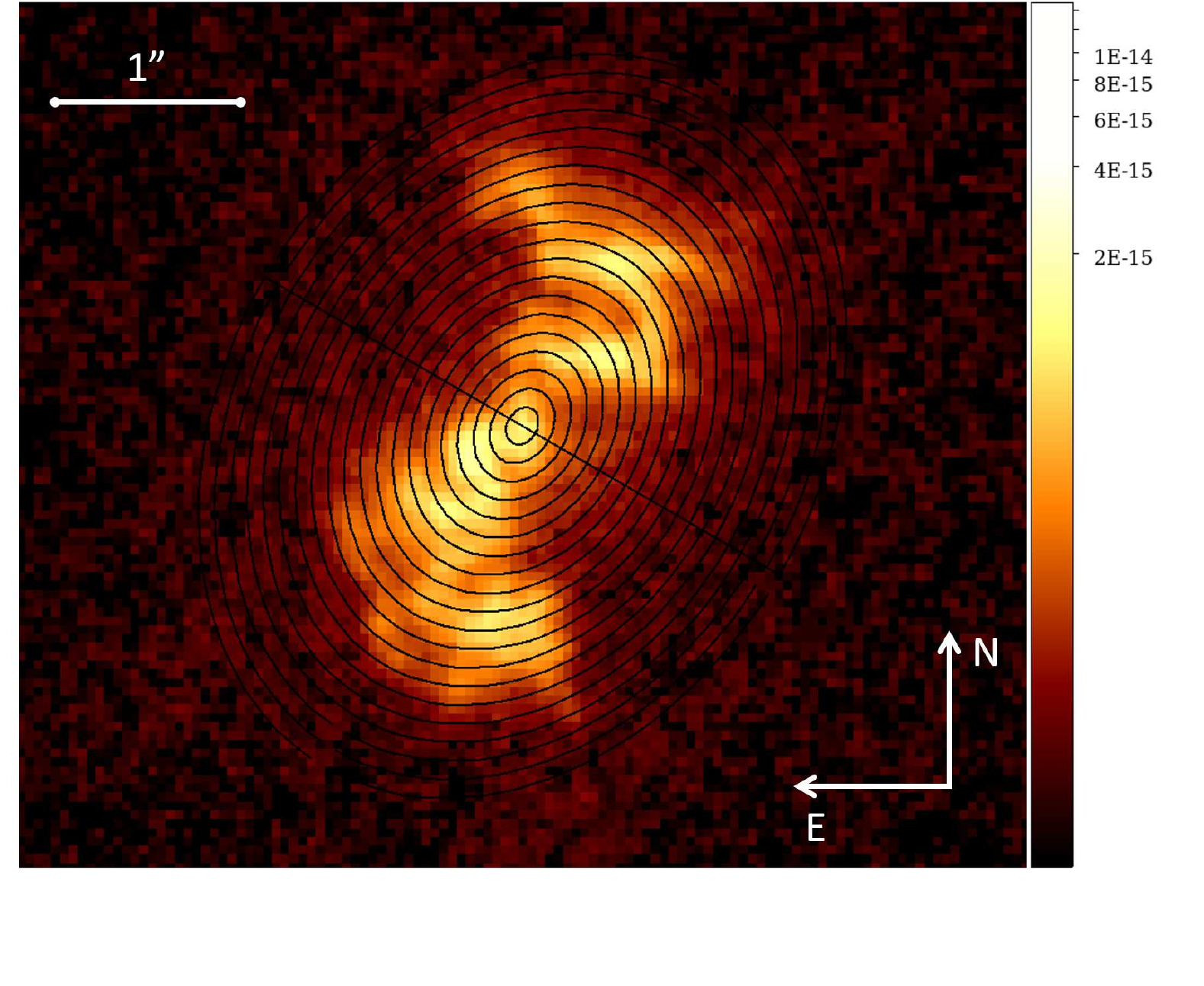}}
\subfigure{
\includegraphics[scale=0.77]{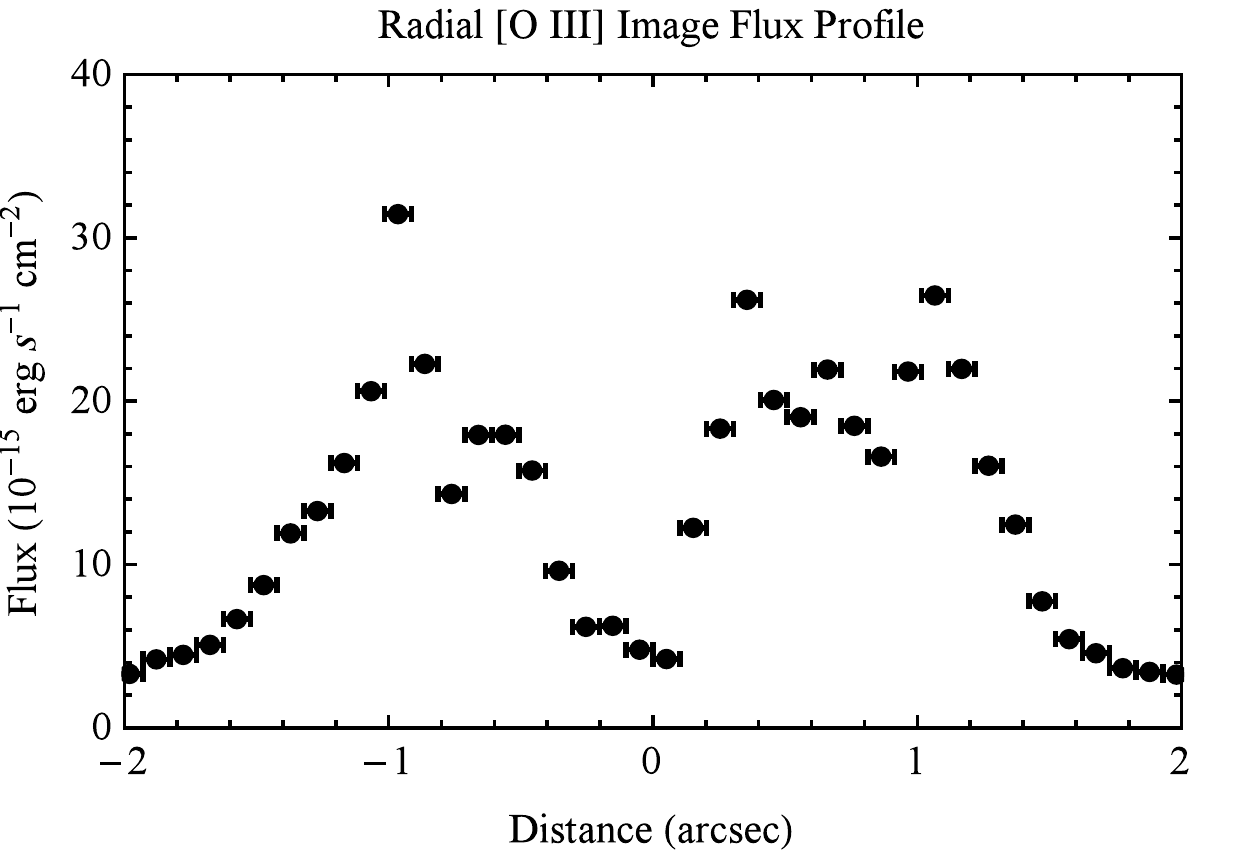}}
\vspace{-4pt}
\caption{The left panel is a portion of the {\it HST} [O~III] image with overlaid elliptical semi-annuli representing rings of constant distance from the nucleus. The color bar gives fluxes in units of erg s$^{-1}$ cm$^{-2}$. The right panel is the azimuthally summed [O~III] semi-annuli fluxes oriented along the major axis of the ellipse, with typical errors smaller than the size of the points. SE is to the left and NW is to the right.}
\label{imaging}
\end{figure*}

The [O~III] image and azimuthally summed flux profile are shown in Figure \ref{imaging}. An error of $\sigma \approx 8.77 \times 10^{-18}$ erg s$^{-1}$ cm$^{-2}$ pixel$^{-1}$ was calculated from line-free regions in the image, with the error in each annulus equal to $\sqrt{N_{pix}}\cdot \sigma$, where $N_{pix}$ is the number of pixels.

\break

\section{Photoionization Models}

Ultimately, our techniques rely on an accurate scale factor for converting [O~III] flux to ionized gas mass. The most accurate method requires detailed photoionization models, as the emission coefficient of the gas will vary across the NLR due to changing physical conditions, such as density. Our process for creating multi-component photoionization models is described in \S5 of Paper I, and we summarize here the most pertinent details.

\setlength{\tabcolsep}{0.06in}
\tabletypesize{\footnotesize}
\begin{deluxetable}{c|c|c|c|c|c|}
\tablenum{6}
\tablecaption{Cloudy Model Input Parameters}
\tablehead{
\colhead{Distance} & \colhead{Comp} & \colhead{Ionization} & \colhead{Column} & \colhead{Number} & \colhead{Dust}\\
\colhead{from} & \colhead{ION} & \colhead{Parameter} & \colhead{Density} & \colhead{Density} & \colhead{Content}\\
\colhead{Nucleus} & \colhead{Name} & \colhead{Log(U)} & \colhead{Log(N$_H$)} & \colhead{Log(n$_H$)} & \colhead{Relative}\\
\colhead{(arcsec)}& \colhead{} & \colhead{(unitless)} & \colhead{(cm$^{-2}$)} & \colhead{(cm$^{-3}$)} & \colhead{to ISM}\\
\colhead{(1)} & \colhead{(2)} & \colhead{(3)} & \colhead{(4)} & \colhead{(5)} & \colhead{(6)}
}
\startdata
2.00	 & 	High	 & 	...	 & 	...	 & 	...	 & 	...	 \\ 
2.00	 & 	Med		 & 	-2.00	 & 	21.20	 & 	1.46	 & 	0.5	 \\ 
2.00	 & 	Low		 & 	-4.00	 & 	19.40	 & 	3.46	 & 	0.5	 \\ 
1.60	 & 	High	 & 	...	 & 	...	 & 	...	 & 	...	 \\ 
1.60	 & 	Med		 & 	-2.00	 & 	21.40	 & 	1.64	 & 	0.5	 \\ 
1.60	 & 	Low		 & 	-3.80	 & 	19.40	 & 	3.44	 & 	0.5	 \\ 
1.20	 & 	High	 & 	...	 & 	...	 & 	...	 & 	...	 \\ 
1.20	 & 	Med		 & 	-2.00	 & 	21.30	 & 	1.86	 & 	0.5	 \\ 
1.20	 & 	Low		 & 	-3.90	 & 	19.60	 & 	3.76	 & 	0.5	 \\ 
0.80	 & 	High	 & 	...	 & 	...	 & 	...	 & 	...	 \\ 
0.80	 & 	Med		 & 	-2.00	 & 	21.40	 & 	2.14	 & 	0.5	 \\ 
0.80	 & 	Low		 & 	-3.20	 & 	20.30	 & 	3.34	 & 	0.5	 \\ 
0.40	 & 	High	 & 	-0.80	 & 	22.20	 & 	1.28	 & 	0.0	 \\ 
0.40	 & 	Med		 & 	-2.80	 & 	20.80	 & 	3.28	 & 	0.5	 \\ 
0.40	 & 	Low		 & 	...	 & 	...	 & 	...	 & 	...	 \\ 
0.00	 & 	High	 & 	-0.90	 & 	22.00	 & 	1.59	 & 	0.0	 \\ 
0.00	 & 	Med		 & 	-2.80	 & 	20.80	 & 	3.49	 & 	0.5	 \\ 
0.00	 & 	Low		 & 	...	 & 	...	 & 	...	 & 	...	 \\ 
-0.40	 & 	High	 & 	-1.40	 & 	21.60	 & 	1.88	 & 	0.0	 \\ 
-0.40	 & 	Med		 & 	-2.90	 & 	20.60	 & 	3.38	 & 	0.5	 \\ 
-0.40	 & 	Low		 & 	...	 & 	...	 & 	...	 & 	...	 \\ 
-0.80	 & 	High	 & 	...	 & 	...	 & 	...	 & 	...	 \\ 
-0.80	 & 	Med		 & 	-2.00	 & 	21.40	 & 	2.14	 & 	0.5	 \\ 
-0.80	 & 	Low		 & 	-3.50	 & 	20.00	 & 	3.64	 & 	0.5	 \\ 
-1.20	 & 	High	 & 	...	 & 	...	 & 	...	 & 	...	 \\ 
-1.20	 & 	Med		 & 	-2.00	 & 	20.90	 & 	1.86	 & 	0.5	 \\ 
-1.20	 & 	Low		 & 	-4.00	 & 	19.70	 & 	3.86	 & 	0.5	 \\ 
-1.60	 & 	High	 & 	...	 & 	...	 & 	...	 & 	...	 \\ 
-1.60	 & 	Med		 & 	-2.00	 & 	21.40	 & 	1.64	 & 	0.5	 \\ 
-1.60	 & 	Low		 & 	-3.70	 & 	19.90	 & 	3.34	 & 	0.5	 \\ 
-2.00	 & 	High	 & 	...	 & 	...	 & 	...	 & 	...	 \\ 
-2.00	 & 	Med		 & 	-2.00	 & 	21.20	 & 	1.46	 & 	0.5	 \\ 
-2.00	 & 	Low		 & 	-4.00	 & 	19.60	 & 	3.46	 & 	0.5 
\enddata
\tablecomments{The Cloudy photoionization model input parameters. The columns are: (1) position, (2) component name, (3) log$_{10}$ ionization parameter, (4) log$_{10}$ column density, (5) log$_{10}$ number density, and (6) dust fraction relative to ISM.}
\end{deluxetable}

\setlength{\tabcolsep}{0.028in}
\tabletypesize{\footnotesize}
\begin{deluxetable}{c|c|c|c|c|c|c|}
\tablenum{7}
\tablecaption{Cloudy Model Output Parameters}
\tablehead{
\colhead{Distance} & \colhead{Comp} & \colhead{Fraction} & \colhead{Log(F$_{H\beta}$)} & \colhead{Cloud} & \colhead{Cloud} & \colhead{Cloud}\\
\colhead{from} & \colhead{ION} & \colhead{of Total} & \colhead{Model} & \colhead{Surface} & \colhead{Model} & \colhead{Model}\\
\colhead{Nucleus} & \colhead{Name} & \colhead{Model} & \colhead{Flux} & \colhead{Area} & \colhead{Thickness} & \colhead{Depth}\\
\colhead{(arcsec)} & \colhead{} & \colhead{} & \colhead{(cgs)} & \colhead{($10^3$ pc$^2$)} & \colhead{(pc)} & \colhead{(pc)}\\
\colhead{(1)} & \colhead{(2)} & \colhead{(3)} & \colhead{(4)} & \colhead{(5)} & \colhead{(6)} & \colhead{(7)}
}
\startdata
2.00	 & 	High & 	...	 & 	...	 & 	...	 & 	...	 & 	... 	 \\ 
2.00	 & 	Med	 & 	0.70	 & 	-2.54	 & 	4292	 & 	18	 & 	2117	 \\ 
2.00	 & 	Low	 & 	0.30	 & 	-2.46	 & 	1506	 & 	$<$0.1	 & 	742	 \\ 
1.60	 & 	High & 	...	 & 	...	 & 	...	 & 	...	 & 	... 	 \\ 
1.60	 & 	Med	 & 	0.70	 & 	-2.35	 & 	3527	 & 	19	 & 	1739	 \\ 
1.60	 & 	Low	 & 	0.30	 & 	-2.28	 & 	1278	 & 	$<$0.1	 & 	630	 \\ 
1.20	 & 	High & 	...	 & 	...	 & 	...	 & 	...	 & 	... 	 \\ 
1.20	 & 	Med	 & 	0.65	 & 	-2.14	 & 	1644	 & 	9	 & 	811	 \\ 
1.20	 & 	Low	 & 	0.35	 & 	-2.05	 & 	714	 & 	$<$0.1	 & 	352	 \\ 
0.80	 & 	High & 	...	 & 	...	 & 	...	 & 	...	 & 	... 	 \\ 
0.80	 & 	Med	 & 	0.40	 & 	-1.75	 & 	347	 & 	6	 & 	171	 \\ 
0.80	 & 	Low	 & 	0.60	 & 	-1.85	 & 	657	 & 	$<$0.1	 & 	324	 \\ 
0.40	 & 	High & 	0.05	 & 	-1.47	 & 	20	 & 	269.6	 & 	10	 \\ 
0.40	 & 	Med	 & 	0.95	 & 	-1.42	 & 	330	 & 	$<$0.1	 & 	163	 \\ 
0.40	 & 	Low	 & 	...	 & 	...	 & 	...	 & 	...	 & 	... 	 \\ 
0.00	 & 	High & 	0.05	 & 	-1.38	 & 	23	 & 	83	 & 	11	 \\ 
0.00	 & 	Med	 & 	0.95	 & 	-1.21	 & 	294	 & 	$<$0.1	 & 	145	 \\ 
0.00	 & 	Low	 & 	...	 & 	...	 & 	...	 & 	...	 & 	... 	 \\ 
-0.40	 & 	High & 	0.10	 & 	-1.44	 & 	49	 & 	17.0	 & 	24	 \\ 
-0.40	 & 	Med	 & 	0.90	 & 	-1.42	 & 	418	 & 	$<$0.1	 & 	206	 \\ 
-0.40	 & 	Low	 & 	...	 & 	...	 & 	...	 & 	...	 & 	... 	 \\ 
-0.80	 & 	High & 	...	 & 	...	 & 	...	 & 	...	 & 	... 	 \\ 
-0.80	 & 	Med	 & 	0.60	 & 	-1.85	 & 	513	 & 	6	 & 	253	 \\ 
-0.80	 & 	Low	 & 	0.40	 & 	-1.75	 & 	272	 & 	$<$0.1	 & 	134	 \\ 
-1.20	 & 	High & 	...	 & 	...	 & 	...	 & 	...	 & 	... 	 \\ 
-1.20	 & 	Med	 & 	0.60	 & 	-2.18	 & 	956	 & 	4	 & 	471	 \\ 
-1.20	 & 	Low	 & 	0.40	 & 	-2.04	 & 	467	 & 	$<$0.1	 & 	230	 \\ 
-1.60	 & 	High & 	...	 & 	...	 & 	...	 & 	...	 & 	... 	 \\ 
-1.60	 & 	Med	 & 	0.65	 & 	-2.35	 & 	1246	 & 	19	 & 	615	 \\ 
-1.60	 & 	Low	 & 	0.35	 & 	-2.25	 & 	534	 & 	$<$0.1	 & 	264	 \\ 
-2.00	 & 	High & 	...	 & 	...	 & 	...	 & 	...	 & 	... 	 \\ 
-2.00	 & 	Med	 & 	0.65	 & 	-2.54	 & 	1149	 & 	18	 & 	567	 \\ 
-2.00	 & 	Low	 & 	0.35	 & 	-2.45	 & 	494	 & 	$<$0.1	 & 	244
\enddata
\tablecomments{The best-fitting Cloudy model output parameters. The columns are: (1) position, (2) component name, (3) fraction of model contributing to the H$\beta$ luminosity, (4) log$_{10}$ H$\beta$ model flux (erg s$^{-1}$ cm$^{-2}$), (5) surface area of the gas in units of $10^3$, (6) gas cloud thickness (N$_\mathrm{H}$/$n_\mathrm{H}$), and (7) depth into the plane of the sky.}
\end{deluxetable}

\subsection{Input Parameters}

We construct models using version 13.04 of the photoionization code Cloudy \citep{ferland2013}. A self-consistent model requires supplying the quantity and energy distribution of photons intercepting a cloud of known composition and geometry. These quantities are encapsulated by the ionization parameter (U), which is the ratio of the number of ionizing photons to atoms at the cloud face \citep[\S 13.6]{osterbrock2006},
\begin{equation}
U = \frac{1}{4 \pi r^2 n_H c} \int_{\nu_0}^{\infty} \frac{L_{\nu}}{h\nu} d\nu,
\end{equation}
\noindent
where $r$ is the radial distance from the AGN, $n_\mathrm{H}$ is the hydrogen number density cm$^{-3}$, and $c$ is the speed of light. The integral is the number of ionizing photons s$^{-1}$, $Q(H) = \int_{\nu_0}^{\infty} (L_{\nu}/h\nu) d\nu$, where $L_{\nu}$ is the luminosity of the AGN as a function of frequency as determined from the spectral energy distribution (SED), $h$ is Planck's constant, and $\nu_0 = 13.6 eV/h$ is the ionization potential of hydrogen \citep[\S 14.3]{osterbrock2006}.

We adopt a typical power-law SED that has worked well in previous studies \citep{kraemer2000a, kraemer2000b}, taking into consideration the X-ray modeling of \cite{gandhi2014}. For $L_{\nu} \propto \nu^{\alpha}$ we adopt slopes of $\alpha = -0.5$ from 1 eV to 13.6 eV, $\alpha = -1.4$ from 13.6 eV to 0.5 keV, $\alpha = -1$ from 0.5 keV to 10 keV, and $\alpha = -0.5$ from 10 keV to 100 keV, with low and high energy cutoffs below 1 eV and above 100 keV, respectively. Normalizing this SED to the 2-10 keV luminosity from \cite{gandhi2014}, $L_{2-10} = 9(\pm3)\cdot10^{43}$ erg s$^{-1}$, we numerically compute the above integral and find $Q(H) = 7.8(\pm2.6) \cdot 10^{54}$ photons s$^{-1}$, or equivalently, Log $Q(H) \approx 54.89$, in approximate agreement with \cite{wilson1988}. We also investigated scaling the 2-10 keV luminosity to our adopted distance, a difference of $\sim$ 22\%, but this did not noticeably improve the model fits.

The gas composition is determined by the elemental abundances, dust content, and corresponding depletions of elements into dust grains. We adopt abundances of $\sim 1.3~Z_{\odot}$, and the exact logarithmic values relative to hydrogen by number for dust free models are: He = -0.96, C = -3.46, N = -3.94, O = -3.20, Ne = -3.96, Na = -5.65, Mg = -4.29, Al = -5.44, Si = -4.38, P = -6.48, S = -4.77, Ar = -5.49, Ca = -5.55, Fe = -4.39, Ni = -5.67. The strong low ionization lines are more easily reproduced by including a dusty component, and for models with a dust level of 0.5 relative to the interstellar medium we accounted for depletion of certain elements in graphite and silicate grains \citep{seab1983, snow1996, collins2009}. The logarithmic abundances relative to hydrogen by number for the dusty models are: He = -0.96, C = -3.63, N = -3.94, O = -3.32, Ne = -3.96, Na = -5.65, Mg = -4.57, Al = -5.70, Si = -4.66, P = -6.48, S = -4.77, Ar = -5.49, Ca = -5.81, Fe = -4.67, Ni = -5.93.

\begin{figure*}
\vspace{-10pt}
\centering
\subfigure{
\includegraphics[scale=0.66]{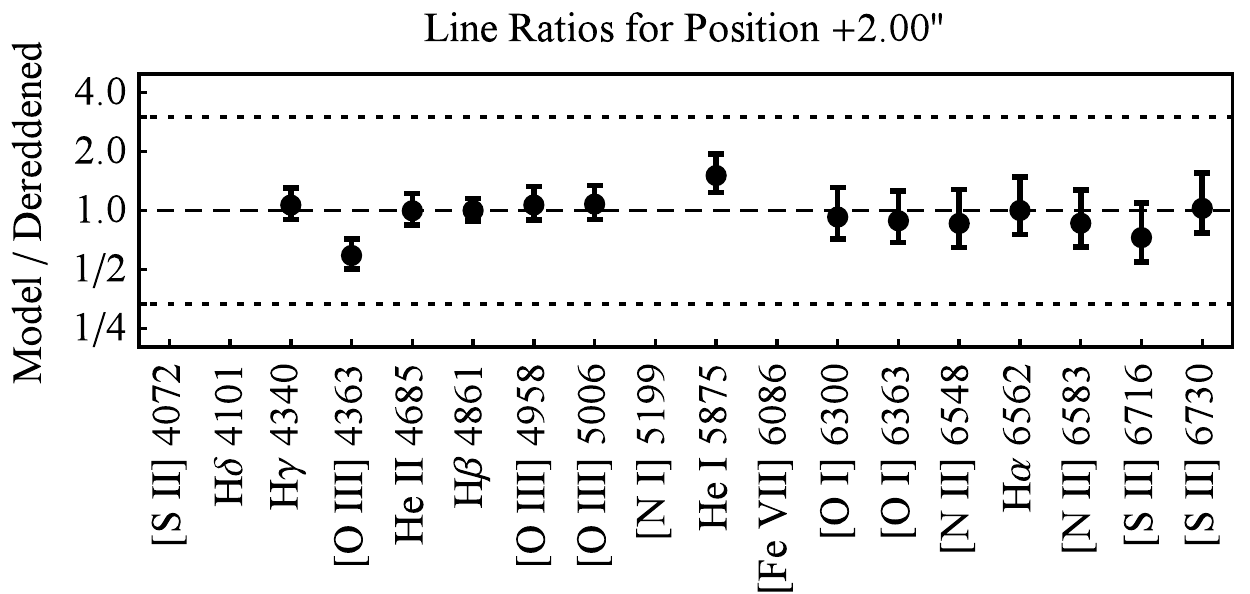}}
\vspace{-10pt}
\subfigure{
\includegraphics[scale=0.66]{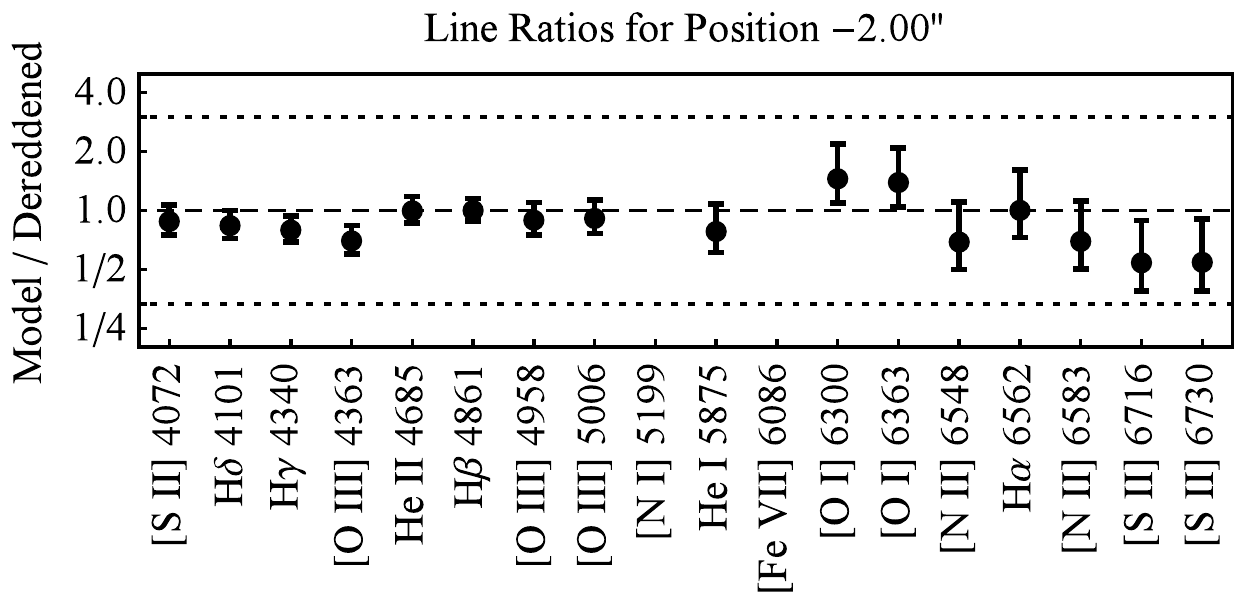}}
\subfigure{
\includegraphics[scale=0.66]{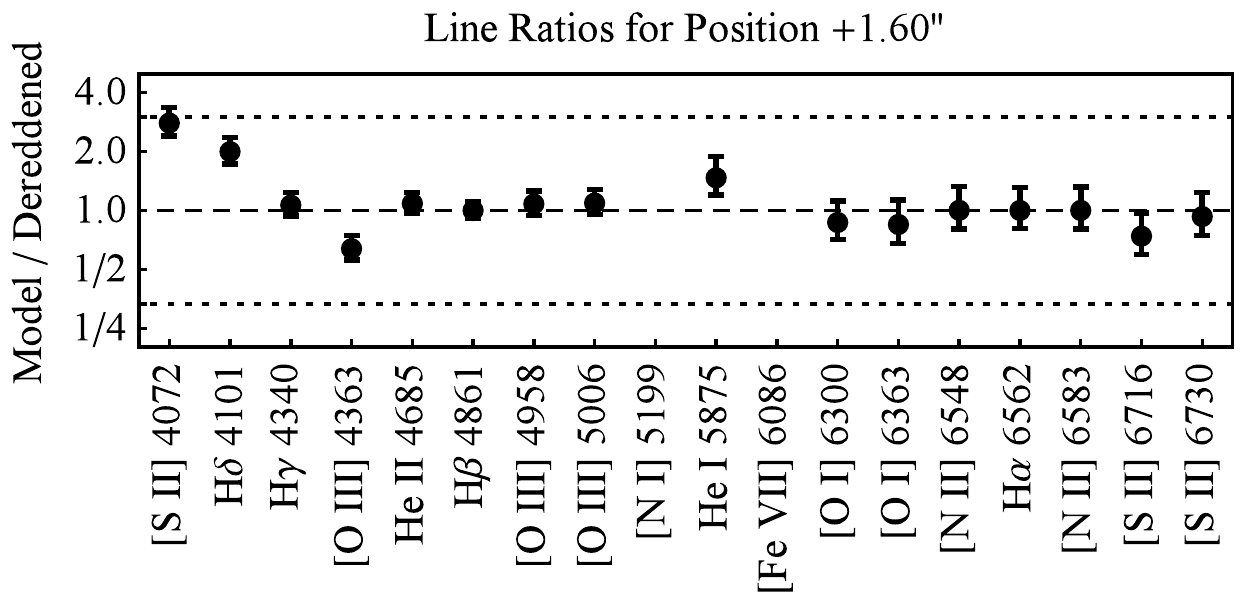}}
\vspace{-10pt}
\subfigure{
\includegraphics[scale=0.66]{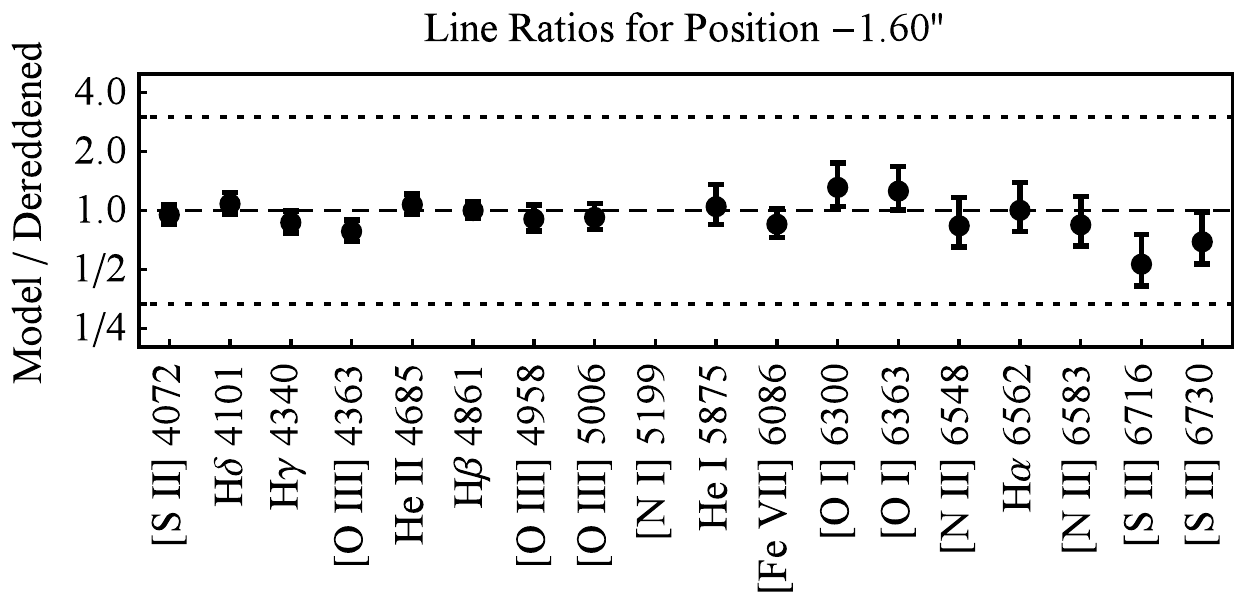}}
\subfigure{
\includegraphics[scale=0.66]{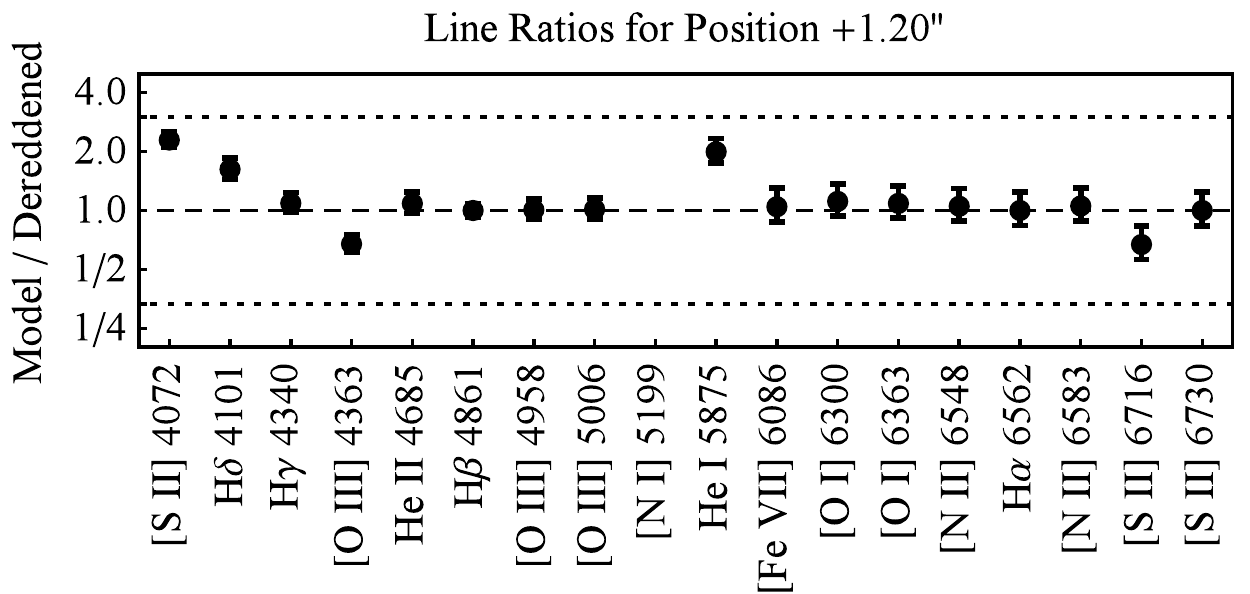}}
\vspace{-10pt}
\subfigure{
\includegraphics[scale=0.66]{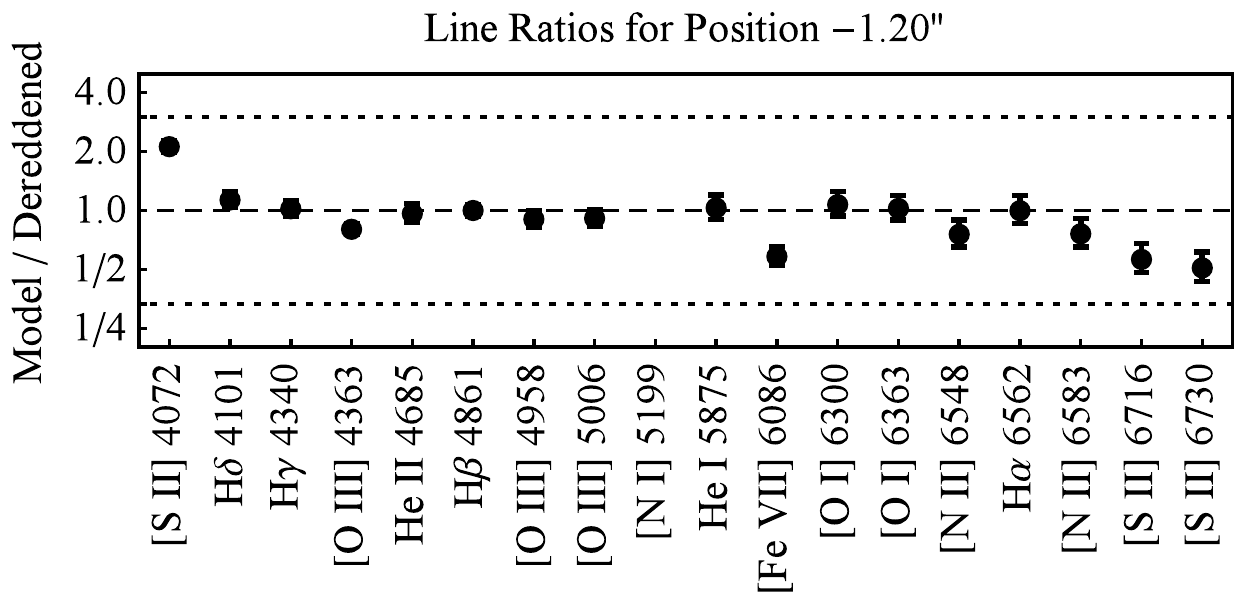}}
\subfigure{
\includegraphics[scale=0.66]{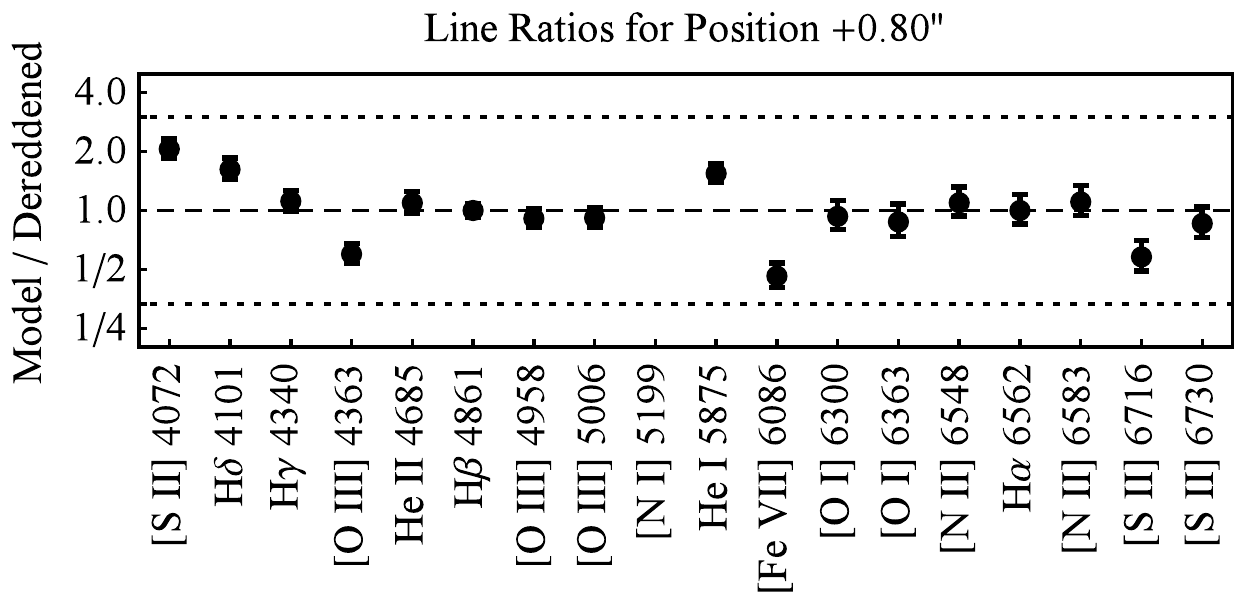}}
\vspace{-10pt}
\subfigure{
\includegraphics[scale=0.66]{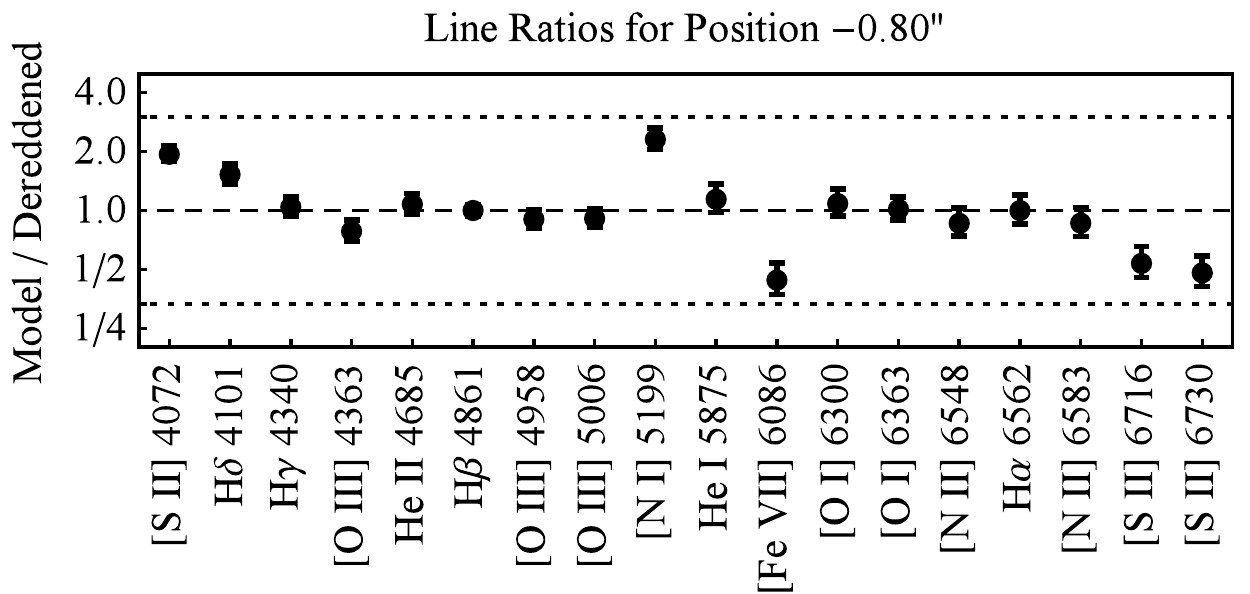}}
\subfigure{
\includegraphics[scale=0.66]{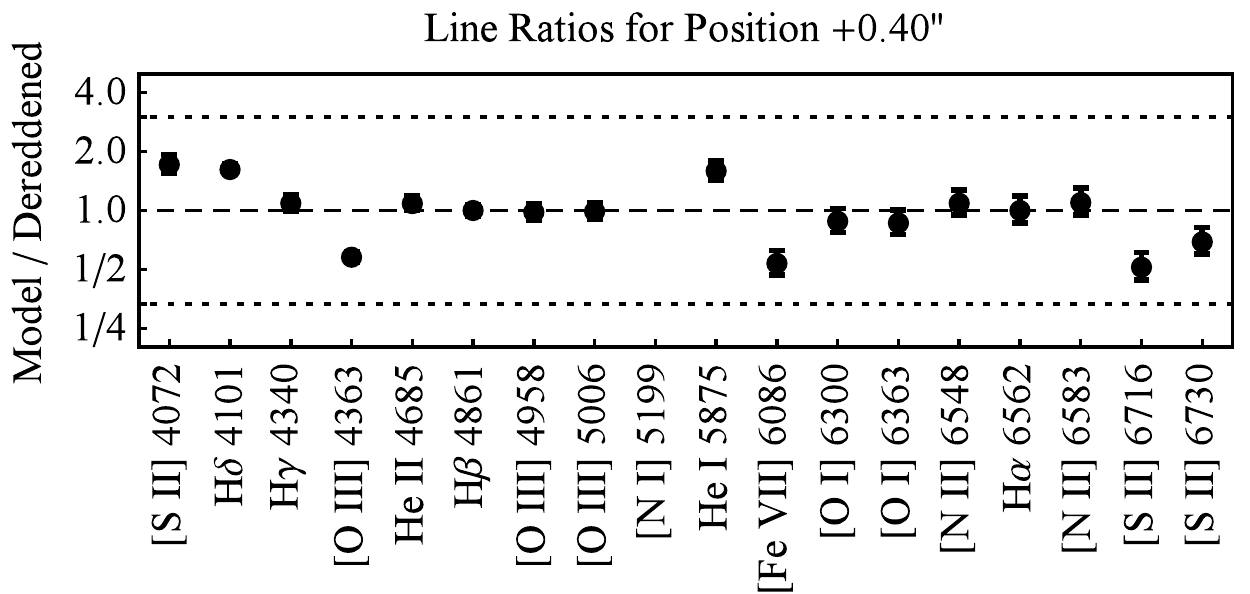}}
\vspace{-12pt}
\subfigure{
\includegraphics[scale=0.66]{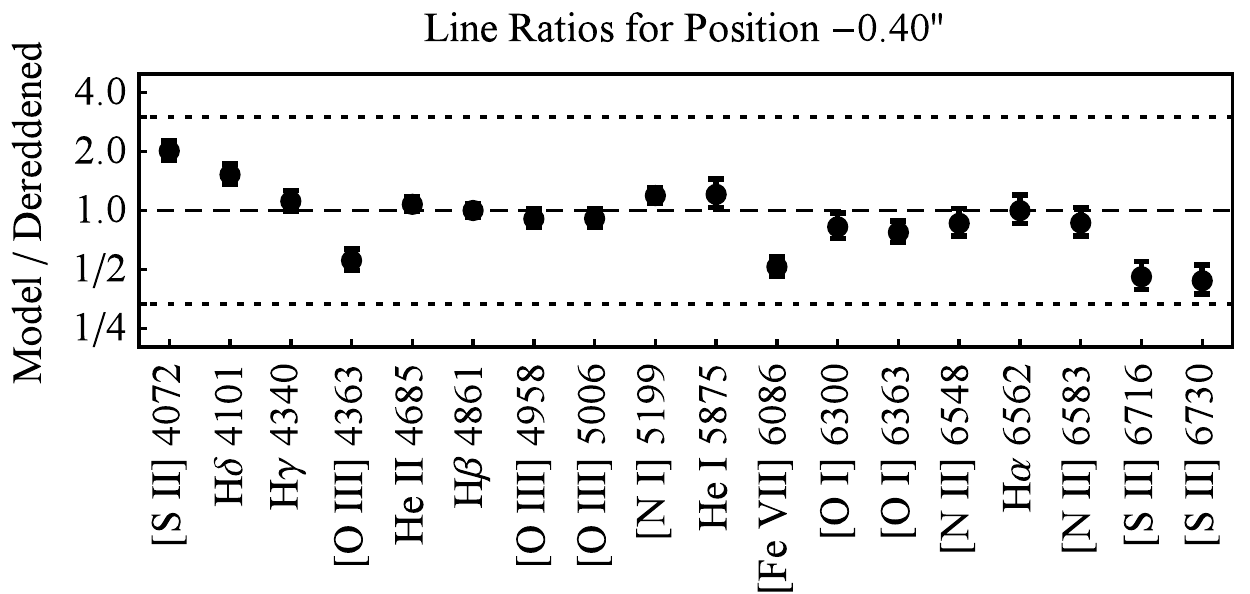}}
\subfigure{
\includegraphics[scale=0.66]{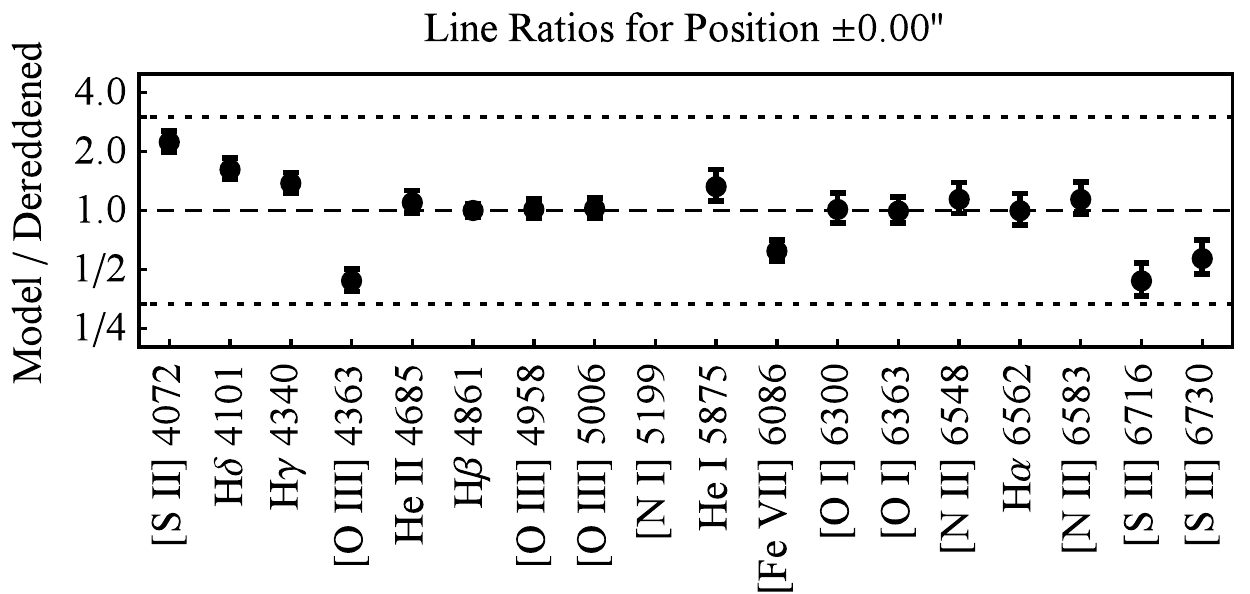}}
\vspace{-8pt}
\caption{The composite model line ratios divided by the dereddened values, where the spatial distance is that from the nucleus to the center of the pixel extraction. The dashed unity lines indicate an exact match, while the dotted lines are factor of three boundaries. The tick marks are logarithmically spaced for even distributions above and below the unity line. Points above this line are overpredicted, while points below are underpredicted by the models.}
\label{models}
\end{figure*}

\subsection{Model Selection}

To account for gas in multiple ionization states with different densities at each location, we use up to three model components referred to as HIGH, MED, and LOW ION. At each location along the slit, the only unknown quantities in Equation 7 are $U$ and $n_\mathrm{H}$, so for a grid of models we choose a range of $U$ values to produce the observed emission and solve for the corresponding density to maintain physical consistency. We then add fractional combinations of the components to create a composite model that matches the H$\beta$ luminosity.

To determine the best model for each location, we use a simple numerical scheme that compares the model line ratios to the dereddened values for all fractional combinations of HIGH, MED, and LOW ION, with an ideal match having a ratio of unity. Similar to Paper I, our limiting criteria for a successful fit were the following. First, the He II ratio that is critically sensitive to the column density, and the [O~III] doublet that determines our flux to mass scaling, must match the observations within 10\%. We then aim to constrain [O~I], [N~II], and [S~II] within 30\%. We loosened this criterion up to a factor of $\sim$ 2 for [S~II] at some locations to find a match. Finally, all remaining lines in the spectra must match within a factor of three.

In contrast to our modeling of Mrk 573, we find that simpler two-component models are able to match the line ratios, in agreement with the general conclusions of \cite{rosario2007}. This could be in part due to the fewer number of emission lines available in the spectra, in combination with the significantly larger APO extraction areas that will tend to blend the conditions of various emission line knots. The models employ pure AGN ionization, and we do not see evidence for shock excitation, in agreement with the findings of \cite{jackson2007}. We will explore this in more detail using {\it Chandra} data in a forthcoming paper (Fischer et al., in prep). The Cloudy model input and output parameters for our best-fitting models are given in Tables 6 and 7, respectively. The predicted emission line ratios for the final composite models are given in Table 8.

 \subsection{Comparison to the Observations}

The comparison of our model and dereddened emission line ratios is presented in Figure \ref{models}. The dashed unity line indicates an exact match, and points between the dotted lines agree to within better than a factor of three. A variety of factors contribute to the observed deviations, such as a poor Gaussian fit, the S/N, the quality of atomic data, and the accuracy of our models. We discuss here systematics discrepancies and those greater than a factor of two for important diagnostic lines at each position.

Qualitatively, the [S~II] $\lambda 4072$ line should be treated with skepticism, as it is an unresolved doublet and at the extreme edge of the spectral coverage. The agreement of the H$\gamma$ line is indicative of a proper reddening correction, while the mild overprediction of H$\delta$ may be due to a noisy continuum. The slight underprediction of [O~III] $\lambda$4363 is a minor concern as it indicates an underprediction of the temperature in more highly ionized zones and may be partially attributed to blending with H$\gamma$.

Finally, there is the general underprediction of [S~II] $\lambda \lambda$6716, 6731, which may indicate the need for more dust, or exposure to a partially-absorbed SED from a closer in absorber as found for the NLRs of Mrk 573 \citep{revalski2018} and Mrk 3 \citep{collins2009}. The issue is most severe near the nucleus, which may also indicate that some of the [S~II] emission arises from the edges of the ionized NLR bicone, at larger distances from the nucleus than those adopted in our models. Specifically, our models use a distance corresponding to the midpoint between the pixel center and the edge of the slit, calculated using Equation (2) for each position (e.g. the central extraction uses a distance of $\pm0\farcs5$, while the pixel covers an area of $2\farcs0$ x $0\farcs4$).

The introduction of additional dust brought the [S~II] emission lines into agreement with the observations for the extractions at larger distances, but degraded the overall fit in some cases and introduced significant scatter into the scale factors. Allowing the low ionization component to be at larger distances near the edge of the slit, rather than in the middle, marginally improved the fits at the $\sim$10--20\% level. Considering the limited constraints we are able to place on the physical conditions in the gas from the small number of emission lines, and the ability of the models to accurately reproduce all other lines in the spectrum, we decided against further fine-tuning. Overall, our models are able to successfully match all emission lines within a factor of three or better at all locations in the NLR.

To check the physical reality of our models, we also derived several quantities at each location. These include the surface areas ($A = L_{\mathrm{H}\beta}/F_{\mathrm{H}\beta}$) and thicknesses ($N_{H}/n_{H}$) of the emitting clouds, which must fit within the slit. Finally, we calculated the depths of the clouds into the plane of the sky by dividing the cloud area by the projected slit width ($\sim$ 2 kpc) to verify that they were within the scale height of a typical disk. It is important to note that each ionized component may not be co-located in the slit, as the emission is spread across 2$\arcsec$ in the spectral direction.

\begin{figure*}
\centering
\subfigure{
\includegraphics[scale=0.46]{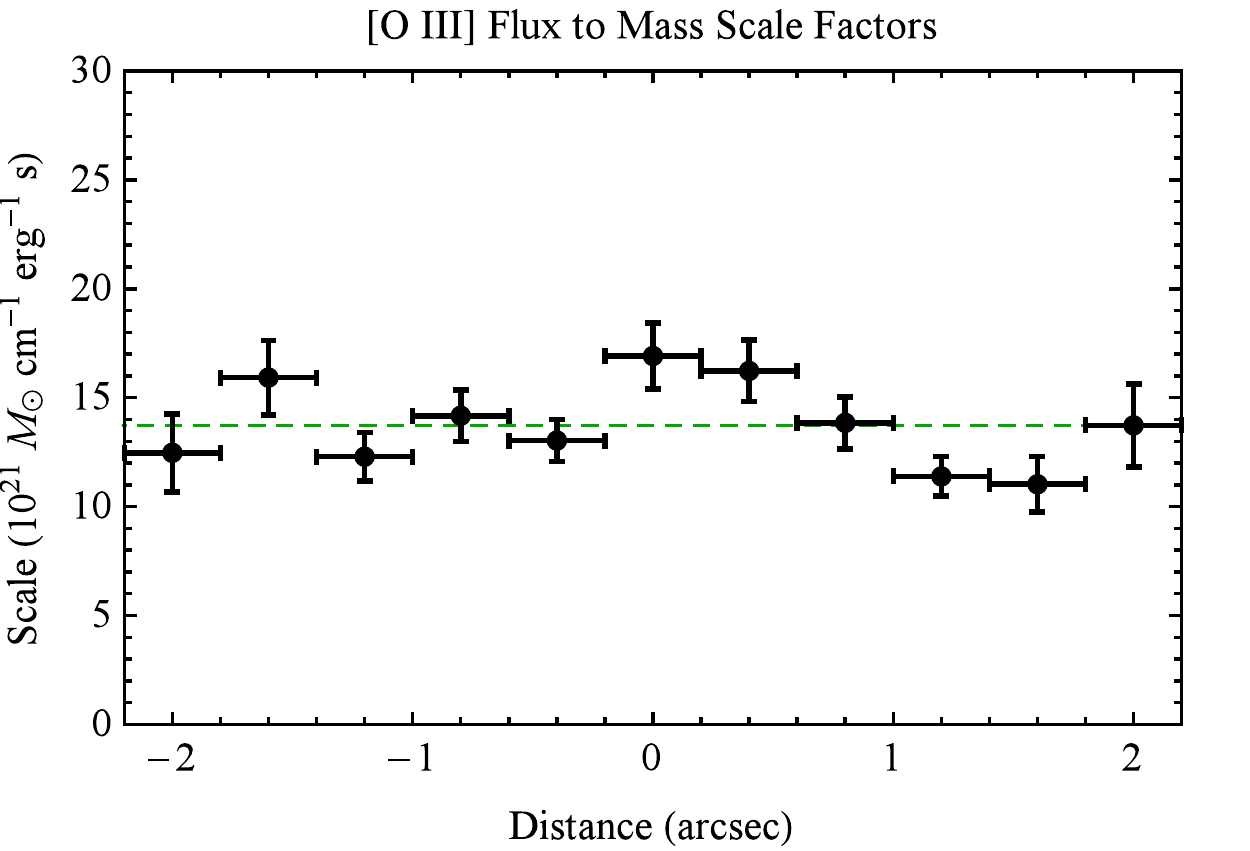}}
\subfigure{
\includegraphics[scale=0.46]{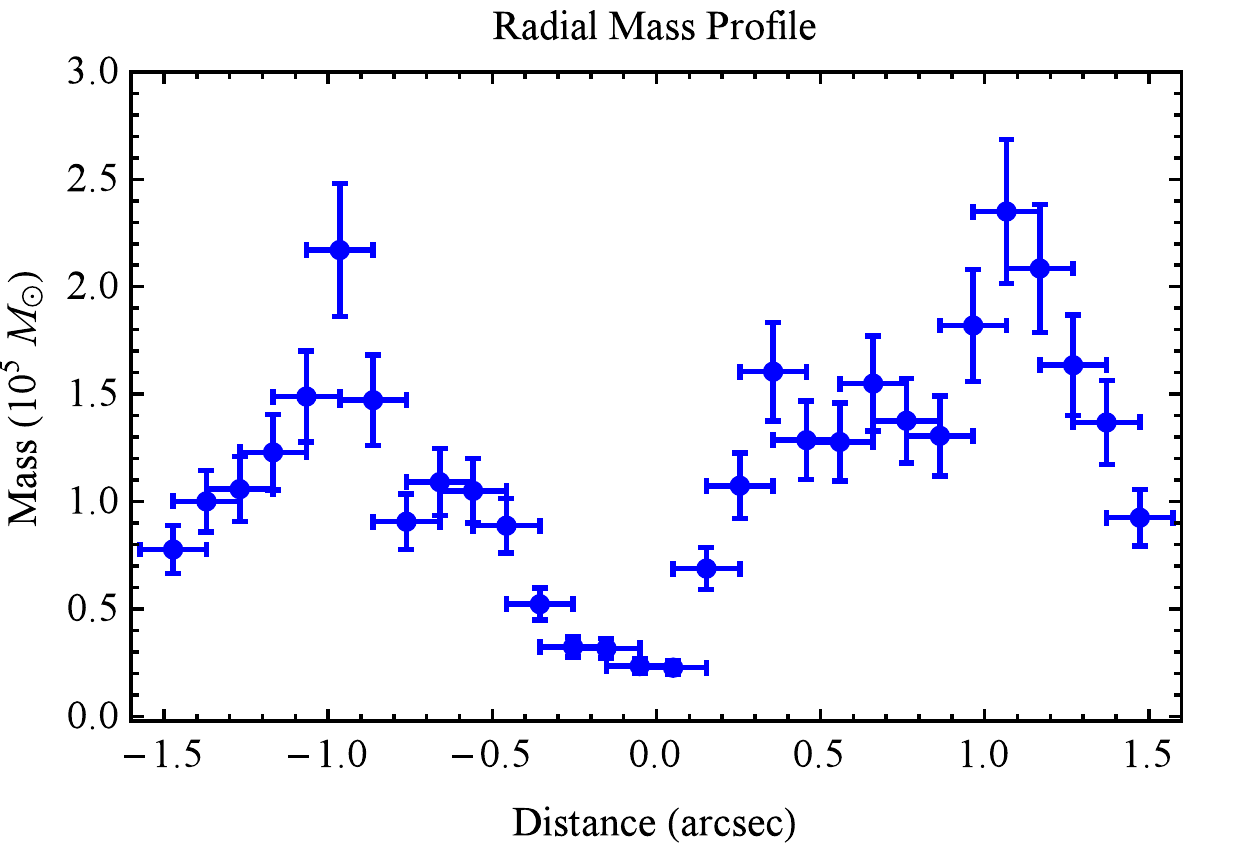}}
\subfigure{
\includegraphics[scale=0.46]{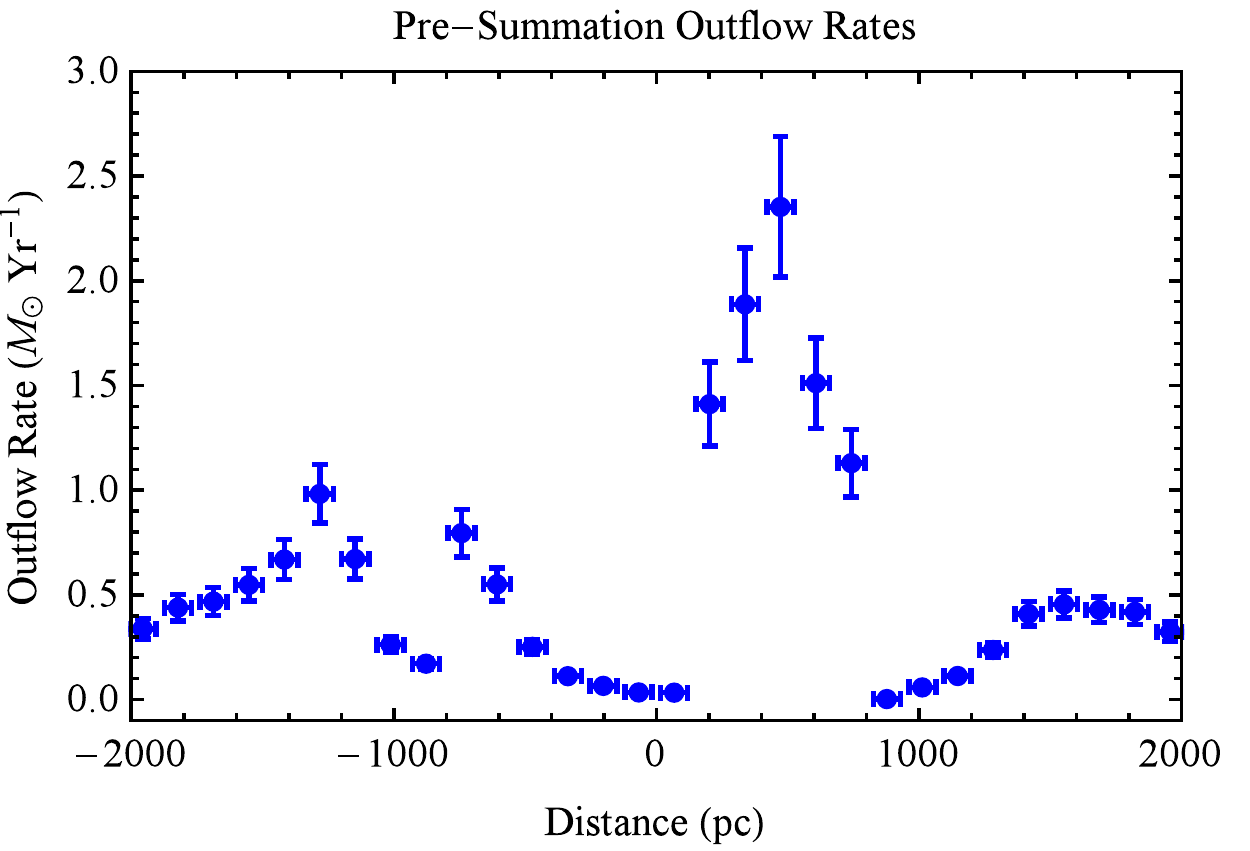}}
\caption{The left panel shows the [O~III] flux to mass scale factors, with the mean indicated by the green dashed line. The middle panel shows the ionized mass profile in units of $10^5 M_{\odot}$ calculated from the total flux in each semi-ellipse. The right panel shows the mass outflow rates assuming that all of the material is in outflow. Distances in arcseconds are observed, and distances in pc are corrected for projection.}
\label{scale}
\end{figure*}

\setlength{\tabcolsep}{0.008in}
\tabletypesize{\scriptsize}
\begin{deluxetable}{l|c|c|c|c|c|c|c|c|c|c|c|}
\tablenum{8}
\tablecaption{Predicted Cloudy Model Emission Line Ratios \vspace{-5pt}}
\tablehead{
\colhead{Line} & \colhead{--2$\farcs$0} & \colhead{--1$\farcs$6} & \colhead{--1$\farcs$2} & \colhead{--0$\farcs$8} & \colhead{--0$\farcs$4} & \colhead{0$\farcs$0} & \colhead{+0$\farcs$4} & \colhead{+0$\farcs$8} & \colhead{+1$\farcs$2} & \colhead{+1$\farcs$6}  & \colhead{+2$\farcs$0}}
\startdata
[S II]	 & 	0.20	 & 	0.17	 & 	0.30	 & 	0.21	 & 	0.18	 & 	0.18	 & 	0.15	 & 	0.19	 & 	0.25	 & 	0.17	 & 	0.18	\\ \relax
H$\delta$	 & 	0.26	 & 	0.26	 & 	0.26	 & 	0.26	 & 	0.26	 & 	0.26	 & 	0.26	 & 	0.26	 & 	0.26	 & 	0.26	 & 	0.26	\\ \relax
H$\gamma$	 & 	0.47	 & 	0.47	 & 	0.47	 & 	0.47	 & 	0.47	 & 	0.47	 & 	0.47	 & 	0.47	 & 	0.47	 & 	0.47	 & 	0.47	\\ \relax
[O III]	 & 	0.13	 & 	0.13	 & 	0.13	 & 	0.13	 & 	0.09	 & 	0.10	 & 	0.09	 & 	0.10	 & 	0.13	 & 	0.13	 & 	0.14	\\ \relax
He II	 & 	0.26	 & 	0.27	 & 	0.26	 & 	0.27	 & 	0.25	 & 	0.25	 & 	0.25	 & 	0.26	 & 	0.26	 & 	0.27	 & 	0.27	\\ \relax
H$\beta$	 & 	1.00	 & 	1.00	 & 	1.00	 & 	1.00	 & 	1.00	 & 	1.00	 & 	1.00	 & 	1.00	 & 	1.00	 & 	1.00	 & 	1.00	\\ \relax
[O III]	 & 	3.44	 & 	3.46	 & 	3.47	 & 	3.49	 & 	3.41	 & 	3.64	 & 	3.52	 & 	3.21	 & 	3.48	 & 	3.67	 & 	3.70	\\ \relax
[O III]	 & 	10.37	 & 	10.41	 & 	10.43	 & 	10.51	 & 	10.26	 & 	10.95	 & 	10.60	 & 	9.68	 & 	10.46	 & 	11.05	 & 	11.15	\\ \relax
[N I]	 & 	0.22	 & 	0.21	 & 	0.12	 & 	0.18	 & 	0.13	 & 	0.13	 & 	0.14	 & 	0.18	 & 	0.20	 & 	0.18	 & 	0.20	\\ \relax
He I	 & 	0.14	 & 	0.14	 & 	0.14	 & 	0.14	 & 	0.15	 & 	0.15	 & 	0.14	 & 	0.14	 & 	0.14	 & 	0.13	 & 	0.14	\\ \relax
[Fe VII]	 & 	0.05	 & 	0.05	 & 	0.05	 & 	0.05	 & 	0.05	 & 	0.05	 & 	0.04	 & 	0.03	 & 	0.05	 & 	0.06	 & 	0.06	\\ \relax
[O I]	 & 	0.52	 & 	0.47	 & 	0.45	 & 	0.48	 & 	0.39	 & 	0.41	 & 	0.38	 & 	0.44	 & 	0.55	 & 	0.43	 & 	0.47	\\ \relax
[O I]	 & 	0.17	 & 	0.15	 & 	0.14	 & 	0.15	 & 	0.12	 & 	0.13	 & 	0.12	 & 	0.14	 & 	0.17	 & 	0.14	 & 	0.15	\\ \relax
[N II]	 & 	0.68	 & 	0.71	 & 	0.67	 & 	0.79	 & 	0.81	 & 	0.76	 & 	0.74	 & 	0.85	 & 	0.73	 & 	0.66	 & 	0.63	\\ \relax
H$\alpha$	 & 	2.90	 & 	2.91	 & 	2.89	 & 	2.90	 & 	2.90	 & 	2.89	 & 	2.90	 & 	2.90	 & 	2.90	 & 	2.90	 & 	2.90	\\ \relax
[N II]	 & 	2.00	 & 	2.09	 & 	1.99	 & 	2.34	 & 	2.39	 & 	2.25	 & 	2.18	 & 	2.52	 & 	2.14	 & 	1.95	 & 	1.85	\\ \relax
[S II]	 & 	0.72	 & 	0.69	 & 	0.50	 & 	0.54	 & 	0.45	 & 	0.39	 & 	0.45	 & 	0.59	 & 	0.58	 & 	0.58	 & 	0.63	\\ \relax
[S II]	 & 	0.88	 & 	0.78	 & 	0.77	 & 	0.68	 & 	0.61	 & 	0.55	 & 	0.58	 & 	0.72	 & 	0.76	 & 	0.68	 & 	0.76		
\enddata
\tablecomments{The predicted Cloudy emission line ratios for our final composite models with the fractional weight of each component given in Table 6. The emission lines are in the same order as Tables 2--5.}
\end{deluxetable}

\section{Calculations}

\subsection{Mass of the Ionized Gas}

Calculating gas masses from our observed spectra and photoionization models is summarized here as described in \S5 of Paper I. For multi-component models, the mass in each component is calculated separately by dividing up the H$\beta$ luminosity, and then the masses are summed. This involves first determining the mass in the slit from
\begin{equation}
M_{slit} = N_H \mu m_p \left(\frac{L(H\beta)}{F(H\beta)_{m}}\right),
\end{equation}
\noindent
where $N_H$ is the model hydrogen column density, $\mu$ is the mean mass per proton ($\sim$ 1.4 for our abundances), $m_p$ is the proton mass, $F(H\beta)_{m}$ is the H$\beta$ model flux, and $L(H\beta)$ is the luminosity of H$\beta$ calculated from the extinction-corrected flux and distance. This establishes a direct relationship describing the number of H$\beta$ photons emitted per unit mass at a specific density. To determine the ionized gas mass at all radii we use available [O~III] imaging, so we derive a scale factor that allows us to calculate mass from observed [O~III] flux rather than H$\beta$ luminosity. Specifically,
\begin{equation}
S = \left(\frac{M_{slit} n_\mathrm{H}}{F_{\lambda 5007}} \right),
\end{equation}
\noindent
where $M_{slit}$ is the ionized mass in the slit from Equation 8, $n_\mathrm{H}$ is the fractional weighted mean hydrogen number density (cm$^{-3}$) for all components, and $F_{\lambda 5007}$ is the extinction-corrected [O~III] emission line flux from our spectra. Figure \ref{scale} shows the mean scale factor, with S $= (1.37 \pm 0.20) \times 10^{22}$ M$_{\odot}$ cm$^{-1}$ erg$^{-1}$ s, and the 1$\sigma$ error corresponding to a fractional uncertainty of 14.3\%. For position $-2\farcs0$ the scale factor was discrepantly low and was replaced with a mean value. We take an average value of the scale factors and then calculate the ionized gas mass for each image flux using
\begin{equation}
M_{ion} = S \left(\frac{F_\mathrm{[O~III]}}{n_\mathrm{H}} \right),
\end{equation}
\noindent
where $F_\mathrm{[O~III]}$ is the image flux in each semi-annulus of width $\delta r$ (Figure \ref{imaging}) and n$_\mathrm{H}$ is the hydrogen number density found by interpolating between our model points.

\subsection{Outflow Parameters}

The goal of our study is to quantify the power and impact of the NLR outflows. This is encapsulated by mass outflow rates, kinetic energy and luminosity, and momenta. The mass outflow rate ($\dot{M}_{out}$) is given by
\begin{equation}
\dot{M}_{out} = \left(\frac{Mv}{\delta r} \right),
\end{equation}
\noindent
where $M$ is the mass in each semi-annulus, $v$ is the deprojected velocity corrected for inclination and position angle on the sky (\S 3.2 of Paper I), and $\delta r$ is the deprojected width for each extraction. Deprojecting the distances results in a bin width that is 31\% larger than the observed value, thus each observed 103 pc bin spans $\delta r \approx$ 135 pc after deprojection. The kinetic energy ($E$), kinetic energy flow rate ($\dot E$), momentum ($p$), and momentum flow rate ($\dot p$) are given by
\begin{equation}
E = \frac{1}{2}{M}_{out} v^2,
\end{equation}
\begin{equation}
\dot E = \frac{1}{2} \dot{M}_{out} v^2,
\end{equation}
\begin{equation}
p = M_{out}v,
\end{equation}
\begin{equation}
\dot p = \dot M_{out}v.
\end{equation}
We do not include contributions from velocity spread, such as turbulence, which would add a $\sigma_v$ term to these expressions. We obtain a single profile for each of these quantities by radially summing the values derived for each of the semi-annuli, and the mass profile and outflow rates for each semi-ellipse can be seen in Figure \ref{scale}.

\begin{figure*}
\centering
\vspace{-5pt}
\subfigure{
\includegraphics[scale=0.95]{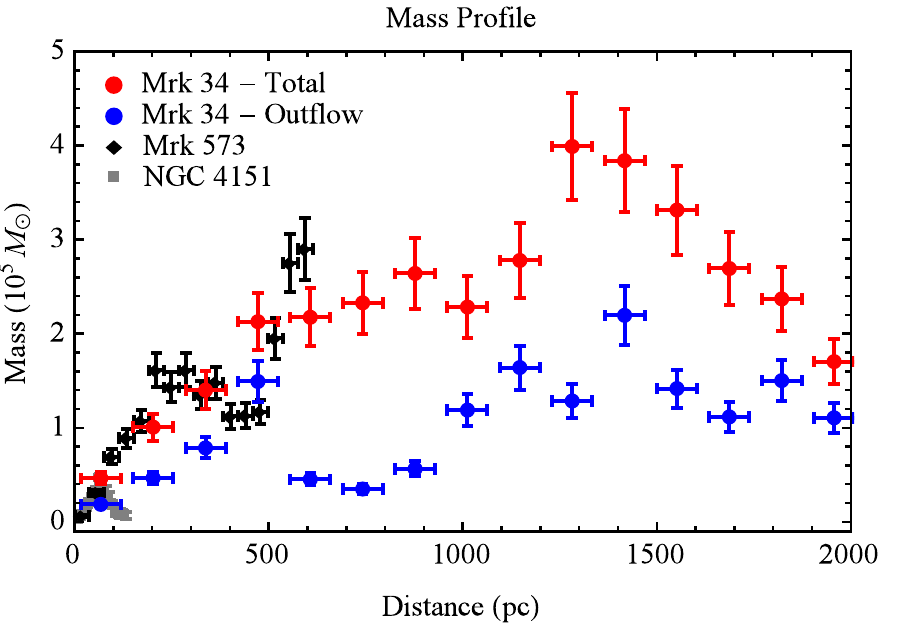}}
\vspace{-10pt}
\subfigure{
\includegraphics[scale=0.95]{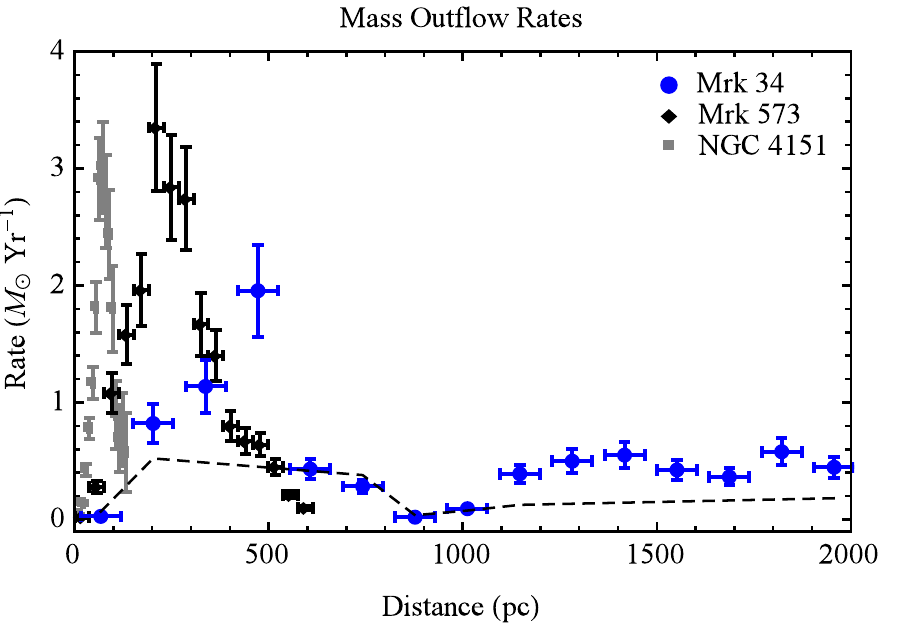}}
\subfigure{
\includegraphics[scale=0.95]{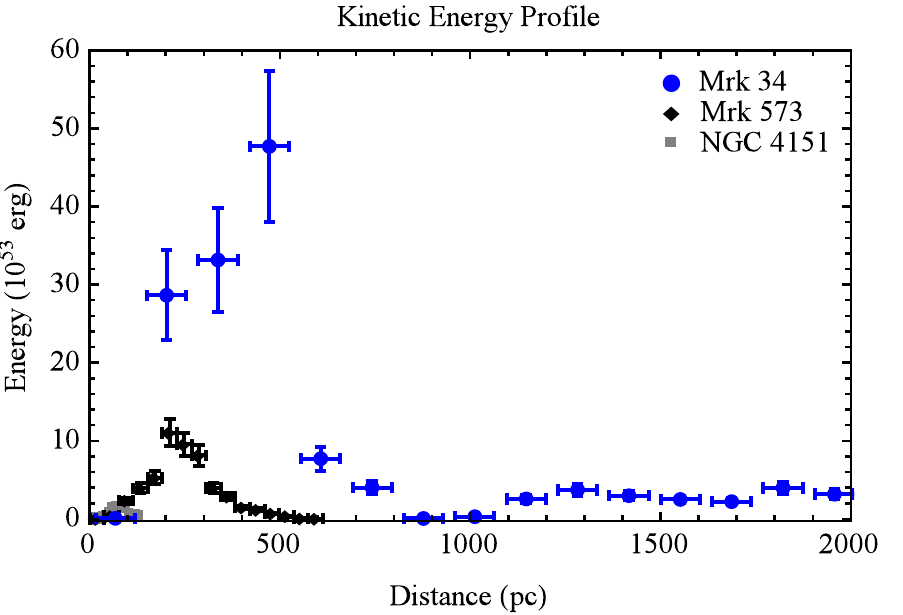}}
\vspace{-10pt}
\subfigure{
\includegraphics[scale=0.95]{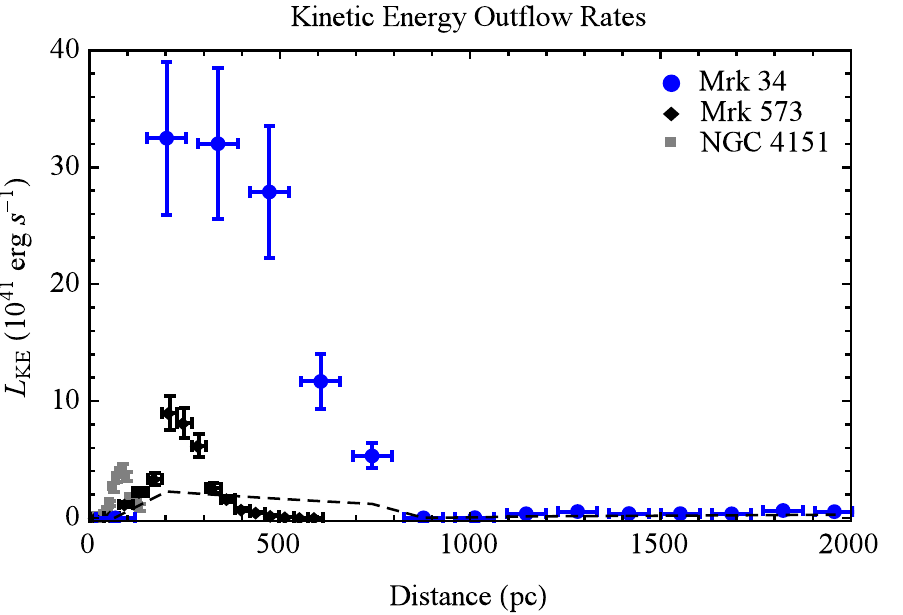}}
\subfigure{
\includegraphics[scale=0.95]{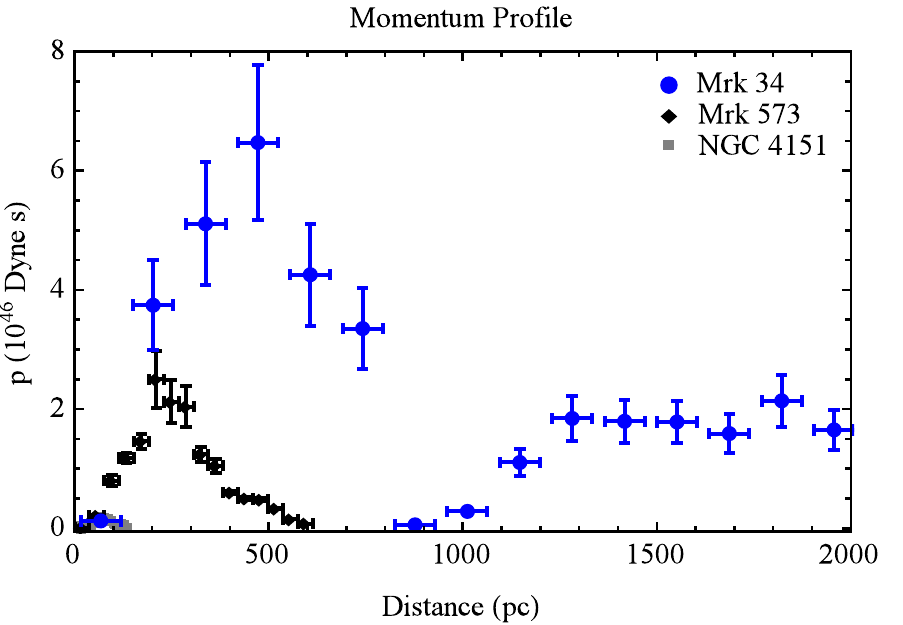}}
\subfigure{
\includegraphics[scale=0.96]{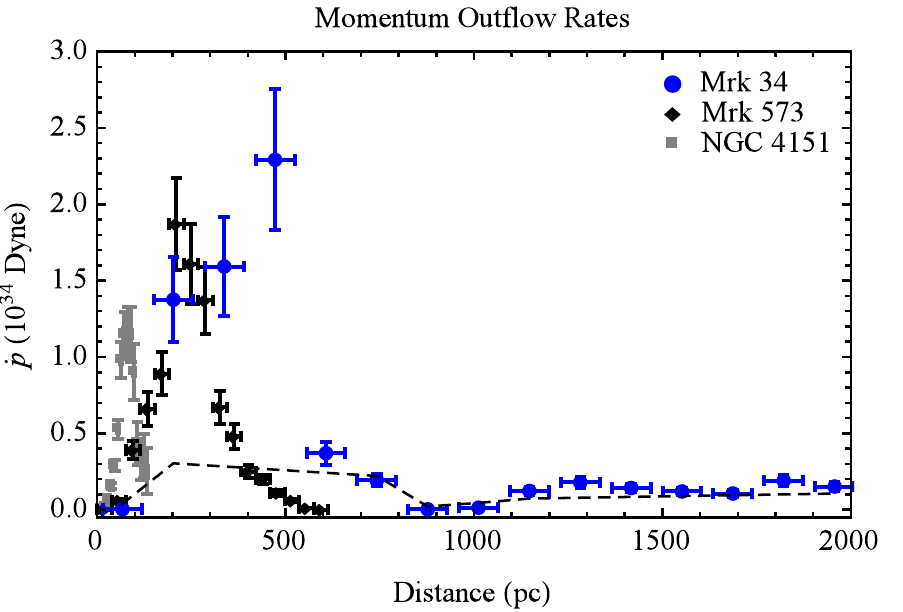}}
\vspace{-5pt}
\caption{Top left to bottom right are the azimuthally summed mass profiles, mass outflow rates, kinetic energy profiles, kinetic energy outflow rates, momentum profiles, and momentum outflow rates for Mrk 34, Mrk 573 \citep{revalski2018}, and NGC 4151 \citep{crenshaw2015}. The red points represent the result that is obtained assuming that all of the mass is in outflow, and the blue points show the net result after multiplying by the fraction of flux in outflow as shown in Figure \ref{kinematics}. The dashed lines represent the profiles that would result from the mass in the center bin ($M \approx 5 \times 10^4 M_{\odot}$) traveling through the velocity profile. Quantities are per bin, and targets have different bin sizes. (\textcolor{red}{{\it See appended erratum.}})}
\label{results}
\end{figure*}

\setlength{\tabcolsep}{0.121in}
\tabletypesize{\small}
\begin{deluxetable*}{c|c|c|c|c|c|c|c|}
\tablenum{9}
\tablecaption{Radial Mass Outflow \& Energetic Results \vspace{-6pt}}
\tablehead{
\colhead{Distance} & \colhead{Velocity} & \colhead{Mass} & \colhead{$\dot{M}$} & \colhead{Energy} & \colhead{$\dot{E}$} & \colhead{Momentum} & \colhead{$\dot{P}$}\\
\colhead{(pc)} & \colhead{(km s$^{-1}$)} & \colhead{(10$^5$ M$_{\odot}$)} & \colhead{(M$_{\odot}$ yr$^{-1}$)} & \colhead{(10$^{53}$ erg)} & \colhead{(10$^{41}$ erg s$^{-1}$)} & \colhead{(10$^{46}$ dyne s)} & \colhead{(10$^{34}$ dyne)}\\
\colhead{(1)} & \colhead{(2)} & \colhead{(3)} & \colhead{(4)} & \colhead{(5)} & \colhead{(6)} & \colhead{(7)} & \colhead{(8)}
}
\startdata
67.5	 & 	191.8	 & 	0.19	$\pm$	0.03	 & 	0.03	$\pm$	0.01	 & 	0.07	$\pm$	0.01	 & 	0.01	$\pm$	0.00	 & 	0.45	$\pm$	0.09	 & 	0.00	$\pm$	0.00	\\
202.4	 & 	2323.5	 & 	0.47	$\pm$	0.07	 & 	0.82	$\pm$	0.17	 & 	28.63	$\pm$	5.78	 & 	32.48	$\pm$	6.55	 & 	4.10	$\pm$	0.83	 & 	1.38	$\pm$	0.28	\\
337.4	 & 	1910.5	 & 	0.79	$\pm$	0.11	 & 	1.14	$\pm$	0.23	 & 	33.15	$\pm$	6.69	 & 	32.01	$\pm$	6.46	 & 	6.51	$\pm$	1.31	 & 	1.59	$\pm$	0.32	\\
472.4	 & 	1729.5	 & 	1.49	$\pm$	0.21	 & 	1.96	$\pm$	0.39	 & 	47.70	$\pm$	9.62	 & 	27.88	$\pm$	5.62	 & 	9.15	$\pm$	1.85	 & 	2.29	$\pm$	0.46	\\
607.3	 & 	1261.4	 & 	0.45	$\pm$	0.06	 & 	0.43	$\pm$	0.09	 & 	7.71	$\pm$	1.56	 & 	11.68	$\pm$	2.36	 & 	6.86	$\pm$	1.38	 & 	0.37	$\pm$	0.07	\\
742.3	 & 	1072.8	 & 	0.35	$\pm$	0.05	 & 	0.28	$\pm$	0.06	 & 	4.02	$\pm$	0.81	 & 	5.30	$\pm$	1.07	 & 	6.83	$\pm$	1.38	 & 	0.19	$\pm$	0.04	\\
877.2	 & 	47.1	 & 	0.56	$\pm$	0.08	 & 	0.02	$\pm$	0.00	 & 	0.05	$\pm$	0.01	 & 	0.00	$\pm$	0.00	 & 	0.51	$\pm$	0.10	 & 	0.00	$\pm$	0.00	\\
1012.2	 & 	102.9	 & 	1.19	$\pm$	0.17	 & 	0.09	$\pm$	0.02	 & 	0.28	$\pm$	0.06	 & 	0.02	$\pm$	0.00	 & 	1.45	$\pm$	0.29	 & 	0.01	$\pm$	0.00	\\
1147.1	 & 	314.7	 & 	1.64	$\pm$	0.23	 & 	0.39	$\pm$	0.08	 & 	2.56	$\pm$	0.52	 & 	0.36	$\pm$	0.07	 & 	8.38	$\pm$	1.69	 & 	0.12	$\pm$	0.02	\\
1282.1	 & 	512.2	 & 	1.28	$\pm$	0.18	 & 	0.50	$\pm$	0.10	 & 	3.72	$\pm$	0.75	 & 	0.54	$\pm$	0.11	 & 	12.83	$\pm$	2.59	 & 	0.18	$\pm$	0.04	\\
1417.0	 & 	330.9	 & 	2.20	$\pm$	0.31	 & 	0.55	$\pm$	0.11	 & 	2.97	$\pm$	0.60	 & 	0.37	$\pm$	0.08	 & 	8.28	$\pm$	1.67	 & 	0.14	$\pm$	0.03	\\
1552.0	 & 	394.8	 & 	1.42	$\pm$	0.20	 & 	0.42	$\pm$	0.09	 & 	2.49	$\pm$	0.50	 & 	0.36	$\pm$	0.07	 & 	7.10	$\pm$	1.43	 & 	0.12	$\pm$	0.02	\\
1687.0	 & 	431.2	 & 	1.12	$\pm$	0.16	 & 	0.36	$\pm$	0.07	 & 	2.21	$\pm$	0.45	 & 	0.35	$\pm$	0.07	 & 	5.77	$\pm$	1.16	 & 	0.11	$\pm$	0.02	\\
1821.9	 & 	509.1	 & 	1.50	$\pm$	0.21	 & 	0.58	$\pm$	0.12	 & 	3.98	$\pm$	0.80	 & 	0.63	$\pm$	0.13	 & 	11.49	$\pm$	2.32	 & 	0.19	$\pm$	0.04	\\
1956.9	 & 	533.2	 & 	1.11	$\pm$	0.16	 & 	0.45	$\pm$	0.09	 & 	3.16	$\pm$	0.64	 & 	0.53	$\pm$	0.11	 & 	8.86	$\pm$	1.79	 & 	0.15	$\pm$	0.03
\vspace{-3pt}
\enddata
\tablecomments{Numerical results for the mass and energetic quantities as a function of radial distance for the outflowing gas component. Columns are (1) deprojected distance from the nucleus, (2) mass-weighted mean velocity, (3) gas mass in units of 10$^{5}$ M$_{\odot}$, (4) mass outflow rates, (5) kinetic energies, (6) kinetic energy outflow rates, (7) momenta, and (8) momenta flow rates. These results, shown in Figure \ref{results}, are the sum of the individual radial profiles calculated for each of the semi-annuli (see Figures \ref{imaging} and \ref{scale}). The value at each distance is the quantity contained within the annulus of width $\delta r$. (\textcolor{red}{{\it See appended erratum.}})}
\end{deluxetable*}

\section{Results}

In Figure \ref{results} and Table 9 we present our mass outflow rates and energetics as functions of distance from the nucleus for Mrk 34. We also show the results for NGC 4151 \citep{crenshaw2015} and Mrk 573 \citep{revalski2018} for comparison. All data are available in machine-readable form by request to M.R. The quantities displayed are the value within each bin of width $\delta r$, and each target has a different bin size such that Mrk 573 and Mrk 34 have a similar total gas mass despite their appearance in Figure \ref{results}.

The outflow reaches a maximum radial extent of $\sim$~2~kpc from the nucleus and contains an ionized gas mass of $M \approx 1.6 \times 10^6 M_{\odot}$. The mass of ionized gas in the rotational component is nearly equal (Figure 10), yielding a total ionized gas mass of $M \approx 3.2 \times 10^6 M_{\odot}$ for the NLR. The total kinetic energy of the outflow over all distances is $E \approx 1.4 \times 10^{55}$ erg. The mass outflow rate reaches a peak value of $\dot M_{out} \approx$ 2.0 $\pm$ 0.4 $M_{\odot}$ yr$^{-1}$ at a distance of 470 pc from the nucleus and then decreases to nearly zero at $\sim$ 900 pc before rising slightly out to distances of $\sim$ 2 kpc. As discussed in \cite{fischer2017} and \S7, points beyond 1.5 kpc represent disturbed kinematics that may not be in radial outflow and should be considered upper limits. Beyond these distances, the observed {\it HST} kinematics are generally consistent with rotation \citep{fischer2018}.

The dashed lines in Figure 10 represent the mass outflow rates and energetics that would be observed if the mass in the central bin ($M \approx 4.6 \times 10^4 M_{\odot}$) propagated through the velocity profile. At 470 pc, where the outflow peaks, this is $\sim$ 5 times smaller than the observed value, indicating that the outflow is likely accelerated in-situ such that material does not originate at small radii and travel large distances, but is actually host galaxy material driven by the nuclear radiation field.

Compared to the AGN bolometric luminosity of Mrk~34, Log(L$_{\mathrm{bol}}$) = 46.2 $\pm$ 0.4 erg s$^{-1}$, the peak kinetic luminosity reaches $\sim$ 0.1--0.5\% of $L_{\mathrm{bol}}$. The photon momentum ($L/c$) from the bolometric luminosity is $\dot{p} \approx 5.3 \times 10^{35}$ Dyne, and the peak momentum flow rate is $\dot{p} \approx 2.3 \times 10^{34}$ Dyne. Thus, the peak outflow momentum rate is $\sim$ 2--11\% of the AGN photon momentum. Interestingly, the masses, velocities, and extraction sizes conspire so that all objects have peak outflow rates $\sim$ 2--3 M$_{\odot}$ yr$^{-1}$. This makes a comparison of the energetics more meaningful.

\section{Discussion}

\subsection{Comparison with Previous Work}

In comparison to Mrk 573 and NGC 4151, the outflows in Mrk 34 reach significantly larger distances from the nucleus. However, on average only half of the gas at each location is participating in the outflow. The increased complexity of the kinematics also corresponds to more uncertainty in the proper correction for projection effects. The clear high velocity separations at $r\leq0\farcs7$ are consistent with our assumptions of radial outflow along the host galaxy disk. At larger radii, the emission line splitting may indicate an ``ablation'' scenario where pure radial outflow transitions into the ablation of gas off of nuclear spiral arms as discussed in \cite{fischer2018}. In this case, the motion is not purely radial and the projection effects are less severe, and the mass outflow rate and energetic points at all radii $>$ 1.5 kpc are upper limits.

Our new APO observations also allow us to probe fainter and more extended emission line knots, including those that fall outside of the narrow {\it HST} slits. These deep exposures reveal evidence of weak, high velocity outflow components that extend to $\sim$ $2\farcs5$, or deprojected distances of $\sim$ 3.1~kpc. While these weak components likely contain little mass, they indicate the presence of outflows at nearly twice the distance that we detected in \cite{fischer2018}. There is also indication from the APO observations that some outflow or ablation continues out to $\sim$ 5 kpc, with velocities and FWHM that are larger than systemic for a number of components, including the brightest one at PA = 163$\degr$. These components are not included in our mass outflow calculations and require high spatial resolution IFU observations to be properly characterized.

It is encouraging for future studies that the emission line ratios derived from the {\it HST} and APO observations demonstrate quantitative agreement, as the ground-based observations may be obtained more easily. This could indicate that less stringent data requirements may suffice for this type of study; however, high spatial resolution observations are ultimately needed to constrain the velocity and mass profiles. In addition, it is possible to create photoionization models that sufficiently constrain the physical conditions in the gas for determining outflow rates and energetics with fewer emission lines than were available for our studies in \cite{crenshaw2015} and \cite{revalski2018}. Ultimately, enough emission lines to constrain the gas ionization parameter, number and column density, temperature, and reddening are needed to create a complete model. At a practical minimum, this may include: H$\gamma$ $\lambda 4340$, [O~III] $\lambda 4363$, He II $\lambda 4686$, [O~III] $\lambda 5007$, [O~I] $\lambda 6300$, H$\alpha$ $\lambda 6563$, [N~II] $\lambda 6584$, and [S~II] $\lambda \lambda 6716, 6731$. However, we caution that constraining the density from [S~II] alone can easily bias gas mass estimates when multi-component models are more appropriate. In the case of Mrk 573, the mean model density is $>$ 1 dex higher near the nucleus, corresponding to significantly less mass than that estimated from [S~II] alone.

Finally, the dichotomy in the derived line ratios, abundances, temperatures, and densities on either side of the nucleus is intriguing. It is difficult to conceive of a physical model for this stark bimodality and may further indicate that one side of the NLR is exposed to a more heavily filtered SED. This would affect the quantities derived from the emission line diagnostics, as the various models generally assume a particular power-law SED. Alternatively, the bimodality could also be due to a small tilt of the NLR relative to the host galaxy, which is supported by the small but systemically higher E(B-V) values to the SE that visually correspond to a strong dust lane in Figure \ref{structure}. Additional possibilities, including variations in dust content and varying the locations of each ionized component within the spectral extraction, are discussed in \S4.3.

\subsection{Comparison with Global Outflow Rates}

We refer to single value mass outflow rates that are calculated from mean conditions across the entire NLR as ``global'' outflow rates. We compare our results with these techniques to explore systematics and better understand the uncertainties that are introduced by various assumptions.

Following the techniques of \cite{nesvadba2006} and \cite{bae2017} we calculate the NLR mass using $M = (9.73\times10^8 M_{\odot}) \times L_{H\alpha, 43} \times n^{-1}_{e,100}$, where $L_{H\alpha, 43}$ is the H$\alpha$ luminosity in units of 10$^{43}$ erg s$^{-1}$, and $n^{-1}_{e,100}$ is the electron density in units of 100 cm$^{-3}$. Here we scale the [O~III] luminosity by the average H$\alpha$/[O~III] ratio as a proxy for $L_{H\alpha}$. Using an average velocity of 1000 km s$^{-1}$, we find $\dot{M}_{out} \approx 3-63$ $M_{\odot}$ yr$^{-1}$ for $n_e = 3000-150$ cm$^{-3}$, which is the range of our derived [S~II] densities. The corresponding NLR mass estimate is $\sim$ 0.6--13 times the value from our models ($\sim$ 3.1$\times$10$^6$ M$_{\odot}$), emphasizing the importance of density choice on the final mass and outflow energetics. A value of $n_e = 100$ cm$^{-3}$ is often adopted in the literature, which may significantly overestimate the mass and energetics of NLR outflows.

Using our findings from Paper I (\S8.1), we no longer consider geometric methods that employ filling factors to yield a reliable estimate of the gas mass, unless the filling factor is derived for individual objects using a physical tracer of the gas mass, such as emission line modeling or narrow band imaging.

\subsection{Implications for Feedback}

The outflows in Mrk 34 reach typical galaxy bulge radii and may deliver important feedback to the host galaxy. The global kinetic energy ($M = 1.6 \times 10^6 M_{\odot}$, $\delta r = 2000$ pc, $V = 1000$ km s$^{-1}$) of the outflow is $\sim$ 0.1\% of the AGN bolometric luminosity (L$_\mathrm{bol} \approx 10^{45.9}$ erg s$^{-1}$), which approaches the 0.5\%-5\% range used in some models of efficient feedback \citep{dimatteo2005, hopkins2010}. This result is consistent with the old nuclear stellar population \citep{gonzalezdelgado2001} and lack of current star formation \citep{wang2007}; however, mass modeling of the galaxy is required to reveal whether or not the outflows are ultimately capable of escaping the gravitational potential of the host galaxy bulge.

Under the idea that the outflows are radiatively driven \citep{fischer2017, wylezalek2018}, the mix of outflow and rotational kinematics at all radii could indicate that the coupling efficiency between the ionizing radiation and gas is very sensitive to the physical conditions in the outflow \citep{zubovas2018}. The lack of strong high ionization lines seen in the spectra as compared with NGC 4151 and Mrk 573 would lend support to this idea, but a detailed study of the gas acceleration through photoionization models is needed.

Finally, a multiwavelength picture incorporating the UV/X-ray and molecular gas phases would further illuminate the details behind the driving mechanisms \citep{cicone2018}. NGC 4151 in particular displays ultra-fast outflows (UFOs), UV/X-ray winds, and NLR outflows \citep{tombesi2010, tombesi2011, tombesi2013, wang2011, crenshaw2012, crenshaw2015}, some of which display similar mass outflow rates or energetics to the NLR outflows. This is discussed further in \S8.3 of Paper I.
\vspace{-1.2em}
\section{Conclusions}

Using long-slit spectroscopy, [O~III] imaging, and photoionization modeling, we determined spatially resolved mass outflow rates and energetics for the NLR of the QSO2 Mrk~34. This is the first spatially resolved outflow profile for a QSO2, following the results of \cite{crenshaw2015, revalski2018}, and \cite{venturi2018} for nearby Seyferts. Our conclusions are the following: 

\begin{enumerate}
\item The outflow contains $M \approx 1.6 \times 10^6 M_{\odot}$ of ionized gas, with a total kinetic energy of $E \approx 1.4 \times 10^{55}$ erg. These are larger than for the NLRs of the lower luminosity galaxies NGC 4151 \citep{crenshaw2015} and Mrk 573 \citep{revalski2018}.

\item The outflows extend to $\sim$~2~kpc, reaching a peak mass outflow rate of $\dot M_{out} \approx$ 2.0 $\pm$ 0.4 $M_{\odot}$ yr$^{-1}$ at a distance of 470 pc from the SMBH, with evidence of disturbed kinematics extending up to $\sim$ 5 kpc. The resulting mass outflow rates are consistent with in-situ acceleration where host galaxy material in the NLR is accelerated by nuclear radiation.

\item The presence of multiple kinematic components indicates that only a portion of the NLR gas is in outflow and that global outflow rate techniques may overestimate $\dot M$ in these cases. Without photoionization modeling, gas masses may also be overestimated.

\item The presence of rotation and outflow at all radii in this more luminous target may indicate that the coupling efficiency between the radiation and gas, and the ability of the AGN to radiatively drive outflows, is sensitive to the physical conditions in the gas, as well as the luminosity and driving timescale.
\end{enumerate}

(\textcolor{red}{{\it See appended erratum for updated values.}})\\

The authors thank the anonymous referee for helpful comments that improved the clarity of this paper. M.R. gratefully acknowledges support from the National Science Foundation through the Graduate Research Fellowship Program (GRFP). This material is based upon work supported by the National Science Foundation Graduate Research Fellowship Program under Grant No. DGE-1550139. Any opinions, findings, and conclusions or recommendations expressed in this material are those of the author(s) and do not necessarily reflect the views of the National Science Foundation. Support for this work was provided by NASA through grant number HST-AR-14290.001-A from the Space Telescope Science Institute, which is operated by AURA, Inc., under NASA contract NAS 5-26555. Basic astrophysics research at the Naval Research Laboratory is supported by 6.1 base money. T.C.F. was supported by an appointment to the NASA Postdoctoral Program at the NASA Goddard Space Flight Center, administered by the Universities Space Research Association under contract with NASA. T.S.-B. acknowledges support from the Brazilian institutions CNPq (Conselho Nacional de Desenvolvimento Cient\'ifico e Tecnol\'ogico) and FAPERGS (Funda\c c\~ao de Amparo \`a Pesquisa do Estado do Rio Grande do Sul). This paper used the photoionization code Cloudy, which can be obtained from \url{http://www.nublado.org} and the Atomic Line List available at \url{http://www.pa.uky.edu/~peter/atomic/}. This research has made use of NASA's Astrophysics Data System.

\facility{{\it Facilities: }HST (STIS, WFPC2, WFC3), ARC (DIS)}

\appendix



In this Appendix we include additional figures and tables that further illuminate the analysis. In Figure 11, we provide the Gaussian profile decomposition for important emission lines at each spatial position in our APO long-slit observations along PA = 163$\degr$. Figure 12 displays the spatially resolved BPT diagrams for each of the four APO long-slit position angles, with the fluxes of all kinematic components added together. These are further divided into the ratios obtained for each individual Gaussian component at each position angle in Figure 13. In Tables 10 and 11 we provide the emission line ratios for each kinematic component measured in the HST and APO spectra, respectively, that were summed to produce the integrated ratios in Tables 2 and 4.

\begin{figure*}[h]
\centering
\subfigure{
\includegraphics[scale=0.34]{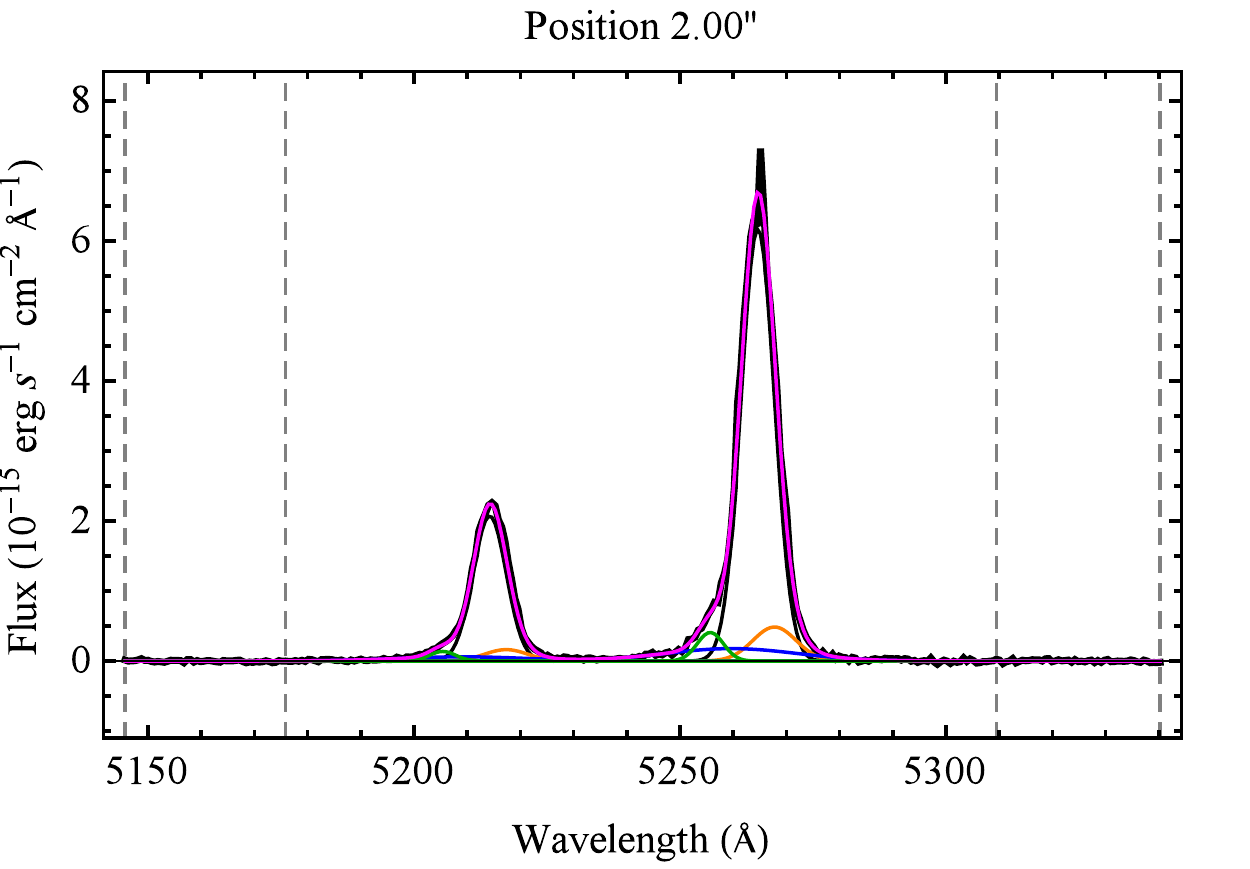}}
\subfigure{
\includegraphics[scale=0.34]{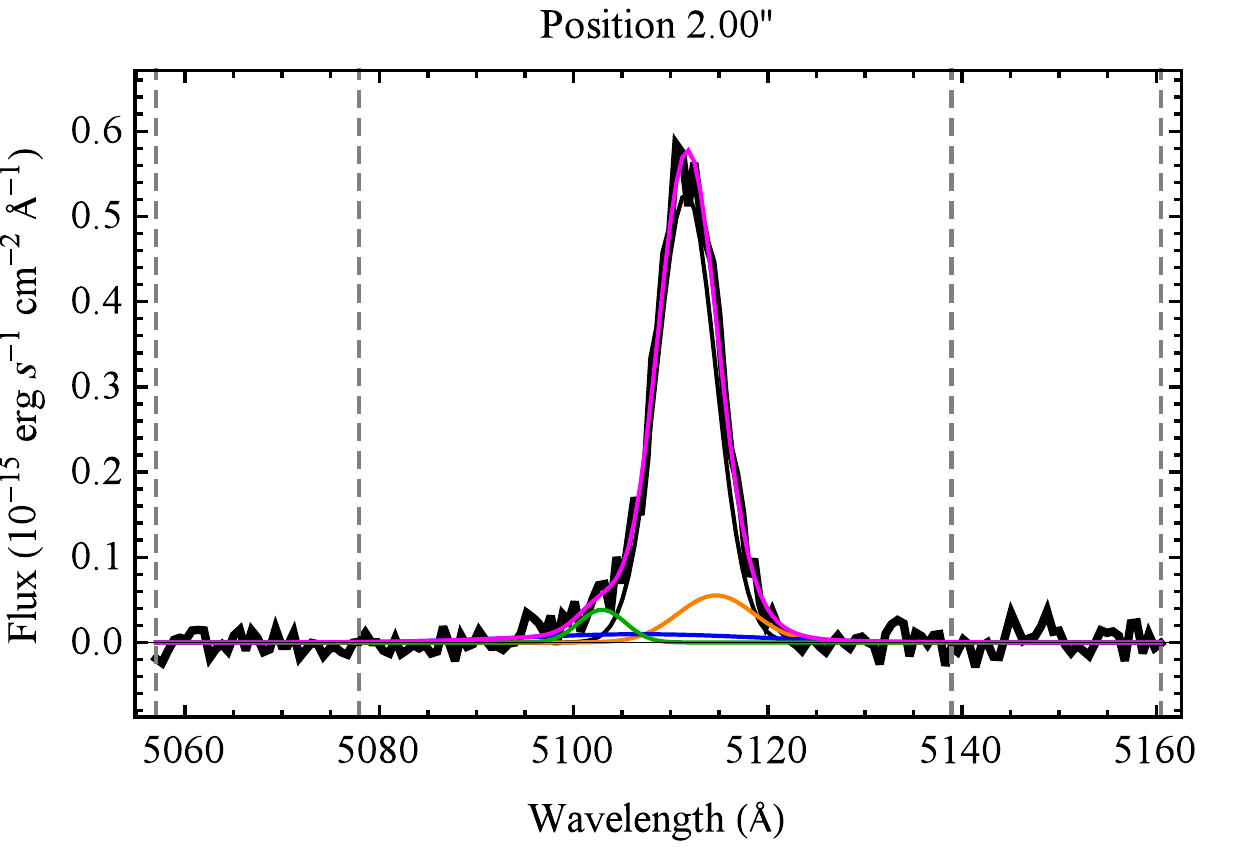}}
\subfigure{
\includegraphics[scale=0.34]{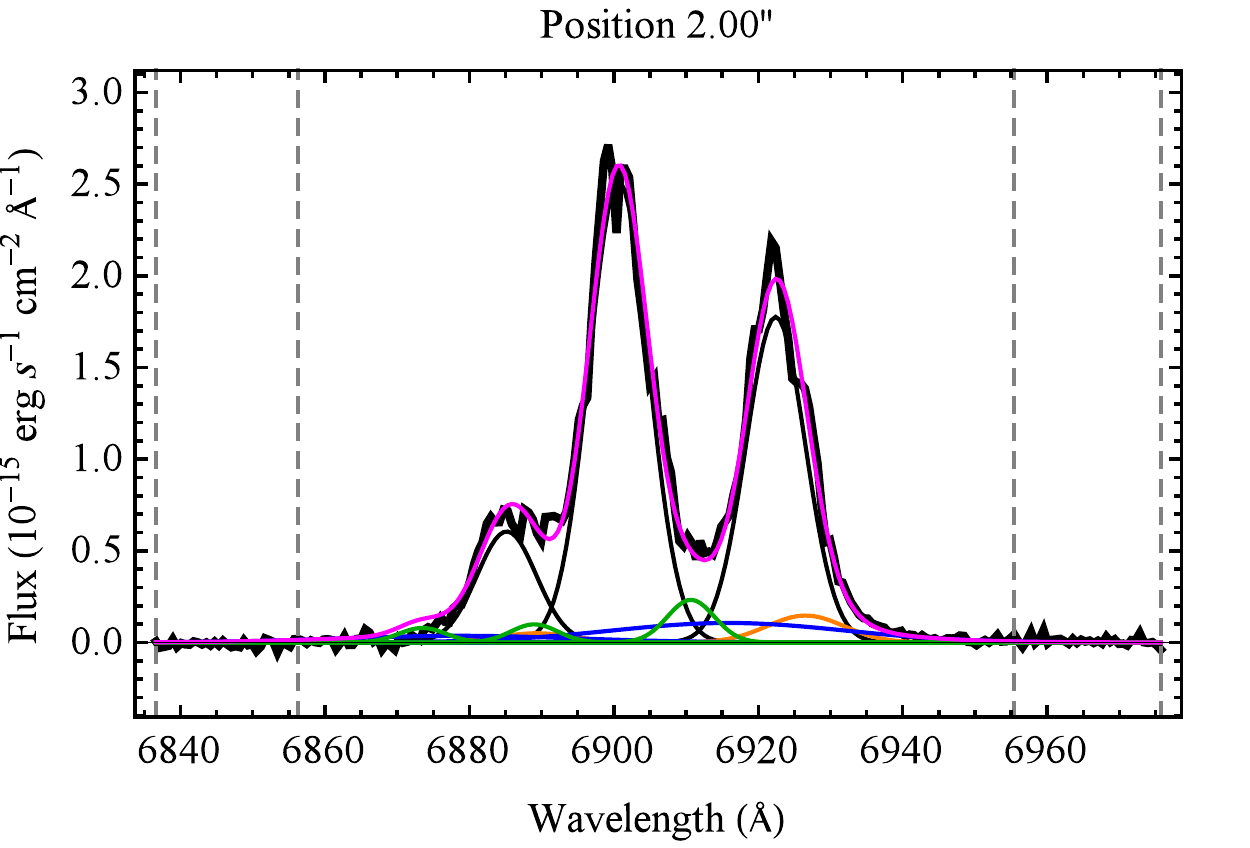}}
\subfigure{
\includegraphics[scale=0.34]{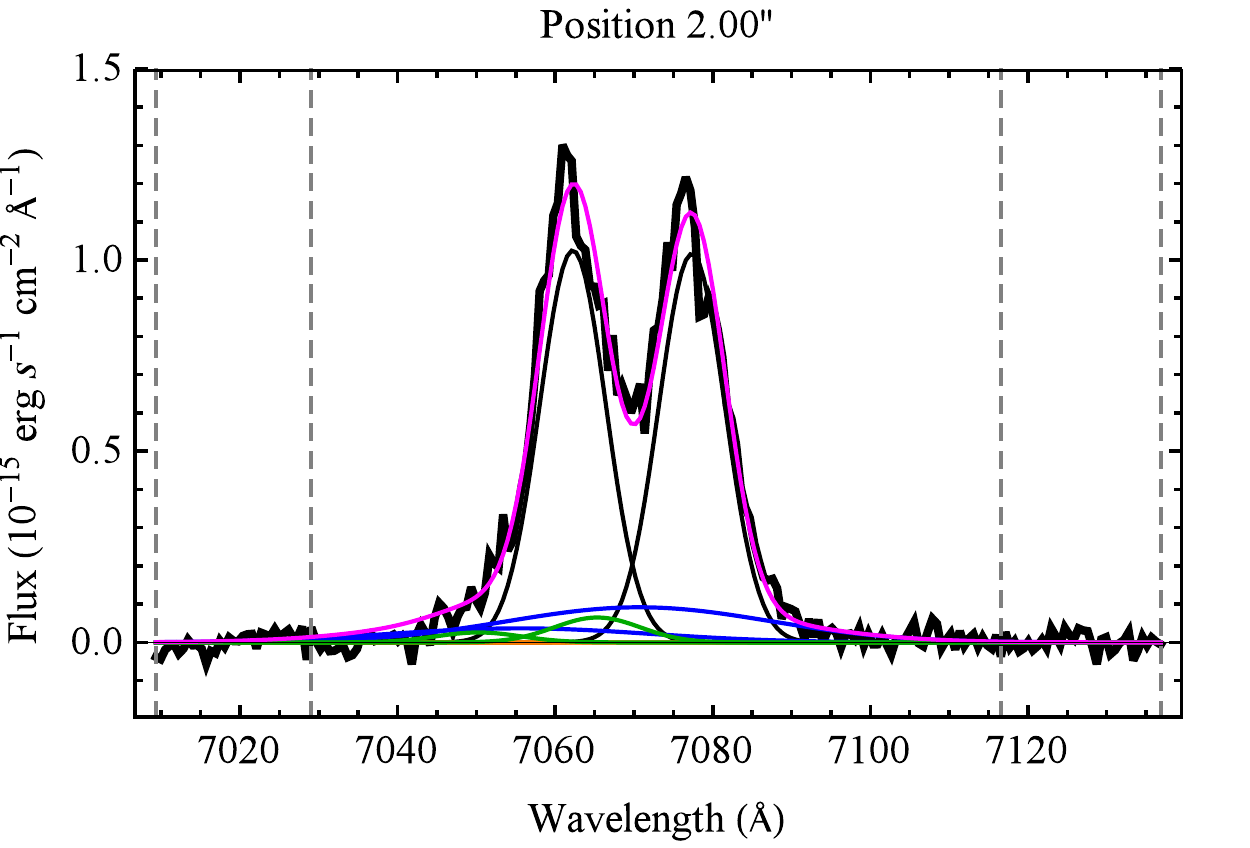}}
\subfigure{
\includegraphics[scale=0.34]{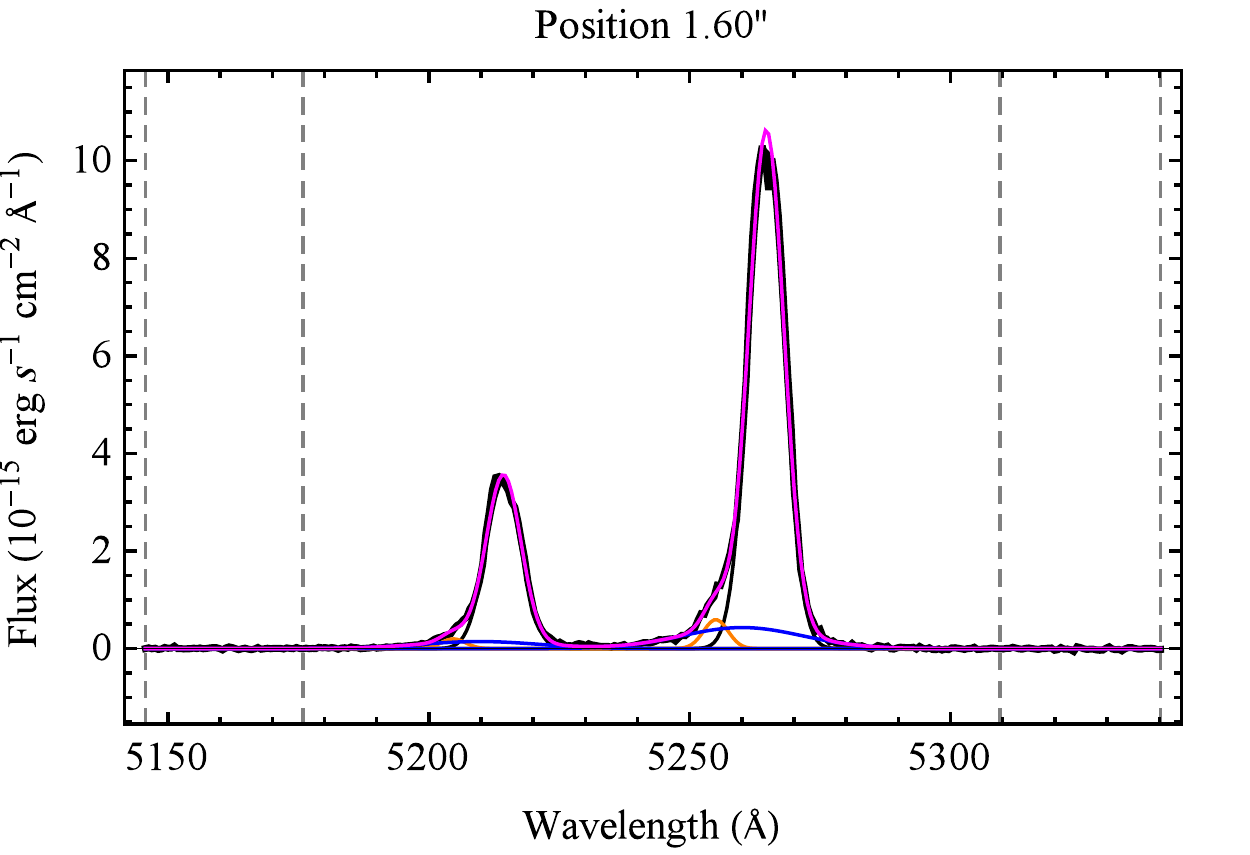}}
\subfigure{
\includegraphics[scale=0.34]{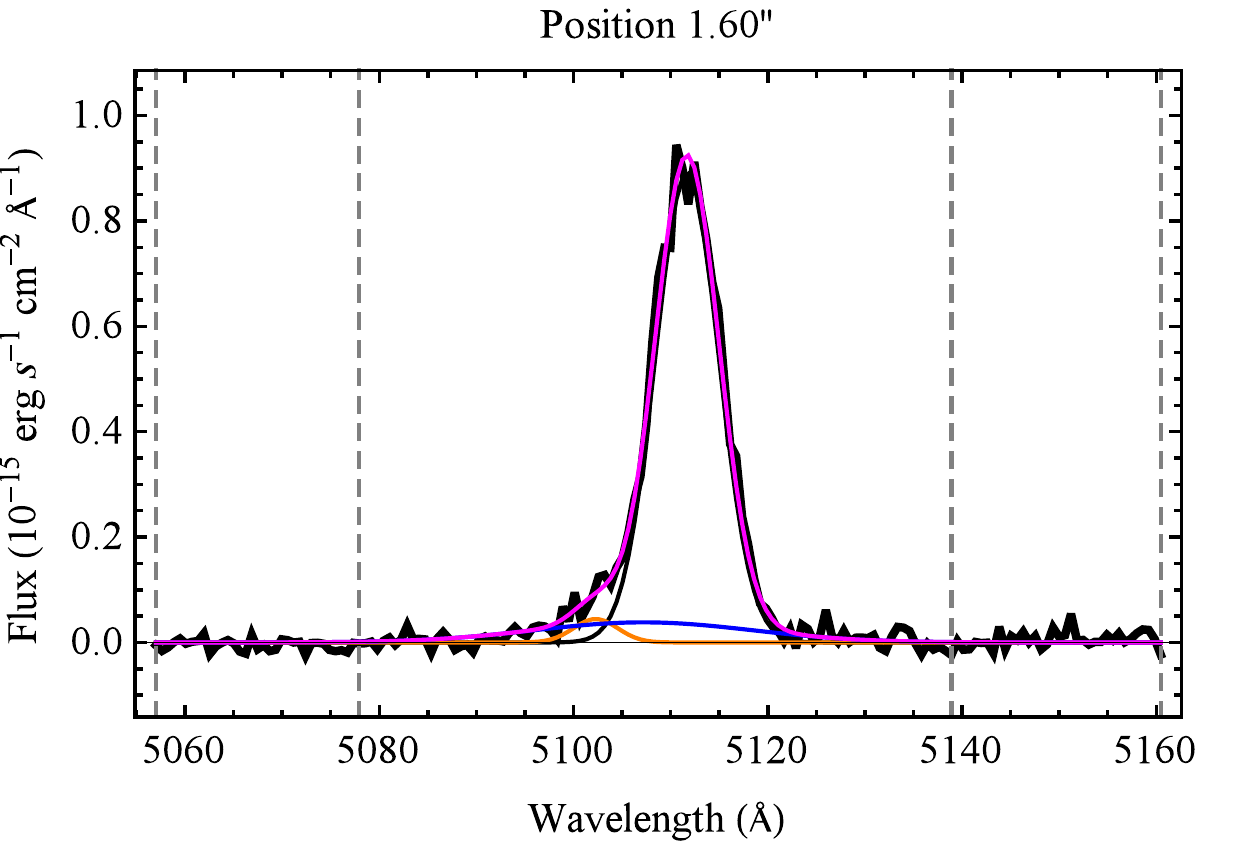}}
\subfigure{
\includegraphics[scale=0.34]{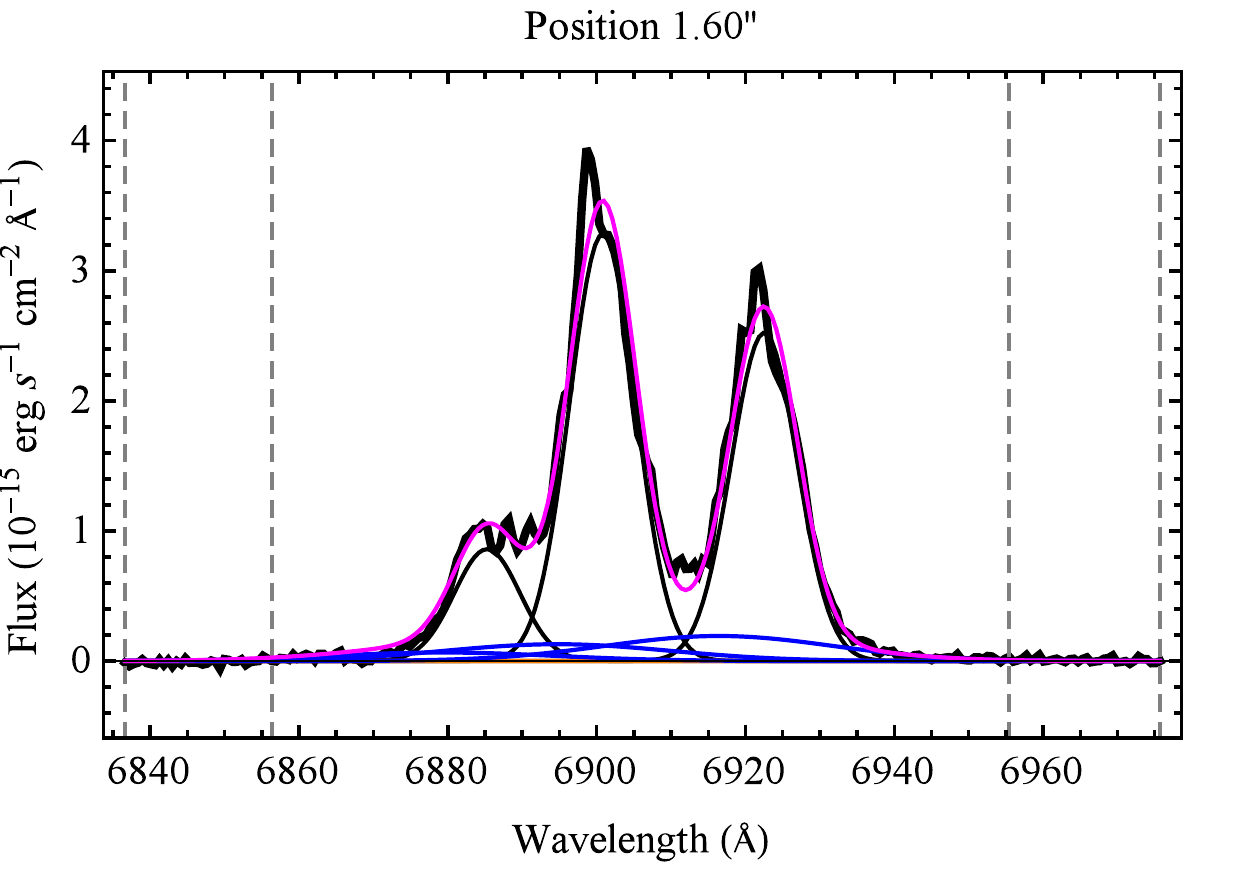}}
\subfigure{
\includegraphics[scale=0.34]{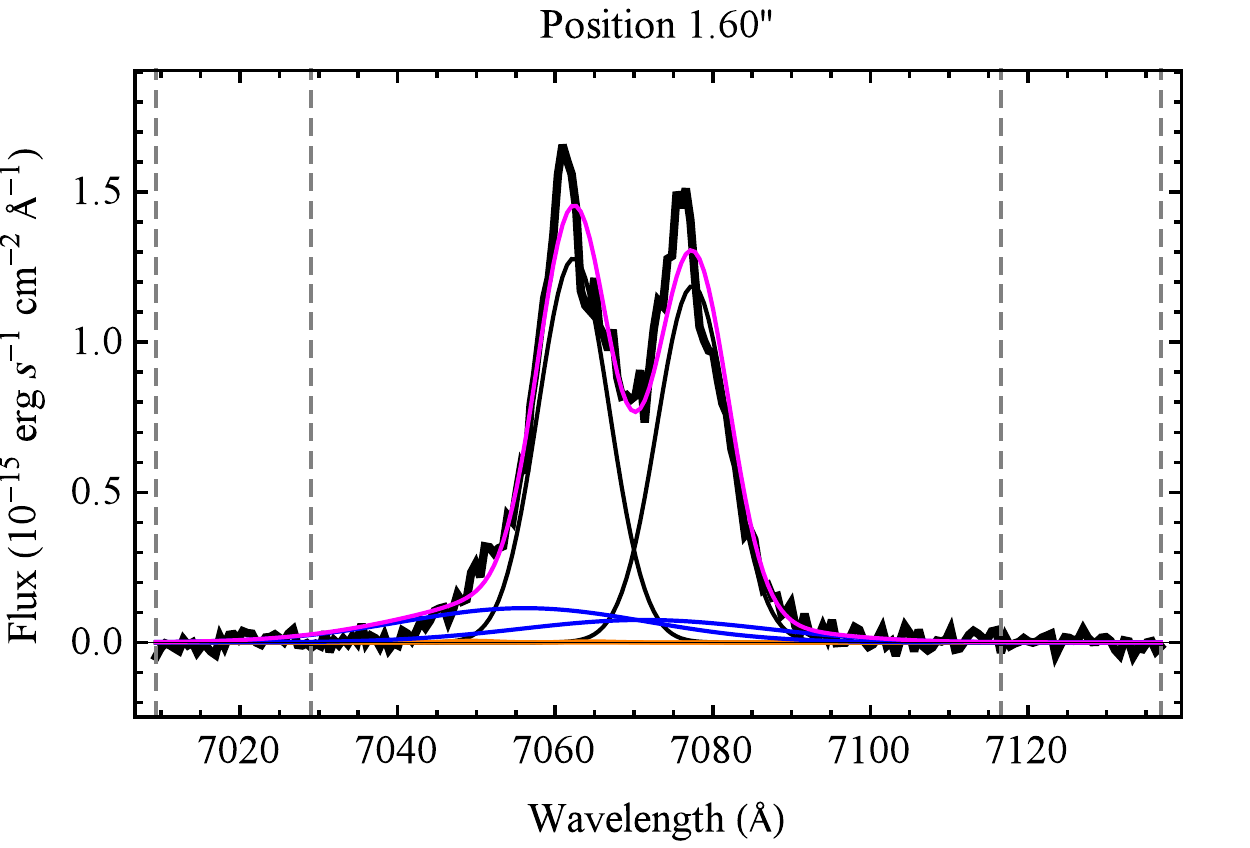}}
\subfigure{
\includegraphics[scale=0.34]{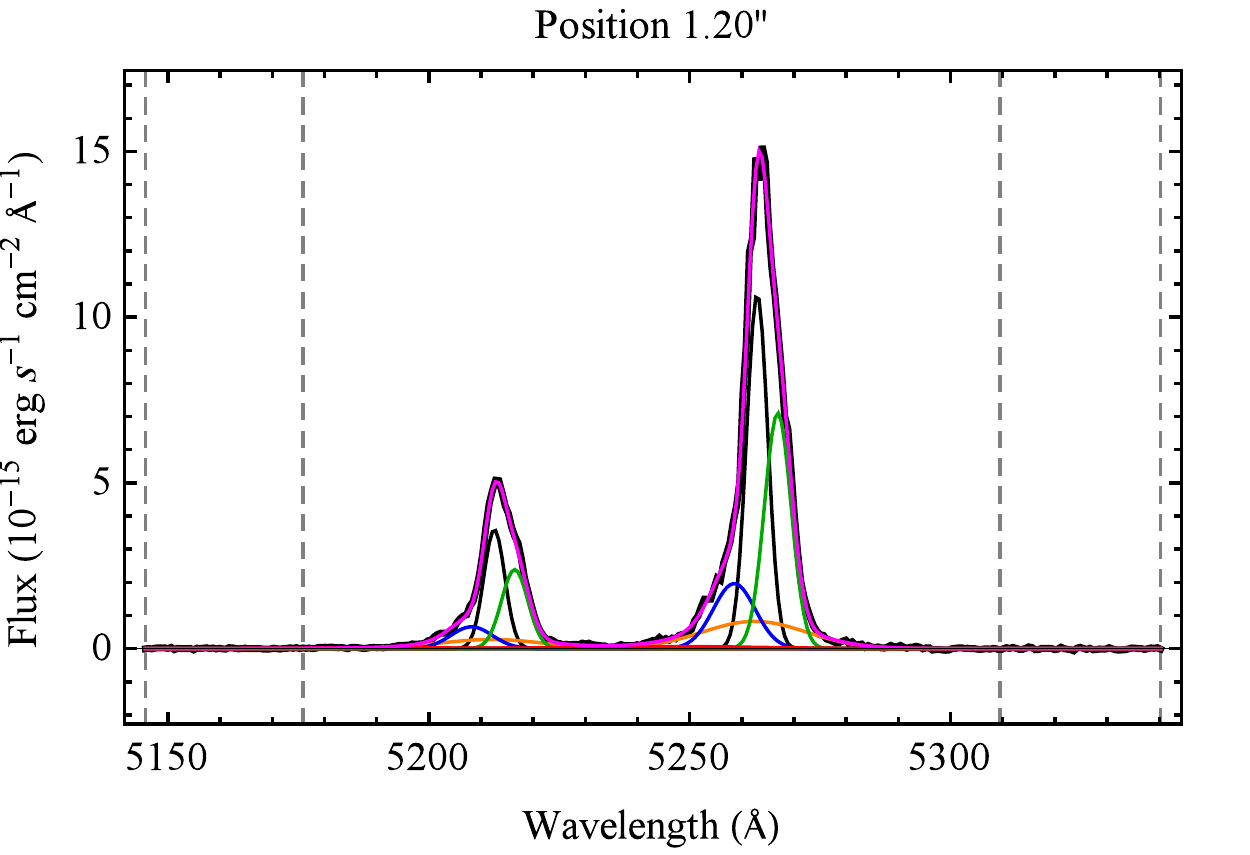}}
\subfigure{
\includegraphics[scale=0.34]{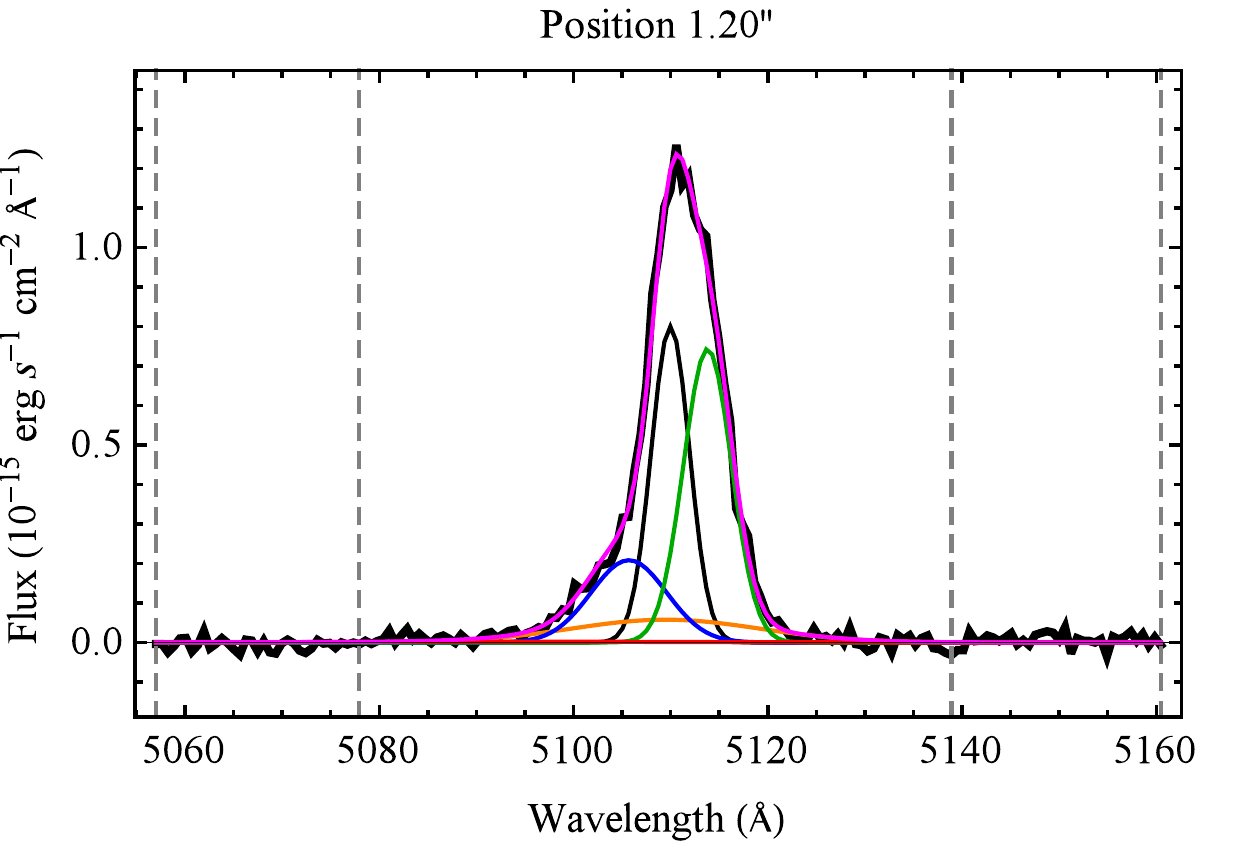}}
\subfigure{
\includegraphics[scale=0.34]{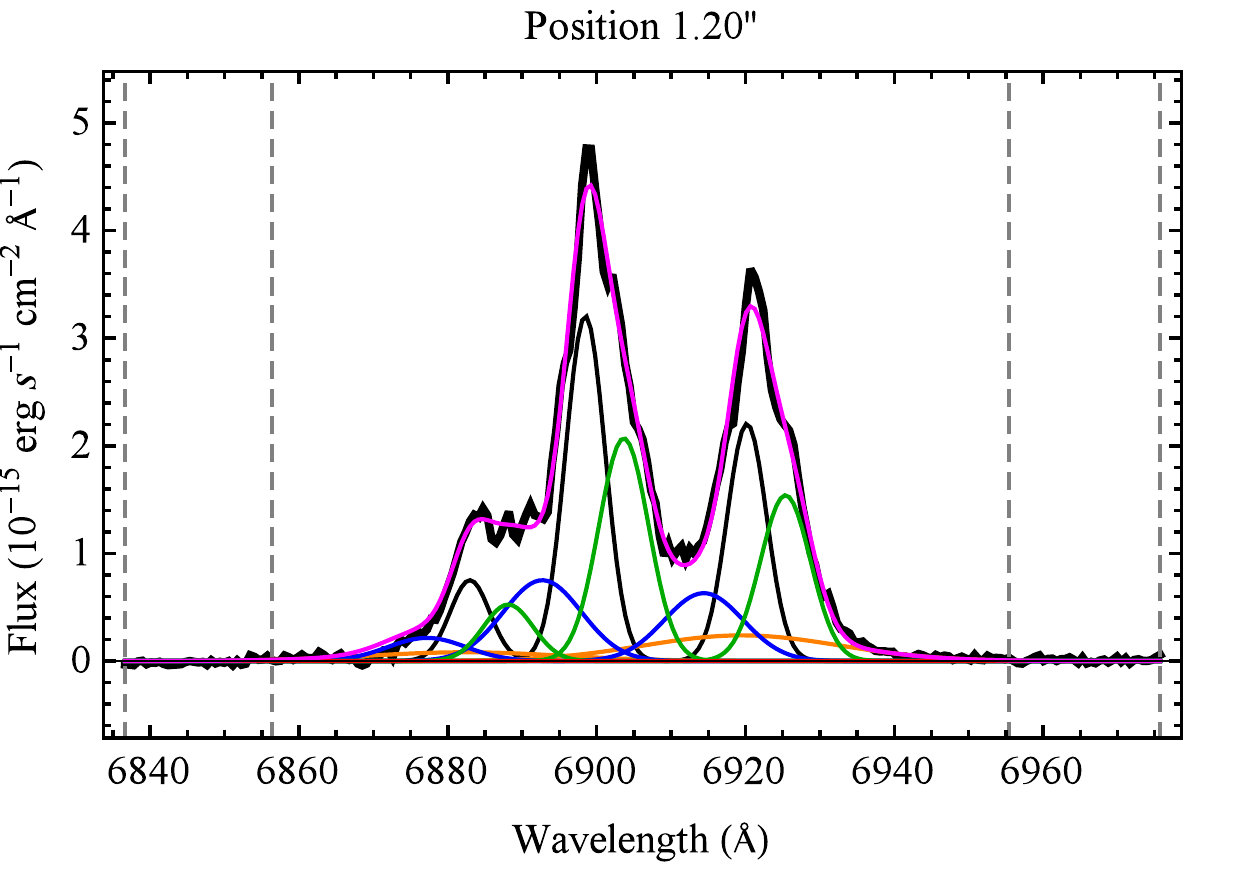}}
\subfigure{
\includegraphics[scale=0.34]{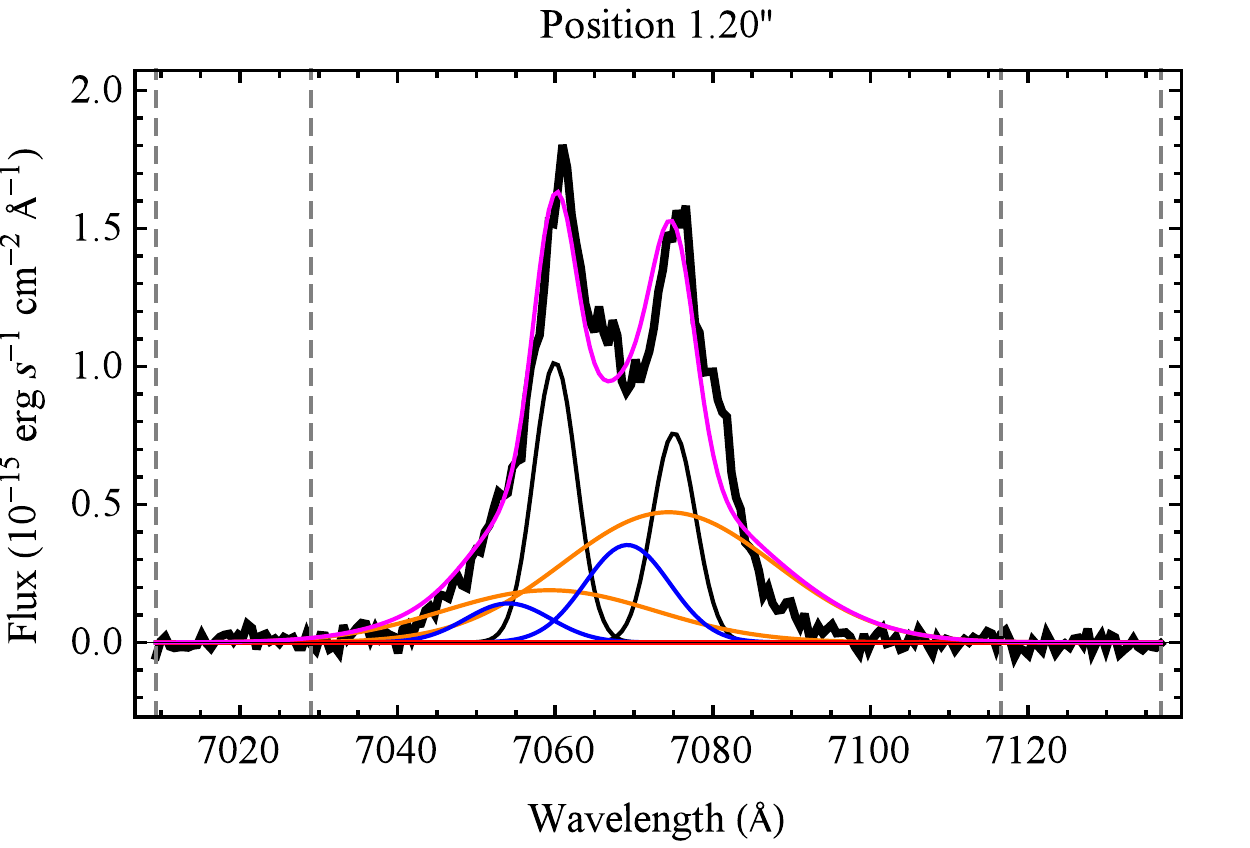}}
\subfigure{
\includegraphics[scale=0.34]{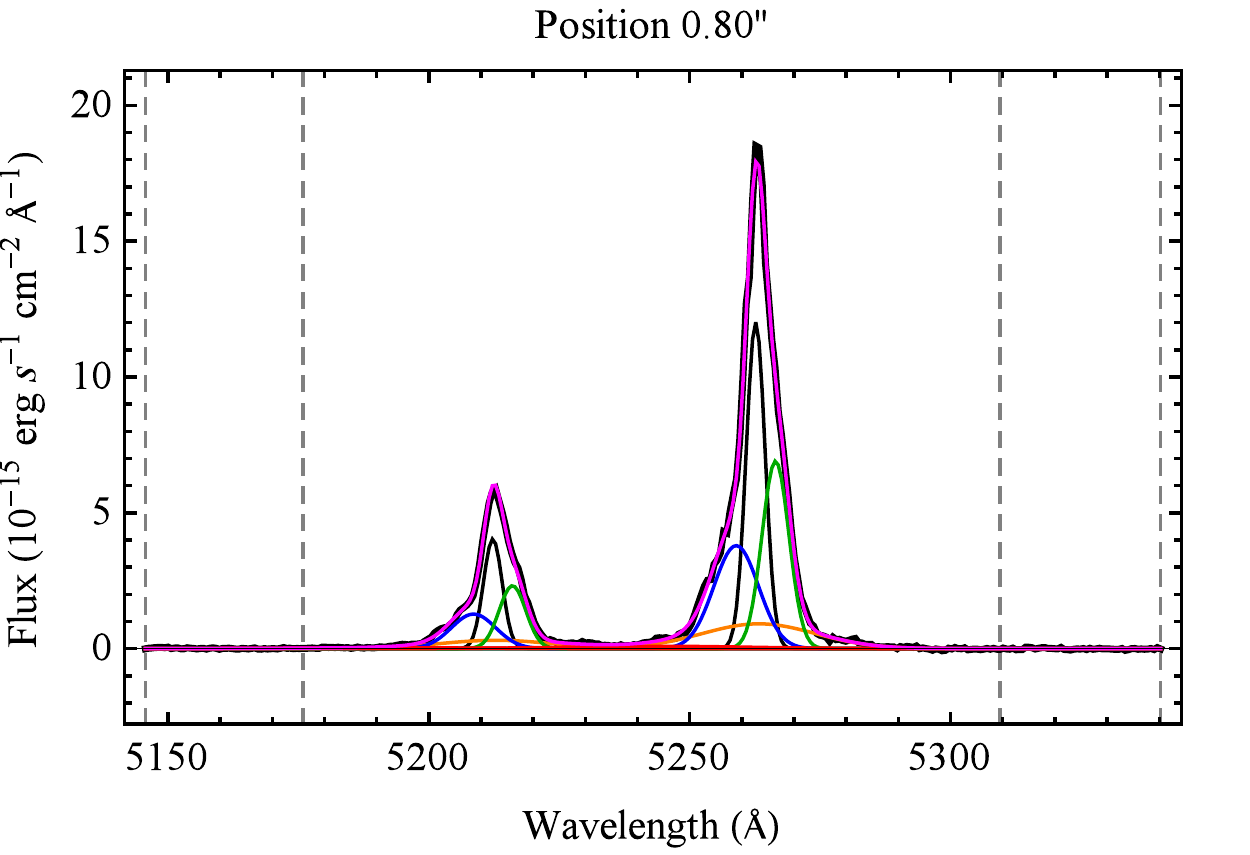}}
\subfigure{
\includegraphics[scale=0.34]{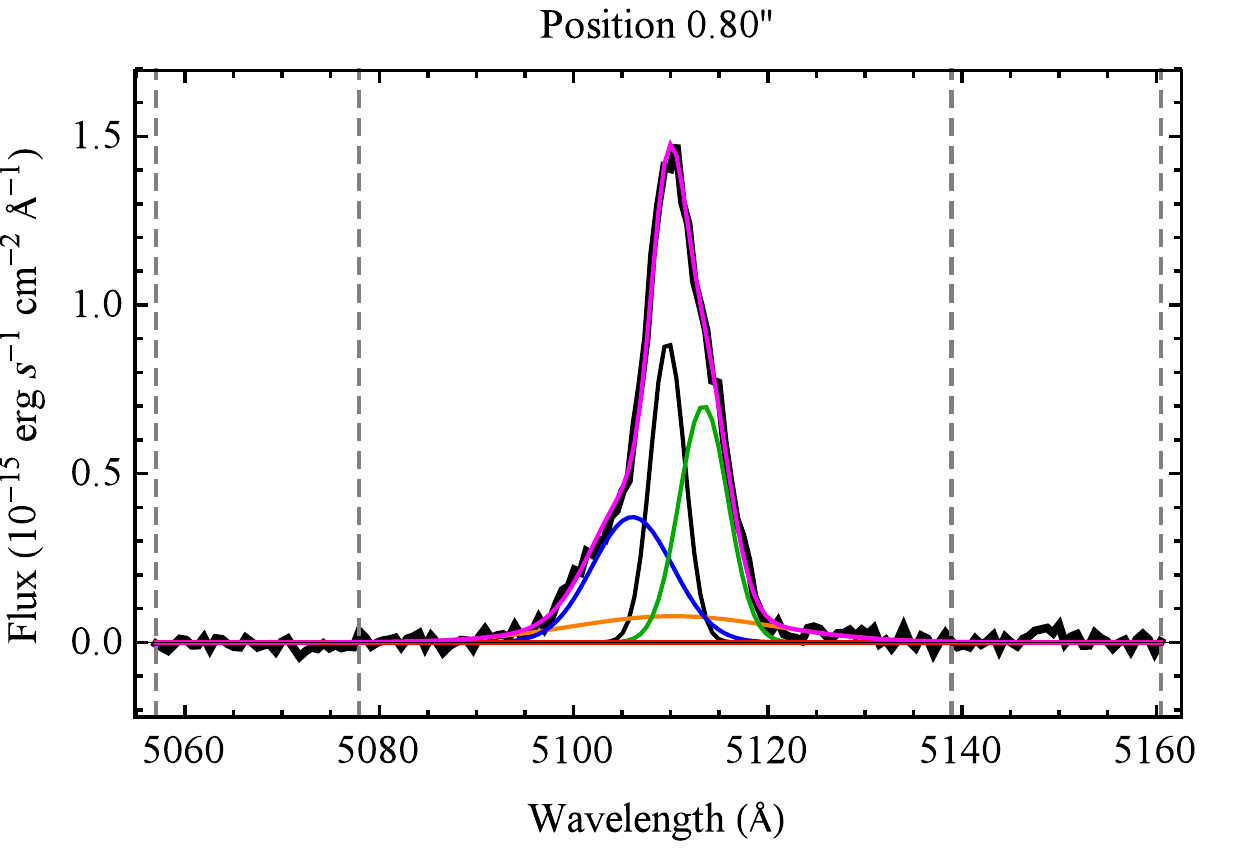}}
\subfigure{
\includegraphics[scale=0.34]{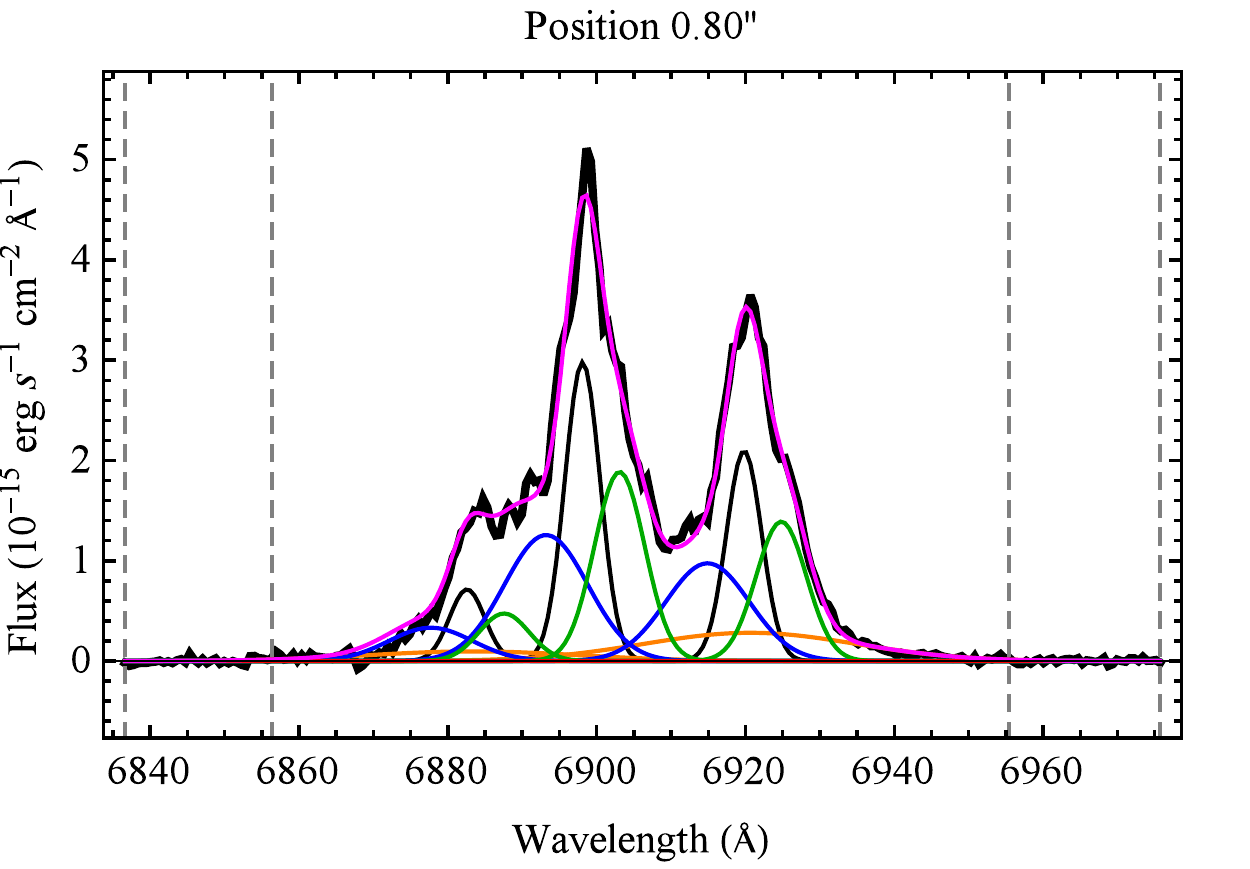}}
\subfigure{
\includegraphics[scale=0.34]{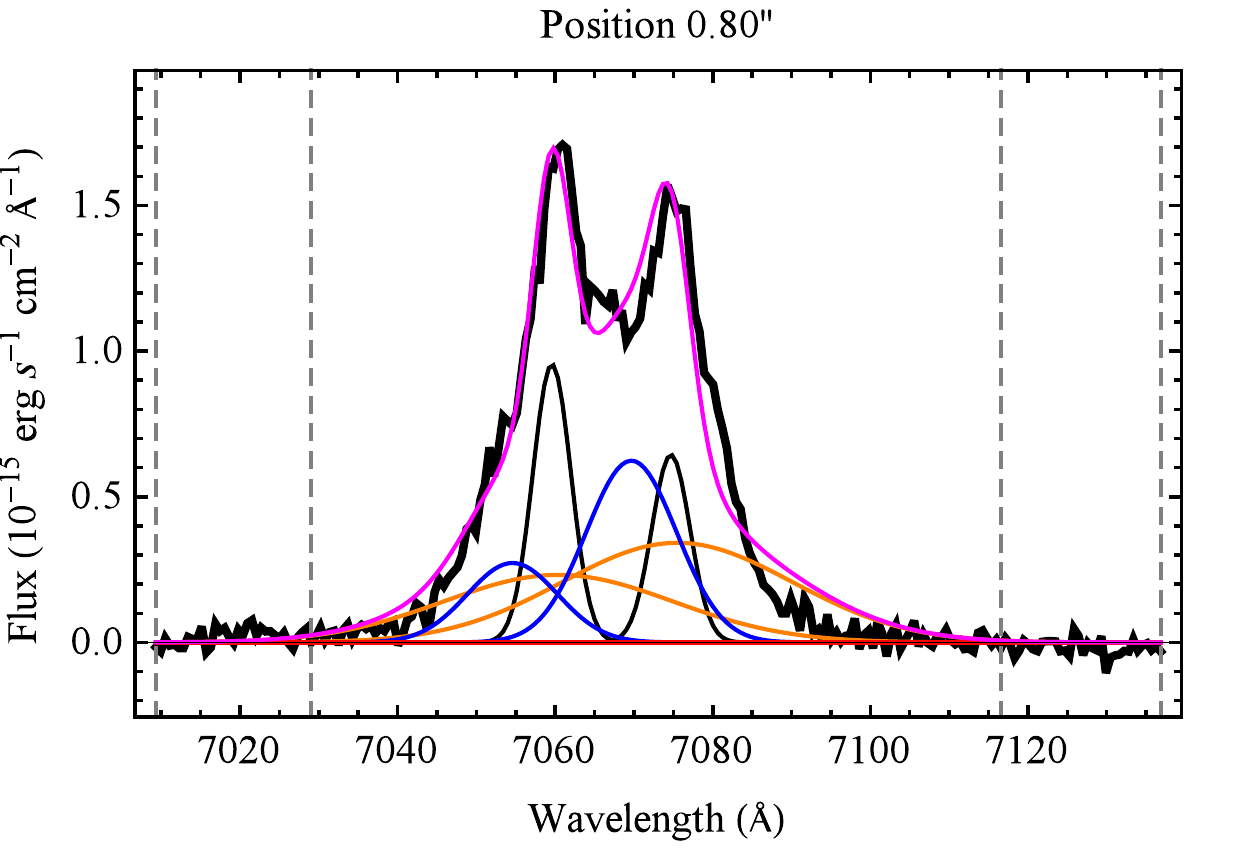}}
\subfigure{
\includegraphics[scale=0.34]{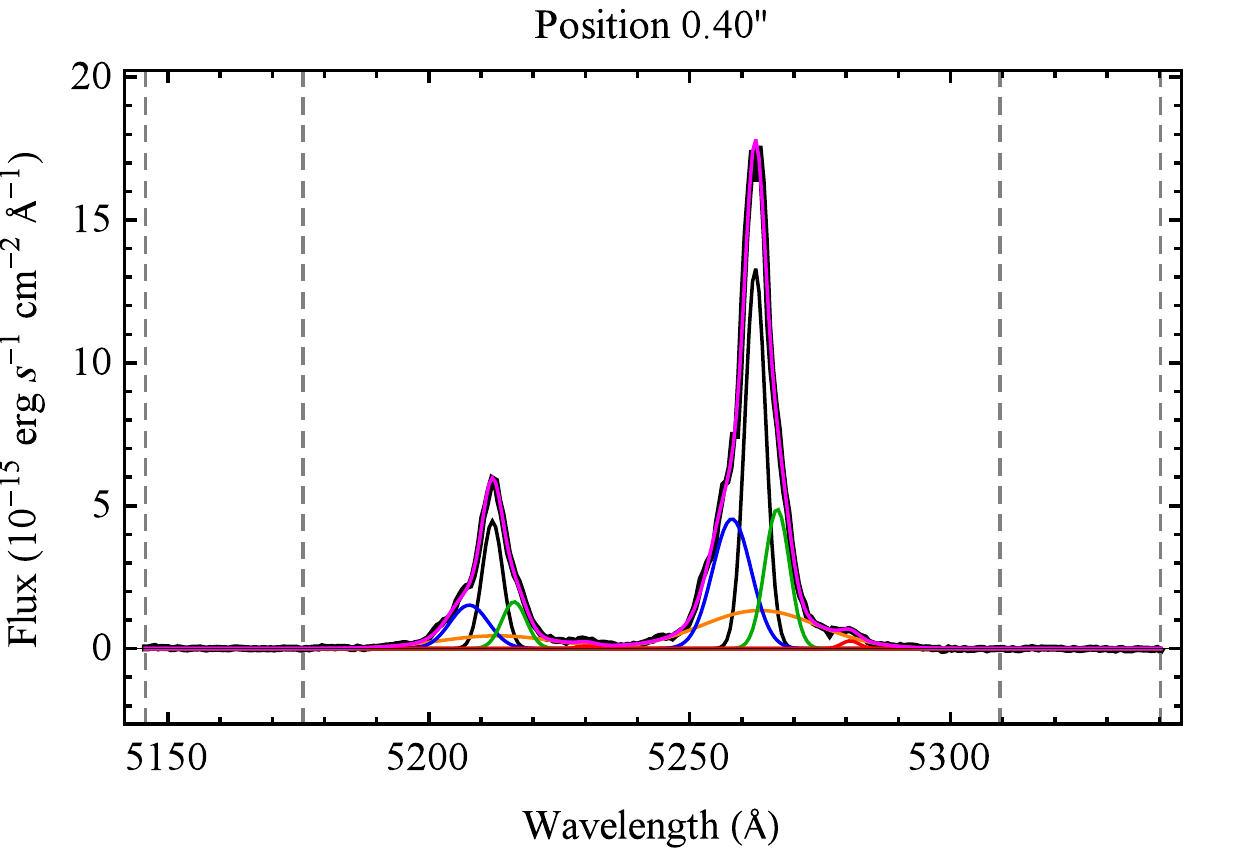}}
\subfigure{
\includegraphics[scale=0.34]{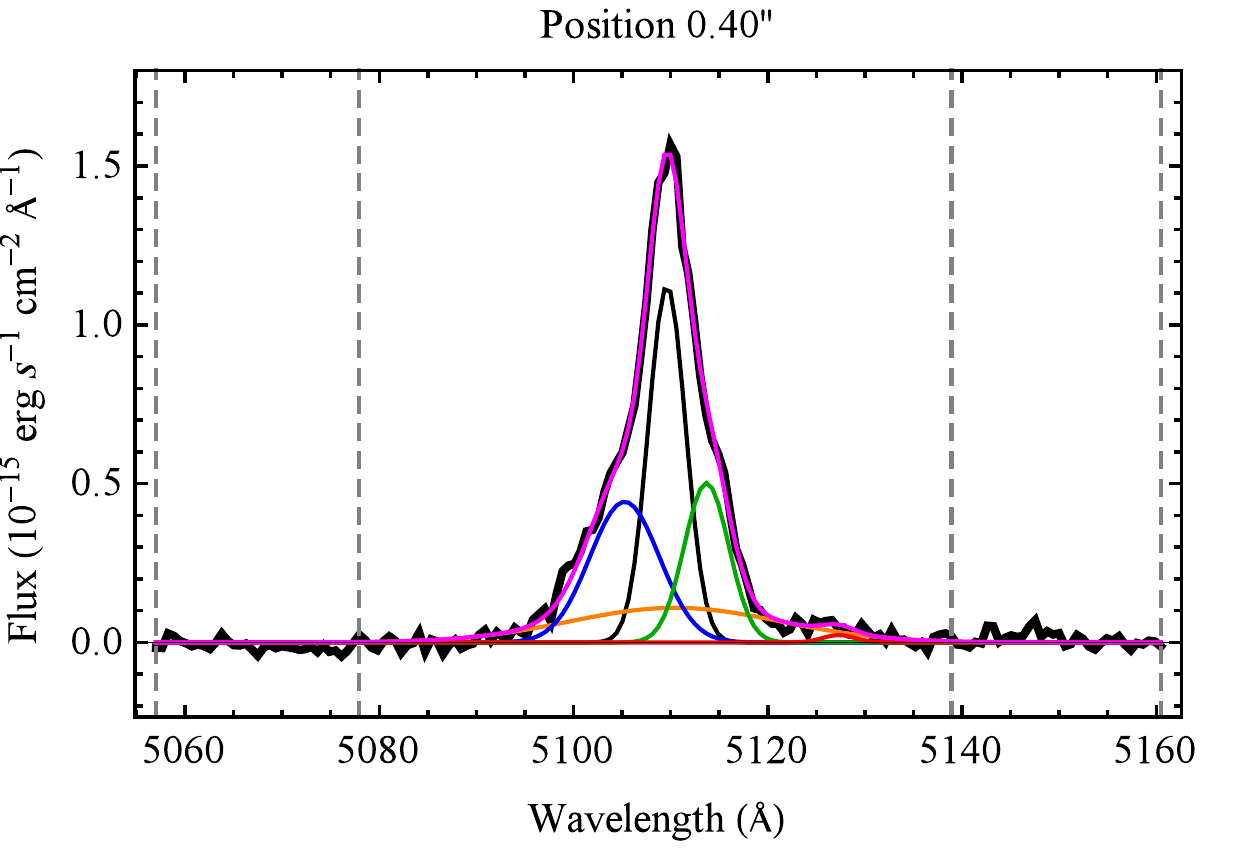}}
\subfigure{
\includegraphics[scale=0.34]{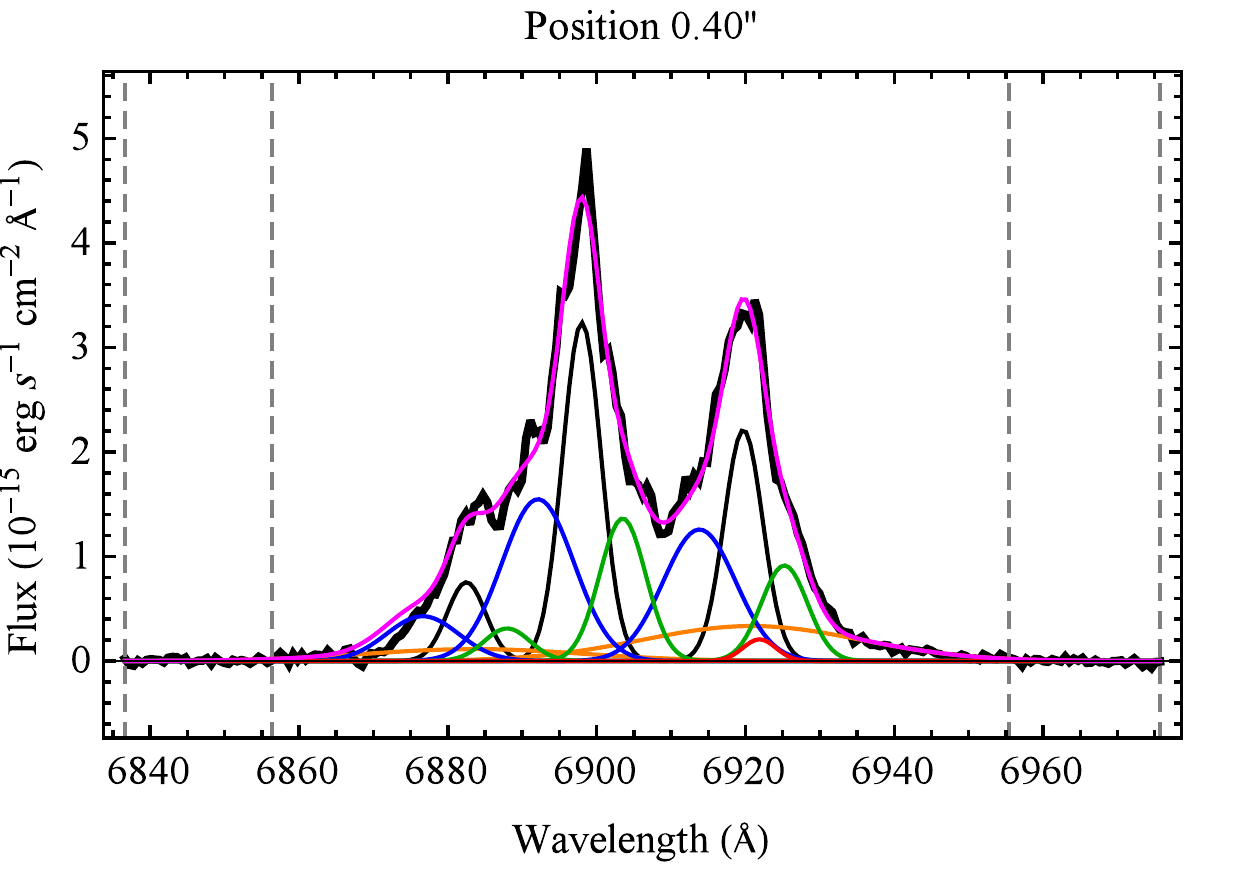}}
\subfigure{
\includegraphics[scale=0.34]{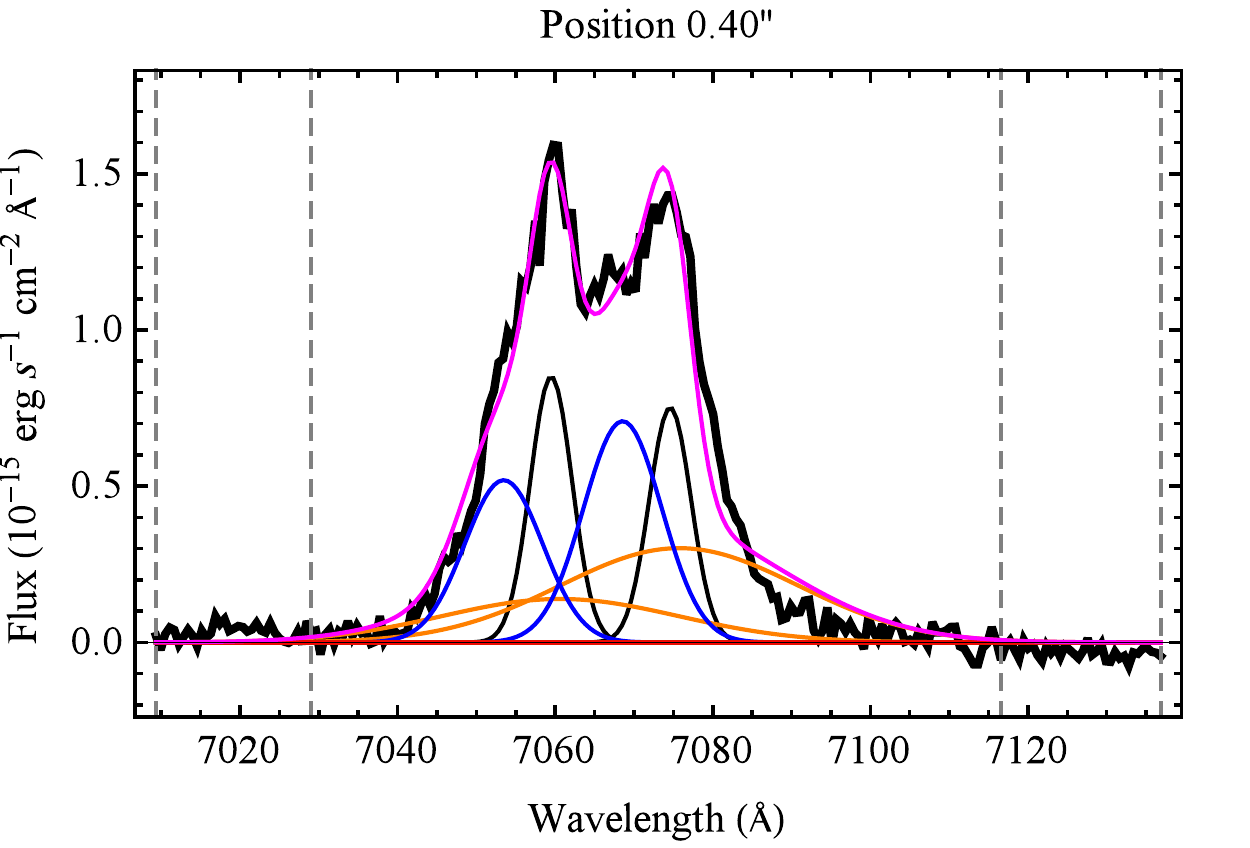}}
\subfigure{
\includegraphics[scale=0.34]{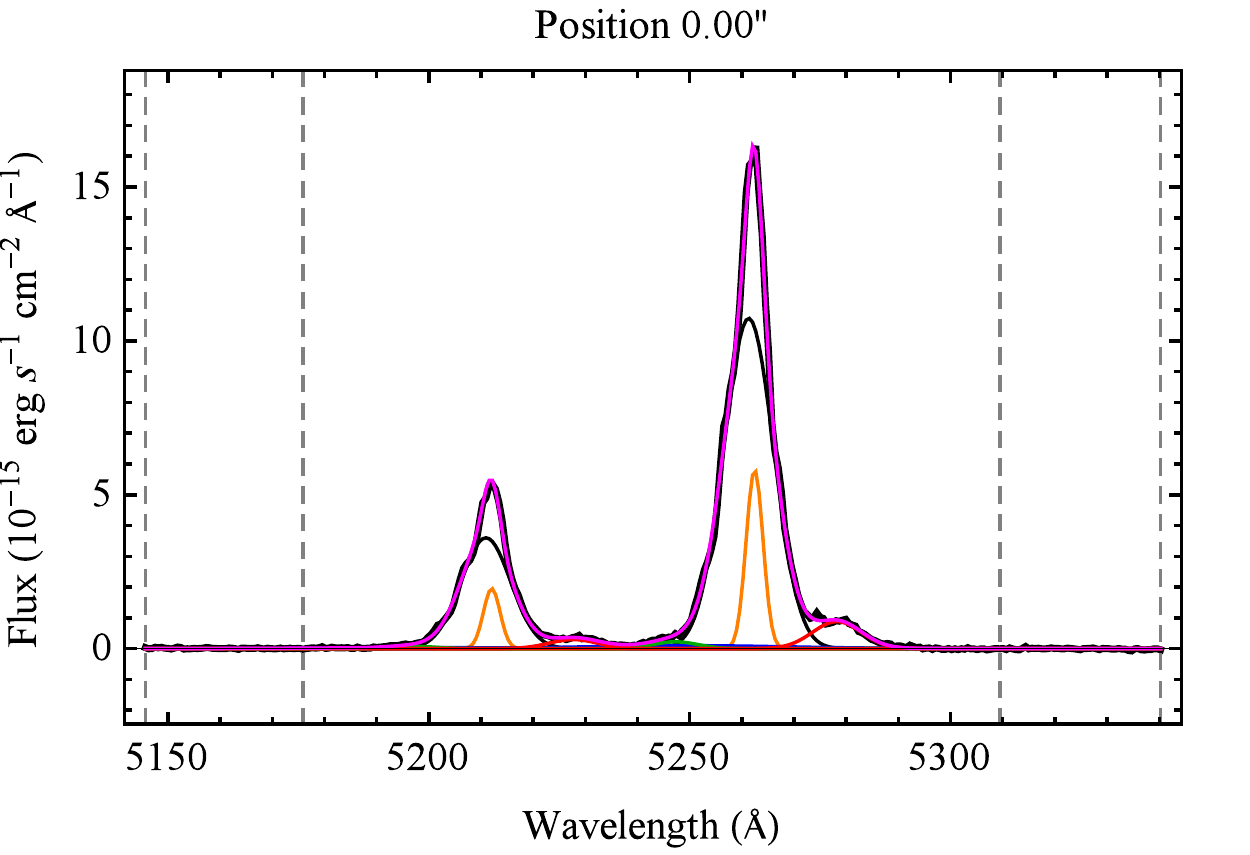}}
\subfigure{
\includegraphics[scale=0.34]{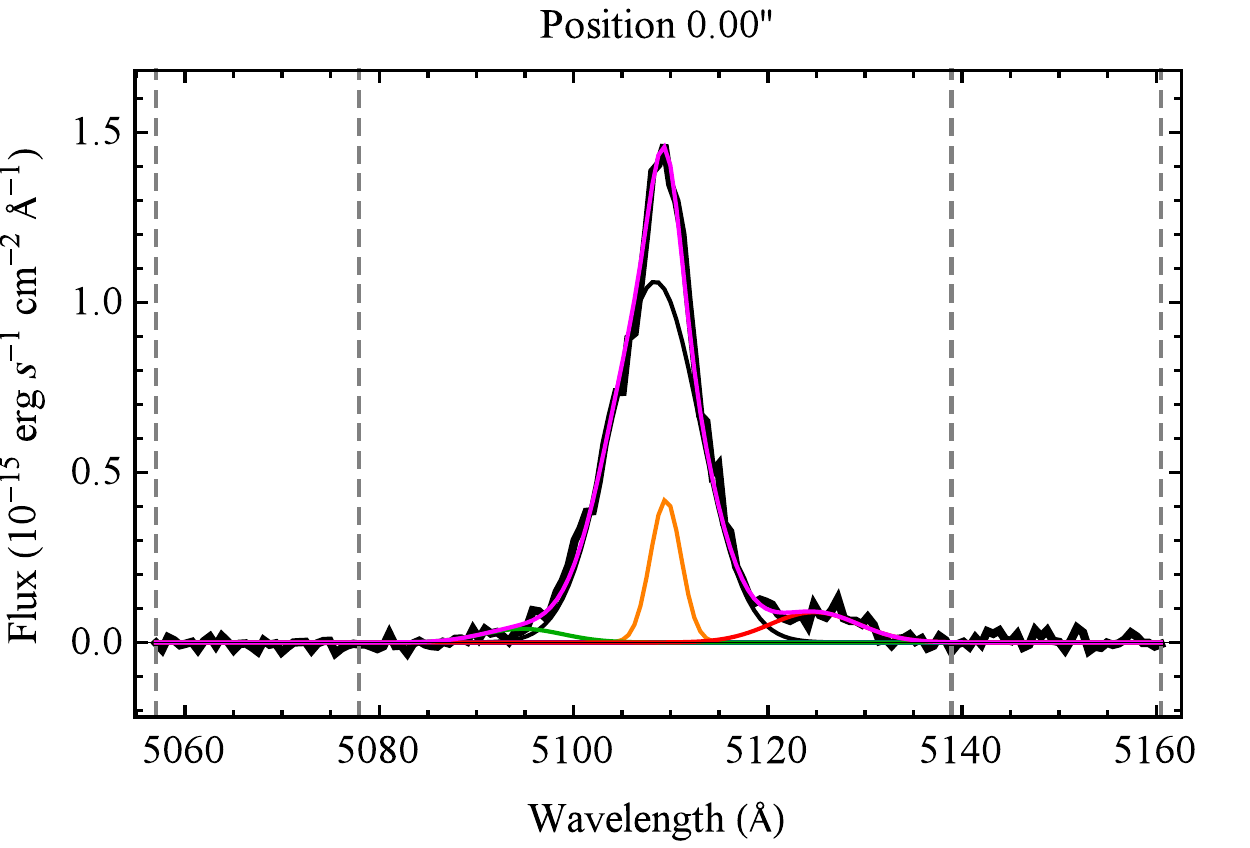}}
\subfigure{
\includegraphics[scale=0.34]{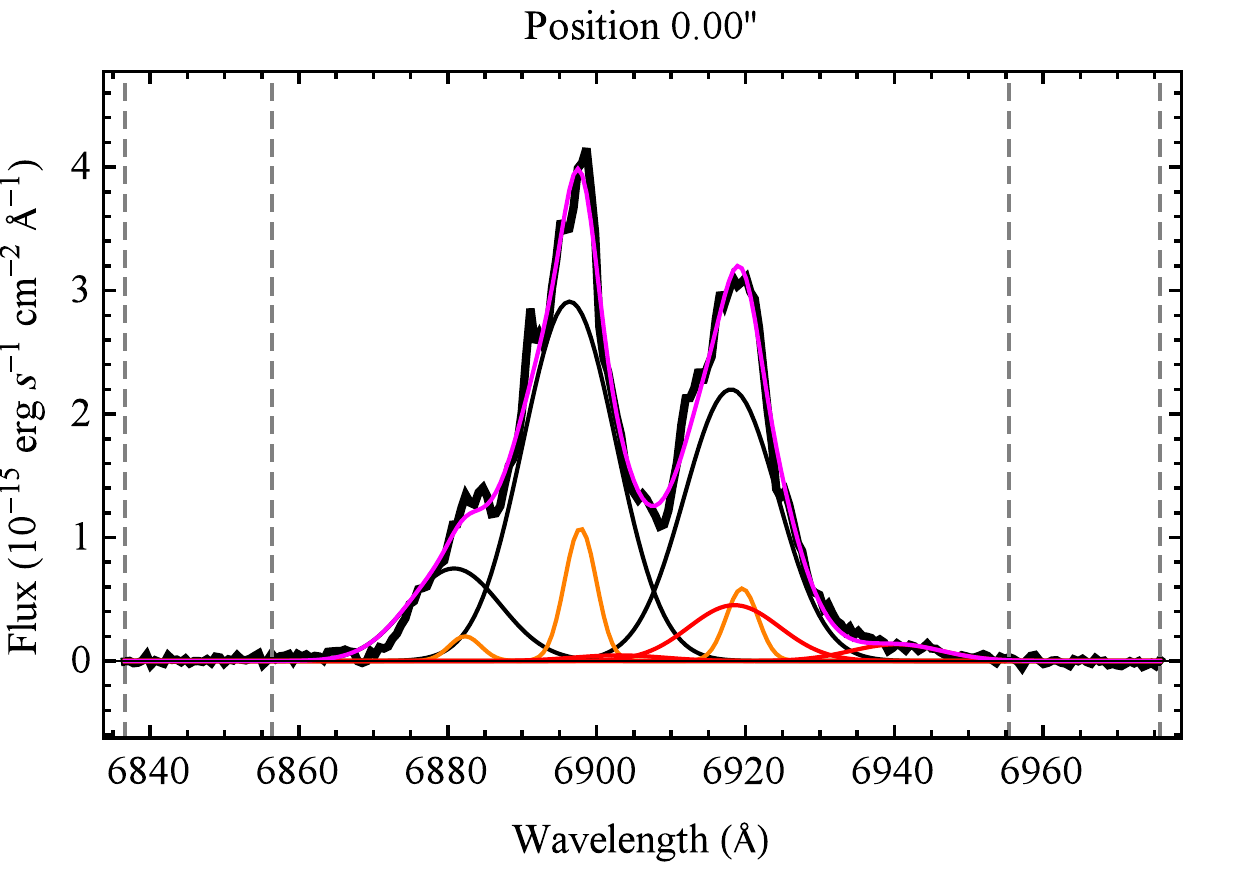}}
\subfigure{
\includegraphics[scale=0.34]{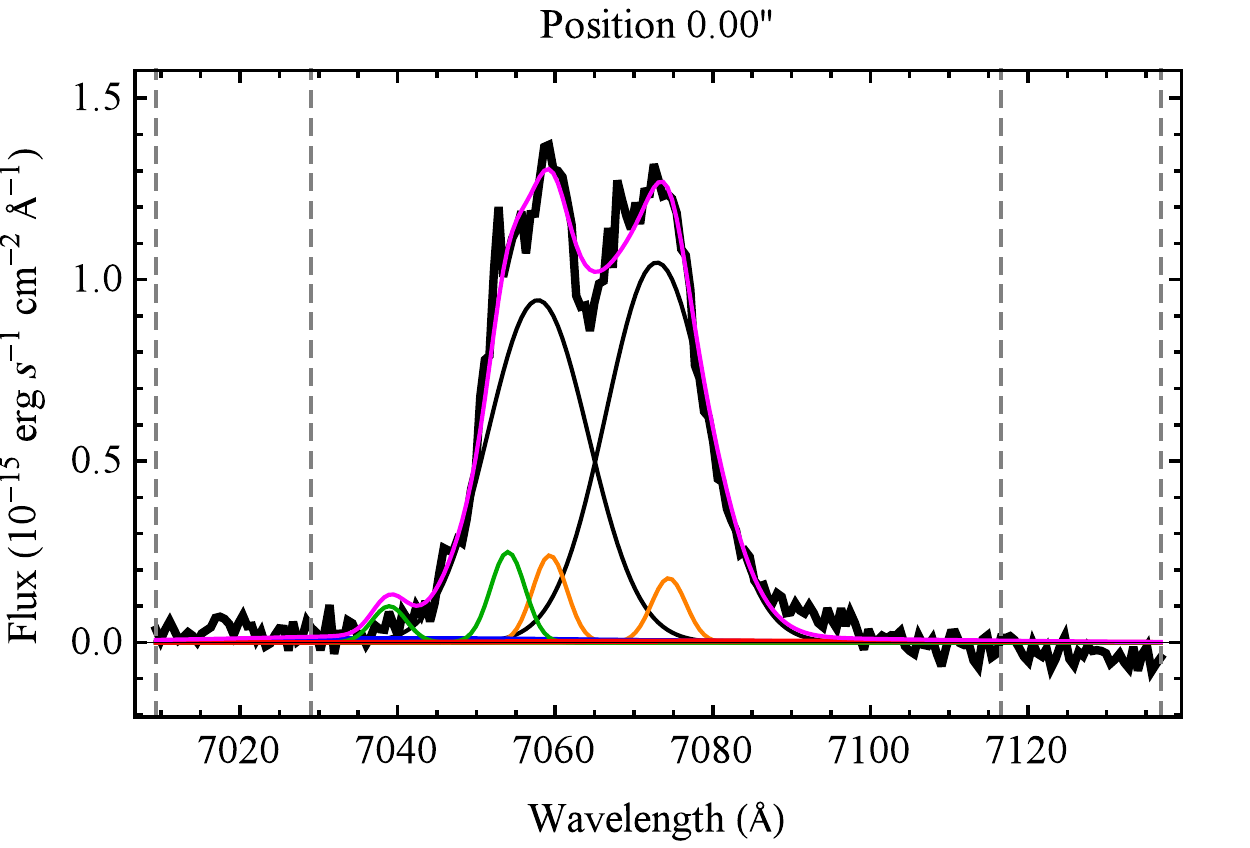}}
\caption{Gaussian fits, from left to right, of the [O~III], H$\beta$, [N~II]+H$\alpha$, and [S~II] emission lines for the APO DIS observations along PA~=~163$\degr$. The components are sorted by peak flux from strongest to weakest: black, orange, blue, green, red, and the total in magenta.}
\end{figure*}

\addtocounter{figure}{-1}

\begin{figure*}
\centering
\subfigure{
\includegraphics[scale=0.34]{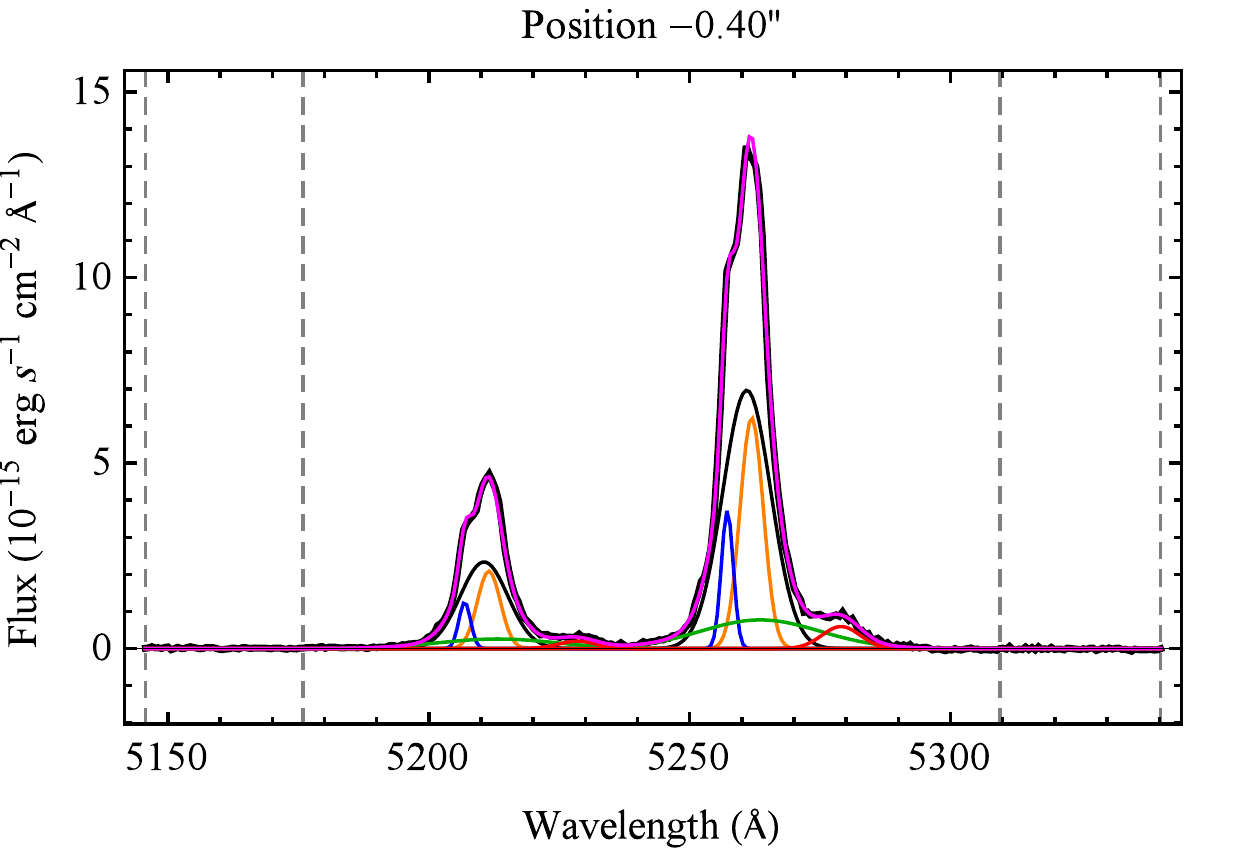}}
\subfigure{
\includegraphics[scale=0.34]{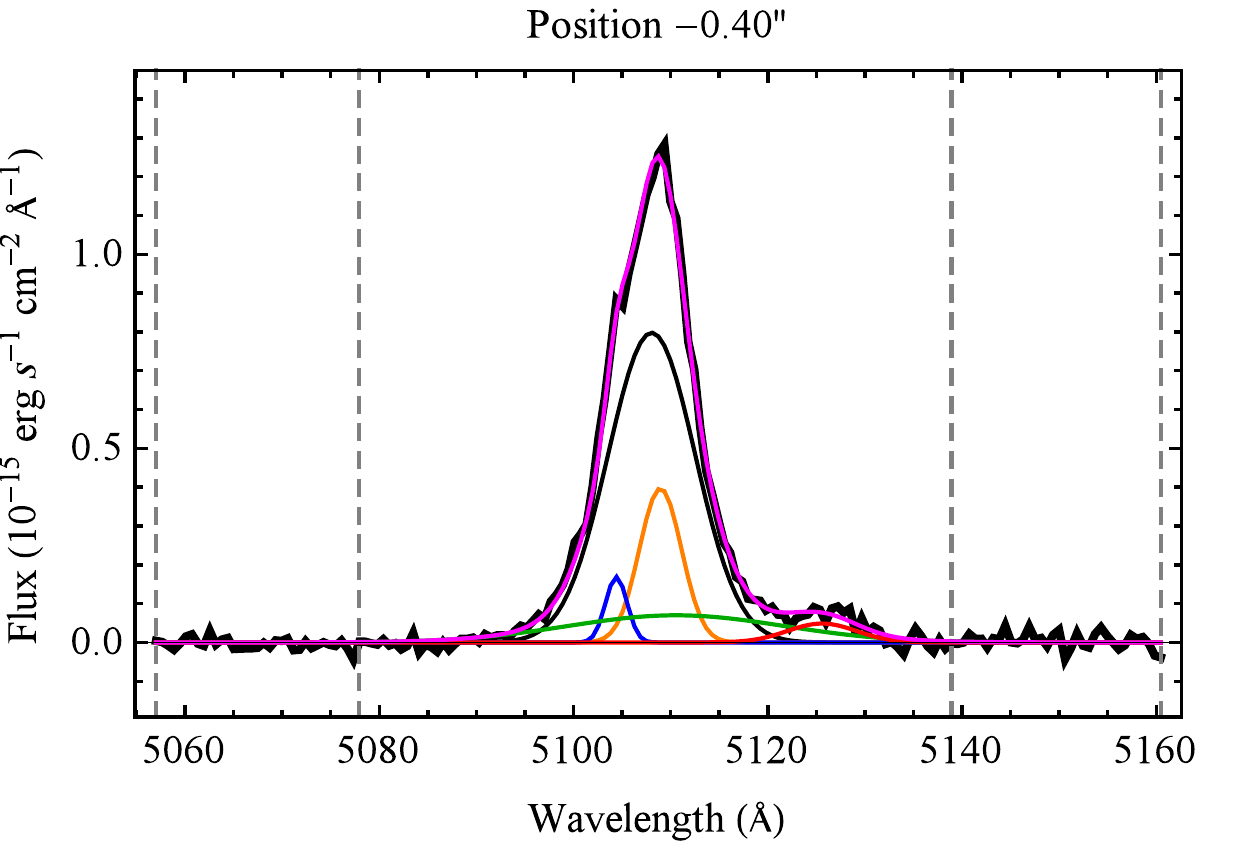}}
\subfigure{
\includegraphics[scale=0.34]{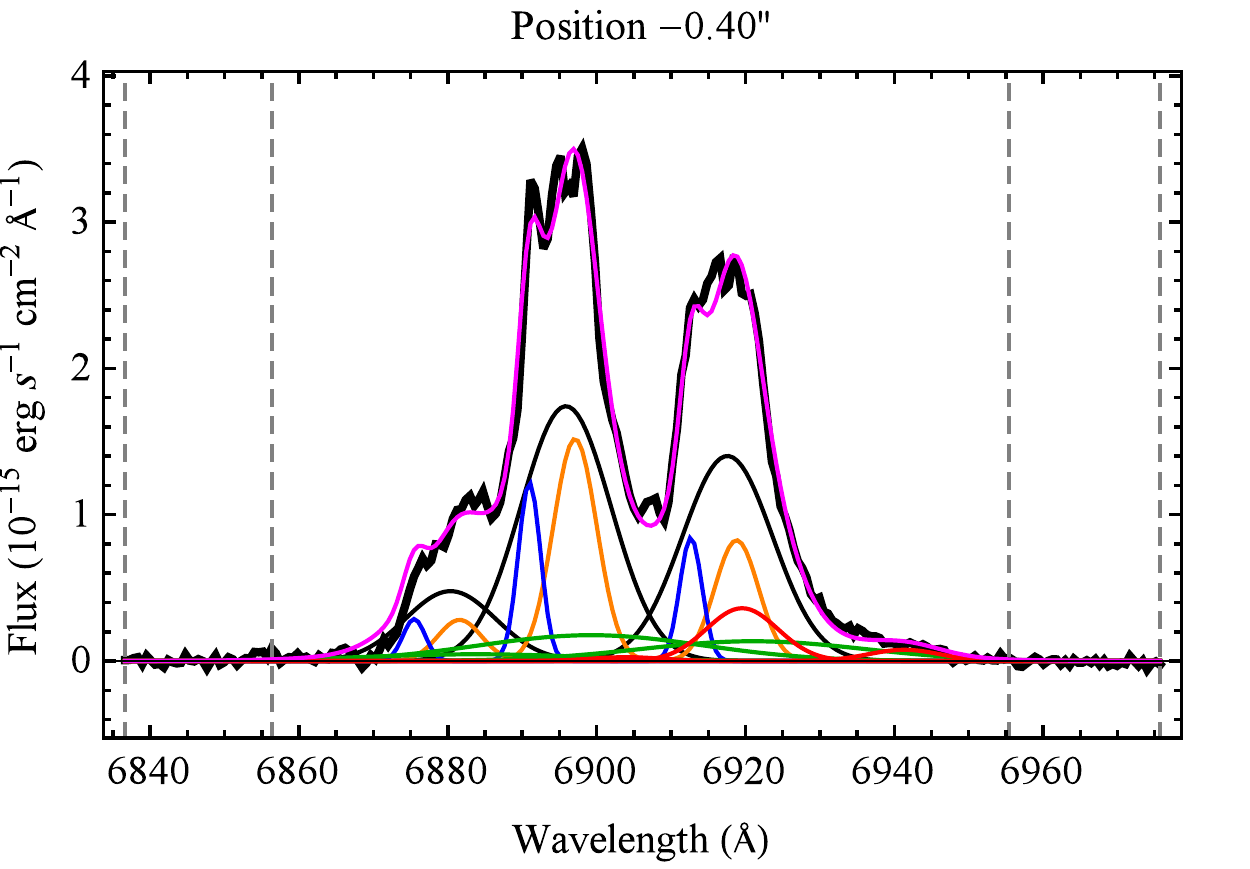}}
\subfigure{
\includegraphics[scale=0.34]{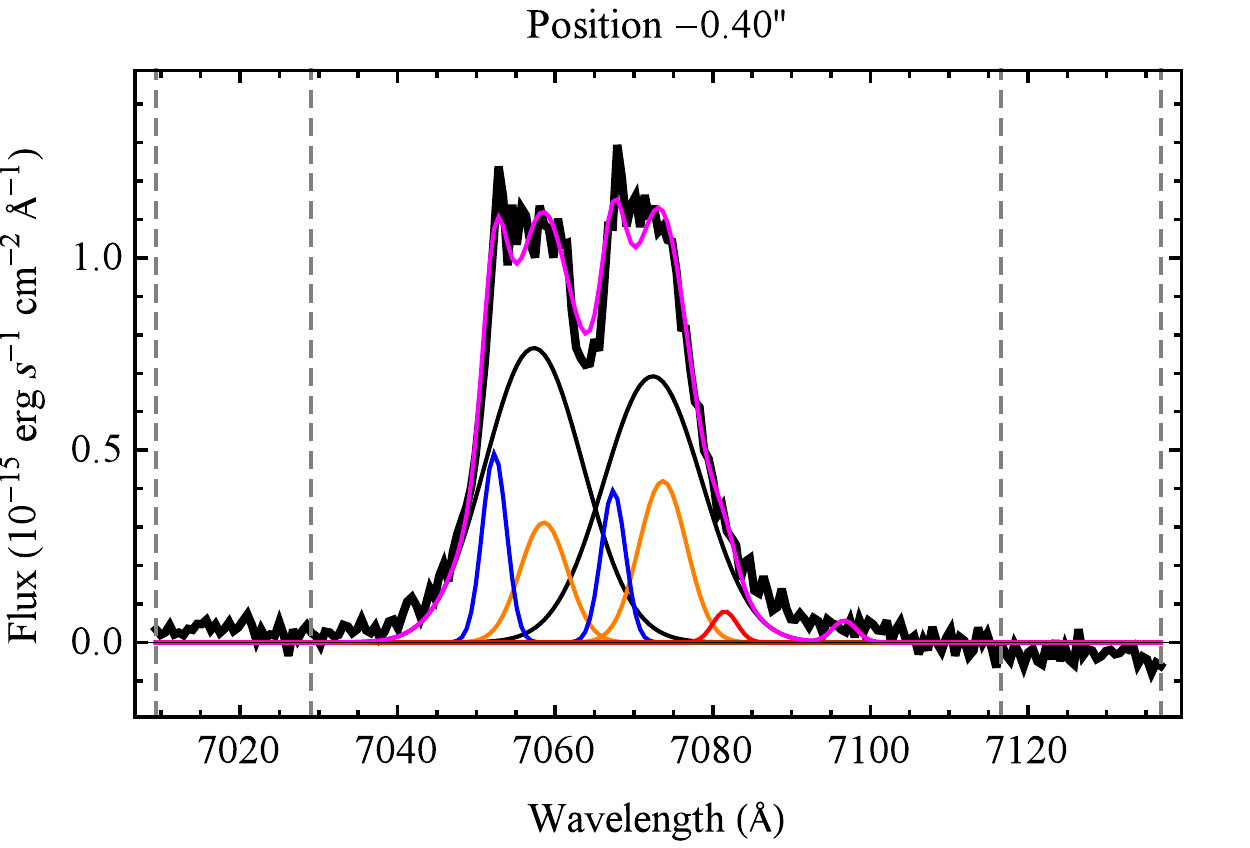}}
\subfigure{
\includegraphics[scale=0.34]{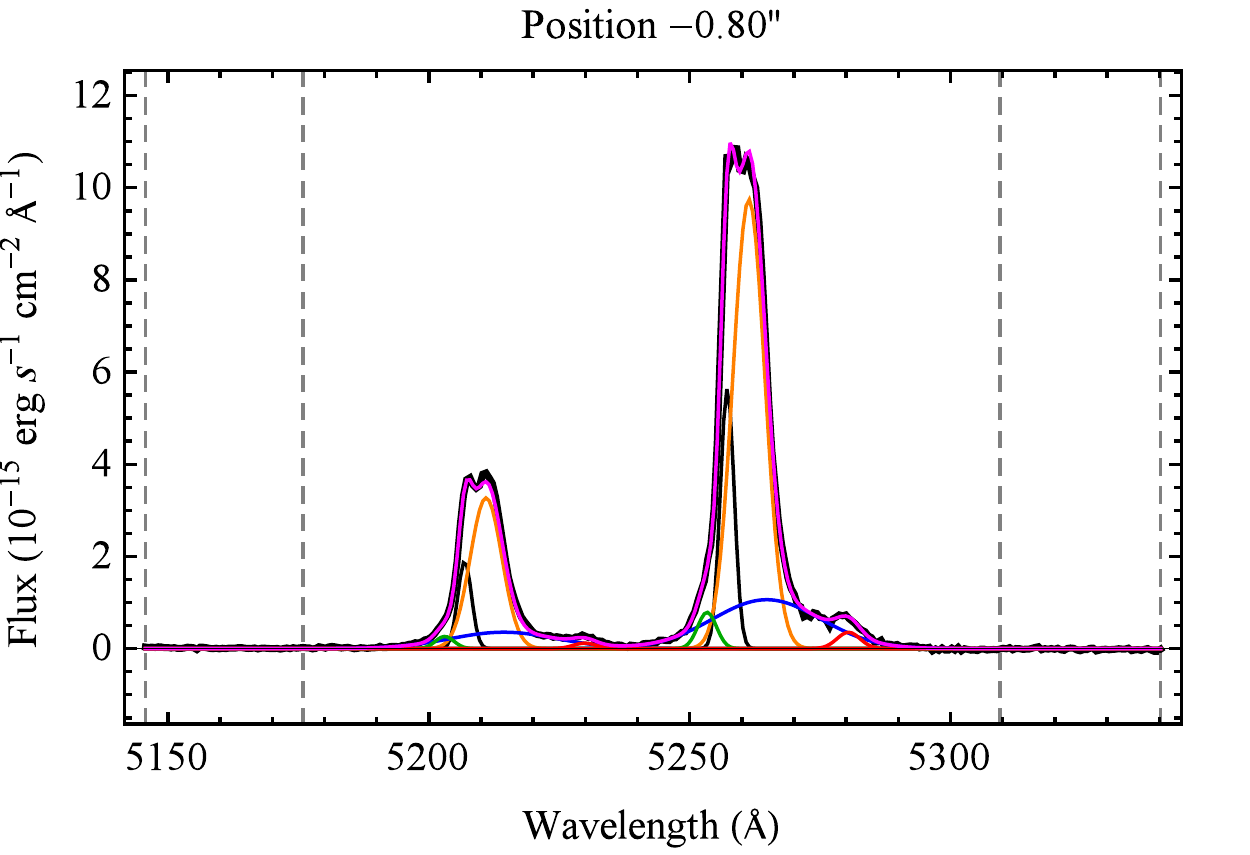}}
\subfigure{
\includegraphics[scale=0.34]{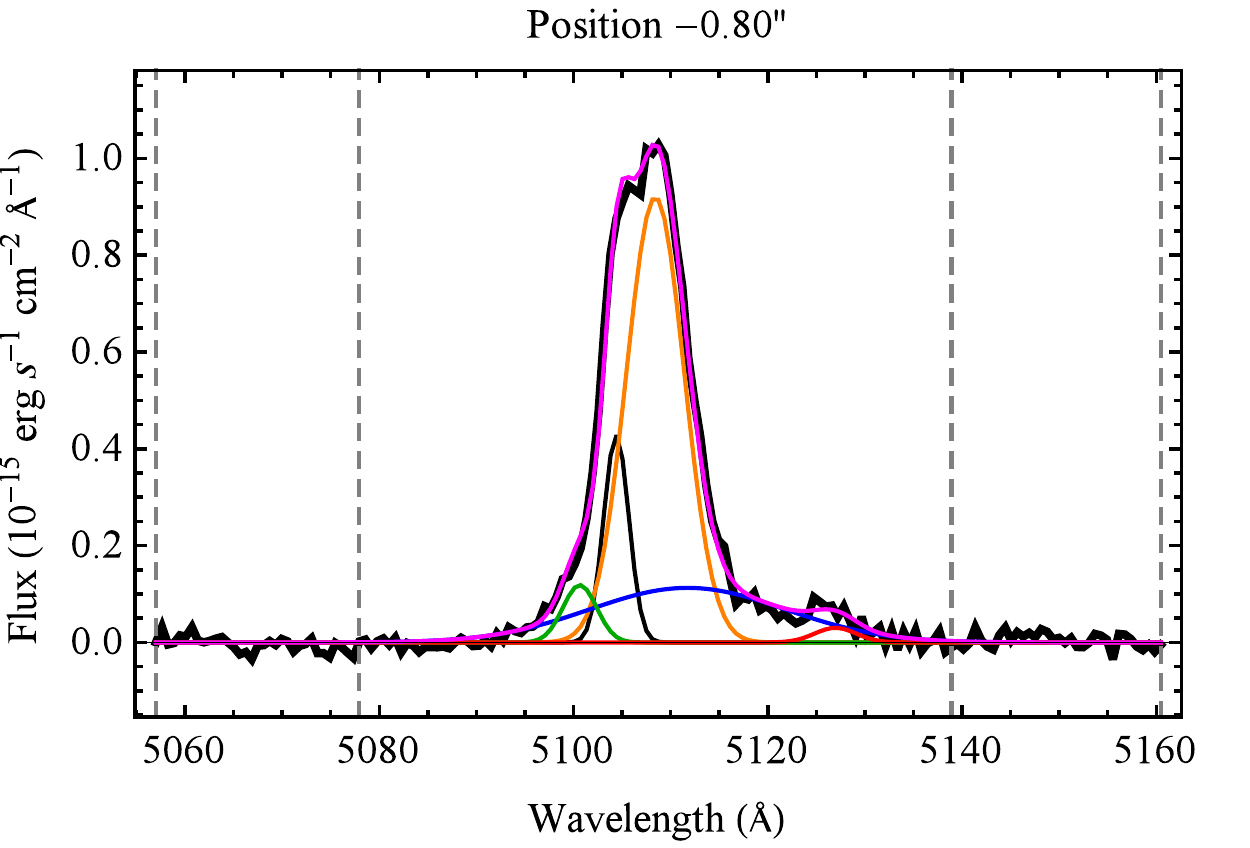}}
\subfigure{
\includegraphics[scale=0.34]{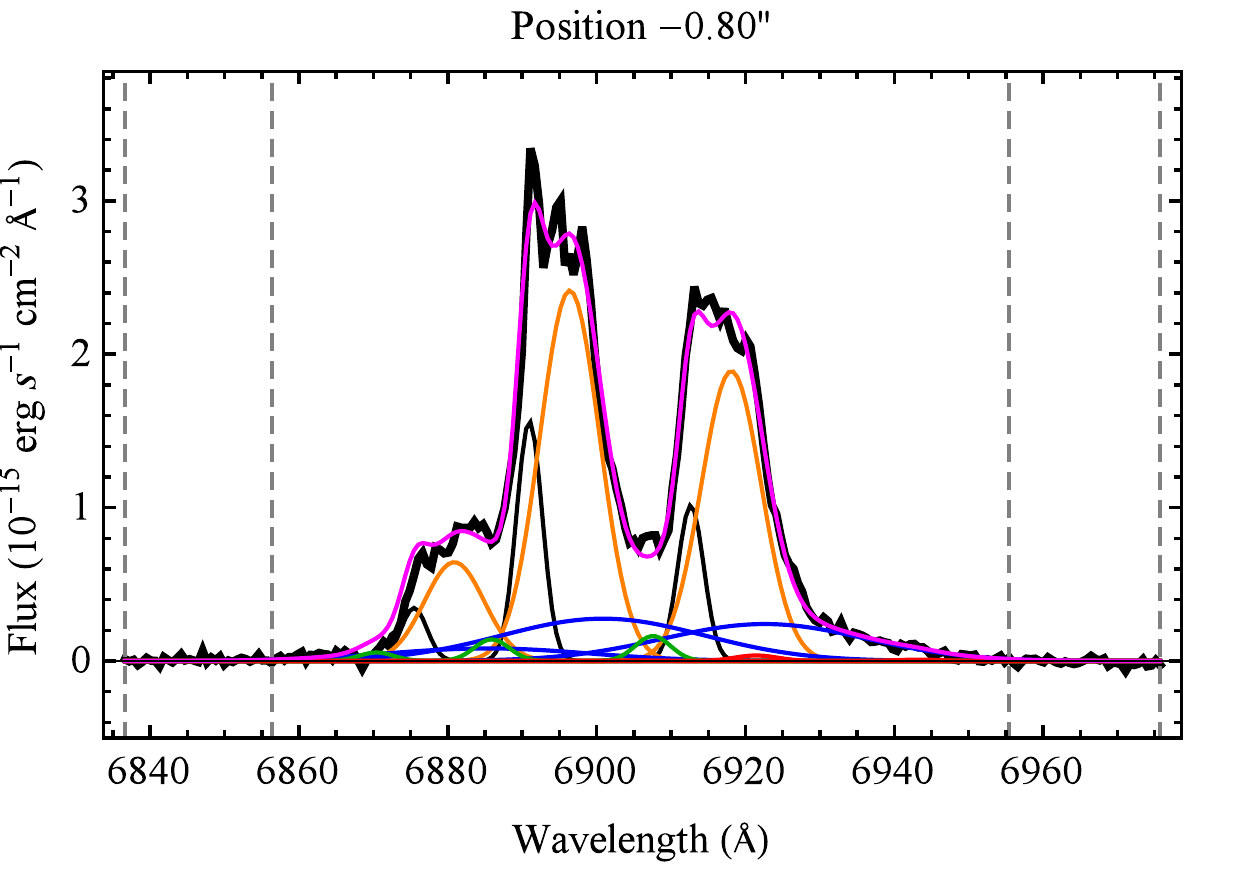}}
\subfigure{
\includegraphics[scale=0.34]{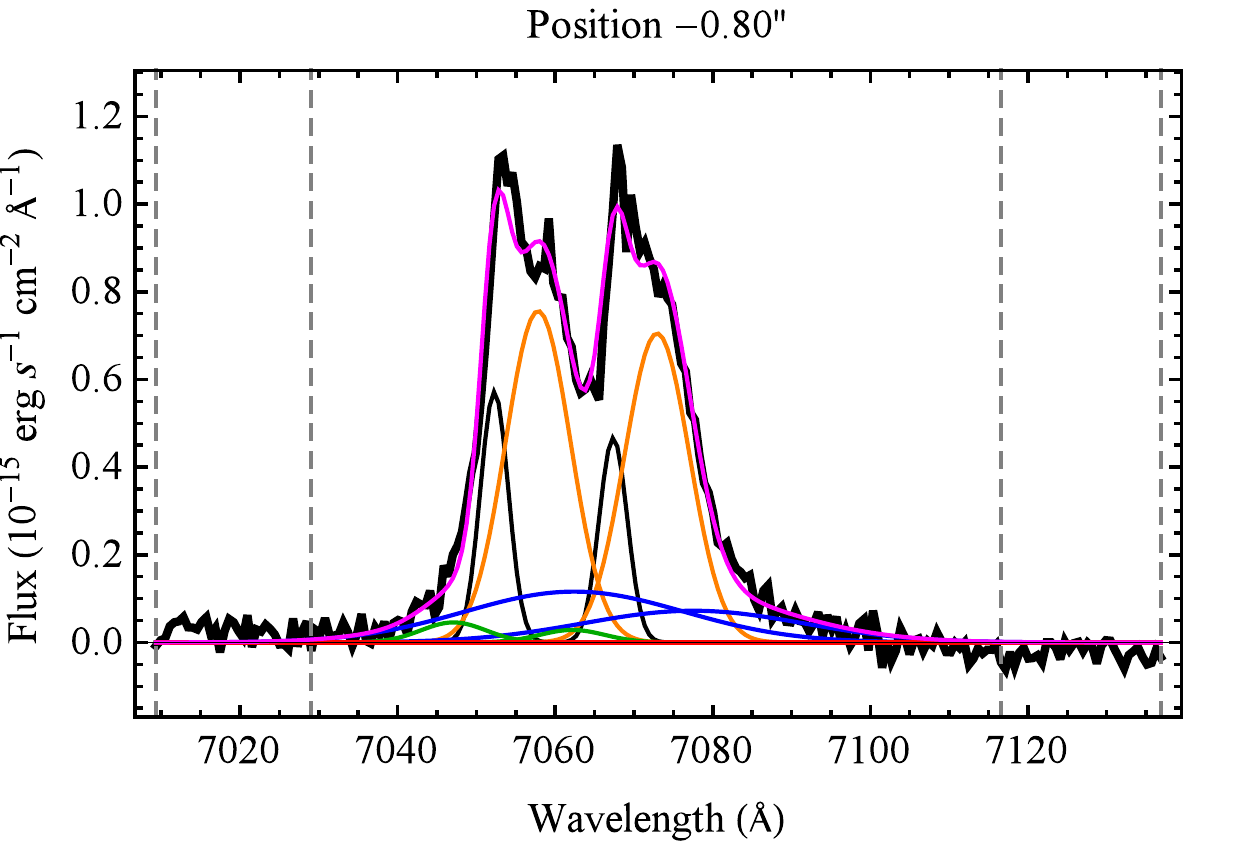}}
\subfigure{
\includegraphics[scale=0.34]{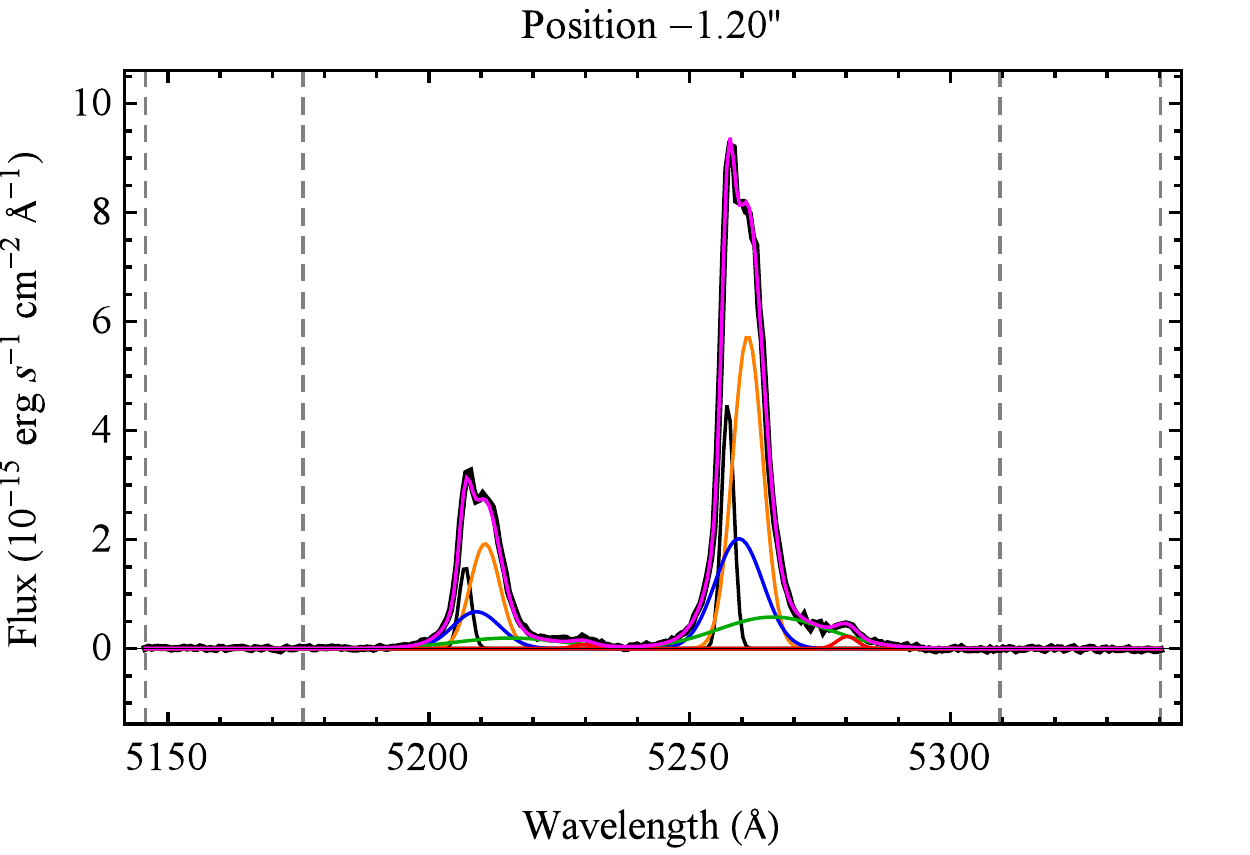}}
\subfigure{
\includegraphics[scale=0.34]{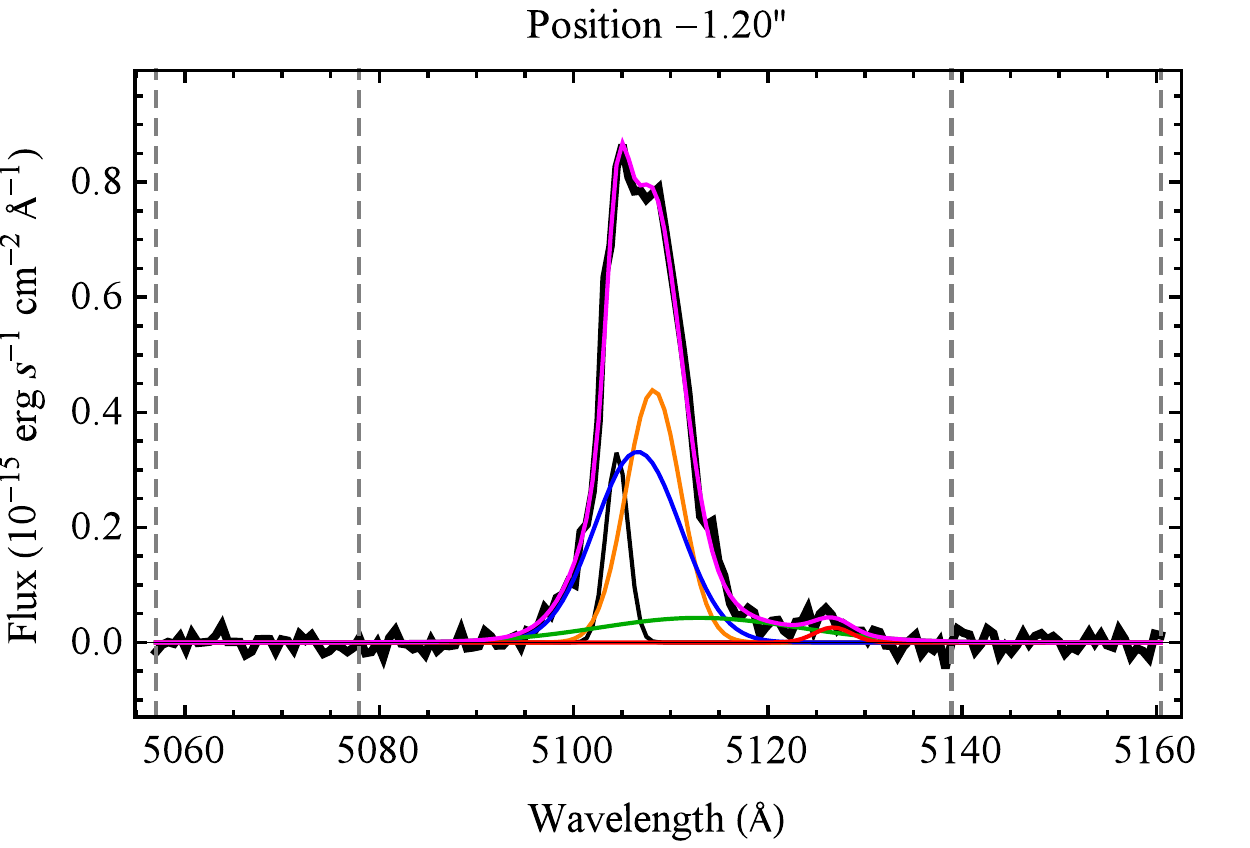}}
\subfigure{
\includegraphics[scale=0.34]{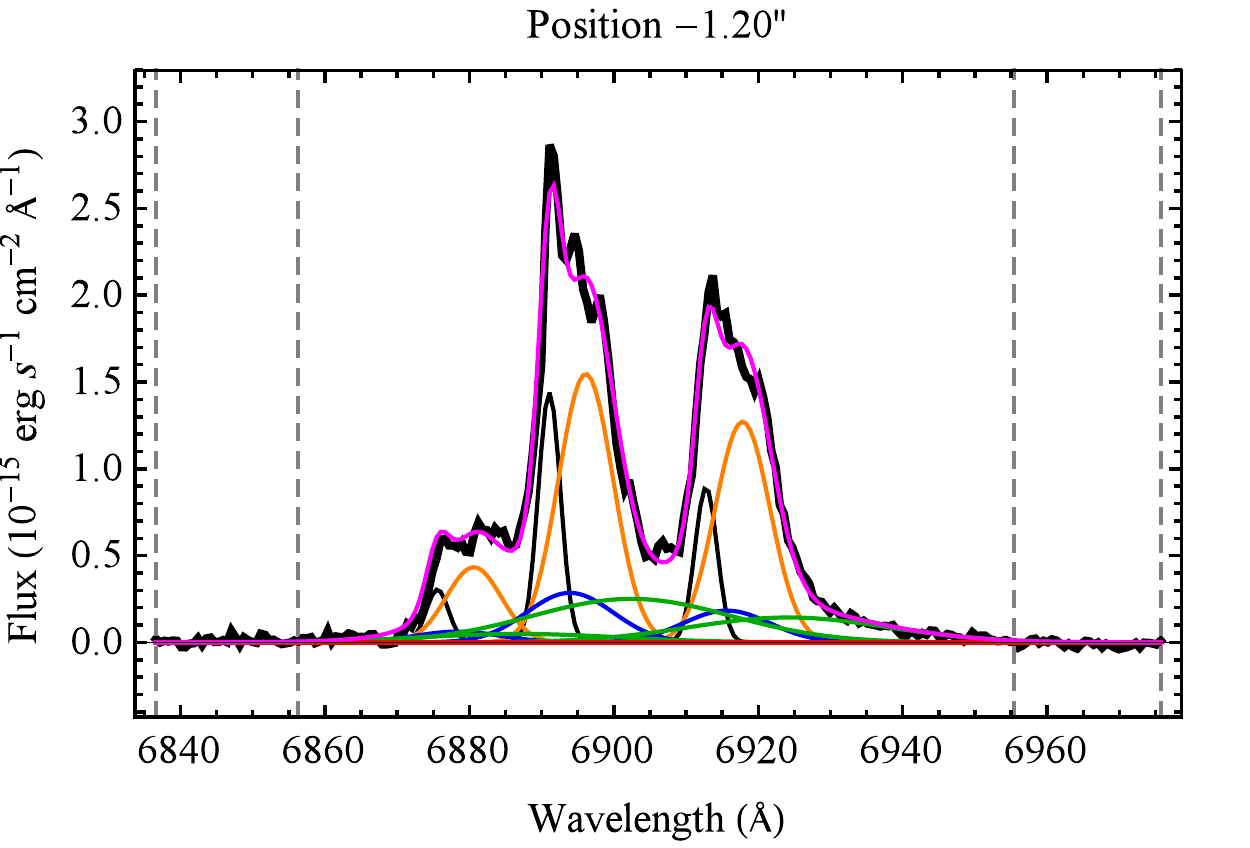}}
\subfigure{
\includegraphics[scale=0.34]{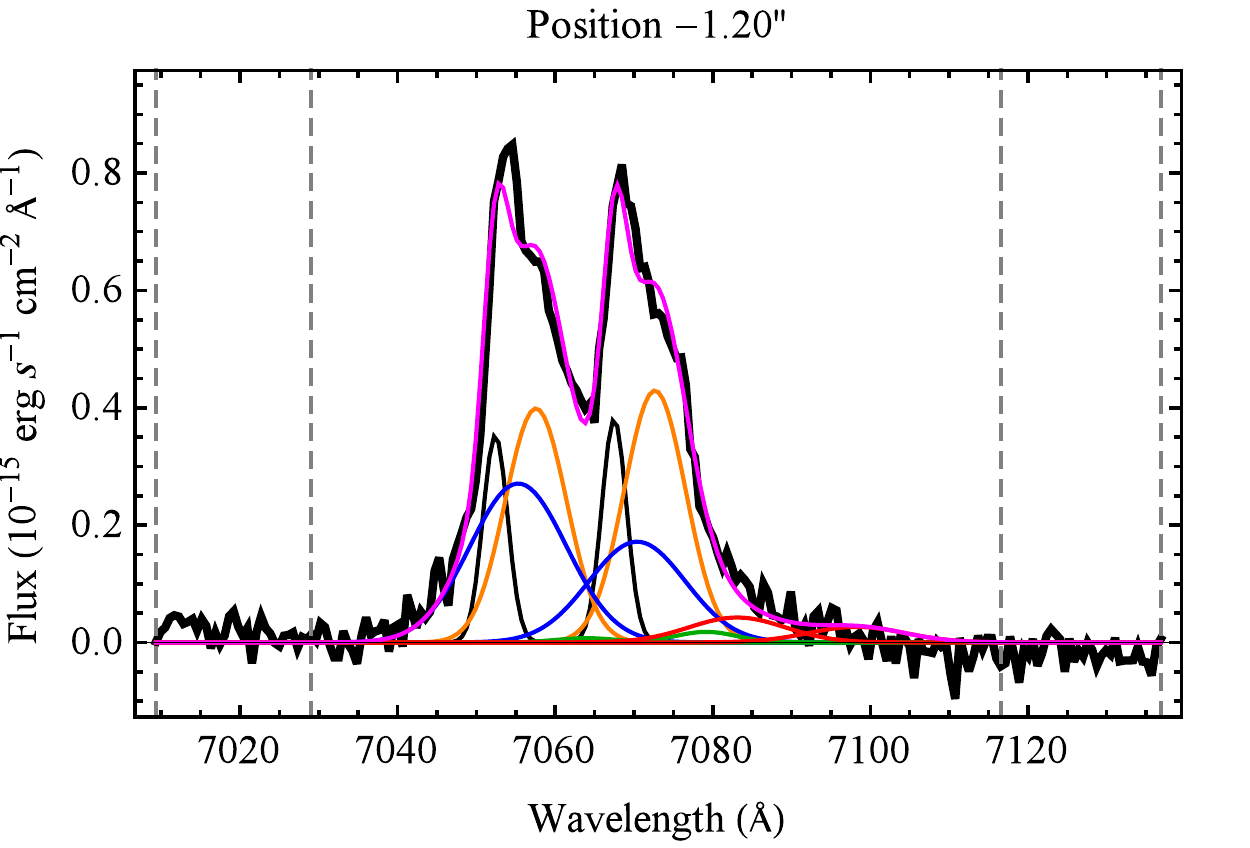}}
\subfigure{
\includegraphics[scale=0.34]{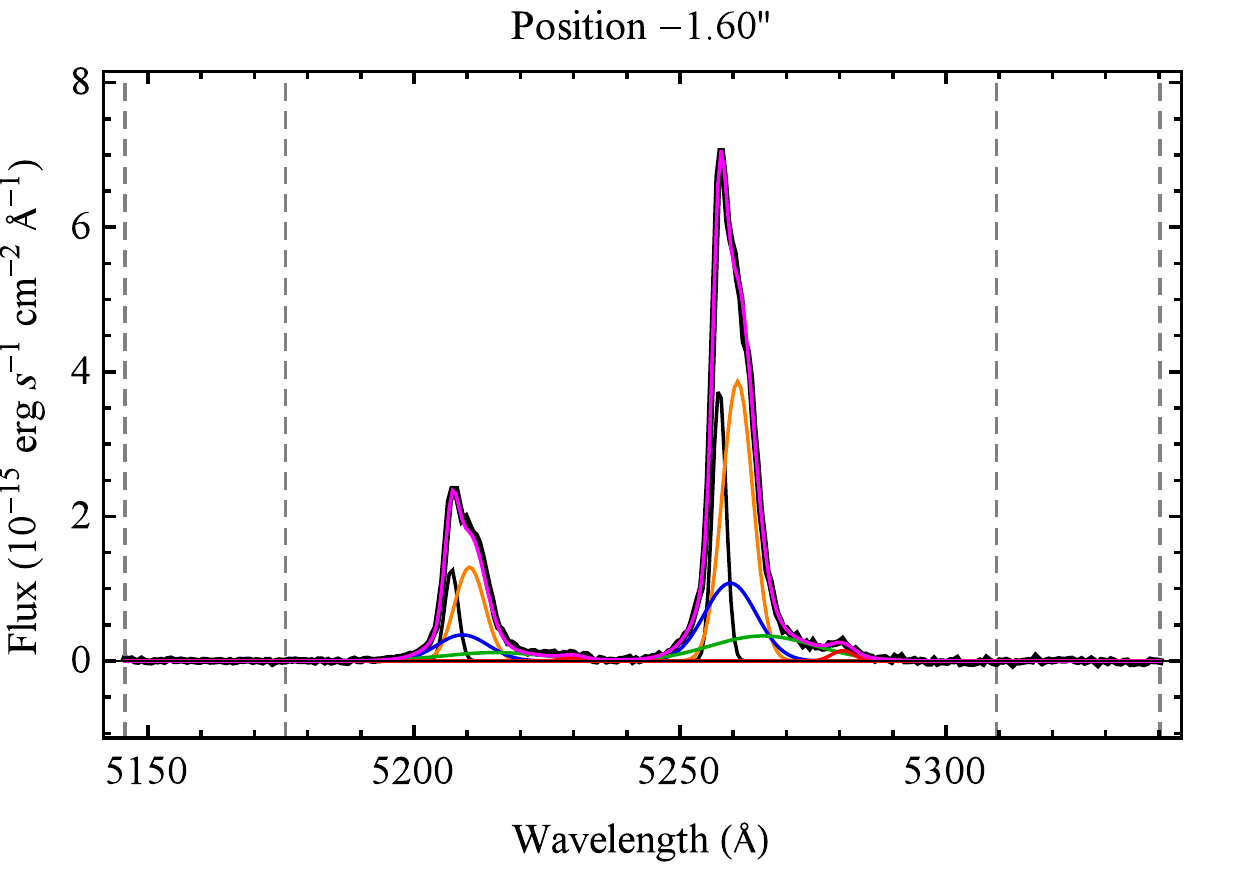}}
\subfigure{
\includegraphics[scale=0.34]{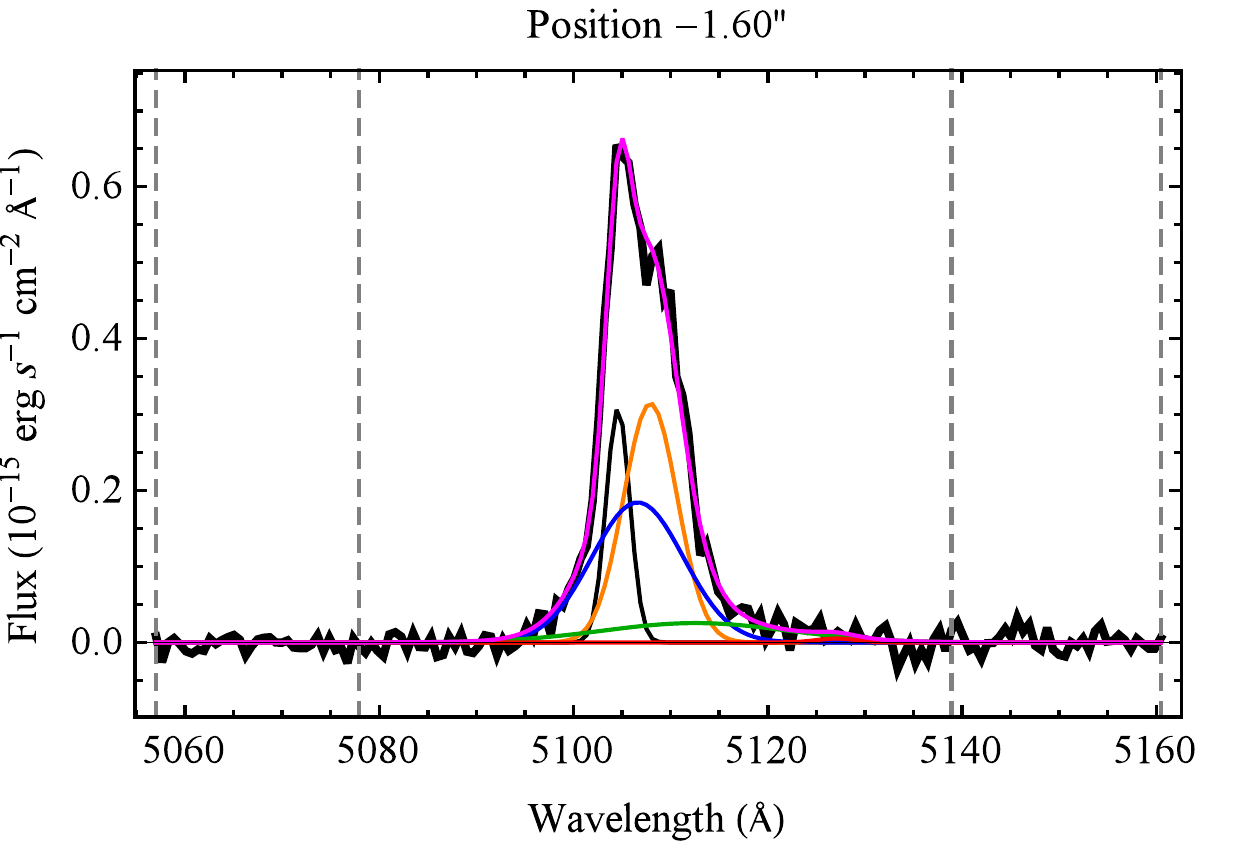}}
\subfigure{
\includegraphics[scale=0.34]{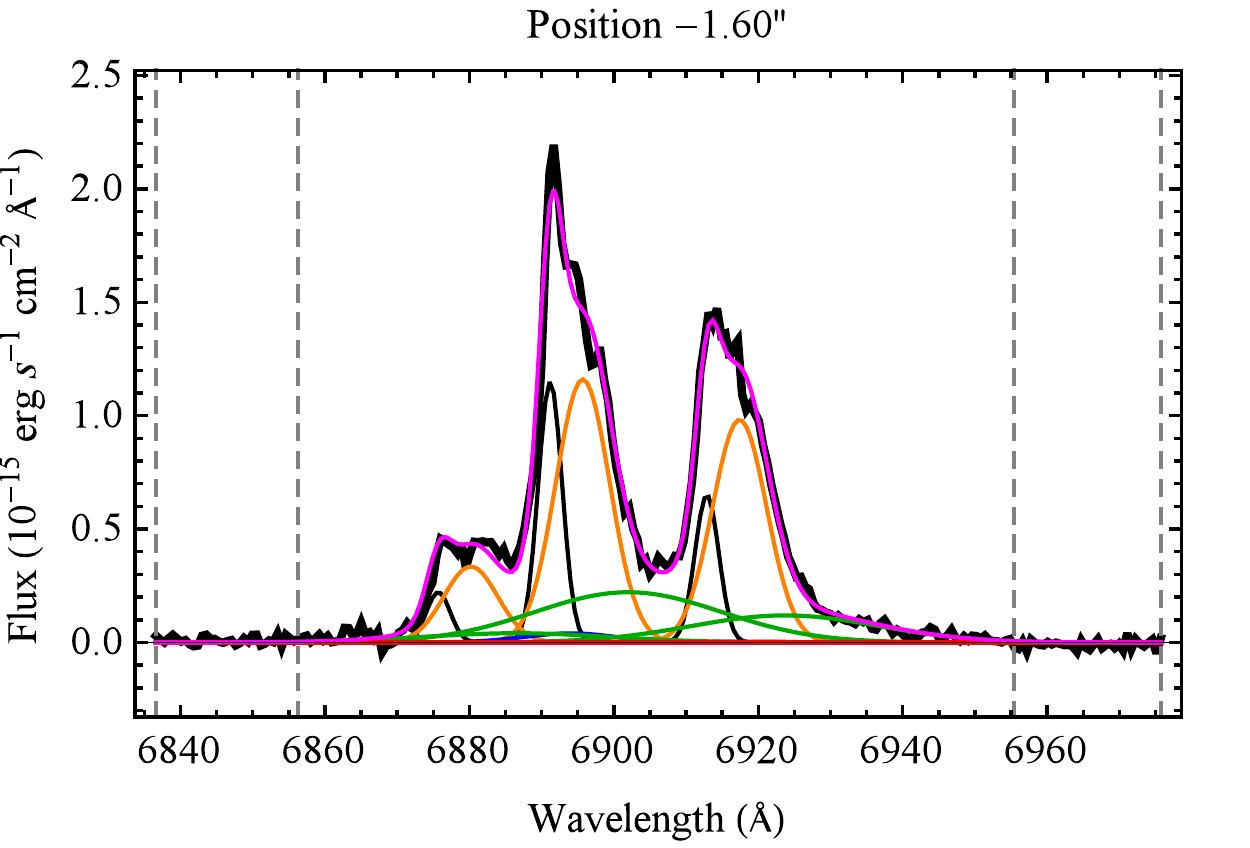}}
\subfigure{
\includegraphics[scale=0.34]{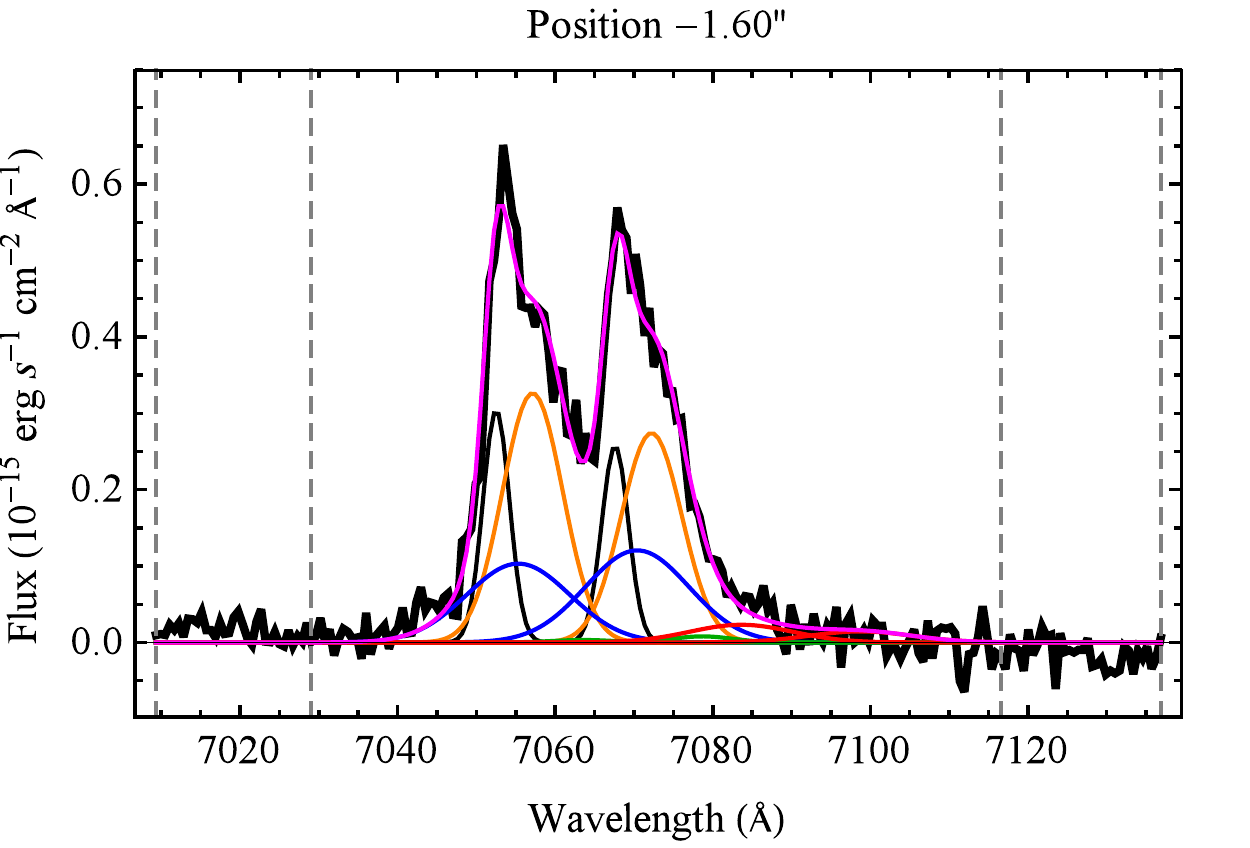}}
\subfigure{
\includegraphics[scale=0.34]{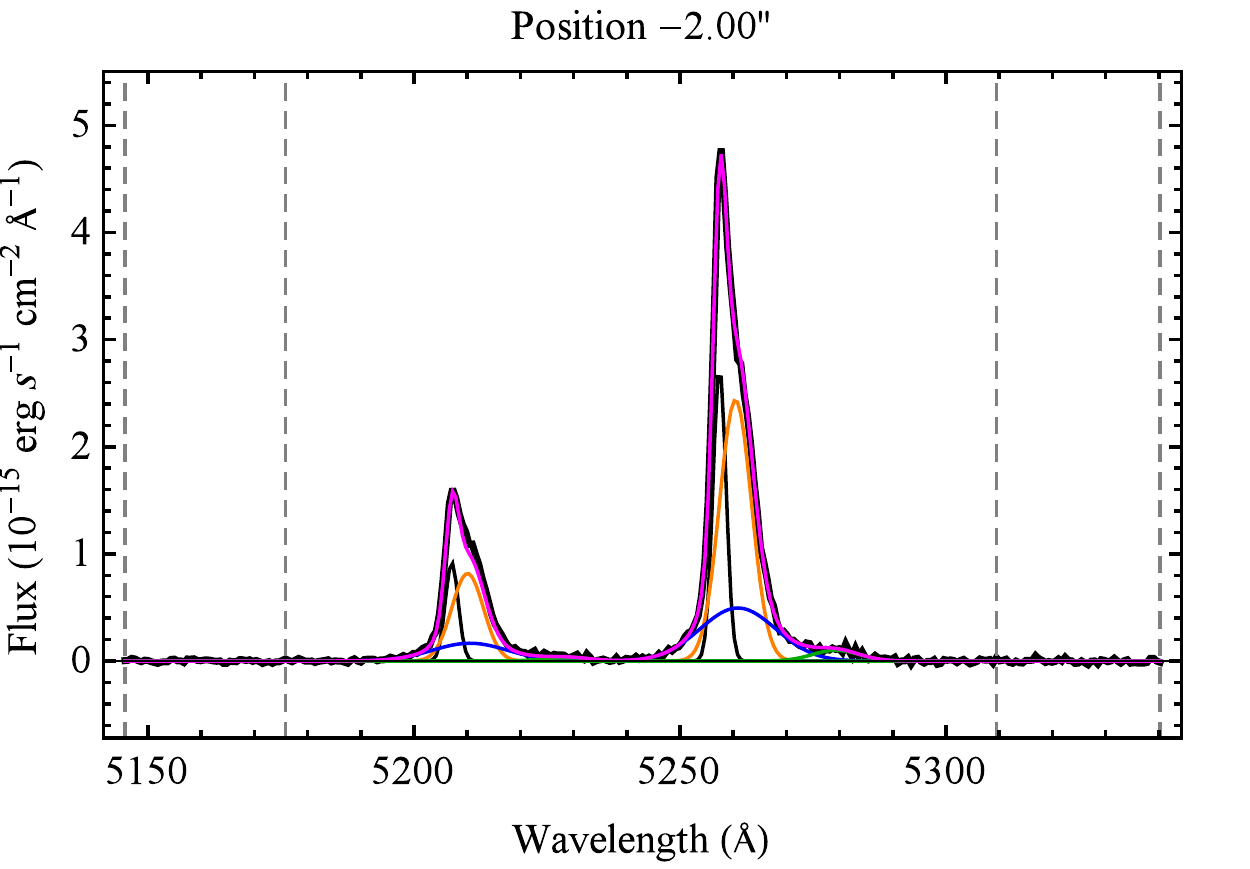}}
\subfigure{
\includegraphics[scale=0.34]{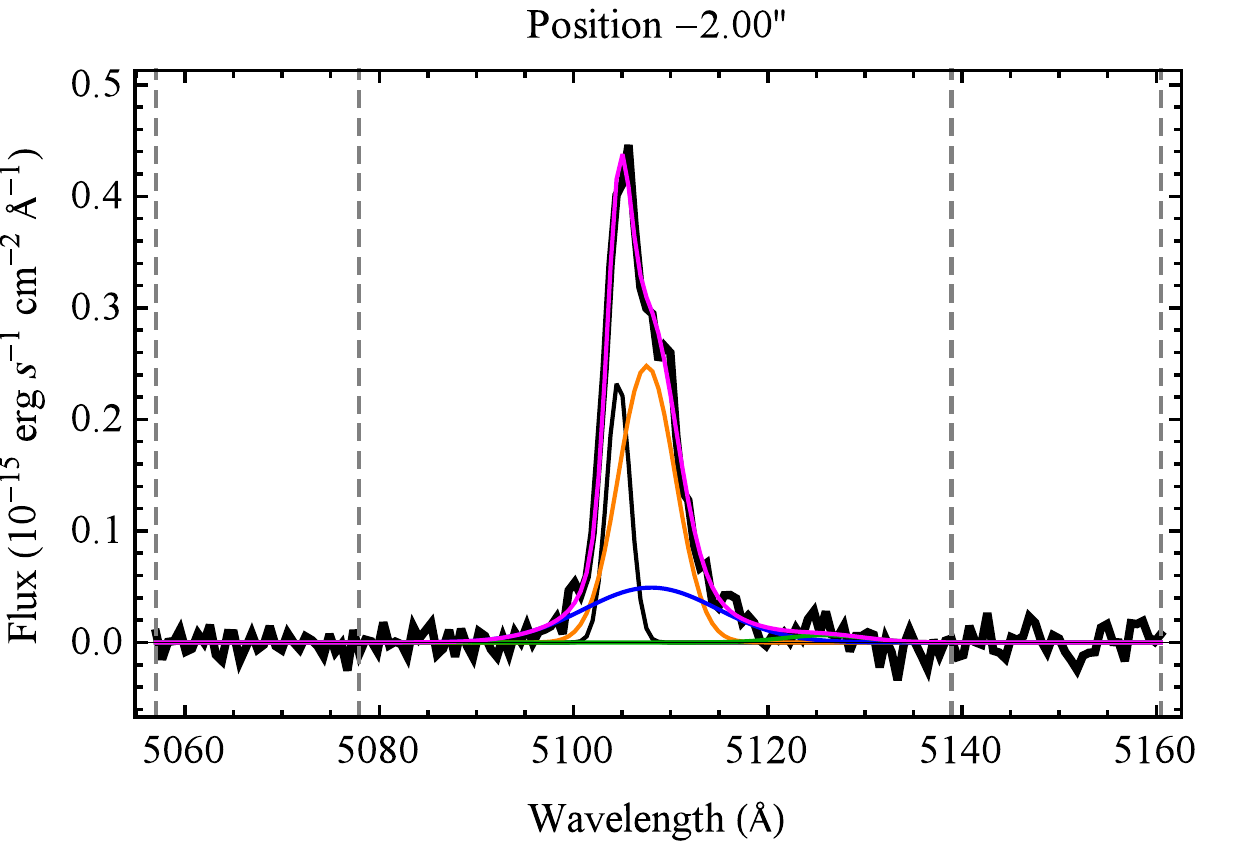}}
\subfigure{
\includegraphics[scale=0.34]{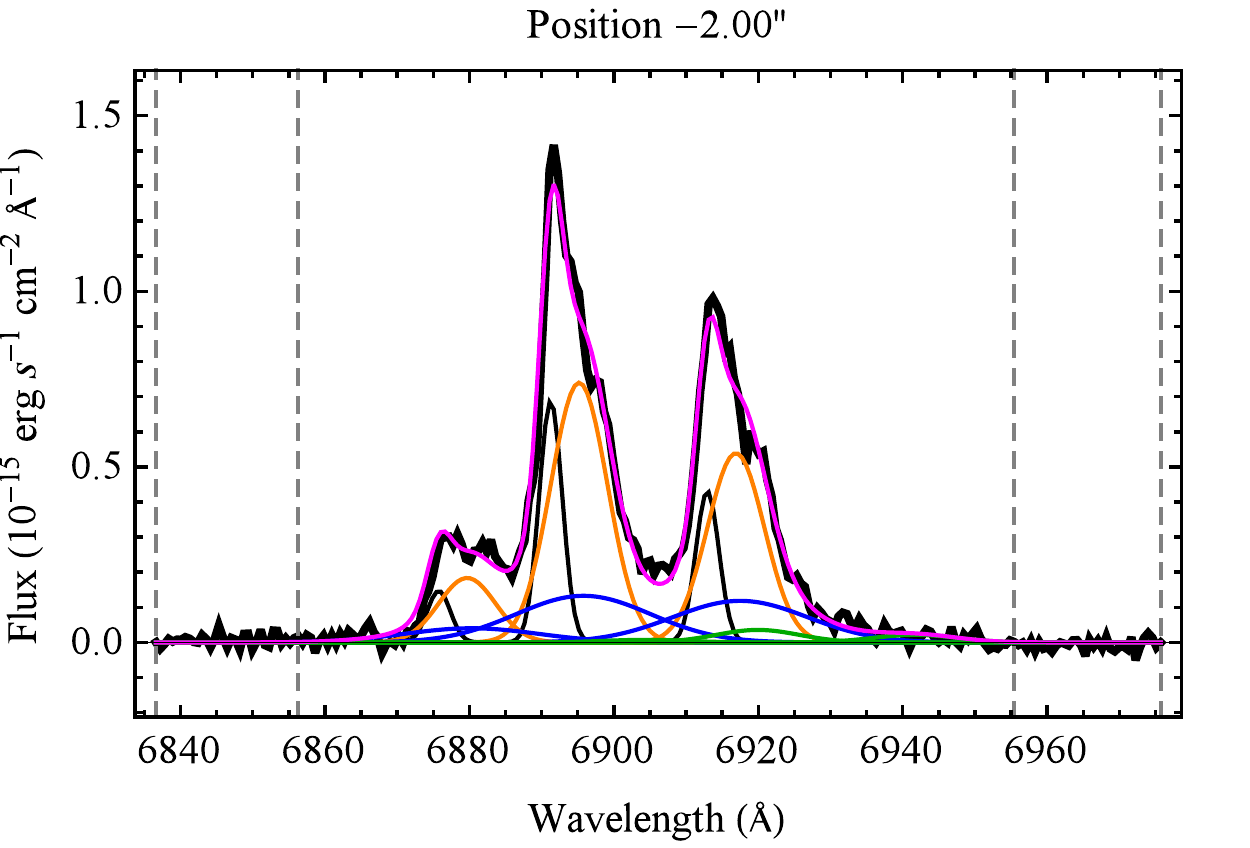}}
\subfigure{
\includegraphics[scale=0.34]{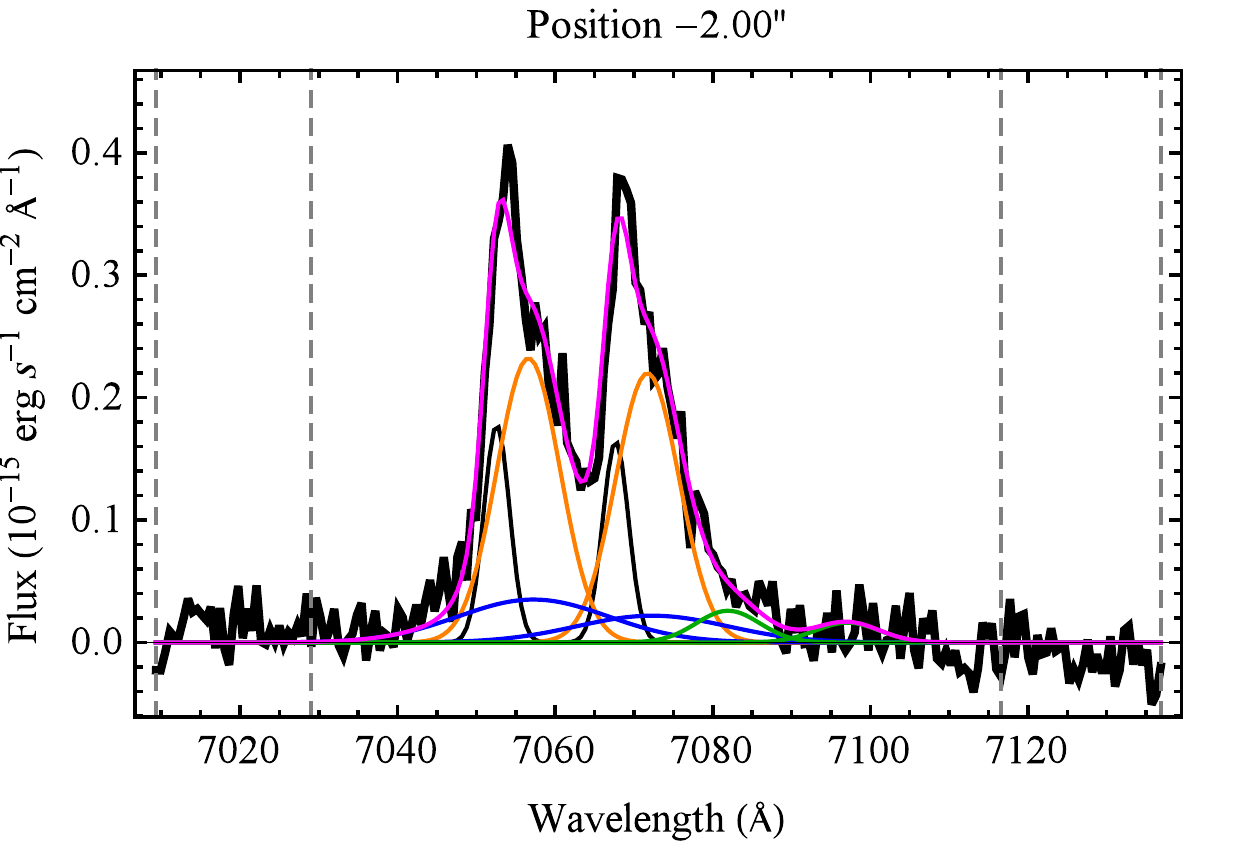}}
\caption{{\it continued.}}
\end{figure*}

\clearpage

\begin{figure*}
\centering
\subfigure{
\includegraphics[scale=0.45]{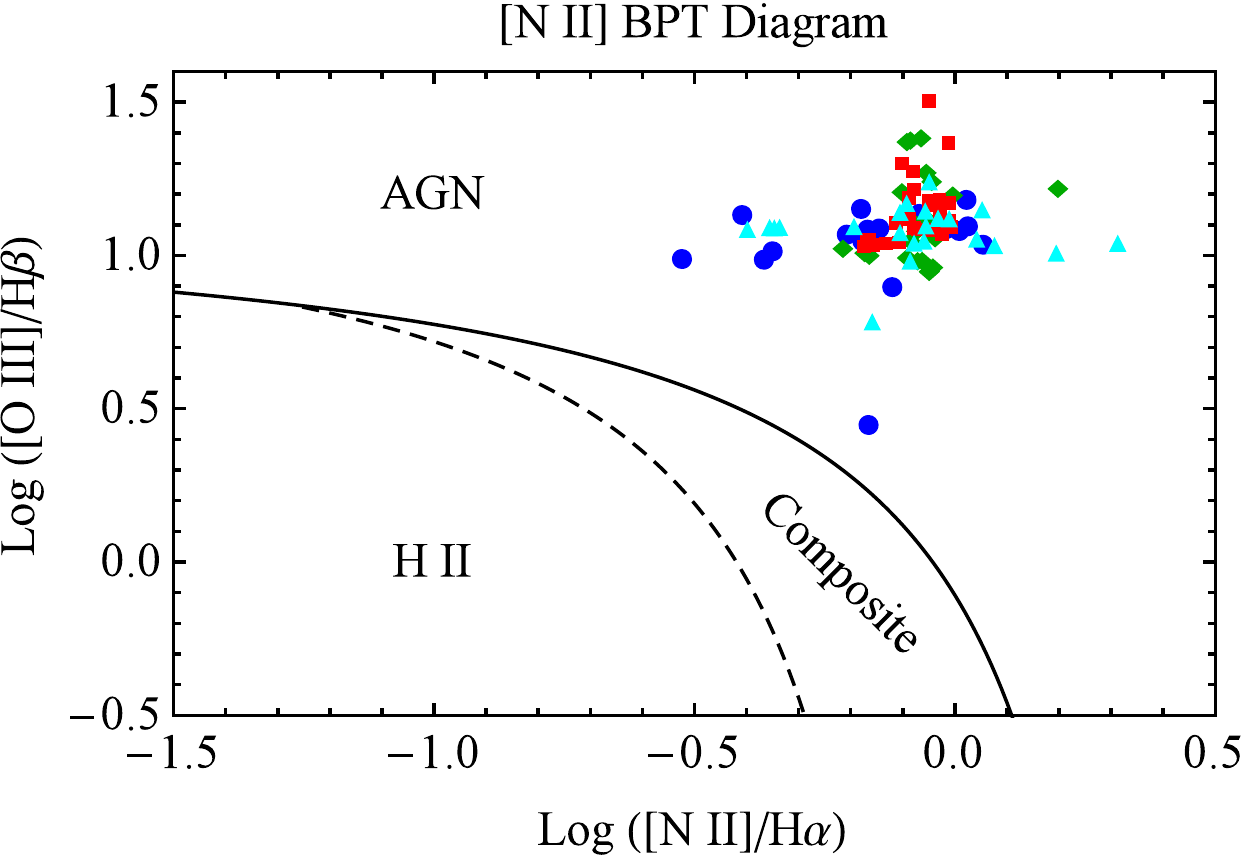}}
\subfigure{
\includegraphics[scale=0.45]{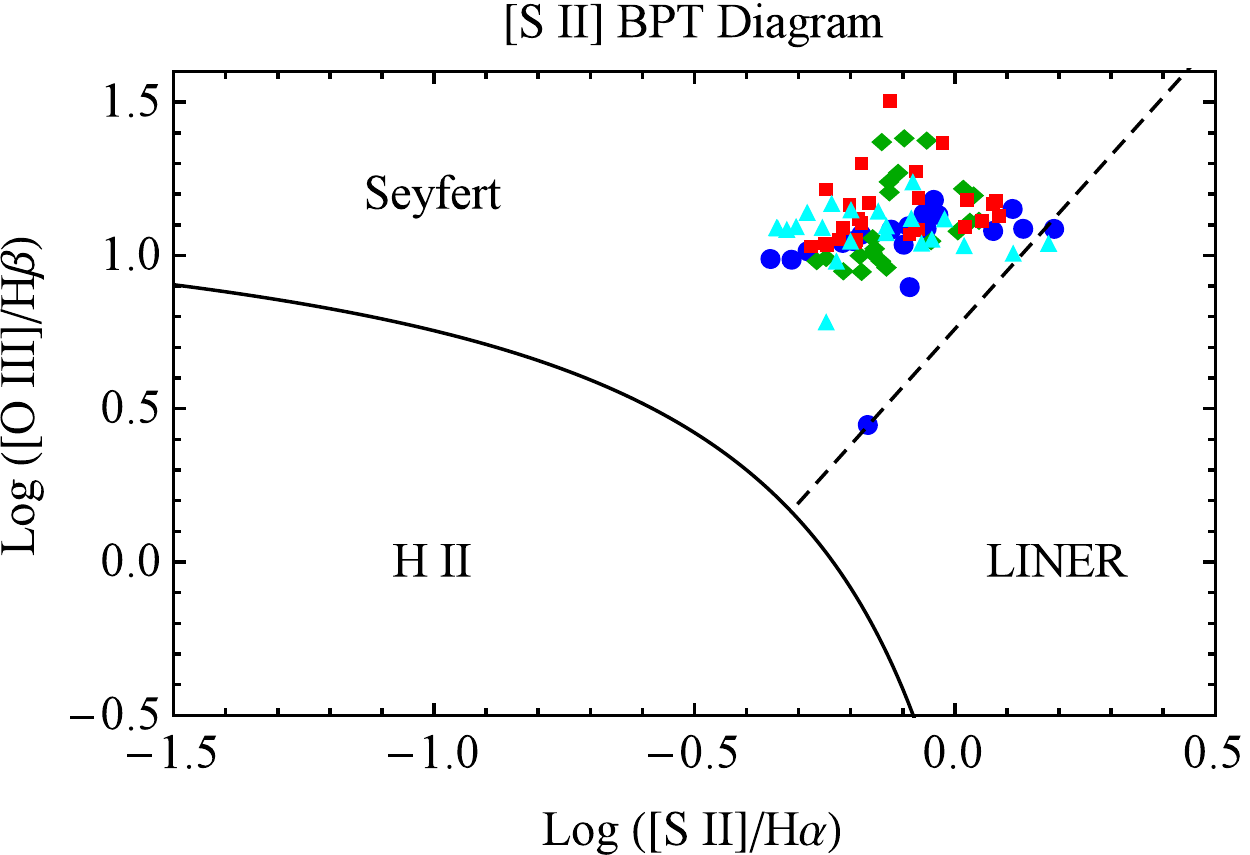}}
\subfigure{
\includegraphics[scale=0.45]{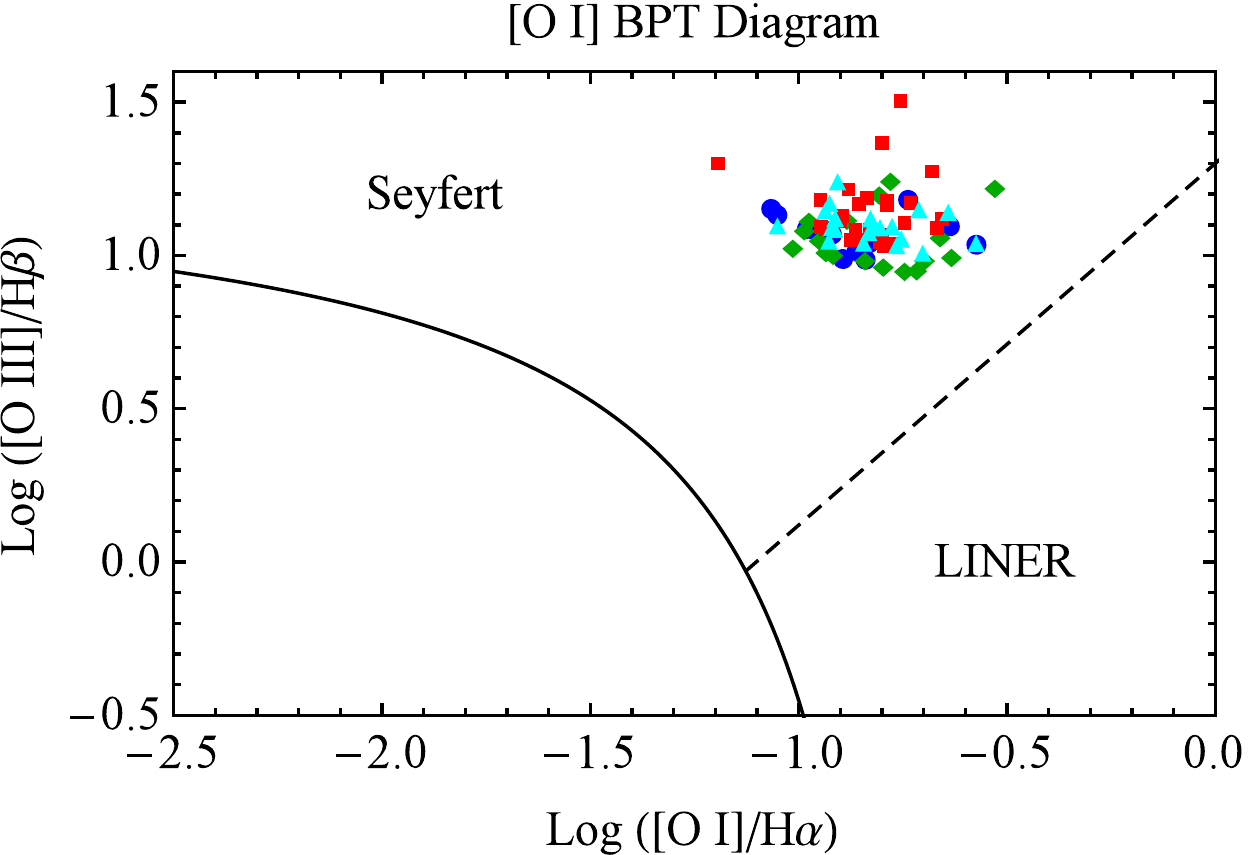}}
\caption{BPT ionization diagrams for [N~II], [S~II], and [O~I] using ratios calculated with the fluxes of all kinematic components summed together. The points reach a maximum radial extent of $\sim 8\arcsec$ from the nucleus, and symbols are color-coded for each of the four APO DIS position angles: 73$\degr$ (blue circles), 118$\degr$ (green diamonds), 163$\degr$ (red squares), 208$\degr$ (cyan triangles). The demarcation lines for distinguishing ionization mechanisms are from \cite{kewley2001, kewley2006, kauffmann2003}.}
\end{figure*}

\begin{figure*}
\centering
\subfigure{
\includegraphics[scale=0.45]{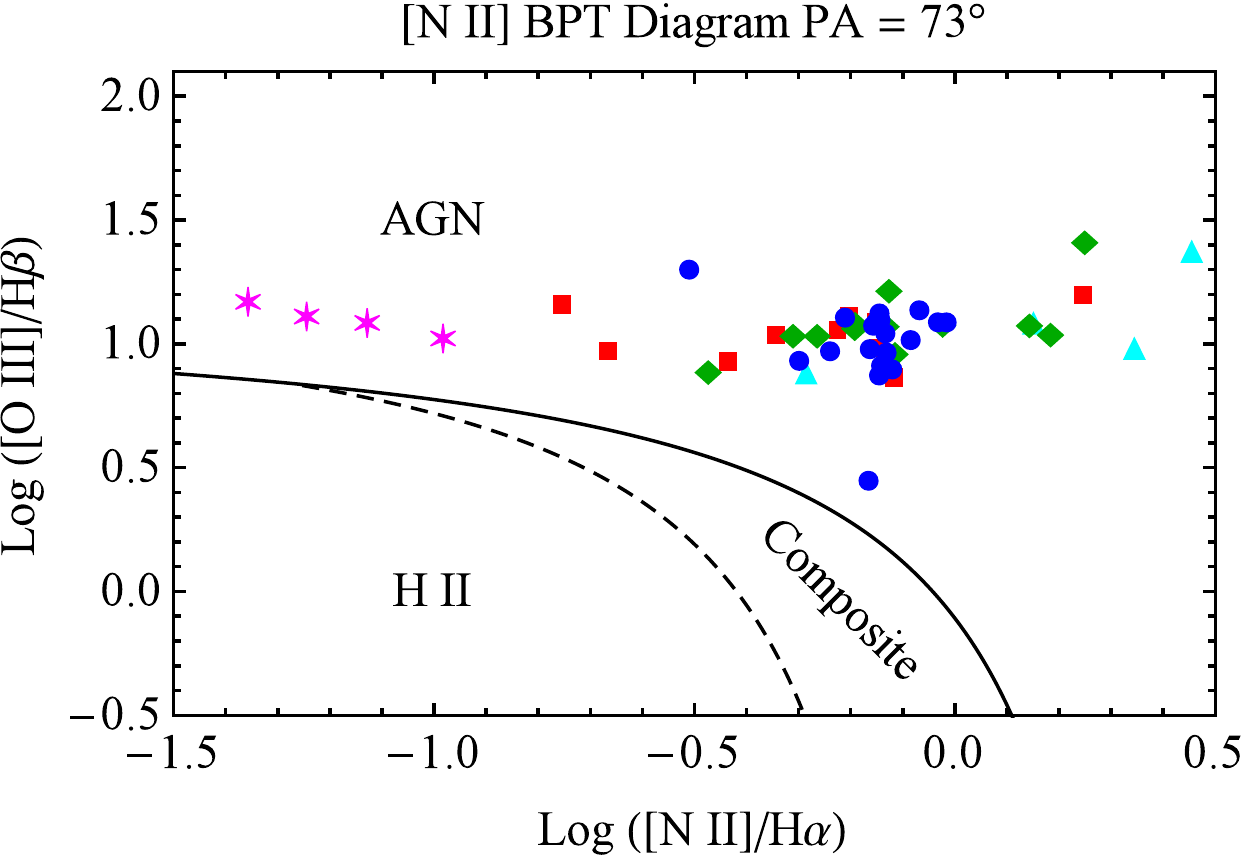}}
\vspace{-4pt}
\subfigure{
\includegraphics[scale=0.45]{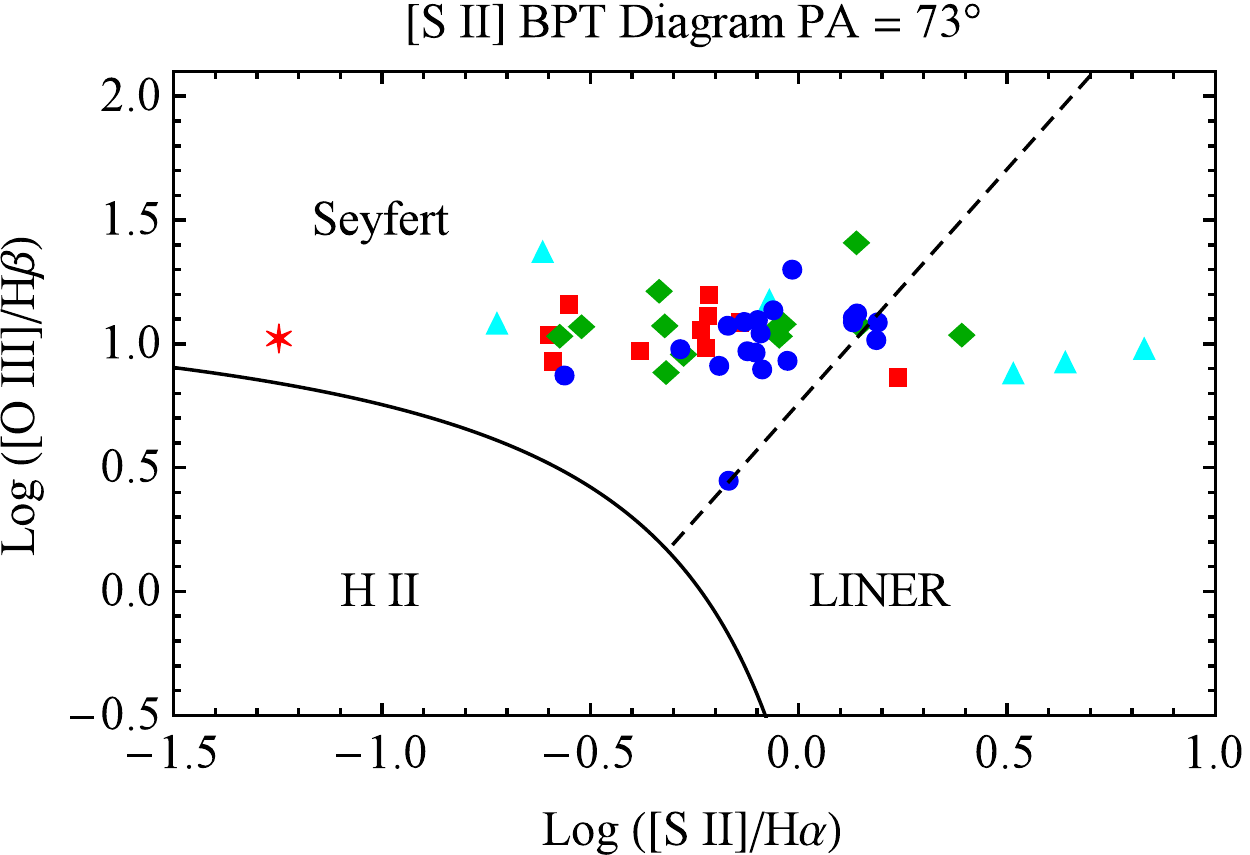}}
\subfigure{
\includegraphics[scale=0.45]{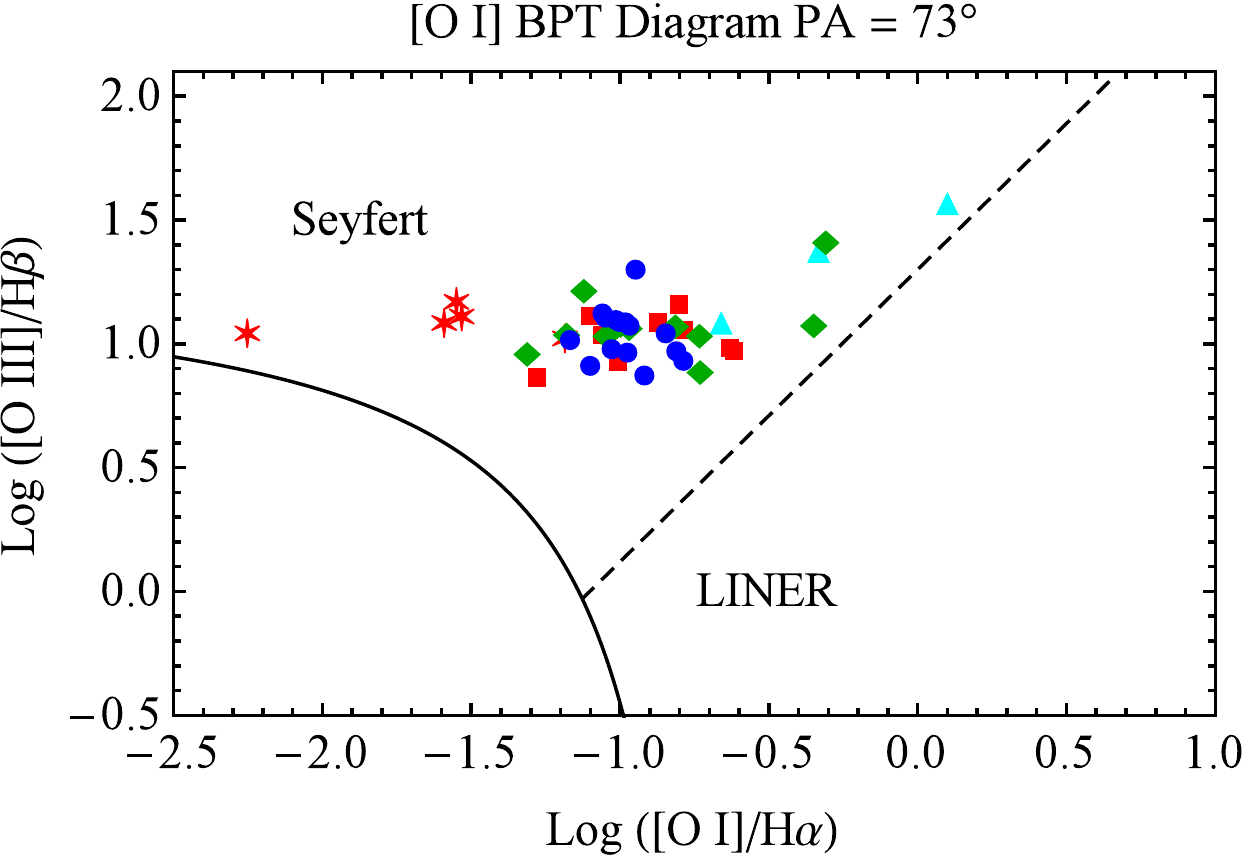}}
\vspace{-4pt}
\subfigure{
\includegraphics[scale=0.45]{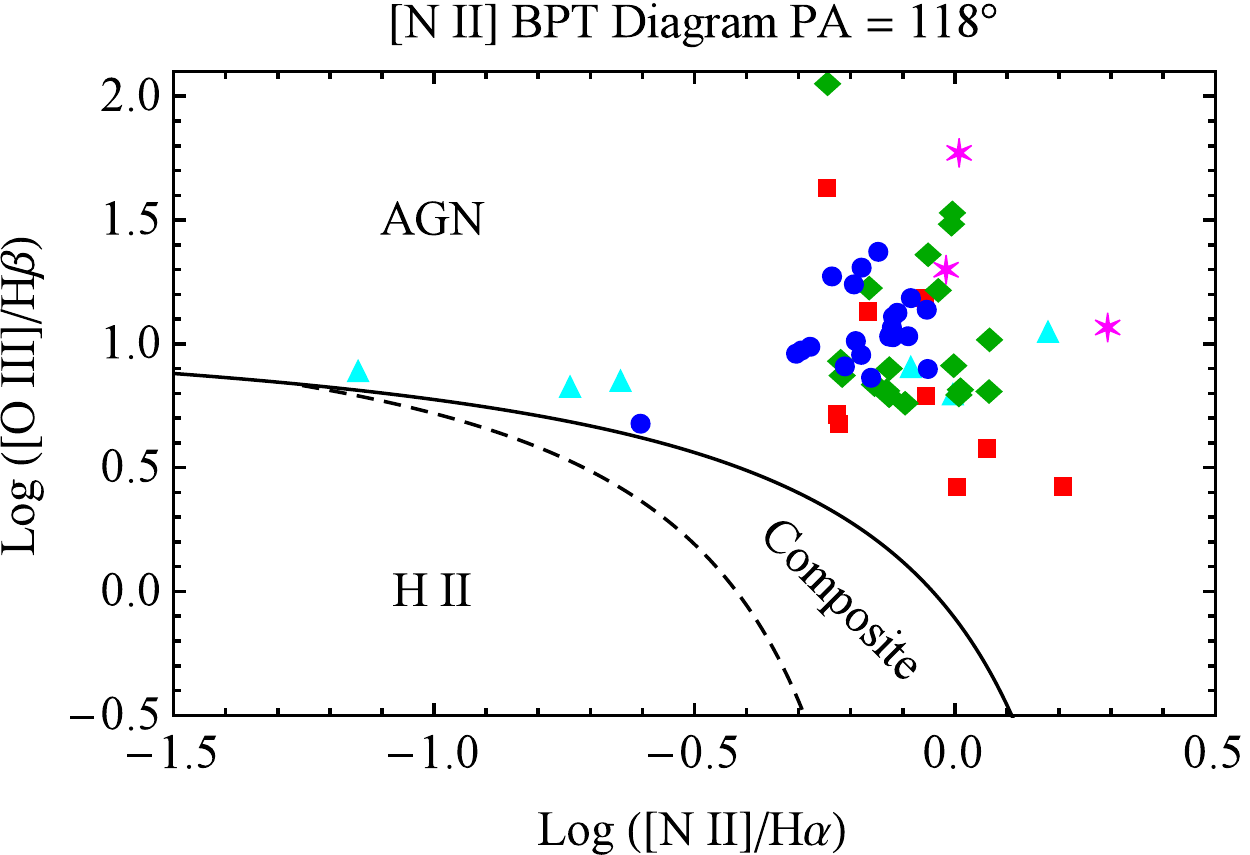}}
\subfigure{
\includegraphics[scale=0.45]{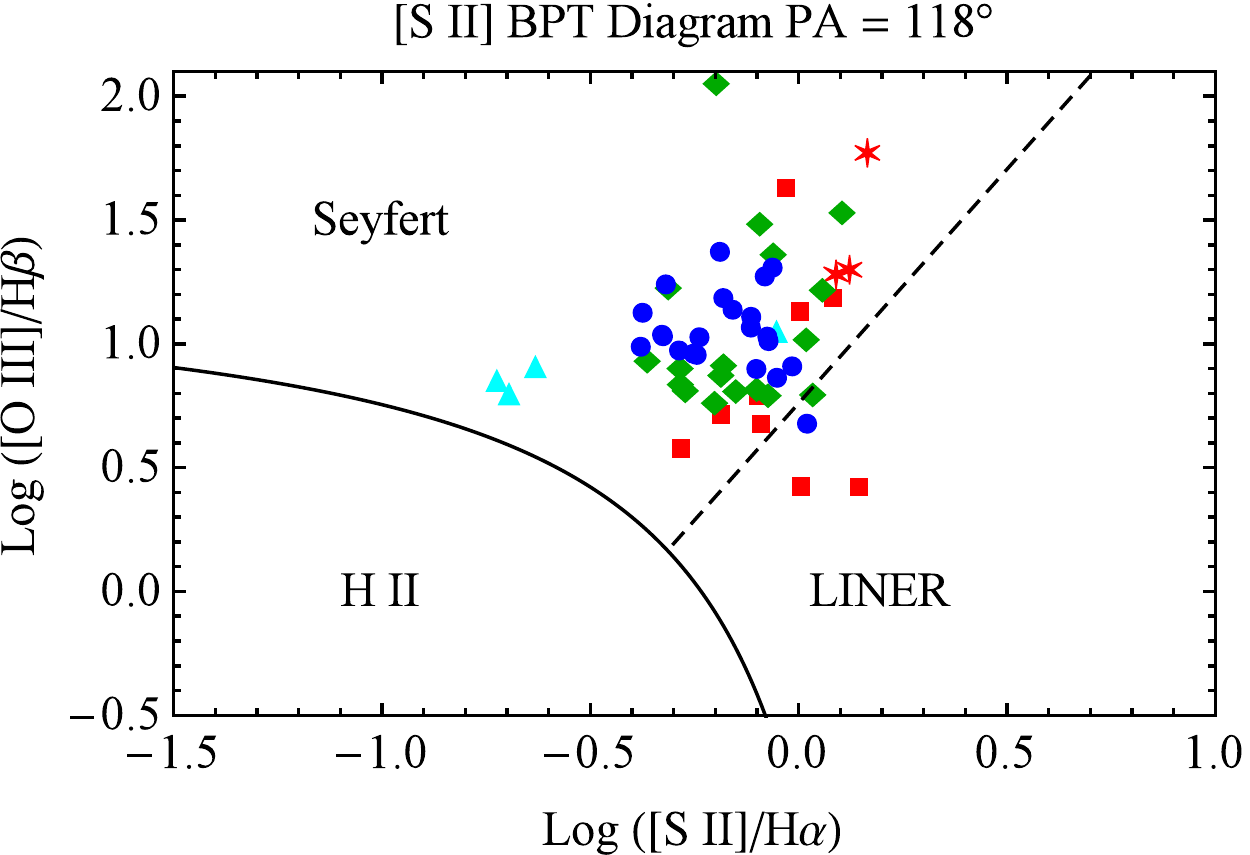}}
\subfigure{
\includegraphics[scale=0.45]{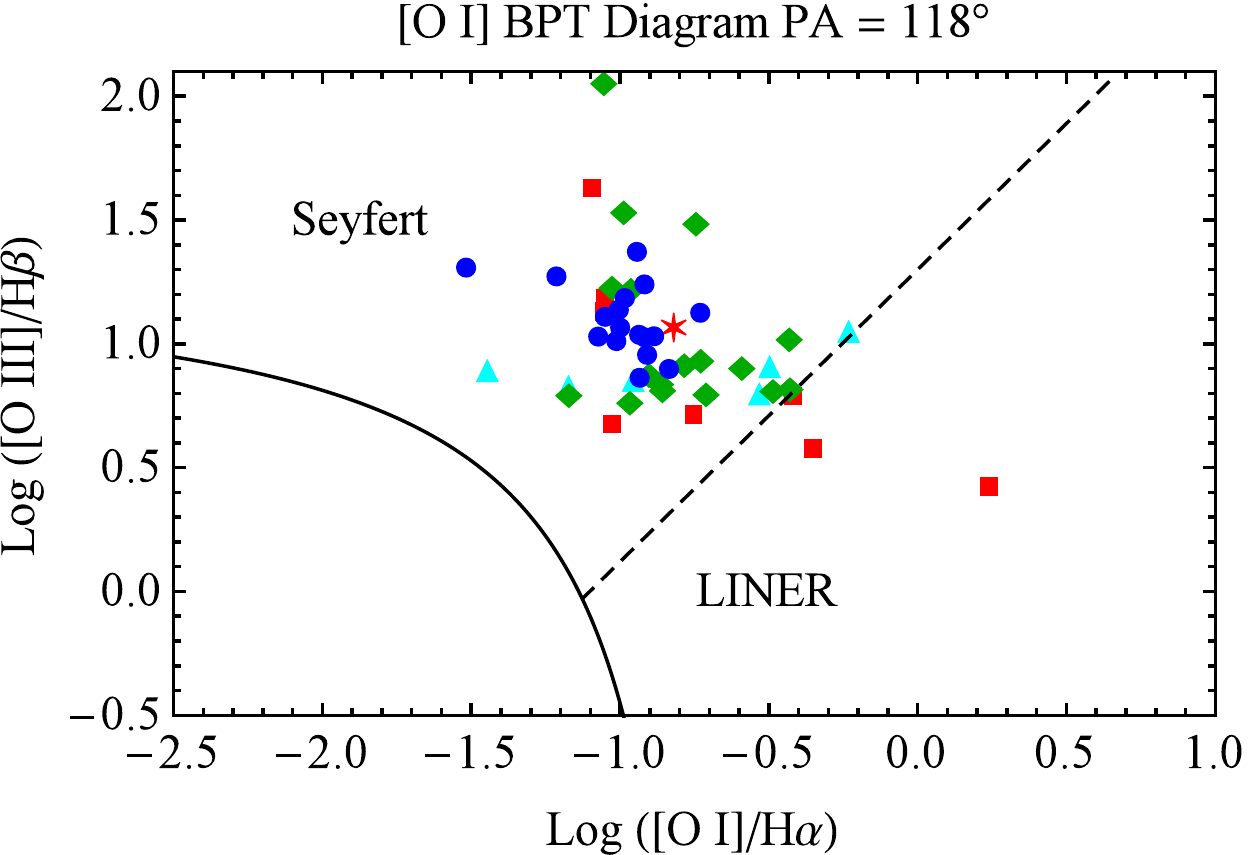}}
\vspace{-4pt}
\subfigure{
\includegraphics[scale=0.45]{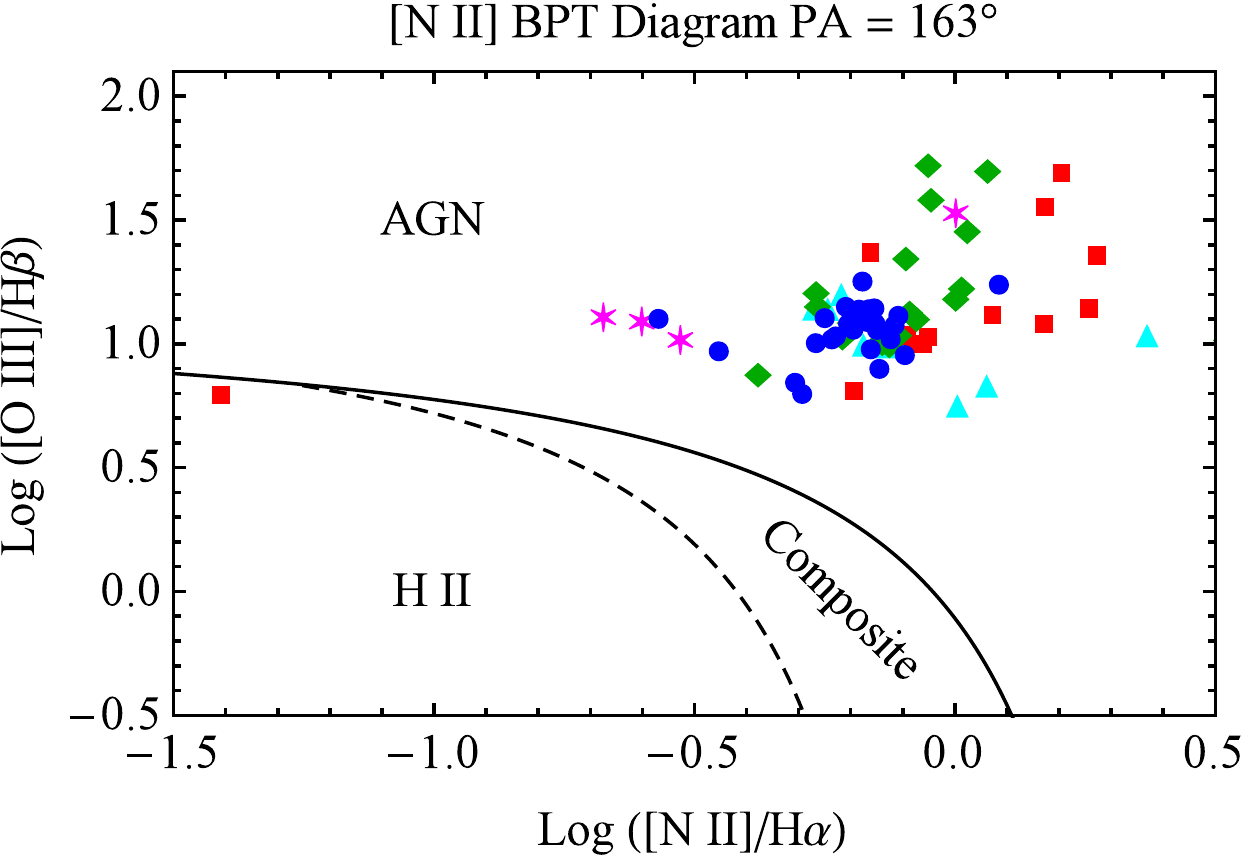}}
\subfigure{
\includegraphics[scale=0.45]{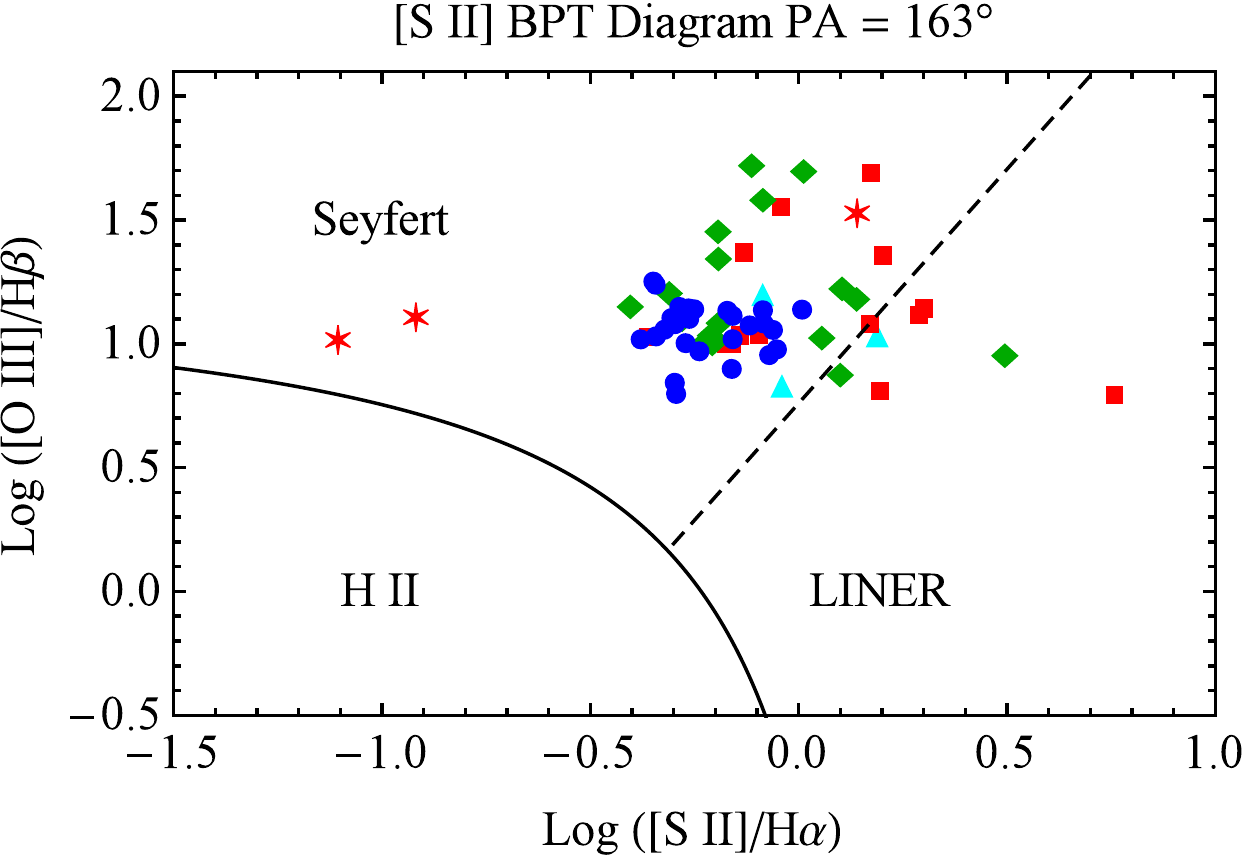}}
\subfigure{
\includegraphics[scale=0.45]{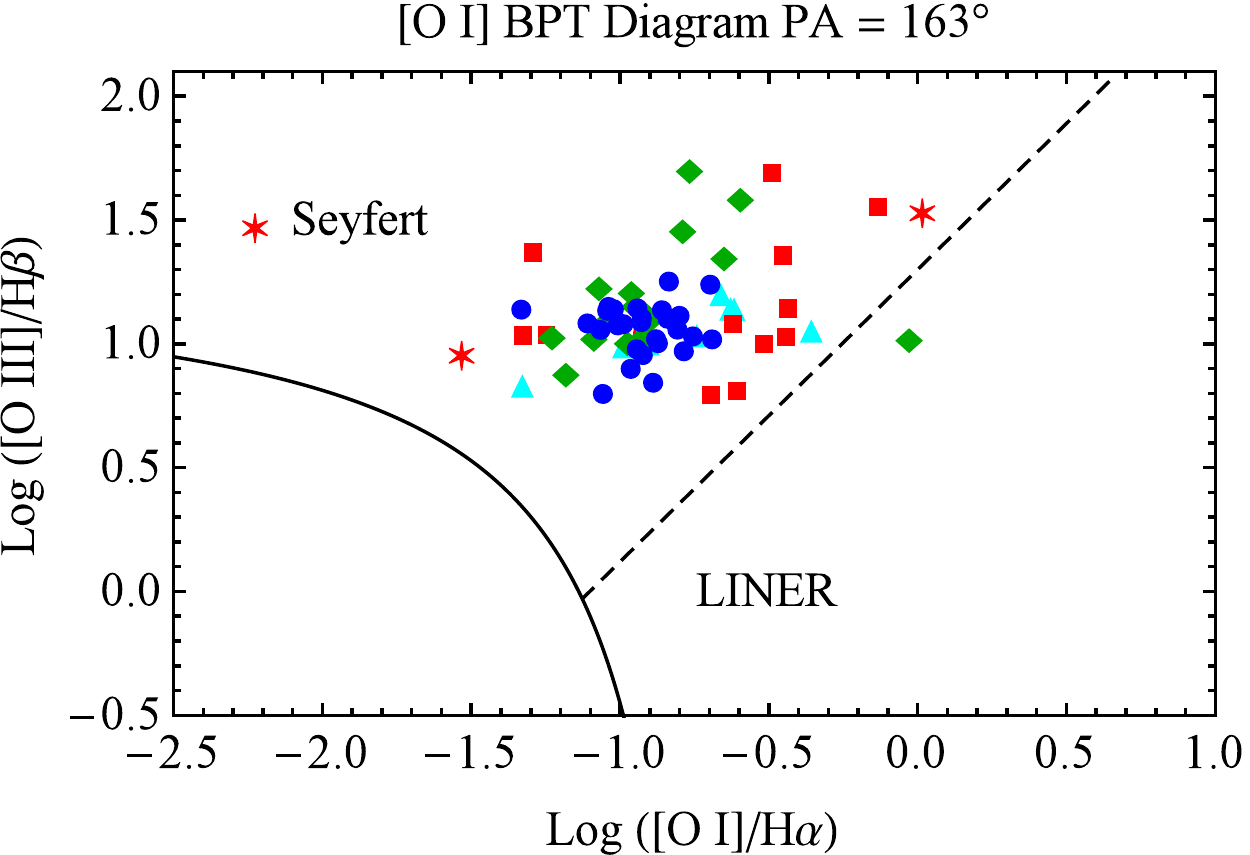}}
\subfigure{
\includegraphics[scale=0.45]{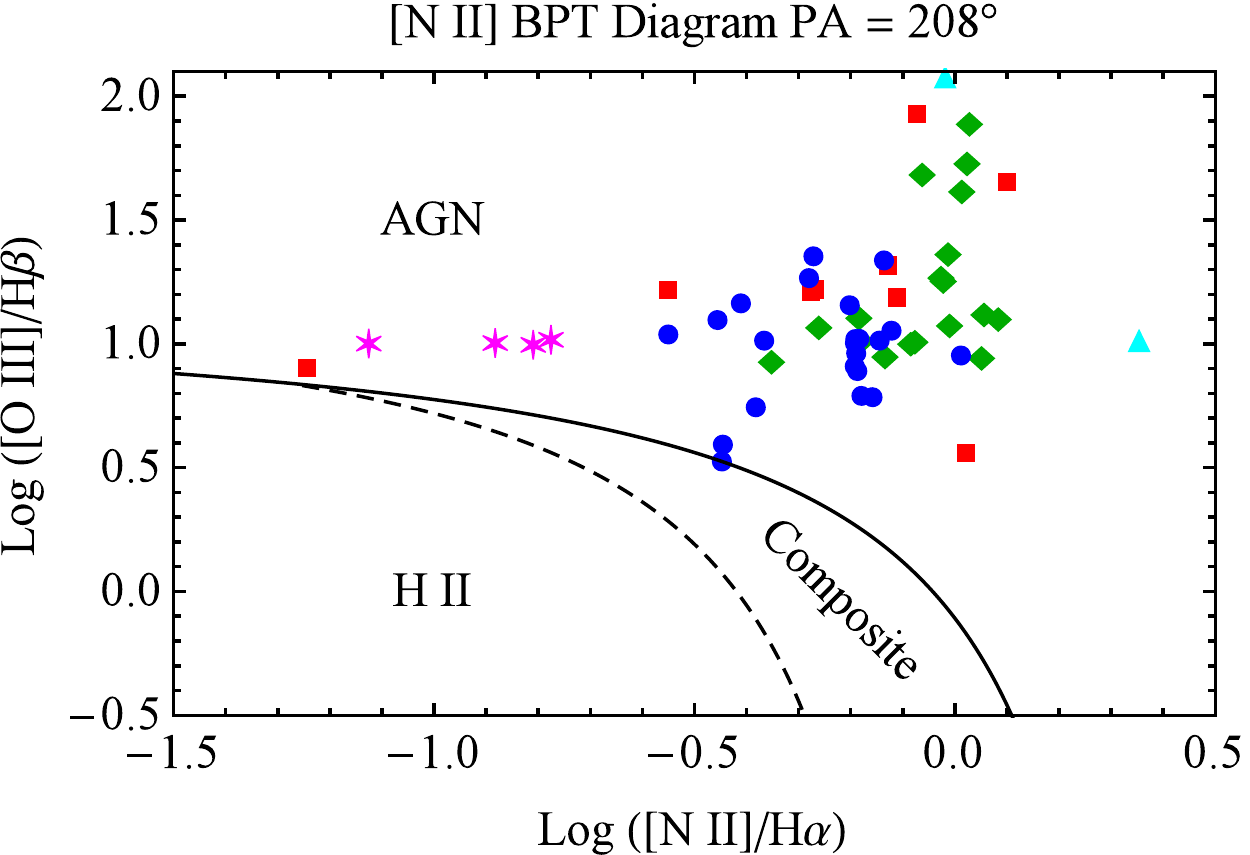}}
\subfigure{
\includegraphics[scale=0.45]{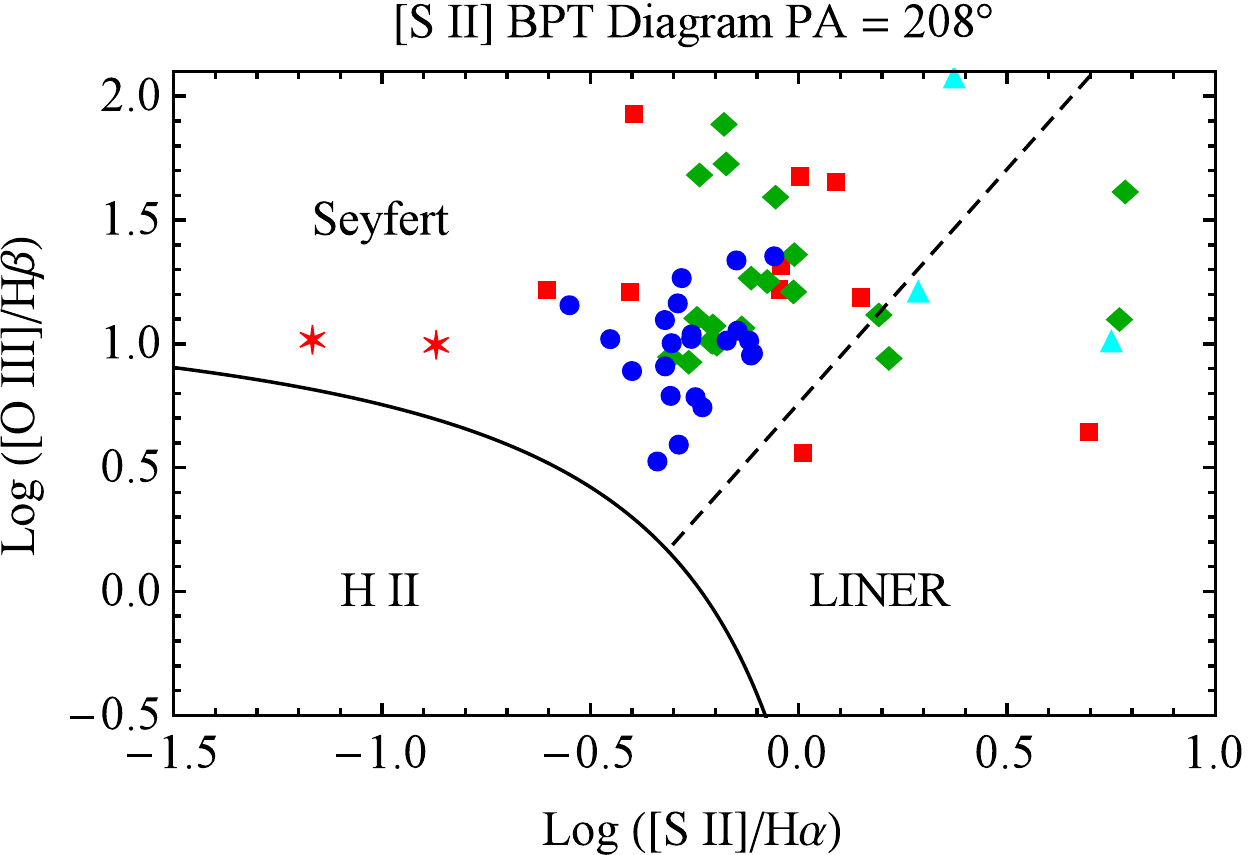}}
\subfigure{
\includegraphics[scale=0.45]{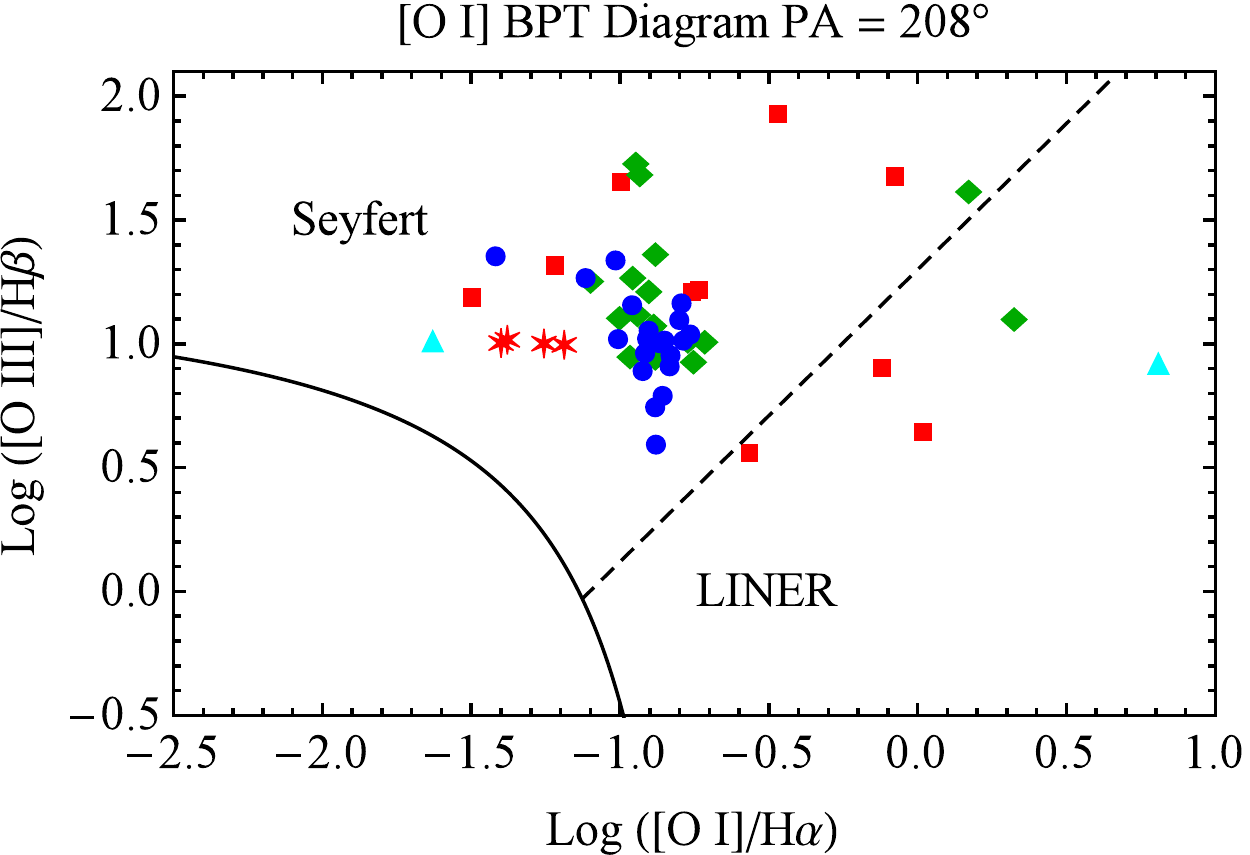}}
\caption{BPT ionization diagrams for [N~II], [S~II], and [O~I] for the individual kinematic components for each position angle. The symbols correspond to the peak flux of the kinematic components from strongest to weakest: blue circles, green diamonds, red squares, cyan triangles, magenta stars.}
\label{...}
\end{figure*}

\clearpage

\setlength{\tabcolsep}{0.008in} 
\tabletypesize{\tiny}
\begin{deluxetable*}{l|c|c|c|c|c|c|c|c|c|c|c|c|c|}
\tablenum{10}
\tablecaption{Observed Emission Line Ratios - Markarian 34 HST STIS Spectrum - Components}
\tablewidth{0pt}
\startdata
\hline
Component \#1 & --0$\farcs$66 & --0$\farcs$56 & --0$\farcs$46 & --0$\farcs$36 & --0$\farcs$25 & --0$\farcs$15 & --0$\farcs$05 & +0$\farcs$05 & +0$\farcs$15 & +0$\farcs$46 & +0$\farcs$56 & +0$\farcs$66\\
\hline
[Ne V] $\lambda$3426	& ... $\pm$ ...	& 0.67 $\pm$ 0.54	& 0.53 $\pm$ 0.45	& 0.72 $\pm$ 0.64	& ... $\pm$ ...	& ... $\pm$ ...	& 0.70 $\pm$ 0.47	& ... $\pm$ ...	& ... $\pm$ ...	& ... $\pm$ ...	& ... $\pm$ ...	& ... $\pm$ ...	\\ \relax
[O II] $\lambda$3728	& ... $\pm$ ...	& 2.31 $\pm$ 1.90	& 2.02 $\pm$ 1.38	& 1.59 $\pm$ 1.03	& ... $\pm$ ...	& 1.28 $\pm$ 1.40	& 1.29 $\pm$ 0.50	& 1.14 $\pm$ 0.47	& 1.95 $\pm$ 4.58	& 1.09 $\pm$ 2.83	& 1.26 $\pm$ 1.38	& 1.96 $\pm$ 1.10	\\ \relax
[Ne III] $\lambda$3870	& ... $\pm$ ...	& 0.99 $\pm$ 0.81	& 0.96 $\pm$ 0.66	& 1.00 $\pm$ 0.68	& 0.46 $\pm$ 1.00	& 0.80 $\pm$ 0.84	& 0.43 $\pm$ 0.21	& 0.33 $\pm$ 0.16	& 0.68 $\pm$ 1.63	& 0.64 $\pm$ 1.70	& 0.54 $\pm$ 0.62	& 0.87 $\pm$ 0.51	\\ \relax
[Ne III] $\lambda$3968	& ... $\pm$ ...	& ... $\pm$ ...	& 0.31 $\pm$ 0.22	& 0.48 $\pm$ 0.31	& 0.49 $\pm$ 1.09	& 0.39 $\pm$ 0.46	& 0.34 $\pm$ 0.17	& 0.37 $\pm$ 0.18	& ... $\pm$ ...	& ... $\pm$ ...	& ... $\pm$ ...	& 0.31 $\pm$ 0.26	\\ \relax
H$\delta$ $\lambda$4102	& ... $\pm$ ...	& ... $\pm$ ...	& 0.25 $\pm$ 0.31	& 0.30 $\pm$ 0.31	& ... $\pm$ ...	& ... $\pm$ ...	& 0.18 $\pm$ 0.21	& ... $\pm$ ...	& 0.52 $\pm$ 1.39	& ... $\pm$ ...	& ... $\pm$ ...	& 0.32 $\pm$ 0.26	\\ \relax
H$\gamma$ $\lambda$4341	& ... $\pm$ ...	& 0.68 $\pm$ 0.58	& 0.46 $\pm$ 0.36	& 0.36 $\pm$ 0.36	& ... $\pm$ ...	& 0.47 $\pm$ 0.52	& 0.20 $\pm$ 0.15	& 0.31 $\pm$ 0.19	& ... $\pm$ ...	& 0.47 $\pm$ 1.24	& 0.33 $\pm$ 0.46	& 0.38 $\pm$ 0.26	\\ \relax
[O III] $\lambda$4364	& ... $\pm$ ...	& ... $\pm$ ...	& ... $\pm$ ...	& ... $\pm$ ...	& ... $\pm$ ...	& ... $\pm$ ...	& ... $\pm$ ...	& ... $\pm$ ...	& ... $\pm$ ...	& 0.47 $\pm$ 1.22	& ... $\pm$ ...	& ... $\pm$ ...	\\ \relax
He II $\lambda$4687	& ... $\pm$ ...	& 0.40 $\pm$ 0.38	& 0.44 $\pm$ 0.34	& 0.32 $\pm$ 0.25	& ... $\pm$ ...	& ... $\pm$ ...	& ... $\pm$ ...	& ... $\pm$ ...	& 0.52 $\pm$ 1.46	& ... $\pm$ ...	& 0.24 $\pm$ 0.27	& 0.36 $\pm$ 0.24	\\ \relax
H$\beta$ $\lambda$4862	& ... $\pm$ ...	& 1.00 $\pm$ 0.27	& 1.00 $\pm$ 0.22	& 1.00 $\pm$ 0.27	& 1.00 $\pm$ 0.70	& 1.00 $\pm$ 0.22	& 1.00 $\pm$ 0.16	& 1.00 $\pm$ 0.21	& 1.00 $\pm$ 0.47	& 1.00 $\pm$ 0.69	& 1.00 $\pm$ 0.42	& 1.00 $\pm$ 0.16	\\ \relax
[O III] $\lambda$4960	& ... $\pm$ ...	& 5.03 $\pm$ 3.98	& 4.60 $\pm$ 3.10	& 4.64 $\pm$ 2.93	& 4.79 $\pm$ 10.43	& 4.34 $\pm$ 4.41	& 3.93 $\pm$ 1.50	& 3.52 $\pm$ 1.34	& 4.90 $\pm$ 10.93	& 2.81 $\pm$ 7.22	& 4.43 $\pm$ 4.74	& 4.42 $\pm$ 2.38	\\ \relax
[O III] $\lambda$5008	& ... $\pm$ ...	& 15.15 $\pm$ 11.98	& 13.85 $\pm$ 9.33	& 13.98 $\pm$ 8.82	& 14.43 $\pm$ 31.39	& 13.05 $\pm$ 13.28	& 11.83 $\pm$ 4.50	& 10.59 $\pm$ 4.03	& 14.74 $\pm$ 32.89	& 8.46 $\pm$ 21.75	& 13.33 $\pm$ 14.25	& 13.30 $\pm$ 7.16	\\ \relax
[O I] $\lambda$6302	& ... $\pm$ ...	& ... $\pm$ ...	& ... $\pm$ ...	& ... $\pm$ ...	& ... $\pm$ ...	& ... $\pm$ ...	& ... $\pm$ ...	& ... $\pm$ ...	& ... $\pm$ ...	& ... $\pm$ ...	& 0.37 $\pm$ 0.82	& ... $\pm$ ...	\\ \relax
[N II] $\lambda$6549	& ... $\pm$ ...	& 1.37 $\pm$ 1.10	& 0.75 $\pm$ 0.52	& 0.83 $\pm$ 0.53	& 0.76 $\pm$ 1.66	& 0.69 $\pm$ 0.71	& 0.71 $\pm$ 0.28	& 0.75 $\pm$ 0.29	& 1.57 $\pm$ 3.51	& ... $\pm$ ...	& 0.81 $\pm$ 0.88	& 0.81 $\pm$ 0.45	\\ \relax
H$\alpha$ $\lambda$6564	& ... $\pm$ ...	& 4.45 $\pm$ 3.55	& 2.91 $\pm$ 2.03	& 2.88 $\pm$ 1.85	& ... $\pm$ ...	& 2.71 $\pm$ 2.77	& 3.98 $\pm$ 1.53	& 4.16 $\pm$ 1.59	& 8.49 $\pm$ 19.00	& 2.70 $\pm$ 6.94	& 2.63 $\pm$ 2.84	& 2.88 $\pm$ 1.60	\\ \relax
[N II] $\lambda$6585	& ... $\pm$ ...	& 4.05 $\pm$ 3.23	& 2.22 $\pm$ 1.55	& 2.44 $\pm$ 1.57	& 2.25 $\pm$ 4.90	& 2.03 $\pm$ 2.08	& 2.11 $\pm$ 0.81	& 2.21 $\pm$ 0.85	& 4.63 $\pm$ 10.35	& 1.63 $\pm$ 4.18	& 2.40 $\pm$ 2.58	& 2.40 $\pm$ 1.33	\\ \relax
[S II] $\lambda$6718	& ... $\pm$ ...	& 0.95 $\pm$ 0.77	& 0.66 $\pm$ 0.46	& 0.68 $\pm$ 0.45	& 0.65 $\pm$ 1.41	& 0.70 $\pm$ 0.74	& 0.82 $\pm$ 0.33	& 0.94 $\pm$ 0.36	& ... $\pm$ ...	& 1.03 $\pm$ 2.71	& 1.28 $\pm$ 1.42	& 0.41 $\pm$ 0.29	\\ \relax
[S II] $\lambda$6732	& ... $\pm$ ...	& 1.38 $\pm$ 1.12	& 0.79 $\pm$ 0.55	& 0.96 $\pm$ 0.64	& 1.39 $\pm$ 3.02	& 0.46 $\pm$ 0.48	& 0.51 $\pm$ 0.21	& 0.80 $\pm$ 0.31	& ... $\pm$ ...	& 0.87 $\pm$ 2.29	& 0.90 $\pm$ 1.00	& 0.53 $\pm$ 0.37	\\ \hline
F(H$\beta$) $\times$10$^{-15}$ & 	0.00 $\pm$ 0.00 & 	0.33 $\pm$ 0.09 & 	0.46 $\pm$ 0.10 & 	0.50 $\pm$ 0.13 & 	0.20 $\pm$ 0.14 & 	0.30 $\pm$ 0.07 & 	0.59 $\pm$ 0.10 & 	0.64 $\pm$ 0.13 & 	0.21 $\pm$ 0.10 & 	0.21 $\pm$ 0.15 & 	0.37 $\pm$ 0.15 & 	0.42 $\pm$ 0.07\\
\hline
\\
\hline
Component \#2 & --0$\farcs$66 & --0$\farcs$56 & --0$\farcs$46 & --0$\farcs$36 & --0$\farcs$25 & --0$\farcs$15 & --0$\farcs$05 & +0$\farcs$05 & +0$\farcs$15 & +0$\farcs$46 & +0$\farcs$56 & +0$\farcs$66\\
\hline
[Ne V] $\lambda$3426	& ... $\pm$ ...	& 0.68 $\pm$ 1.15	& ... $\pm$ ...	& 1.14 $\pm$ 3.32	& 1.90 $\pm$ 3.05	& ... $\pm$ ...	& ... $\pm$ ...	& ... $\pm$ ...	& ... $\pm$ ...	& ... $\pm$ ...	& ... $\pm$ ...	& ... $\pm$ ...	\\ \relax
[O II] $\lambda$3728	& 1.50 $\pm$ 1.60	& 1.20 $\pm$ 2.04	& ... $\pm$ ...	& 1.66 $\pm$ 4.73	& 2.43 $\pm$ 3.88	& ... $\pm$ ...	& ... $\pm$ ...	& 2.02 $\pm$ 3.62	& ... $\pm$ ...	& 1.20 $\pm$ 4.08	& 1.50 $\pm$ 3.17	& ... $\pm$ ...	\\ \relax
[Ne III] $\lambda$3870	& 0.82 $\pm$ 0.85	& 0.76 $\pm$ 1.29	& ... $\pm$ ...	& 0.71 $\pm$ 2.04	& 1.20 $\pm$ 1.93	& ... $\pm$ ...	& ... $\pm$ ...	& 0.53 $\pm$ 0.96	& ... $\pm$ ...	& 0.99 $\pm$ 3.38	& 0.76 $\pm$ 1.62	& ... $\pm$ ...	\\ \relax
[Ne III] $\lambda$3968	& 0.24 $\pm$ 0.26	& 0.86 $\pm$ 1.46	& ... $\pm$ ...	& ... $\pm$ ...	& ... $\pm$ ...	& ... $\pm$ ...	& ... $\pm$ ...	& ... $\pm$ ...	& ... $\pm$ ...	& 0.44 $\pm$ 1.53	& 0.54 $\pm$ 1.19	& ... $\pm$ ...	\\ \relax
H$\delta$ $\lambda$4102	& ... $\pm$ ...	& 0.49 $\pm$ 0.94	& ... $\pm$ ...	& ... $\pm$ ...	& ... $\pm$ ...	& ... $\pm$ ...	& ... $\pm$ ...	& ... $\pm$ ...	& ... $\pm$ ...	& ... $\pm$ ...	& ... $\pm$ ...	& ... $\pm$ ...	\\ \relax
H$\gamma$ $\lambda$4341	& 0.31 $\pm$ 0.40	& ... $\pm$ ...	& ... $\pm$ ...	& ... $\pm$ ...	& 0.48 $\pm$ 0.79	& ... $\pm$ ...	& ... $\pm$ ...	& ... $\pm$ ...	& ... $\pm$ ...	& ... $\pm$ ...	& ... $\pm$ ...	& ... $\pm$ ...	\\ \relax
[O III] $\lambda$4364	& ... $\pm$ ...	& 0.21 $\pm$ 0.36	& ... $\pm$ ...	& ... $\pm$ ...	& ... $\pm$ ...	& ... $\pm$ ...	& ... $\pm$ ...	& ... $\pm$ ...	& ... $\pm$ ...	& 0.30 $\pm$ 1.02	& ... $\pm$ ...	& ... $\pm$ ...	\\ \relax
He II $\lambda$4687	& 0.35 $\pm$ 0.43	& 0.57 $\pm$ 1.02	& ... $\pm$ ...	& ... $\pm$ ...	& 0.85 $\pm$ 1.37	& ... $\pm$ ...	& ... $\pm$ ...	& ... $\pm$ ...	& ... $\pm$ ...	& 0.65 $\pm$ 2.28	& 0.74 $\pm$ 1.58	& ... $\pm$ ...	\\ \relax
H$\beta$ $\lambda$4862	& 1.00 $\pm$ 0.21	& 1.00 $\pm$ 0.58	& ... $\pm$ ...	& 1.00 $\pm$ 1.22	& 1.00 $\pm$ 0.51	& ... $\pm$ ...	& ... $\pm$ ...	& 1.00 $\pm$ 0.98	& ... $\pm$ ...	& 1.00 $\pm$ 0.90	& 1.00 $\pm$ 0.82	& ... $\pm$ ...	\\ \relax
[O III] $\lambda$4960	& 3.99 $\pm$ 4.12	& 5.80 $\pm$ 9.82	& ... $\pm$ ...	& 3.43 $\pm$ 9.74	& 4.39 $\pm$ 6.93	& ... $\pm$ ...	& ... $\pm$ ...	& 1.40 $\pm$ 2.49	& ... $\pm$ ...	& 4.48 $\pm$ 15.09	& 4.18 $\pm$ 8.75	& ... $\pm$ ...	\\ \relax
[O III] $\lambda$5008	& 12.01 $\pm$ 12.39	& 17.46 $\pm$ 29.57	& ... $\pm$ ...	& 10.32 $\pm$ 29.33	& 13.20 $\pm$ 20.86	& ... $\pm$ ...	& ... $\pm$ ...	& 4.22 $\pm$ 7.51	& ... $\pm$ ...	& 13.48 $\pm$ 45.43	& 12.58 $\pm$ 26.32	& ... $\pm$ ...	\\ \relax
[O I] $\lambda$6302	& ... $\pm$ ...	& ... $\pm$ ...	& ... $\pm$ ...	& ... $\pm$ ...	& ... $\pm$ ...	& ... $\pm$ ...	& ... $\pm$ ...	& ... $\pm$ ...	& ... $\pm$ ...	& ... $\pm$ ...	& ... $\pm$ ...	& ... $\pm$ ...	\\ \relax
[N II] $\lambda$6549	& ... $\pm$ ...	& ... $\pm$ ...	& ... $\pm$ ...	& 2.20 $\pm$ 6.26	& 0.84 $\pm$ 1.32	& ... $\pm$ ...	& ... $\pm$ ...	& ... $\pm$ ...	& ... $\pm$ ...	& ... $\pm$ ...	& ... $\pm$ ...	& ... $\pm$ ...	\\ \relax
H$\alpha$ $\lambda$6564	& 1.88 $\pm$ 1.95	& 1.48 $\pm$ 2.51	& ... $\pm$ ...	& ... $\pm$ ...	& 7.01 $\pm$ 11.10	& ... $\pm$ ...	& ... $\pm$ ...	& 2.76 $\pm$ 4.92	& ... $\pm$ ...	& 3.63 $\pm$ 12.23	& 1.99 $\pm$ 4.17	& ... $\pm$ ...	\\ \relax
[N II] $\lambda$6585	& ... $\pm$ ...	& ... $\pm$ ...	& ... $\pm$ ...	& 6.49 $\pm$ 18.47	& 2.47 $\pm$ 3.90	& ... $\pm$ ...	& ... $\pm$ ...	& 2.65 $\pm$ 4.72	& ... $\pm$ ...	& 2.58 $\pm$ 8.71	& 1.31 $\pm$ 2.74	& ... $\pm$ ...	\\ \relax
[S II] $\lambda$6718	& 0.52 $\pm$ 0.56	& ... $\pm$ ...	& ... $\pm$ ...	& ... $\pm$ ...	& 0.86 $\pm$ 1.37	& ... $\pm$ ...	& ... $\pm$ ...	& ... $\pm$ ...	& ... $\pm$ ...	& ... $\pm$ ...	& ... $\pm$ ...	& ... $\pm$ ...	\\ \relax
[S II] $\lambda$6732	& ... $\pm$ ...	& ... $\pm$ ...	& ... $\pm$ ...	& ... $\pm$ ...	& 0.83 $\pm$ 1.31	& ... $\pm$ ...	& ... $\pm$ ...	& ... $\pm$ ...	& ... $\pm$ ...	& ... $\pm$ ...	& ... $\pm$ ...	& ... $\pm$ ...	\\ \hline
F(H$\beta$) $\times$10$^{-15}$ & 	0.28 $\pm$ 0.06 & 	0.16 $\pm$ 0.09 & 	0.00 $\pm$ 0.00 & 	0.11 $\pm$ 0.13 & 	0.28 $\pm$ 0.14 & 	0.00 $\pm$ 0.00 & 	0.00 $\pm$ 0.00 & 	0.14 $\pm$ 0.13 & 	0.00 $\pm$ 0.00 & 	0.16 $\pm$ 0.15 & 	0.19 $\pm$ 0.15 & 	0.00 $\pm$ 0.00
\enddata
\tablecomments{Same as in Table 2, but for the individual rotational (\#1) and outflow (\#2) components.}
\end{deluxetable*}

\setlength{\tabcolsep}{0.03in} 
\tabletypesize{\tiny}
\begin{deluxetable*}{|l|c|c|c|c|c|c|c|c|c|c|c|c|c|c|c|c|c|c|}
\tablenum{11}
\tablecaption{Observed Emission Line Ratios - Markarian 34 APO DIS Spectrum - Components}
\tablewidth{0pt}
\startdata
\hline
Component \#1 & --2$\farcs$0 &--1$\farcs$6 & --1$\farcs$2 & --0$\farcs$8 & --0$\farcs$4 & 0$\farcs$0 & +0$\farcs$4 & +0$\farcs$8 & +1$\farcs$2 & +1$\farcs$6  & +2$\farcs$0\\
\hline
[S II] $\lambda$4072	& ... $\pm$ ...	& ... $\pm$ ...	& ... $\pm$ ...	& 0.13 $\pm$ 0.01	& 0.08 $\pm$ 0.01	& 0.07 $\pm$ 0.01	& ... $\pm$ ...	& ... $\pm$ ...	& ... $\pm$ ...	& ... $\pm$ ...	& ... $\pm$ ...	\\ \relax
H$\delta$ $\lambda$4101	& ... $\pm$ ...	& 0.13 $\pm$ 0.03	& 0.15 $\pm$ 0.02	& 0.20 $\pm$ 0.02	& 0.14 $\pm$ 0.01	& 0.14 $\pm$ 0.02	& 0.15 $\pm$ 0.01	& 0.17 $\pm$ 0.02	& 0.20 $\pm$ 0.02	& 0.18 $\pm$ 0.05	& 0.20 $\pm$ 0.06	\\ \relax
H$\gamma$ $\lambda$4340	& 0.26 $\pm$ 0.05	& 0.19 $\pm$ 0.02	& 0.18 $\pm$ 0.01	& 0.34 $\pm$ 0.02	& 0.36 $\pm$ 0.01	& 0.32 $\pm$ 0.01	& 0.34 $\pm$ 0.01	& 0.33 $\pm$ 0.01	& 0.34 $\pm$ 0.01	& 0.35 $\pm$ 0.03	& 0.39 $\pm$ 0.04	\\ \relax
[O III] $\lambda$4363	& ... $\pm$ ...	& ... $\pm$ ...	& ... $\pm$ ...	& 0.16 $\pm$ 0.01	& 0.12 $\pm$ 0.00	& 0.10 $\pm$ 0.00	& 0.11 $\pm$ 0.00	& 0.12 $\pm$ 0.00	& 0.12 $\pm$ 0.00	& 0.14 $\pm$ 0.01	& 0.16 $\pm$ 0.02	\\ \relax
He II $\lambda$4685	& 0.18 $\pm$ 0.05	& 0.26 $\pm$ 0.03	& 0.33 $\pm$ 0.03	& 0.20 $\pm$ 0.01	& 0.22 $\pm$ 0.01	& 0.22 $\pm$ 0.01	& 0.26 $\pm$ 0.01	& 0.25 $\pm$ 0.01	& 0.28 $\pm$ 0.02	& 0.24 $\pm$ 0.02	& 0.26 $\pm$ 0.03	\\ \relax
H$\beta$ $\lambda$4861	& 1.00 $\pm$ 0.26	& 1.00 $\pm$ 0.13	& 1.00 $\pm$ 0.15	& 1.00 $\pm$ 0.14	& 1.00 $\pm$ 0.02	& 1.00 $\pm$ 0.02	& 1.00 $\pm$ 0.04	& 1.00 $\pm$ 0.06	& 1.00 $\pm$ 0.04	& 1.00 $\pm$ 0.06	& 1.00 $\pm$ 0.06	\\ \relax
[O III] $\lambda$4958	& 3.98 $\pm$ 0.62	& 4.21 $\pm$ 0.36	& 4.67 $\pm$ 0.32	& 4.51 $\pm$ 0.21	& 2.99 $\pm$ 0.03	& 3.46 $\pm$ 0.04	& 4.06 $\pm$ 0.06	& 4.61 $\pm$ 0.08	& 4.57 $\pm$ 0.08	& 3.94 $\pm$ 0.12	& 4.01 $\pm$ 0.14	\\ \relax
[O III] $\lambda$5006	& 11.99 $\pm$ 1.87	& 12.67 $\pm$ 1.08	& 14.06 $\pm$ 0.96	& 13.56 $\pm$ 0.63	& 8.99 $\pm$ 0.09	& 10.42 $\pm$ 0.12	& 12.21 $\pm$ 0.18	& 13.87 $\pm$ 0.25	& 13.77 $\pm$ 0.23	& 11.84 $\pm$ 0.36	& 12.07 $\pm$ 0.41	\\ \relax
[N I] $\lambda$5199	& ... $\pm$ ...	& ... $\pm$ ...	& ... $\pm$ ...	& ... $\pm$ ...	& ... $\pm$ ...	& ... $\pm$ ...	& 0.12 $\pm$ 0.03	& 0.15 $\pm$ 0.05	& 0.12 $\pm$ 0.04	& ... $\pm$ ...	& ... $\pm$ ...	\\ \relax
He I $\lambda$5875	& ... $\pm$ ...	& 0.12 $\pm$ 0.02	& ... $\pm$ ...	& 0.15 $\pm$ 0.01	& 0.10 $\pm$ 0.01	& 0.14 $\pm$ 0.01	& 0.13 $\pm$ 0.01	& 0.15 $\pm$ 0.01	& 0.21 $\pm$ 0.03	& 0.19 $\pm$ 0.04	& 0.28 $\pm$ 0.03	\\ \relax
[Fe VII] $\lambda$6086	& ... $\pm$ ...	& ... $\pm$ ...	& ... $\pm$ ...	& ... $\pm$ ...	& 0.08 $\pm$ 0.01	& 0.09 $\pm$ 0.02	& 0.07 $\pm$ 0.01	& 0.09 $\pm$ 0.02	& 0.07 $\pm$ 0.01	& 0.08 $\pm$ 0.04	& 0.09 $\pm$ 0.05	\\ \relax
[O I] $\lambda$6300	& 0.41 $\pm$ 0.10	& 0.60 $\pm$ 0.07	& 0.55 $\pm$ 0.05	& 0.45 $\pm$ 0.03	& 0.36 $\pm$ 0.02	& 0.50 $\pm$ 0.03	& 0.46 $\pm$ 0.02	& 0.52 $\pm$ 0.03	& 0.52 $\pm$ 0.03	& 0.50 $\pm$ 0.08	& 0.51 $\pm$ 0.07	\\ \relax
[O I] $\lambda$6363	& 0.14 $\pm$ 0.03	& 0.20 $\pm$ 0.02	& 0.18 $\pm$ 0.02	& 0.15 $\pm$ 0.01	& 0.12 $\pm$ 0.01	& 0.17 $\pm$ 0.01	& 0.15 $\pm$ 0.01	& 0.17 $\pm$ 0.01	& 0.17 $\pm$ 0.01	& 0.17 $\pm$ 0.03	& 0.17 $\pm$ 0.02	\\ \relax
[N II] $\lambda$6548	& 0.84 $\pm$ 0.14	& 0.97 $\pm$ 0.08	& 1.24 $\pm$ 0.09	& 1.09 $\pm$ 0.05	& 0.81 $\pm$ 0.02	& 0.95 $\pm$ 0.02	& 0.90 $\pm$ 0.02	& 1.08 $\pm$ 0.03	& 1.27 $\pm$ 0.03	& 1.30 $\pm$ 0.07	& 1.55 $\pm$ 0.08	\\ \relax
H$\alpha$ $\lambda$6562	& 3.94 $\pm$ 0.64	& 5.04 $\pm$ 0.44	& 5.89 $\pm$ 0.41	& 4.92 $\pm$ 0.25	& 2.95 $\pm$ 0.06	& 3.70 $\pm$ 0.07	& 3.89 $\pm$ 0.09	& 4.51 $\pm$ 0.10	& 5.41 $\pm$ 0.12	& 4.98 $\pm$ 0.27	& 6.44 $\pm$ 0.32	\\ \relax
[N II] $\lambda$6583	& 2.48 $\pm$ 0.40	& 2.85 $\pm$ 0.25	& 3.66 $\pm$ 0.26	& 3.21 $\pm$ 0.16	& 2.38 $\pm$ 0.05	& 2.81 $\pm$ 0.05	& 2.67 $\pm$ 0.06	& 3.18 $\pm$ 0.07	& 3.73 $\pm$ 0.08	& 3.84 $\pm$ 0.21	& 4.56 $\pm$ 0.23	\\ \relax
[S II] $\lambda$6716	& 1.04 $\pm$ 0.18	& 1.36 $\pm$ 0.12	& 1.48 $\pm$ 0.11	& 1.84 $\pm$ 0.10	& 1.33 $\pm$ 0.03	& 1.23 $\pm$ 0.03	& 1.05 $\pm$ 0.04	& 1.48 $\pm$ 0.07	& 1.75 $\pm$ 0.08	& 1.99 $\pm$ 0.17	& 2.69 $\pm$ 0.18	\\ \relax
[S II] $\lambda$6730	& 0.97 $\pm$ 0.17	& 1.15 $\pm$ 0.10	& 1.60 $\pm$ 0.12	& 1.51 $\pm$ 0.08	& 1.20 $\pm$ 0.03	& 1.37 $\pm$ 0.03	& 0.93 $\pm$ 0.03	& 1.00 $\pm$ 0.04	& 1.31 $\pm$ 0.06	& 1.85 $\pm$ 0.15	& 2.67 $\pm$ 0.18	\\ \hline
F(H$\beta$) $\times$10$^{-15}$ & 	0.70 $\pm$ 0.18 & 	0.95 $\pm$ 0.12 & 	0.94 $\pm$ 0.15 & 	1.39 $\pm$ 0.19 & 	8.97 $\pm$ 0.19 & 	12.33 $\pm$ 0.30 & 	5.42 $\pm$ 0.23 & 	3.93 $\pm$ 0.22 & 	3.96 $\pm$ 0.16 & 	7.32 $\pm$ 0.42 & 	4.02 $\pm$ 0.24\\
\hline
\\
\hline
Component \#2 & --2$\farcs$0 &--1$\farcs$6 & --1$\farcs$2 & --0$\farcs$8 & --0$\farcs$4 & 0$\farcs$0 & +0$\farcs$4 & +0$\farcs$8 & +1$\farcs$2 & +1$\farcs$6  & +2$\farcs$0\\
\hline
[S II] $\lambda$4072	& ... $\pm$ ...	& ... $\pm$ ...	& ... $\pm$ ...	& ... $\pm$ ...	& ... $\pm$ ...	& ... $\pm$ ...	& ... $\pm$ ...	& ... $\pm$ ...	& ... $\pm$ ...	& ... $\pm$ ...	& ... $\pm$ ...	\\ \relax
H$\delta$ $\lambda$4101	& ... $\pm$ ...	& ... $\pm$ ...	& ... $\pm$ ...	& 0.14 $\pm$ 0.01	& 0.16 $\pm$ 0.02	& 0.12 $\pm$ 0.02	& ... $\pm$ ...	& ... $\pm$ ...	& ... $\pm$ ...	& ... $\pm$ ...	& ... $\pm$ ...	\\ \relax
H$\gamma$ $\lambda$4340	& 0.26 $\pm$ 0.04	& 0.17 $\pm$ 0.01	& 0.21 $\pm$ 0.01	& 0.33 $\pm$ 0.01	& 0.27 $\pm$ 0.01	& 0.38 $\pm$ 0.02	& 0.67 $\pm$ 0.03	& 0.71 $\pm$ 0.03	& ... $\pm$ ...	& ... $\pm$ ...	& ... $\pm$ ...	\\ \relax
[O III] $\lambda$4363	& ... $\pm$ ...	& 0.17 $\pm$ 0.01	& 0.13 $\pm$ 0.01	& 0.10 $\pm$ 0.00	& ... $\pm$ ...	& ... $\pm$ ...	& ... $\pm$ ...	& ... $\pm$ ...	& ... $\pm$ ...	& ... $\pm$ ...	& ... $\pm$ ...	\\ \relax
He II $\lambda$4685	& 0.26 $\pm$ 0.06	& 0.29 $\pm$ 0.02	& 0.31 $\pm$ 0.02	& 0.25 $\pm$ 0.01	& 0.30 $\pm$ 0.02	& 0.33 $\pm$ 0.02	& ... $\pm$ ...	& ... $\pm$ ...	& ... $\pm$ ...	& ... $\pm$ ...	& ... $\pm$ ...	\\ \relax
H$\beta$ $\lambda$4861	& 1.00 $\pm$ 0.10	& 1.00 $\pm$ 0.06	& 1.00 $\pm$ 0.05	& 1.00 $\pm$ 0.03	& 1.00 $\pm$ 0.09	& 1.00 $\pm$ 0.18	& 1.00 $\pm$ 0.08	& 1.00 $\pm$ 0.11	& 1.00 $\pm$ 0.11	& 1.00 $\pm$ 1.65	& 1.00 $\pm$ 0.44	\\ \relax
[O III] $\lambda$4958	& 3.36 $\pm$ 0.21	& 4.21 $\pm$ 0.16	& 4.48 $\pm$ 0.10	& 3.63 $\pm$ 0.04	& 5.37 $\pm$ 0.14	& 4.74 $\pm$ 0.22	& 4.19 $\pm$ 0.09	& 4.01 $\pm$ 0.12	& 4.87 $\pm$ 0.18	& 4.56 $\pm$ 1.51	& 3.01 $\pm$ 0.56	\\ \relax
[O III] $\lambda$5006	& 10.12 $\pm$ 0.65	& 12.67 $\pm$ 0.48	& 13.47 $\pm$ 0.30	& 10.91 $\pm$ 0.13	& 16.16 $\pm$ 0.41	& 14.26 $\pm$ 0.67	& 12.60 $\pm$ 0.28	& 12.08 $\pm$ 0.37	& 14.67 $\pm$ 0.55	& 13.73 $\pm$ 4.55	& 9.05 $\pm$ 1.67	\\ \relax
[N I] $\lambda$5199	& ... $\pm$ ...	& ... $\pm$ ...	& ... $\pm$ ...	& ... $\pm$ ...	& ... $\pm$ ...	& ... $\pm$ ...	& ... $\pm$ ...	& ... $\pm$ ...	& ... $\pm$ ...	& ... $\pm$ ...	& ... $\pm$ ...	\\ \relax
He I $\lambda$5875	& 0.13 $\pm$ 0.06	& ... $\pm$ ...	& 0.10 $\pm$ 0.01	& 0.13 $\pm$ 0.01	& 0.22 $\pm$ 0.02	& ... $\pm$ ...	& ... $\pm$ ...	& ... $\pm$ ...	& ... $\pm$ ...	& ... $\pm$ ...	& ... $\pm$ ...	\\ \relax
[Fe VII] $\lambda$6086	& ... $\pm$ ...	& 0.11 $\pm$ 0.03	& 0.08 $\pm$ 0.02	& 0.07 $\pm$ 0.01	& 0.10 $\pm$ 0.02	& ... $\pm$ ...	& ... $\pm$ ...	& ... $\pm$ ...	& ... $\pm$ ...	& ... $\pm$ ...	& ... $\pm$ ...	\\ \relax
[O I] $\lambda$6300	& 0.43 $\pm$ 0.09	& 0.64 $\pm$ 0.05	& 0.58 $\pm$ 0.04	& 0.44 $\pm$ 0.02	& 0.57 $\pm$ 0.04	& 0.39 $\pm$ 0.03	& 1.14 $\pm$ 0.06	& 1.34 $\pm$ 0.08	& 2.02 $\pm$ 0.13	& ... $\pm$ ...	& 1.00 $\pm$ 0.23	\\ \relax
[O I] $\lambda$6363	& 0.14 $\pm$ 0.03	& 0.21 $\pm$ 0.02	& 0.19 $\pm$ 0.01	& 0.15 $\pm$ 0.01	& 0.19 $\pm$ 0.01	& ... $\pm$ ...	& ... $\pm$ ...	& ... $\pm$ ...	& ... $\pm$ ...	& ... $\pm$ ...	& ... $\pm$ ...	\\ \relax
[N II] $\lambda$6548	& 1.00 $\pm$ 0.08	& 1.43 $\pm$ 0.06	& 1.33 $\pm$ 0.04	& 0.94 $\pm$ 0.02	& 0.96 $\pm$ 0.03	& 0.64 $\pm$ 0.03	& 1.41 $\pm$ 0.04	& 1.66 $\pm$ 0.06	& 1.90 $\pm$ 0.08	& ... $\pm$ ...	& 1.22 $\pm$ 0.23	\\ \relax
H$\alpha$ $\lambda$6562	& 4.03 $\pm$ 0.31	& 4.98 $\pm$ 0.21	& 4.76 $\pm$ 0.14	& 3.54 $\pm$ 0.08	& 5.17 $\pm$ 0.16	& 3.45 $\pm$ 0.17	& ... $\pm$ ...	& ... $\pm$ ...	& ... $\pm$ ...	& ... $\pm$ ...	& ... $\pm$ ...	\\ \relax
[N II] $\lambda$6583	& 2.94 $\pm$ 0.23	& 4.23 $\pm$ 0.18	& 3.92 $\pm$ 0.11	& 2.78 $\pm$ 0.06	& 2.82 $\pm$ 0.09	& 1.90 $\pm$ 0.09	& 4.15 $\pm$ 0.12	& 4.90 $\pm$ 0.17	& 5.60 $\pm$ 0.23	& ... $\pm$ ...	& 3.59 $\pm$ 0.68	\\ \relax
[S II] $\lambda$6716	& 1.29 $\pm$ 0.13	& 1.43 $\pm$ 0.07	& 1.26 $\pm$ 0.04	& 1.14 $\pm$ 0.03	& 1.09 $\pm$ 0.04	& 0.79 $\pm$ 0.04	& 1.76 $\pm$ 0.07	& 4.12 $\pm$ 0.21	& 4.53 $\pm$ 0.27	& ... $\pm$ ...	& ... $\pm$ ...	\\ \relax
[S II] $\lambda$6730	& 1.23 $\pm$ 0.12	& 1.21 $\pm$ 0.06	& 1.36 $\pm$ 0.05	& 1.06 $\pm$ 0.03	& 1.46 $\pm$ 0.05	& 0.58 $\pm$ 0.03	& 3.83 $\pm$ 0.15	& 6.08 $\pm$ 0.31	& 11.33 $\pm$ 0.67	& ... $\pm$ ...	& ... $\pm$ ...	\\ \hline
F(H$\beta$) $\times$10$^{-15}$ & 	1.81 $\pm$ 0.18 & 	2.18 $\pm$ 0.12 & 	3.09 $\pm$ 0.15 & 	6.86 $\pm$ 0.19 & 	2.16 $\pm$ 0.19 & 	1.66 $\pm$ 0.30 & 	2.97 $\pm$ 0.23 & 	2.06 $\pm$ 0.22 & 	1.40 $\pm$ 0.16 & 	0.25 $\pm$ 0.42 & 	0.55 $\pm$ 0.24\\
\hline
\\
\hline
Component \#3 & --2$\farcs$0 &--1$\farcs$6 & --1$\farcs$2 & --0$\farcs$8 & --0$\farcs$4 & 0$\farcs$0 & +0$\farcs$4 & +0$\farcs$8 & +1$\farcs$2 & +1$\farcs$6  & +2$\farcs$0\\
\hline
[S II] $\lambda$4072	& ... $\pm$ ...	& ... $\pm$ ...	& ... $\pm$ ...	& ... $\pm$ ...	& ... $\pm$ ...	& ... $\pm$ ...	& 0.20 $\pm$ 0.02	& 0.16 $\pm$ 0.02	& 0.46 $\pm$ 0.05	& ... $\pm$ ...	& ... $\pm$ ...	\\ \relax
H$\delta$ $\lambda$4101	& ... $\pm$ ...	& 0.23 $\pm$ 0.04	& 0.21 $\pm$ 0.03	& ... $\pm$ ...	& 0.28 $\pm$ 0.04	& ... $\pm$ ...	& 0.15 $\pm$ 0.01	& 0.23 $\pm$ 0.02	& 0.27 $\pm$ 0.03	& ... $\pm$ ...	& ... $\pm$ ...	\\ \relax
H$\gamma$ $\lambda$4340	& 0.69 $\pm$ 0.12	& 0.58 $\pm$ 0.03	& 0.45 $\pm$ 0.02	& 0.55 $\pm$ 0.02	& ... $\pm$ ...	& ... $\pm$ ...	& 0.37 $\pm$ 0.01	& 0.45 $\pm$ 0.02	& 0.55 $\pm$ 0.02	& 1.07 $\pm$ 0.13	& ... $\pm$ ...	\\ \relax
[O III] $\lambda$4363	& ... $\pm$ ...	& ... $\pm$ ...	& ... $\pm$ ...	& ... $\pm$ ...	& ... $\pm$ ...	& ... $\pm$ ...	& 0.14 $\pm$ 0.01	& 0.18 $\pm$ 0.01	& ... $\pm$ ...	& ... $\pm$ ...	& ... $\pm$ ...	\\ \relax
He II $\lambda$4685	& ... $\pm$ ...	& ... $\pm$ ...	& ... $\pm$ ...	& ... $\pm$ ...	& 0.28 $\pm$ 0.03	& ... $\pm$ ...	& 0.20 $\pm$ 0.01	& 0.25 $\pm$ 0.01	& 0.25 $\pm$ 0.02	& ... $\pm$ ...	& ... $\pm$ ...	\\ \relax
H$\beta$ $\lambda$4861	& 1.00 $\pm$ 0.22	& 1.00 $\pm$ 0.06	& 1.00 $\pm$ 0.04	& 1.00 $\pm$ 0.07	& 1.00 $\pm$ 0.40	& ... $\pm$ ...	& 1.00 $\pm$ 0.06	& 1.00 $\pm$ 0.06	& 1.00 $\pm$ 0.08	& 1.00 $\pm$ 0.39	& ... $\pm$ ...	\\ \relax
[O III] $\lambda$4958	& 3.44 $\pm$ 0.45	& 2.00 $\pm$ 0.08	& 2.08 $\pm$ 0.04	& 3.22 $\pm$ 0.08	& 7.55 $\pm$ 0.84	& ... $\pm$ ...	& 3.51 $\pm$ 0.06	& 3.49 $\pm$ 0.06	& 3.22 $\pm$ 0.09	& 3.87 $\pm$ 0.32	& ... $\pm$ ...	\\ \relax
[O III] $\lambda$5006	& 10.37 $\pm$ 1.34	& 6.02 $\pm$ 0.23	& 6.27 $\pm$ 0.12	& 9.69 $\pm$ 0.23	& 22.72 $\pm$ 2.54	& ... $\pm$ ...	& 10.55 $\pm$ 0.19	& 10.51 $\pm$ 0.19	& 9.71 $\pm$ 0.26	& 11.65 $\pm$ 0.97	& ... $\pm$ ...	\\ \relax
[N I] $\lambda$5199	& ... $\pm$ ...	& ... $\pm$ ...	& ... $\pm$ ...	& ... $\pm$ ...	& ... $\pm$ ...	& ... $\pm$ ...	& ... $\pm$ ...	& ... $\pm$ ...	& ... $\pm$ ...	& ... $\pm$ ...	& ... $\pm$ ...	\\ \relax
He I $\lambda$5875	& ... $\pm$ ...	& ... $\pm$ ...	& ... $\pm$ ...	& ... $\pm$ ...	& 0.33 $\pm$ 0.04	& ... $\pm$ ...	& 0.26 $\pm$ 0.02	& 0.24 $\pm$ 0.02	& ... $\pm$ ...	& ... $\pm$ ...	& ... $\pm$ ...	\\ \relax
[Fe VII] $\lambda$6086	& ... $\pm$ ...	& ... $\pm$ ...	& ... $\pm$ ...	& ... $\pm$ ...	& ... $\pm$ ...	& ... $\pm$ ...	& ... $\pm$ ...	& ... $\pm$ ...	& ... $\pm$ ...	& ... $\pm$ ...	& ... $\pm$ ...	\\ \relax
[O I] $\lambda$6300	& 1.34 $\pm$ 0.32	& ... $\pm$ ...	& 0.29 $\pm$ 0.02	& 1.02 $\pm$ 0.06	& 0.50 $\pm$ 0.06	& ... $\pm$ ...	& 0.27 $\pm$ 0.01	& 0.22 $\pm$ 0.01	& ... $\pm$ ...	& 1.11 $\pm$ 0.20	& ... $\pm$ ...	\\ \relax
[O I] $\lambda$6363	& ... $\pm$ ...	& ... $\pm$ ...	& ... $\pm$ ...	& ... $\pm$ ...	& ... $\pm$ ...	& ... $\pm$ ...	& ... $\pm$ ...	& ... $\pm$ ...	& ... $\pm$ ...	& ... $\pm$ ...	& ... $\pm$ ...	\\ \relax
[N II] $\lambda$6548	& 1.11 $\pm$ 0.15	& ... $\pm$ ...	& 0.25 $\pm$ 0.01	& 0.98 $\pm$ 0.03	& 2.29 $\pm$ 0.26	& ... $\pm$ ...	& 1.30 $\pm$ 0.03	& 1.20 $\pm$ 0.03	& 1.39 $\pm$ 0.04	& 2.32 $\pm$ 0.22	& ... $\pm$ ...	\\ \relax
H$\alpha$ $\lambda$6562	& 3.65 $\pm$ 0.50	& ... $\pm$ ...	& 1.16 $\pm$ 0.03	& 3.31 $\pm$ 0.10	& 9.74 $\pm$ 1.10	& ... $\pm$ ...	& 4.72 $\pm$ 0.12	& 4.56 $\pm$ 0.11	& 4.87 $\pm$ 0.15	& 4.57 $\pm$ 0.43	& ... $\pm$ ...	\\ \relax
[N II] $\lambda$6583	& 3.26 $\pm$ 0.45	& ... $\pm$ ...	& 0.75 $\pm$ 0.02	& 2.90 $\pm$ 0.09	& 6.76 $\pm$ 0.76	& ... $\pm$ ...	& 3.85 $\pm$ 0.09	& 3.55 $\pm$ 0.08	& 4.10 $\pm$ 0.13	& 6.83 $\pm$ 0.65	& ... $\pm$ ...	\\ \relax
[S II] $\lambda$6716	& ... $\pm$ ...	& 0.77 $\pm$ 0.04	& 1.13 $\pm$ 0.04	& 1.42 $\pm$ 0.05	& 4.02 $\pm$ 0.46	& ... $\pm$ ...	& 1.62 $\pm$ 0.06	& 1.01 $\pm$ 0.04	& 0.94 $\pm$ 0.05	& 4.14 $\pm$ 0.47	& ... $\pm$ ...	\\ \relax
[S II] $\lambda$6730	& ... $\pm$ ...	& 0.91 $\pm$ 0.04	& 0.72 $\pm$ 0.02	& 0.89 $\pm$ 0.03	& 3.24 $\pm$ 0.37	& ... $\pm$ ...	& 2.21 $\pm$ 0.08	& 2.32 $\pm$ 0.10	& 2.35 $\pm$ 0.12	& 2.72 $\pm$ 0.31	& ... $\pm$ ...	\\ \hline
F(H$\beta$) $\times$10$^{-15}$ & 	0.85 $\pm$ 0.18 & 	2.18 $\pm$ 0.12 & 	3.67 $\pm$ 0.15 & 	2.84 $\pm$ 0.19 & 	0.47 $\pm$ 0.19 & 	0.00 $\pm$ 0.00 & 	3.99 $\pm$ 0.23 & 	3.87 $\pm$ 0.22 & 	2.03 $\pm$ 0.16 & 	1.08 $\pm$ 0.42 & 	0.00 $\pm$ 0.00\\
\hline
\\
\hline
Component \#4 & --2$\farcs$0 &--1$\farcs$6 & --1$\farcs$2 & --0$\farcs$8 & --0$\farcs$4 & 0$\farcs$0 & +0$\farcs$4 & +0$\farcs$8 & +1$\farcs$2 & +1$\farcs$6  & +2$\farcs$0\\
\hline
[S II] $\lambda$4072	& ... $\pm$ ...	& ... $\pm$ ...	& ... $\pm$ ...	& 0.41 $\pm$ 0.06	& ... $\pm$ ...	& ... $\pm$ ...	& ... $\pm$ ...	& ... $\pm$ ...	& ... $\pm$ ...	& ... $\pm$ ...	& ... $\pm$ ...	\\ \relax
H$\delta$ $\lambda$4101	& ... $\pm$ ...	& ... $\pm$ ...	& ... $\pm$ ...	& ... $\pm$ ...	& ... $\pm$ ...	& ... $\pm$ ...	& 0.11 $\pm$ 0.01	& 0.14 $\pm$ 0.01	& 0.13 $\pm$ 0.01	& ... $\pm$ ...	& ... $\pm$ ...	\\ \relax
H$\gamma$ $\lambda$4340	& ... $\pm$ ...	& ... $\pm$ ...	& ... $\pm$ ...	& 0.55 $\pm$ 0.07	& ... $\pm$ ...	& ... $\pm$ ...	& 0.29 $\pm$ 0.01	& 0.32 $\pm$ 0.01	& 0.32 $\pm$ 0.01	& ... $\pm$ ...	& ... $\pm$ ...	\\ \relax
[O III] $\lambda$4363	& ... $\pm$ ...	& ... $\pm$ ...	& ... $\pm$ ...	& ... $\pm$ ...	& ... $\pm$ ...	& ... $\pm$ ...	& ... $\pm$ ...	& 0.10 $\pm$ 0.00	& 0.09 $\pm$ 0.00	& ... $\pm$ ...	& ... $\pm$ ...	\\ \relax
He II $\lambda$4685	& ... $\pm$ ...	& ... $\pm$ ...	& ... $\pm$ ...	& ... $\pm$ ...	& ... $\pm$ ...	& ... $\pm$ ...	& 0.26 $\pm$ 0.01	& 0.25 $\pm$ 0.01	& 0.23 $\pm$ 0.02	& ... $\pm$ ...	& ... $\pm$ ...	\\ \relax
H$\beta$ $\lambda$4861	& ... $\pm$ ...	& 1.00 $\pm$ 0.20	& 1.00 $\pm$ 0.13	& 1.00 $\pm$ 0.37	& 1.00 $\pm$ 0.10	& 1.00 $\pm$ 0.72	& 1.00 $\pm$ 0.08	& 1.00 $\pm$ 0.05	& 1.00 $\pm$ 0.03	& ... $\pm$ ...	& 1.00 $\pm$ 1.04	\\ \relax
[O III] $\lambda$4958	& ... $\pm$ ...	& 4.65 $\pm$ 0.61	& 4.64 $\pm$ 0.27	& 2.27 $\pm$ 0.28	& 3.78 $\pm$ 0.11	& 1.89 $\pm$ 0.34	& 3.33 $\pm$ 0.08	& 3.37 $\pm$ 0.06	& 3.27 $\pm$ 0.05	& ... $\pm$ ...	& 3.63 $\pm$ 1.58	\\ \relax
[O III] $\lambda$5006	& ... $\pm$ ...	& 14.00 $\pm$ 1.85	& 13.97 $\pm$ 0.82	& 6.84 $\pm$ 0.85	& 11.39 $\pm$ 0.32	& 5.68 $\pm$ 1.04	& 10.01 $\pm$ 0.23	& 10.13 $\pm$ 0.17	& 9.85 $\pm$ 0.15	& ... $\pm$ ...	& 10.92 $\pm$ 4.74	\\ \relax
[N I] $\lambda$5199	& ... $\pm$ ...	& ... $\pm$ ...	& ... $\pm$ ...	& ... $\pm$ ...	& ... $\pm$ ...	& ... $\pm$ ...	& ... $\pm$ ...	& ... $\pm$ ...	& ... $\pm$ ...	& ... $\pm$ ...	& ... $\pm$ ...	\\ \relax
He I $\lambda$5875	& ... $\pm$ ...	& ... $\pm$ ...	& ... $\pm$ ...	& ... $\pm$ ...	& ... $\pm$ ...	& ... $\pm$ ...	& ... $\pm$ ...	& 0.09 $\pm$ 0.01	& 0.09 $\pm$ 0.01	& ... $\pm$ ...	& ... $\pm$ ...	\\ \relax
[Fe VII] $\lambda$6086	& ... $\pm$ ...	& ... $\pm$ ...	& ... $\pm$ ...	& ... $\pm$ ...	& ... $\pm$ ...	& ... $\pm$ ...	& 0.17 $\pm$ 0.03	& 0.08 $\pm$ 0.01	& 0.09 $\pm$ 0.02	& ... $\pm$ ...	& ... $\pm$ ...	\\ \relax
[O I] $\lambda$6300	& ... $\pm$ ...	& 2.80 $\pm$ 0.42	& 1.94 $\pm$ 0.16	& ... $\pm$ ...	& 1.52 $\pm$ 0.09	& ... $\pm$ ...	& 0.44 $\pm$ 0.02	& 0.46 $\pm$ 0.02	& 0.39 $\pm$ 0.02	& ... $\pm$ ...	& ... $\pm$ ...	\\ \relax
[O I] $\lambda$6363	& ... $\pm$ ...	& ... $\pm$ ...	& ... $\pm$ ...	& ... $\pm$ ...	& ... $\pm$ ...	& ... $\pm$ ...	& 0.15 $\pm$ 0.01	& 0.15 $\pm$ 0.01	& 0.13 $\pm$ 0.01	& ... $\pm$ ...	& ... $\pm$ ...	\\ \relax
[N II] $\lambda$6548	& ... $\pm$ ...	& 2.15 $\pm$ 0.29	& 1.54 $\pm$ 0.09	& 0.63 $\pm$ 0.08	& ... $\pm$ ...	& ... $\pm$ ...	& 0.84 $\pm$ 0.02	& 0.91 $\pm$ 0.02	& 0.95 $\pm$ 0.02	& ... $\pm$ ...	& 2.77 $\pm$ 1.21	\\ \relax
H$\alpha$ $\lambda$6562	& ... $\pm$ ...	& 11.74 $\pm$ 1.56	& 7.92 $\pm$ 0.49	& 1.61 $\pm$ 0.20	& 3.41 $\pm$ 0.11	& ... $\pm$ ...	& 3.67 $\pm$ 0.10	& 3.63 $\pm$ 0.08	& 3.76 $\pm$ 0.08	& ... $\pm$ ...	& 3.47 $\pm$ 1.51	\\ \relax
[N II] $\lambda$6583	& ... $\pm$ ...	& 6.34 $\pm$ 0.84	& 4.55 $\pm$ 0.28	& 1.86 $\pm$ 0.23	& 2.62 $\pm$ 0.09	& ... $\pm$ ...	& 2.47 $\pm$ 0.07	& 2.68 $\pm$ 0.06	& 2.81 $\pm$ 0.06	& ... $\pm$ ...	& 8.18 $\pm$ 3.57	\\ \relax
[S II] $\lambda$6716	& ... $\pm$ ...	& ... $\pm$ ...	& ... $\pm$ ...	& ... $\pm$ ...	& ... $\pm$ ...	& 1.31 $\pm$ 0.24	& ... $\pm$ ...	& ... $\pm$ ...	& ... $\pm$ ...	& ... $\pm$ ...	& ... $\pm$ ...	\\ \relax
[S II] $\lambda$6730	& ... $\pm$ ...	& ... $\pm$ ...	& ... $\pm$ ...	& ... $\pm$ ...	& ... $\pm$ ...	& 3.27 $\pm$ 0.60	& ... $\pm$ ...	& ... $\pm$ ...	& ... $\pm$ ...	& ... $\pm$ ...	& 3.86 $\pm$ 1.69	\\ \hline
F(H$\beta$) $\times$10$^{-15}$ & 	0.00 $\pm$ 0.00 & 	0.61 $\pm$ 0.12 & 	1.10 $\pm$ 0.15 & 	0.52 $\pm$ 0.19 & 	1.95 $\pm$ 0.19 & 	0.42 $\pm$ 0.30 & 	2.89 $\pm$ 0.23 & 	4.41 $\pm$ 0.22 & 	4.62 $\pm$ 0.16 & 	0.00 $\pm$ 0.00 & 	0.23 $\pm$ 0.24\\
\hline
\\
\hline
Component \#5 & --2$\farcs$0 &--1$\farcs$6 & --1$\farcs$2 & --0$\farcs$8 & --0$\farcs$4 & 0$\farcs$0 & +0$\farcs$4 & +0$\farcs$8 & +1$\farcs$2 & +1$\farcs$6  & +2$\farcs$0\\
\hline
[S II] $\lambda$4072	& ... $\pm$ ...	& ... $\pm$ ...	& ... $\pm$ ...	& ... $\pm$ ...	& ... $\pm$ ...	& ... $\pm$ ...	& ... $\pm$ ...	& ... $\pm$ ...	& ... $\pm$ ...	& ... $\pm$ ...	& ... $\pm$ ...	\\ \relax
H$\delta$ $\lambda$4101	& ... $\pm$ ...	& ... $\pm$ ...	& ... $\pm$ ...	& ... $\pm$ ...	& ... $\pm$ ...	& ... $\pm$ ...	& ... $\pm$ ...	& ... $\pm$ ...	& ... $\pm$ ...	& ... $\pm$ ...	& ... $\pm$ ...	\\ \relax
H$\gamma$ $\lambda$4340	& ... $\pm$ ...	& ... $\pm$ ...	& ... $\pm$ ...	& ... $\pm$ ...	& ... $\pm$ ...	& ... $\pm$ ...	& ... $\pm$ ...	& ... $\pm$ ...	& ... $\pm$ ...	& ... $\pm$ ...	& ... $\pm$ ...	\\ \relax
[O III] $\lambda$4363	& ... $\pm$ ...	& ... $\pm$ ...	& ... $\pm$ ...	& ... $\pm$ ...	& ... $\pm$ ...	& ... $\pm$ ...	& ... $\pm$ ...	& ... $\pm$ ...	& ... $\pm$ ...	& ... $\pm$ ...	& ... $\pm$ ...	\\ \relax
He II $\lambda$4685	& ... $\pm$ ...	& ... $\pm$ ...	& ... $\pm$ ...	& ... $\pm$ ...	& ... $\pm$ ...	& ... $\pm$ ...	& ... $\pm$ ...	& ... $\pm$ ...	& ... $\pm$ ...	& ... $\pm$ ...	& ... $\pm$ ...	\\ \relax
H$\beta$ $\lambda$4861	& ... $\pm$ ...	& ... $\pm$ ...	& ... $\pm$ ...	& ... $\pm$ ...	& 1.00 $\pm$ 0.44	& 1.00 $\pm$ 0.30	& ... $\pm$ ...	& ... $\pm$ ...	& ... $\pm$ ...	& ... $\pm$ ...	& ... $\pm$ ...	\\ \relax
[O III] $\lambda$4958	& ... $\pm$ ...	& ... $\pm$ ...	& ... $\pm$ ...	& ... $\pm$ ...	& 4.18 $\pm$ 0.51	& 3.38 $\pm$ 0.26	& ... $\pm$ ...	& ... $\pm$ ...	& ... $\pm$ ...	& ... $\pm$ ...	& ... $\pm$ ...	\\ \relax
[O III] $\lambda$5006	& ... $\pm$ ...	& ... $\pm$ ...	& ... $\pm$ ...	& ... $\pm$ ...	& 12.57 $\pm$ 1.52	& 10.18 $\pm$ 0.77	& ... $\pm$ ...	& ... $\pm$ ...	& ... $\pm$ ...	& ... $\pm$ ...	& ... $\pm$ ...	\\ \relax
[N I] $\lambda$5199	& ... $\pm$ ...	& ... $\pm$ ...	& ... $\pm$ ...	& ... $\pm$ ...	& ... $\pm$ ...	& ... $\pm$ ...	& ... $\pm$ ...	& ... $\pm$ ...	& ... $\pm$ ...	& ... $\pm$ ...	& ... $\pm$ ...	\\ \relax
He I $\lambda$5875	& ... $\pm$ ...	& ... $\pm$ ...	& ... $\pm$ ...	& ... $\pm$ ...	& ... $\pm$ ...	& ... $\pm$ ...	& ... $\pm$ ...	& ... $\pm$ ...	& ... $\pm$ ...	& ... $\pm$ ...	& ... $\pm$ ...	\\ \relax
[Fe VII] $\lambda$6086	& ... $\pm$ ...	& ... $\pm$ ...	& ... $\pm$ ...	& ... $\pm$ ...	& ... $\pm$ ...	& ... $\pm$ ...	& ... $\pm$ ...	& ... $\pm$ ...	& ... $\pm$ ...	& ... $\pm$ ...	& ... $\pm$ ...	\\ \relax
[O I] $\lambda$6300	& ... $\pm$ ...	& ... $\pm$ ...	& ... $\pm$ ...	& ... $\pm$ ...	& ... $\pm$ ...	& 0.78 $\pm$ 0.07	& ... $\pm$ ...	& ... $\pm$ ...	& ... $\pm$ ...	& ... $\pm$ ...	& ... $\pm$ ...	\\ \relax
[O I] $\lambda$6363	& ... $\pm$ ...	& ... $\pm$ ...	& ... $\pm$ ...	& ... $\pm$ ...	& ... $\pm$ ...	& ... $\pm$ ...	& ... $\pm$ ...	& ... $\pm$ ...	& ... $\pm$ ...	& ... $\pm$ ...	& ... $\pm$ ...	\\ \relax
[N II] $\lambda$6548	& ... $\pm$ ...	& ... $\pm$ ...	& ... $\pm$ ...	& ... $\pm$ ...	& ... $\pm$ ...	& ... $\pm$ ...	& ... $\pm$ ...	& ... $\pm$ ...	& ... $\pm$ ...	& ... $\pm$ ...	& ... $\pm$ ...	\\ \relax
H$\alpha$ $\lambda$6562	& ... $\pm$ ...	& ... $\pm$ ...	& ... $\pm$ ...	& ... $\pm$ ...	& 10.00 $\pm$ 1.22	& 6.87 $\pm$ 0.53	& ... $\pm$ ...	& ... $\pm$ ...	& ... $\pm$ ...	& ... $\pm$ ...	& ... $\pm$ ...	\\ \relax
[N II] $\lambda$6583	& ... $\pm$ ...	& ... $\pm$ ...	& ... $\pm$ ...	& ... $\pm$ ...	& 2.13 $\pm$ 0.26	& 2.06 $\pm$ 0.16	& ... $\pm$ ...	& ... $\pm$ ...	& ... $\pm$ ...	& ... $\pm$ ...	& ... $\pm$ ...	\\ \relax
[S II] $\lambda$6716	& ... $\pm$ ...	& ... $\pm$ ...	& ... $\pm$ ...	& ... $\pm$ ...	& 0.72 $\pm$ 0.09	& ... $\pm$ ...	& ... $\pm$ ...	& ... $\pm$ ...	& ... $\pm$ ...	& ... $\pm$ ...	& ... $\pm$ ...	\\ \relax
[S II] $\lambda$6730	& ... $\pm$ ...	& ... $\pm$ ...	& ... $\pm$ ...	& ... $\pm$ ...	& ... $\pm$ ...	& ... $\pm$ ...	& ... $\pm$ ...	& ... $\pm$ ...	& ... $\pm$ ...	& ... $\pm$ ...	& ... $\pm$ ...	\\ \hline
F(H$\beta$) $\times$10$^{-15}$ & 	0.00 $\pm$ 0.00 & 	0.00 $\pm$ 0.00 & 	0.00 $\pm$ 0.00 & 	0.00 $\pm$ 0.00 & 	0.43 $\pm$ 0.19 & 	1.01 $\pm$ 0.30 & 	0.00 $\pm$ 0.00 & 	0.00 $\pm$ 0.00 & 	0.00 $\pm$ 0.00 & 	0.00 $\pm$ 0.00 & 	0.00 $\pm$ 0.00
\enddata
\tablecomments{Same as in Table 4, but for the individual kinematic components, sorted from strongest to weakest peak flux.}
\end{deluxetable*}

\clearpage
\section*{Erratum: ``Quantifying Feedback from Narrow Line Region Outflows in Nearby Active Galaxies. II. Spatially Resolved Mass Outflow Rates for the QSO2 Markarian 34$^{\dagger \star}$'' (2018, \apj, 867, 88)}
\section*{Erratum}

In the published article, the ionized gas mass profile, mass outflow rates, and outflow energetics presented in Table 9 and Figures 9 and 10 were underestimated due to a calculation error \citep{revalski2018b}. This arose from adopting a mean density law from our photoionization models where each component's density was weighted by its contribution to the luminosity rather than its contribution to the mass. This caused the average gas density at each location to be weighted towards higher values, corresponding to an underestimation of the gas mass at each radius. We have corrected this error by calculating the mass in each high, medium, and low ionization component individually, and then summing their masses.\par
\vspace{\baselineskip}
This erratum contains corrected values for the results presented in Table 9 and Figures 9 and 10 of the published article. The result of this correction is that the total ionized gas mass summed over all radii increases by a factor of $\sim 10$, with the majority located at larger radii. This trend with radius is due to the fact that the photoionization models are dominated by a single medium ionization component at smaller radii ($R \lesssim 1$ kpc) such that the luminosity-weighted mean density is closer to the mass-weighted mean density at those locations.\par
\vspace{\baselineskip}
The total ionized gas mass contained within the APO long-slit observations that we used to construct our photoionization models is $M \approx 3.2 \times 10^8 M_{\odot}$, which may be confirmed using the values provided in Tables 6 and 7 in conjunction with Equation (8) in the published article. These models span a radial extent of $\pm 2\farcs2$ from the nucleus, while the {\it HST} kinematics allowed us to derive an outflow velocity law extending to $\pm 1\farcs5$. Inside of this radius, the correct ionized gas mass is $M \approx 3.3 \times 10^7 M_{\odot}$. We determined that half of the material is outflowing, leading to an outflow gas mass of $M \approx 1.6 \times 10^7 M_{\odot}$. These values are a factor of $\sim 10$ higher than those quoted in the published article.\par
\vspace{\baselineskip}
The numerical values of several quantities listed in the abstract and throughout the discussion and conclusions require revision. The outflow contains a total ionized gas mass of $M \approx 1.6 \times 10^7 M_{\odot}$. The peak mass outflow rate is $\dot M_{out} \approx$ 12.5 $\pm$ 2.4 $M_{\odot}$ yr$^{-1}$ at a distance of 470 pc from the nucleus, with a spatially integrated kinetic energy of $E \approx 1.0 \times 10^{56}$ erg. The central bin mass is $M \approx 2.1 \times 10^5 M_{\odot}$, with $M \approx 8.4 \times 10^4 M_{\odot}$ of that outflowing. The peak momentum flow rate is $\dot{p} \approx 1.5 \times 10^{35}$ dyne, which is $\sim 28\%$ of the active galactic nucleus (AGN) photon momentum. The peak kinetic luminosity reaches $\sim$ $0.1-0.3$\% of the bolometric luminosity, which is log($L_{\mathrm{bol}}$) = 46.2 $\pm$ 0.4 erg s$^{-1}$.\par
\vspace{\baselineskip}
In \S 7.2, using the [S~II] line ratios to determine the gas density results in a NLR gas mass estimate that is $\sim$ $0.06-1.24$ times the value from our models ($\sim$ $3.3 \times 10^7$ M$_{\odot}$). In \S 8, the first conclusion point should be updated with the correct mass estimate of $M \approx 1.6 \times 10^7 M_{\odot}$ and outflow kinetic energy of $E \approx 1.0 \times 10^{56}$ erg. The second conclusion point should be updated with the correct peak mass outflow rate of $\dot M_{out} \approx$ 12.5 $\pm$ 2.4 $M_{\odot}$ yr$^{-1}$, with the implication that the peak rate is no longer similar to the AGN in our previous studies \citep{crenshaw2015, revalski2018a}.\par
\vspace{\baselineskip}
Overall, the discussion and conclusions in the published article are correct. While the total ionized gas mass is larger by a factor of $\sim 10$ than originally reported, the majority of this gas is at larger radii than those displaying outflow kinematics. As noted above, the enclosed ionized gas mass at $R \leq 1\farcs5$ is $M \approx 3.3 \times 10^7 M_{\odot}$, while extending out to $R \leq 2\farcs2$ encompasses $M \approx 3.2 \times 10^8 M_{\odot}$. This indicates that there is an immense amount of ionized gas in the ENLR, which is important for future studies that compare the amount of mass in different gas phases. Finally, it is worthwhile to note that this issue did not affect our results for Mrk 573 \citep{revalski2018a}, as that analysis did not require interpolation of the model densities.\par

\clearpage

\setcounter{figure}{8}    
\begin{figure*}
\vspace{4\baselineskip}
\centering
\subfigure{
\includegraphics[scale=0.46]{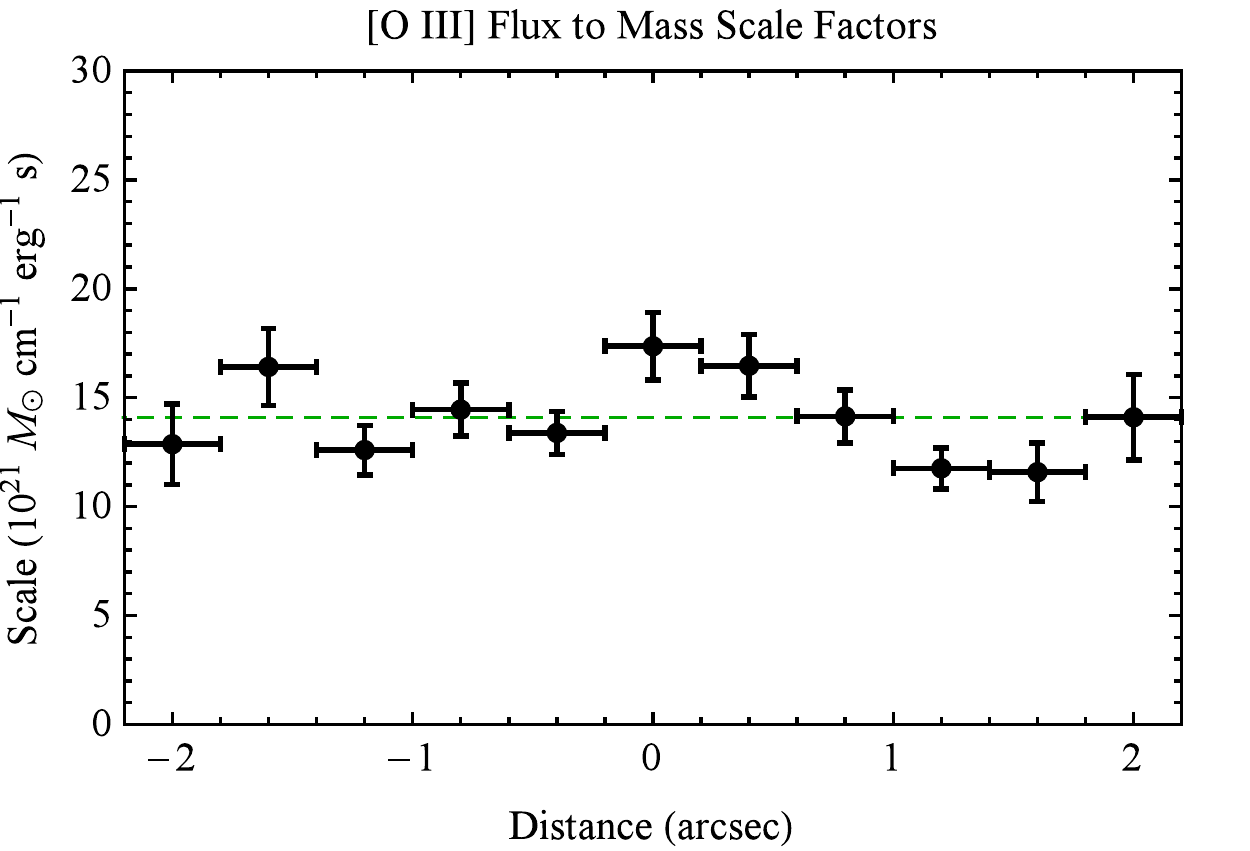}}
\subfigure{
\includegraphics[scale=0.46]{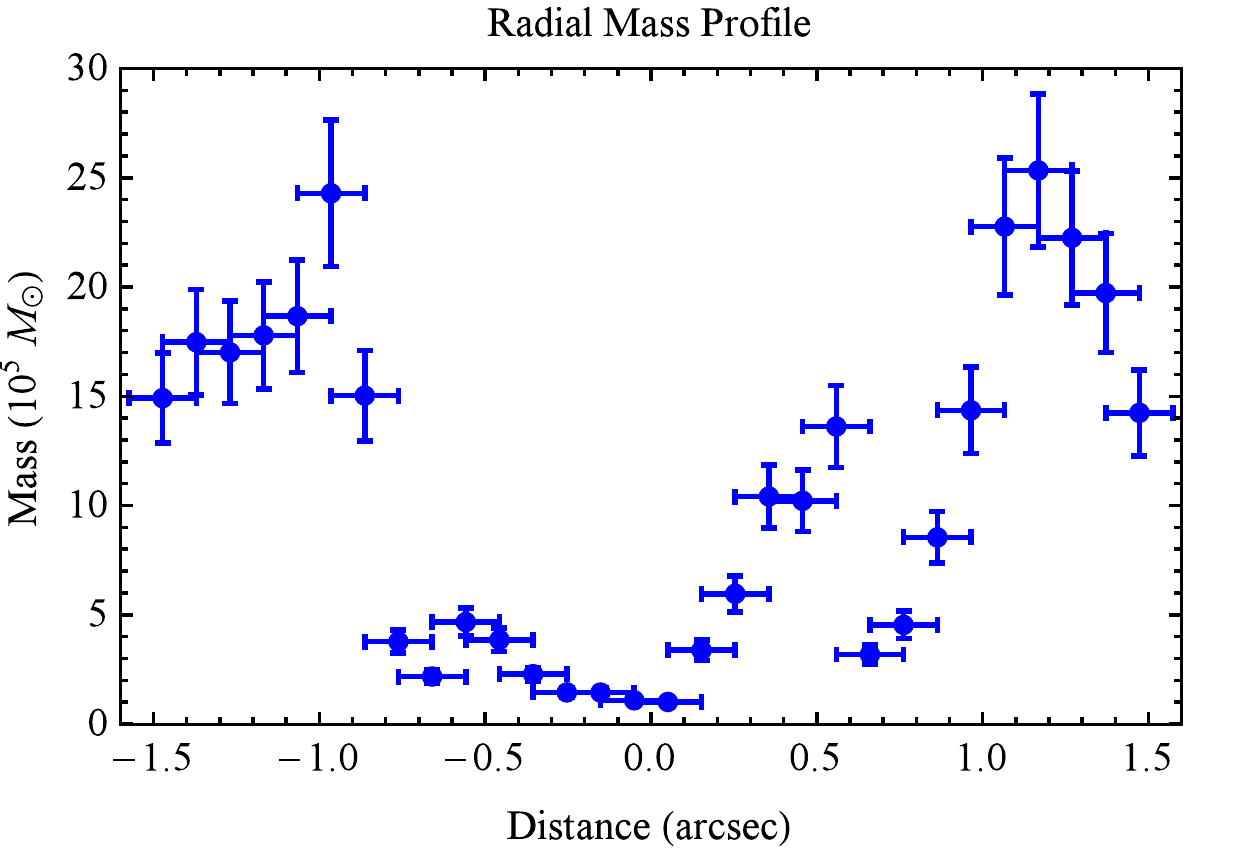}}
\subfigure{
\includegraphics[scale=0.46]{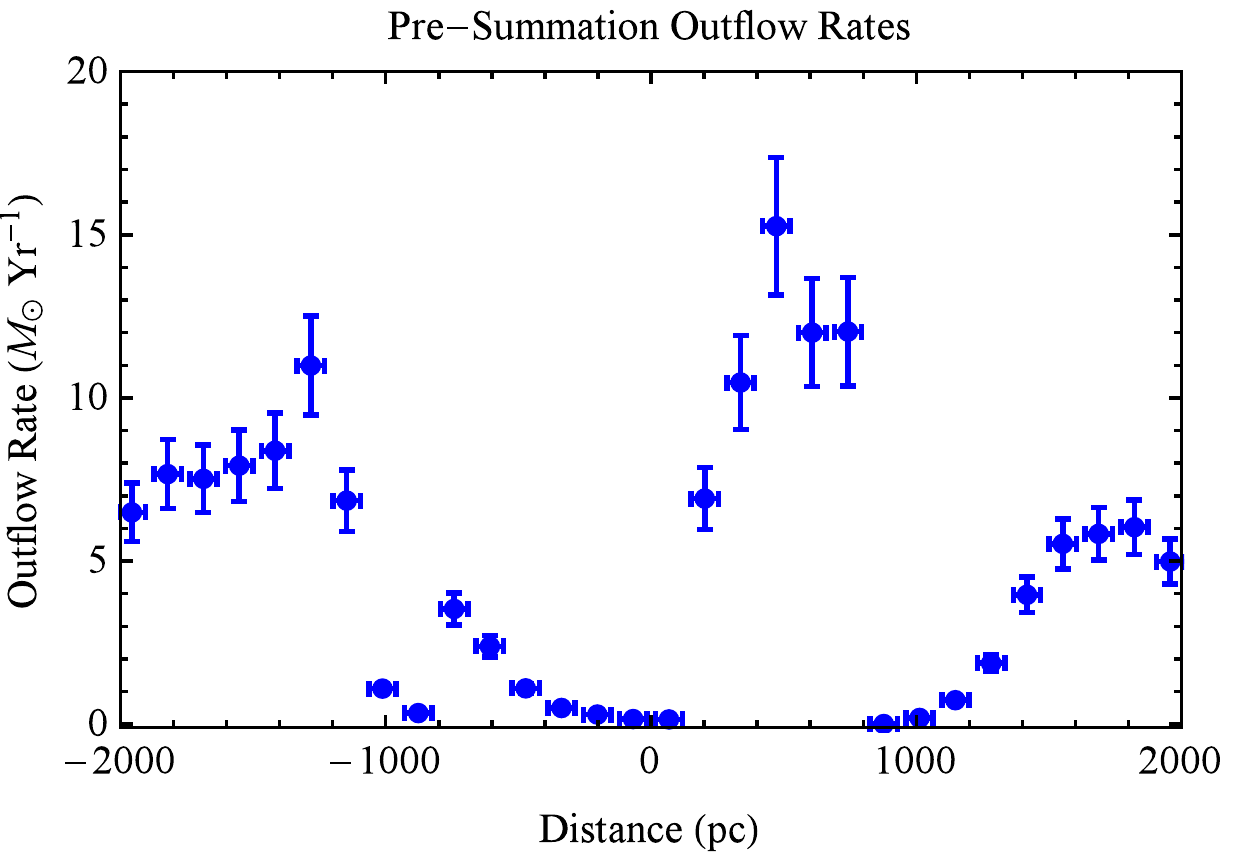}}
\caption{The corrected center and right panels for Figure 9. The center panel shows the ionized gas mass profile in units of $10^5 M_{\odot}$ calculated from the total flux in each semi-ellipse. The right panel shows the mass outflow rates assuming that all of the material is in outflow. Distances in arcseconds are the observed values, while distances in pc are corrected for projection.}
\label{fig9}
\end{figure*}

\begin{figure}
\vspace{4\baselineskip}
\centering
\subfigure{
\includegraphics[width=0.49\textwidth]{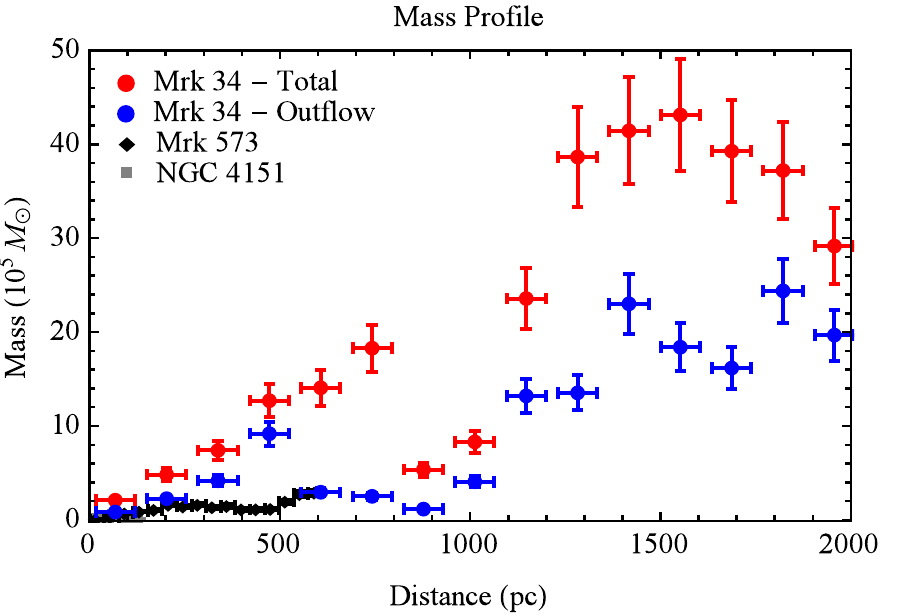}}
\subfigure{
\includegraphics[width=0.49\textwidth]{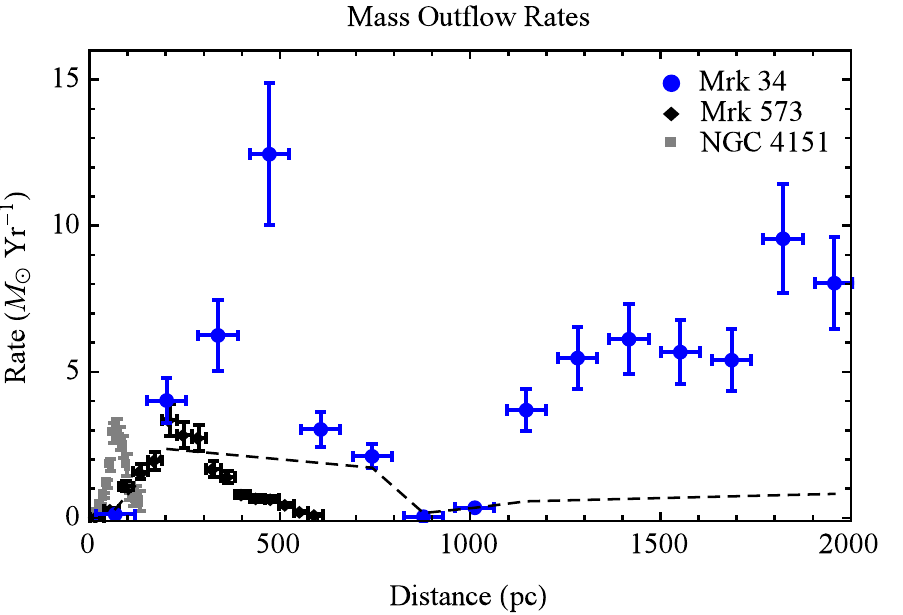}}
\subfigure{
\includegraphics[width=0.49\textwidth]{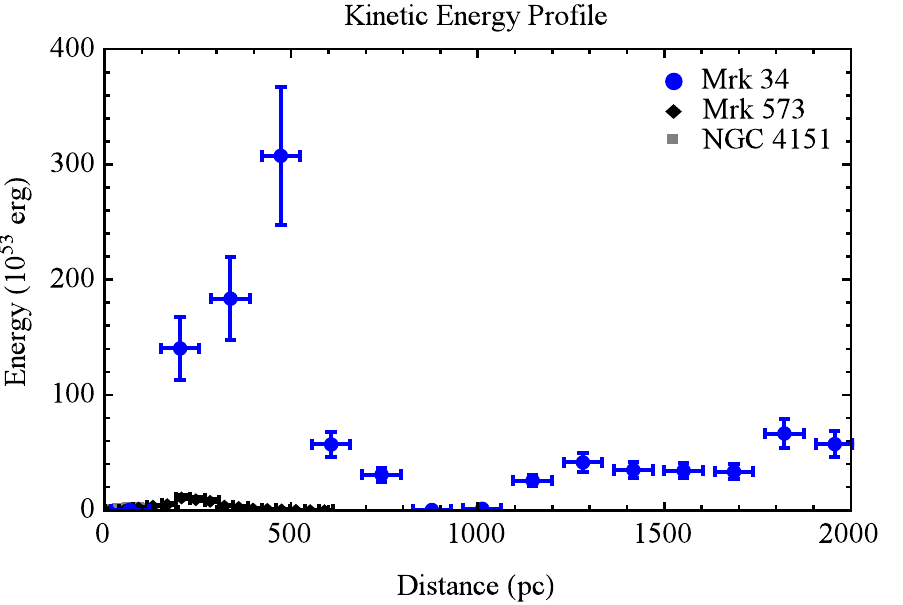}}
\subfigure{
\includegraphics[width=0.49\textwidth]{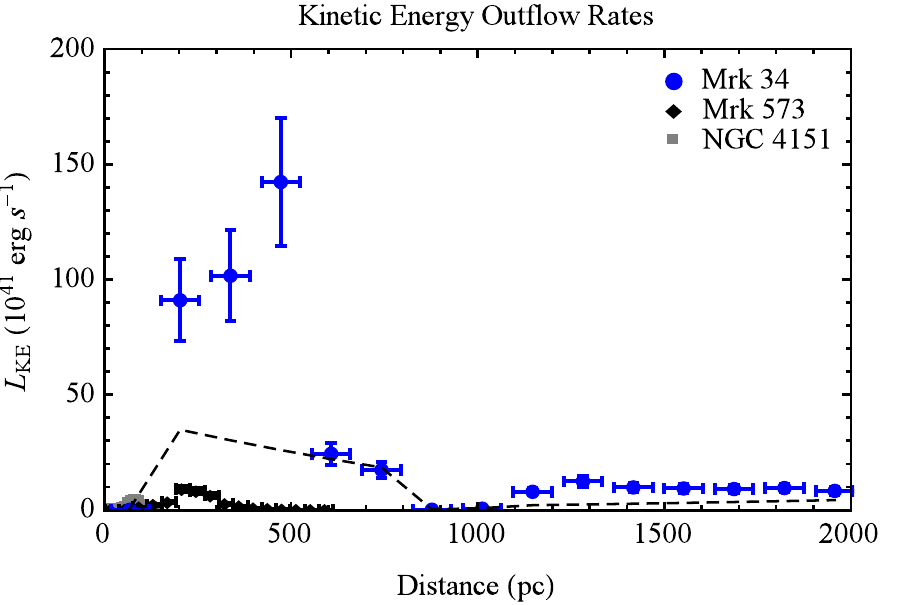}}
\subfigure{
\includegraphics[width=0.49\textwidth]{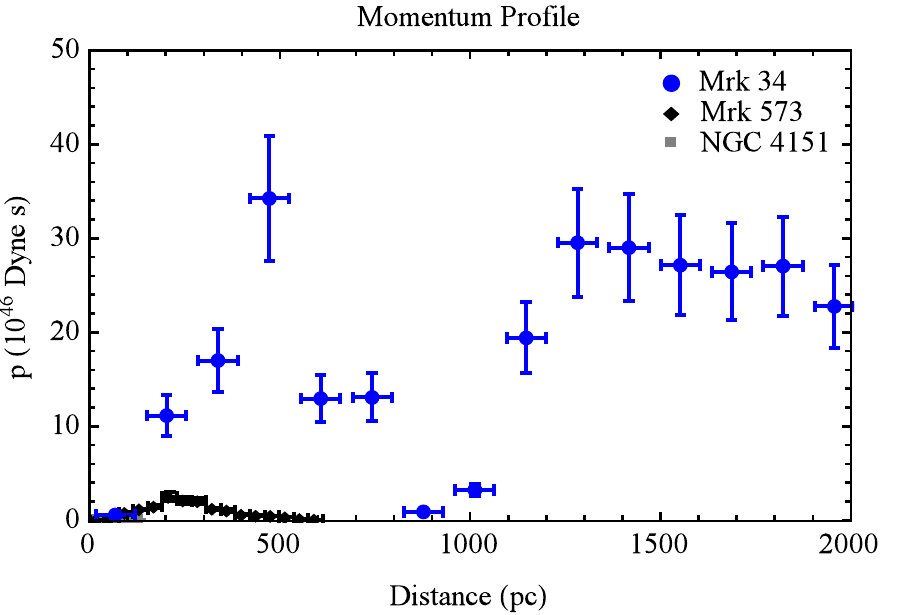}}
\subfigure{
\includegraphics[width=0.49\textwidth]{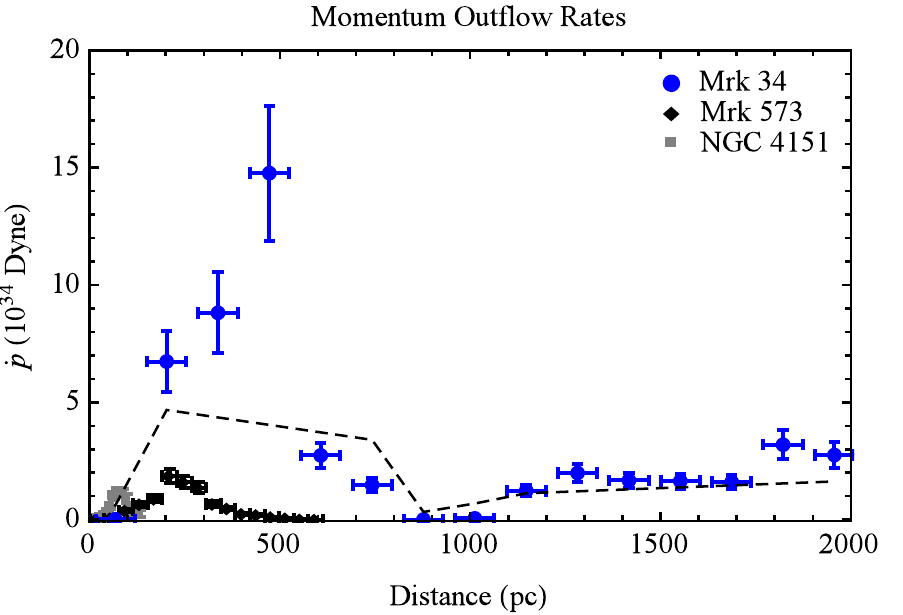}}
\caption{The corrected panels for Figure 10. Top left to bottom right are the azimuthally summed mass profiles, mass outflow rates, kinetic energy profiles, kinetic energy outflow rates, momentum profiles, and momentum outflow rates for Mrk 34, Mrk 573 \citep{revalski2018a}, and NGC 4151 \citep{crenshaw2015}. The red points represent the result that is obtained assuming that all of the mass is in outflow, and the blue points show the net result after multiplying by the fraction of flux in outflow as shown in Figure 4. The dashed lines represent the profiles that would result from the mass in the center bin ($M \approx 2.1 \times 10^5 M_{\odot}$) traveling through the velocity profile. Quantities are per bin, and targets have different bin sizes.}
\end{figure}

\setlength{\tabcolsep}{0.1in}
\tabletypesize{\small}
\begin{deluxetable*}{c|c|c|c|c|c|c|c|}
\tablenum{9}
\tablecaption{Radial Mass Outflow \& Energetic Results \vspace{-6pt}}
\tablehead{
\colhead{Distance} & \colhead{Velocity} & \colhead{Mass} & \colhead{$\dot{M}$} & \colhead{Energy} & \colhead{$\dot{E}$} & \colhead{Momentum} & \colhead{$\dot{P}$}\\
\colhead{(pc)} & \colhead{(km s$^{-1}$)} & \colhead{(10$^5$ M$_{\odot}$)} & \colhead{(M$_{\odot}$ yr$^{-1}$)} & \colhead{(10$^{53}$ erg)} & \colhead{(10$^{41}$ erg s$^{-1}$)} & \colhead{(10$^{46}$ dyne s)} & \colhead{(10$^{34}$ dyne)}\\
\colhead{(1)} & \colhead{(2)} & \colhead{(3)} & \colhead{(4)} & \colhead{(5)} & \colhead{(6)} & \colhead{(7)} & \colhead{(8)}
}
\startdata
67.5	 & 	191.8	 & 	0.84	$\pm$	0.12	 & 	0.12	$\pm$	0.02	 & 	0.31	$\pm$	0.06	 & 	0.03	$\pm$	0.01	 & 	0.57	$\pm$	0.11	 & 	0.01	$\pm$	0.01	\\
202.4	 & 	2347.9	 & 	2.26	$\pm$	0.31	 & 	4.02	$\pm$	0.78	 & 	140.31	$\pm$	27.37	 & 	90.98	$\pm$	17.75	 & 	11.14	$\pm$	2.17	 & 	6.74	$\pm$	1.31	\\
337.4	 & 	1976.6	 & 	4.17	$\pm$	0.57	 & 	6.24	$\pm$	1.22	 & 	183.52	$\pm$	35.80	 & 	101.63	$\pm$	19.83	 & 	17.00	$\pm$	3.32	 & 	8.81	$\pm$	1.72	\\
472.4	 & 	1789.2	 & 	9.18	$\pm$	1.27	 & 	12.45	$\pm$	2.43	 & 	307.48	$\pm$	59.98	 & 	142.32	$\pm$	27.76	 & 	34.25	$\pm$	6.68	 & 	14.77	$\pm$	2.88	\\
607.3	 & 	1358.2	 & 	2.94	$\pm$	0.41	 & 	3.03	$\pm$	0.59	 & 	56.98	$\pm$	11.11	 & 	24.33	$\pm$	4.75	 & 	12.95	$\pm$	2.53	 & 	2.74	$\pm$	0.53	\\
742.3	 & 	1108.4	 & 	2.51	$\pm$	0.35	 & 	2.11	$\pm$	0.41	 & 	30.77	$\pm$	6.00	 & 	17.32	$\pm$	3.38	 & 	13.07	$\pm$	2.55	 & 	1.48	$\pm$	0.29	\\
877.2	 & 	46.0	 & 	1.15	$\pm$	0.16	 & 	0.04	$\pm$	0.01	 & 	0.10	$\pm$	0.02	 & 	0.05	$\pm$	0.01	 & 	0.90	$\pm$	0.18	 & 	0.01	$\pm$	0.01	\\
1012.2	 & 	113.0	 & 	4.06	$\pm$	0.56	 & 	0.35	$\pm$	0.07	 & 	1.14	$\pm$	0.22	 & 	0.50	$\pm$	0.10	 & 	3.24	$\pm$	0.63	 & 	0.05	$\pm$	0.01	\\
1147.1	 & 	368.6	 & 	13.20	$\pm$	1.82	 & 	3.69	$\pm$	0.72	 & 	25.63	$\pm$	5.00	 & 	7.84	$\pm$	1.53	 & 	19.41	$\pm$	3.79	 & 	1.23	$\pm$	0.24	\\
1282.1	 & 	533.5	 & 	13.53	$\pm$	1.87	 & 	5.47	$\pm$	1.07	 & 	41.37	$\pm$	8.07	 & 	12.37	$\pm$	2.41	 & 	29.54	$\pm$	5.76	 & 	1.99	$\pm$	0.39	\\
1417.0	 & 	350.7	 & 	23.01	$\pm$	3.17	 & 	6.11	$\pm$	1.19	 & 	34.82	$\pm$	6.79	 & 	9.72	$\pm$	1.90	 & 	29.01	$\pm$	5.66	 & 	1.67	$\pm$	0.33	\\
1552.0	 & 	407.1	 & 	18.39	$\pm$	2.54	 & 	5.67	$\pm$	1.11	 & 	34.24	$\pm$	6.68	 & 	9.27	$\pm$	1.81	 & 	27.14	$\pm$	5.29	 & 	1.64	$\pm$	0.32	\\
1687.0	 & 	440.4	 & 	16.17	$\pm$	2.23	 & 	5.40	$\pm$	1.05	 & 	33.36	$\pm$	6.51	 & 	9.03	$\pm$	1.76	 & 	26.43	$\pm$	5.16	 & 	1.60	$\pm$	0.31	\\
1821.9	 & 	517.0	 & 	24.38	$\pm$	3.36	 & 	9.55	$\pm$	1.86	 & 	66.48	$\pm$	12.97	 & 	9.45	$\pm$	1.84	 & 	27.05	$\pm$	5.28	 & 	3.19	$\pm$	0.62	\\
1956.9	 & 	538.7	 & 	19.68	$\pm$	2.71	 & 	8.03	$\pm$	1.57	 & 	57.31	$\pm$	11.18	 & 	8.21	$\pm$	1.60	 & 	22.76	$\pm$	4.44	 & 	2.75	$\pm$	0.54
\vspace{-3pt}
\enddata
\tablecomments{The corrected numerical results for the mass and energetic quantities as a function of radial distance for the outflowing gas component. Columns are (1) deprojected distance from the nucleus, (2) mass-weighted mean velocity, (3) gas mass in units of 10$^{5}$ M$_{\odot}$, (4) mass outflow rates, (5) kinetic energies, (6) kinetic energy outflow rates, (7) momenta, and (8) momenta flow rates. These results, shown in Figure 10, are the sum of the individual radial profiles calculated for each of the semi-annuli. The value at each distance is the quantity contained within the annulus of width $\delta r$.}
\end{deluxetable*}



\begin{thebibliography}{}
\bibitem[Asplund et al.(2009)]{asplund2009} Asplund, M., Grevesse, N., Sauval, A.~J., \& Scott, P.\ 2009, \araa, 47, 481

\bibitem[Bae et al.(2017)]{bae2017} Bae, H.-J., Woo, J.-H., Karouzos, M., et al.\ 2017, \apj, 837, 91 

\bibitem[Baldwin et al.(1981)]{baldwin1981} Baldwin, J.~A., Phillips, M.~M., \& Terlevich, R.\ 1981, \pasp, 93, 5

\bibitem[Batiste et al.(2017)]{batiste2017} Batiste, M., Bentz, M.~C., Raimundo, S.~I., Vestergaard, M., \& Onken, C.~A.\ 2017, \apjl, 838, L10

\bibitem[Baum et al.(1993)]{baum1993} Baum, S.~A., O'Dea, C.~P., Dallacassa, D., de Bruyn, A.~G., \& Pedlar, A.\ 1993, \apj, 419, 553

\bibitem[Bischetti et al.(2017)]{bischetti2017} Bischetti, M., Piconcelli, E., Vietri, G., et al.\ 2017, \aap, 598, A122 

\bibitem[Buchner et al.(2014)]{buchner2014} Buchner, J., Georgakakis, A., Nandra, K., et al.\ 2014, \aap, 564, A125

\bibitem[Castro et al.(2017)]{castro2017} Castro, C.~S., Dors, O.~L., Cardaci, M.~V., \& H{\"a}gele, G.~F.\ 2017, \mnras, 467, 1507

\bibitem[Cicone et al.(2018)]{cicone2018} Cicone, C., Brusa, M., Ramos Almeida, C., et al.\ 2018, Nature Astronomy, 2, 176

\bibitem[Ciotti \& Ostriker(2001)]{ciotti2001} Ciotti, L., \& Ostriker, J.~P.\ 2001, \apj, 551, 131

\bibitem[Collins et al.(2009)]{collins2009} Collins, N.~R., Kraemer, S.~B., Crenshaw, D.~M., Bruhweiler, F.~C., \& Mel{\'e}ndez, M.\ 2009, \apj, 694, 765 

\bibitem[Cresci \& Maiolino(2018)]{cresci2018} Cresci, G., \& Maiolino, R.\ 2018, Nature Astronomy, 2, 179

\bibitem[Crenshaw \& Kraemer(2012)]{crenshaw2012} Crenshaw, D.~M., \& Kraemer, S.~B.\ 2012, \apj, 753, 75

\bibitem[Crenshaw et al.(2015)]{crenshaw2015} Crenshaw, D.~M., Fischer, T.~C., Kraemer, S.~B., \& Schmitt, H.~R.\ 2015, \apj, 799, 83

\bibitem[Di Matteo et al.(2005)]{dimatteo2005} Di Matteo, T., Springel, V., \& Hernquist, L.\ 2005, \nat, 433, 604

\bibitem[Dressel(2012)]{wfc3ihb} Dressel, L.\ 2012, Wide Field Camera 3, HST Instrument Handbook,

\bibitem[Falcke et al.(1998)]{falcke1998} Falcke, H., Wilson, A.~S., \& Simpson, C.\ 1998, \apj, 502, 199

\bibitem[Ferrarese \& Merritt(2000)]{ferrarese2000} Ferrarese, L., \& Merritt, D.\ 2000, \apjl, 539, L9 

\bibitem[Fiore et al.(2017)]{fiore2017} Fiore, F., Feruglio, C., Shankar, F., et al.\ 2017, \aap, 601, A143 

\bibitem[Fischer et al.(2013)]{fischer2013} Fischer, T.~C., Crenshaw, D.~M., Kraemer, S.~B., \& Schmitt, H.~R.\ 2013, \apjs, 209, 1

\bibitem[Fischer et al.(2014)]{fischer2014} Fischer, T.~C., Crenshaw, D.~M., Kraemer, S.~B., Schmitt, H.~R., \& Turner, T.~J.\ 2014, \apj, 785, 25 

\bibitem[Fischer et al.(2017)]{fischer2017} Fischer, T.~C., Machuca, C., Diniz, M.~R., et al.\ 2017, \apj, 834, 30

\bibitem[Fischer et al.(2018)]{fischer2018} Fischer, T.~C., Kraemer, S.~B., Schmitt, H.~R., et al.\ 2018, \apj, 856, 102

\bibitem[Ferland et al.(2013)]{ferland2013} Ferland, G.~J., Porter, R.~L., van Hoof, P.~A.~M., et al.\ 2013, RMxAA, 49, 137 

\bibitem[Feroz \& Hobson(2008)]{feroz2008} Feroz, F., \& Hobson, M.~P.\ 2008, \mnras, 384, 449

\bibitem[Feroz et al.(2009)]{feroz2009} Feroz, F., Hobson, M.~P., \& Bridges, M.\ 2009, \mnras, 398, 1601

\bibitem[Feroz et al.(2013)]{feroz2013} Feroz, F., Hobson, M.~P., Cameron, E., \& Pettitt, A.~N.\ 2013, arXiv:1306.2144

\bibitem[F{\"o}rster Schreiber et al.(2014)]{forster2014} F{\"o}rster Schreiber, N.~M., Genzel, R., Newman, S.~F., et al.\ 2014, \apj, 787, 38 

\bibitem[Gandhi et al.(2014)]{gandhi2014} Gandhi, P., Lansbury, G.~B., Alexander, D.~M., et al.\ 2014, \apj, 792, 117

\bibitem[Gebhardt et al.(2000)]{gebhardt2000} Gebhardt, K., Bender, R., Bower, G., et al.\ 2000, \apjl, 539, L13

\bibitem[Genzel et al.(2014)]{genzel2014} Genzel, R., F{\"o}rster Schreiber, N.~M., Rosario, D., et al.\ 2014, \apj, 796, 7 

\bibitem[Gonz{\'a}lez Delgado et al.(2001)]{gonzalezdelgado2001} Gonz{\'a}lez Delgado, R.~M., Heckman, T., \& Leitherer, C.\ 2001, \apj, 546, 845 

\bibitem[Haniff et al.(1988)]{haniff1988} Haniff, C.~A., Wilson, A.~S., \& Ward, M.~J.\ 1988, \apj, 334, 104

\bibitem[Harrison et al.(2014)]{harrison2014} Harrison, C.~M., Alexander, D.~M., Mullaney, J.~R., \& Swinbank, A.~M.\ 2014, \mnras, 441, 3306 

\bibitem[Harrison(2017)]{harrison2017} Harrison, C.~M.\ 2017, Nature Astronomy, 1, 0165

\bibitem[Harrison et al.(2018)]{harrison2018} Harrison, C.~M., Costa, T., Tadhunter, C.~N., et al.\ 2018, Nature Astronomy, 2, 198

\bibitem[Heckman et al.(2004)]{heckman2004} Heckman, T.~M., Kauffmann, G., Brinchmann, J., et al.\ 2004, \apj, 613, 109 

\bibitem[Heckman et al.(2005)]{heckman2005} Heckman, T.~M., Ptak, A., Hornschemeier, A., \& Kauffmann, G.\ 2005, \apj, 634, 161

\bibitem[Heckman \& Best(2014)]{heckman2014} Heckman, T.~M., \& Best, P.~N.\ 2014, \araa, 52, 589 

\bibitem[Henkel et al.(2005)]{henkel2005} Henkel, C., Peck, A.~B., Tarchi, A., et al.\ 2005, \aap, 436, 75

\bibitem[Hopkins \& Elvis(2010)]{hopkins2010} Hopkins, P.~F., \& Elvis, M.\ 2010, \mnras, 401, 7 

\bibitem[Hopkins et al.(2005)]{hopkins2005} Hopkins, P.~F., Hernquist, L., Cox, T.~J., et al.\ 2005, \apj, 630, 705 

\bibitem[Jackson \& Beswick(2007)]{jackson2007} Jackson, N., \& Beswick, R.~J.\ 2007, \mnras, 376, 719

\bibitem[Joye \& Mandel(2003)]{joye2003} Joye, W.~A., \& Mandel, E.\ 2003, Astronomical Data Analysis Software and Systems XII, 295, 489

\bibitem[Kakkad et al.(2016)]{kakkad2016} Kakkad, D., Mainieri, V., Padovani, P., et al.\ 2016, \aap, 592, A148

\bibitem[Kang \& Woo(2018)]{kang2018} Kang, D., \& Woo, J.-H.\ 2018, arXiv:1807.08356

\bibitem[Karouzos et al.(2016)]{karouzos2016} Karouzos, M., Woo, J.-H., \& Bae, H.-J.\ 2016, \apj, 833, 171 

\bibitem[Kauffmann et al.(2003)]{kauffmann2003} Kauffmann, G., Heckman, T.~M., Tremonti, C., et al.\ 2003, \mnras, 346, 1055 

\bibitem[Kewley et al.(2001)]{kewley2001} Kewley, L.~J., Dopita, M.~A., Sutherland, R.~S., Heisler, C.~A., \& Trevena, J.\ 2001, \apj, 556, 121

\bibitem[Kewley et al.(2006)]{kewley2006} Kewley, L.~J., Groves, B., Kauffmann, G., \& Heckman, T.\ 2006, \mnras, 372, 961 

\bibitem[Kormendy \& Ho(2013)]{kormendy2013} Kormendy, J., \& Ho, L.~C.\ 2013, \araa, 51, 511 

\bibitem[Koski(1978)]{koski1978} Koski, A.~T.\ 1978, \apj, 223, 56

\bibitem[Kraemer \& Crenshaw(2000a)]{kraemer2000a} Kraemer, S.~B., \& Crenshaw, D.~M.\ 2000, \apj, 532, 256

\bibitem[Kraemer \& Crenshaw(2000b)]{kraemer2000b} Kraemer, S.~B., \& Crenshaw, D.~M.\ 2000, \apj, 544, 763

\bibitem[Leung et al.(2017)]{leung2017} Leung, G.~C.~K., Coil, A.~L., Azadi, M., et al.\ 2017, \apj, 849, 48 

\bibitem[Liu et al.(2017)]{liu2017} Liu, Z.~W., Zhang, J.~S., Henkel, C., et al.\ 2017, \mnras, 466, 1608

\bibitem[McElroy et al.(2015)]{mcelroy2015} McElroy, R., Croom, S.~M., Pracy, M., et al.\ 2015, \mnras, 446, 2186

\bibitem[McMaster \& et al.(2008)]{wfpc2ihb} McMaster, M., \& et al.\ 2008, Wide Field and Planetary Camera 2, HST Instrument Handbook

\bibitem[Nagar \& Wilson(1999)]{nagar1999} Nagar, N.~M., \& Wilson, A.~S.\ 1999, \apj, 516, 97

\bibitem[Nair \& Abraham(2010)]{nair2010} Nair, P.~B., \& Abraham, R.~G.\ 2010, \apjs, 186, 427

\bibitem[Nesvadba et al.(2006)]{nesvadba2006} Nesvadba, N.~P.~H., Lehnert, M.~D., Eisenhauer, F., et al.\ 2006, \apj, 650, 693 

\bibitem[Netzer(2009)]{netzer2009} Netzer, H.\ 2009, \mnras, 399, 1907 

\bibitem[Oh et al.(2011)]{oh2011} Oh, K., Sarzi, M., Schawinski, K., \& Yi, S.~K.\ 2011, \apjs, 195, 13 

\bibitem[Oke(1990)]{oke1990} Oke, J.~B.\ 1990, \aj, 99, 1621

\bibitem[Osterbrock \& Ferland(2006)]{osterbrock2006} Osterbrock, D.~E., \& Ferland, G.~J.\ 2006, Astrophysics of gaseous nebulae and active galactic nuclei, 2nd.~ed.~by D.E.~Osterbrock and G.J.~Ferland.~Sausalito, CA: University Science Books, 2006

\bibitem[Peterson(1997)]{peterson1997} Peterson, B.~M.\ 1997, An introduction to active galactic nuclei, Publisher: Cambridge, New York Cambridge University Press, 1997 Physical description xvi, 238 p.~ISBN 0521473489

\bibitem[Revalski et al.(2018)]{revalski2018} Revalski, M., Crenshaw, D.~M., Kraemer, S.~B., et al.\ 2018, \apj, 856, 46, Paper I

\bibitem[Reyes et al.(2008)]{reyes2008} Reyes, R., Zakamska, N.~L., Strauss, M.~A., et al.\ 2008, \aj, 136, 2373

\bibitem[Riley(2017)]{stisihb} Riley, A.\ 2017, Space Telescope Imaging Spectrograph Instrument Handbook

\bibitem[Rosario(2007)]{rosario2007} Rosario, D.~J.~V.\ 2007, Ph.D.~Thesis

\bibitem[Rosario et al.(2008)]{rosario2008} Rosario, D.~J., Whittle, M., Nelson, C.~H., \& Wilson, A.~S.\ 2008, \memsai, 79, 1217

\bibitem[Savage \& Mathis(1979)]{savage1979} Savage, B.~D., \& Mathis, J.~S.\ 1979, \araa, 17, 73

\bibitem[Seab \& Shull(1983)]{seab1983} Seab, C.~G., \& Shull, J.~M.\ 1983, \apj, 275, 652 

\bibitem[Snow \& Witt(1996)]{snow1996} Snow, T.~P., \& Witt, A.~N.\ 1996, \apjl, 468, L65 

\bibitem[Stoklasov{\'a} et al.(2009)]{stoklasov2009} Stoklasov{\'a}, I., Ferruit, P., Emsellem, E., et al.\ 2009, \aap, 500, 1287

\bibitem[Storchi-Bergmann et al.(1998)]{storchibergmann1998} Storchi-Bergmann, T., Schmitt, H.~R., Calzetti, D., \& Kinney, A.~L.\ 1998, \aj, 115, 909 

\bibitem[Tody(1986)]{tody1986} Tody, D.\ 1986, \procspie, 627, 733

\bibitem[Tody(1993)]{tody1993} Tody, D.\ 1993, Astronomical Data Analysis Software and Systems II, 52, 173

\bibitem[Tombesi et al.(2010)]{tombesi2010} Tombesi, F., Cappi, M., Reeves, J.~N., et al.\ 2010, \aap, 521, A57

\bibitem[Tombesi et al.(2011)]{tombesi2011} Tombesi, F., Cappi, M., Reeves, J.~N., et al.\ 2011, \apj, 742, 44

\bibitem[Tombesi et al.(2013)]{tombesi2013} Tombesi, F., Cappi, M., Reeves, J.~N., et al.\ 2013, \mnras, 430, 1102

\bibitem[Ulvestad \& Wilson(1984)]{ulvestad1984} Ulvestad, J.~S., \& Wilson, A.~S.\ 1984, \apj, 278, 544

\bibitem[Unger et al.(1987)]{unger1987} Unger, S.~W., Pedlar, A., Axon, D.~J., et al.\ 1987, \mnras, 228, 671

\bibitem[van Dokkum(2001)]{vandokkum2001} van Dokkum, P.~G.\ 2001, \pasp, 113, 1420

\bibitem[Veilleux \& Osterbrock(1987)]{veilleux1987} Veilleux, S., \& Osterbrock, D.~E.\ 1987, \apjs, 63, 295

\bibitem[Venturi et al.(2018)]{venturi2018} Venturi, G., Nardini, E., Marconi, A., et al.\ 2018, arXiv:1809.01206 

\bibitem[Villar-Mart{\'{\i}}n et al.(2016)]{villar2016} Villar-Mart{\'{\i}}n, M., Arribas, S., Emonts, B., et al.\ 2016, \mnras, 460, 130 

\bibitem[Wang et al.(2007)]{wang2007} Wang, J.-M., Chen, Y.-M., Yan, C.-S., Hu, C., \& Bian, W.-H.\ 2007, \apjl, 661, L143

\bibitem[Wang et al.(2011)]{wang2011} Wang, J., Fabbiano, G., Elvis, M., et al.\ 2011, \apj, 742, 23 

\bibitem[Whittle et al.(1988)]{whittle1988} Whittle, M., Pedlar, A., Meurs, E.~J.~A., et al.\ 1988, \apj, 326, 125

\bibitem[Wilson et al.(1988)]{wilson1988} Wilson, A.~S., Ward, M.~J., \& Haniff, C.~A.\ 1988, \apj, 334, 121

\bibitem[Wylezalek \& Morganti(2018)]{wylezalek2018} Wylezalek, D., \& Morganti, R.\ 2018, Nature Astronomy, 2, 181

\bibitem[Zubovas(2018)]{zubovas2018} Zubovas, K.\ 2018, \mnras, 479, 3189
\end{thebibliography}

\begin{thebibliography}{}
\bibitem[Crenshaw et al.(2015)]{crenshaw2015} Crenshaw, D.~M., Fischer, T.~C., Kraemer, S.~B., \& Schmitt, H.~R.\ 2015, \apj, 799, 83

\bibitem[Revalski et al.(2018a)]{revalski2018a} Revalski, M., Crenshaw, D.~M., Kraemer, S.~B., et al.\ 2018a, \apj, 856, 46

\bibitem[Revalski et al.(2018b)]{revalski2018b} Revalski, M., Dashtamirova, D., Crenshaw, D.~M., et al.\ 2018b, \apj, 867, 88
\end{thebibliography}
\end{document}